\documentclass[final,5p,twocolumn]{elsarticle}

\usepackage{amssymb}
\usepackage{amsmath}
\usepackage{algorithm}
\usepackage{enumerate}
\usepackage{algpseudocode}
\usepackage{float}
\usepackage[colorlinks=true,linkcolor=black, citecolor=blue, urlcolor=blue]{hyperref}
\usepackage{tikz}
\usetikzlibrary{shapes.geometric, arrows, calc}
\usepackage{breqn}
\usepackage{bbm}
\usepackage{cleveref}
\usepackage{mathtools}
\usepackage{balance}
\usepackage{caption}
\usepackage{subcaption}
\usepackage{xfrac}
\usepackage{multirow}
\usepackage{booktabs}

\newcommand{\E}{\mathbb{E}}

\newcommand{\Var}{\mbox{Var}}
\newcommand{\Cov}{\mbox{Cov}}

\newcommand{\argmin}{\mathrm{argmin}}

\journal{Physica A: Statistical Mechanics and its Applications}

\begin{document}

\begin{frontmatter}

\title{A simple learning agent interacting with an agent-based market model}

\author[uct-sta]{Matthew Dicks}
\ead{matthew.dicks@alumni.uct.ac.za}
\author[uct-sta]{Andrew Paskaramoorthy}
\ead{andrew.paskaramoorthy@uct.ac.za}
\author[uct-sta]{Tim Gebbie\corref{cor1}}
\ead{tim.gebbie@uct.ac.za}
\address[uct-sta]{Department of Statistical Sciences, University of Cape Town, Rondebosch 7700, South Africa}
\cortext[cor1]{Corresponding author}

\begin{abstract}
 We consider the learning dynamics of a single reinforcement learning optimal execution trading agent when it interacts with an event-driven agent-based financial market model. Trading takes place asynchronously through a matching engine in event time. The optimal execution agent is considered at different levels of initial order sizes and differently sized state spaces. The resulting impact on the agent-based model and market is considered using a calibration approach that explores changes in the empirical stylised facts and price impact curves. Convergence, volume trajectory and action trace plots are used to visualise the learning dynamics. The smaller state space agents had the number of states they visited converge much faster than the larger state space agents, and they were able to start learning to trade intuitively using the spread and volume states. We find that the moments of the model are robust to the impact of the learning agents, except for the Hurst exponent, which was lowered by the introduction of strategic order-splitting. The introduction of the learning agent preserves the shape of the price impact curves but can reduce the trade-sign auto-correlations and increase the micro-price volatility when the trading volumes increase.
\end{abstract}


\end{frontmatter}


\section*{Introduction} \label{sec:Introduction}

Much of the empirical economic and finance literature suffers from a multiple assumption problem with data sets that are often static samples of the system that do not include the impact of system feedback. Often, the key assumptions used to specify agents, and hence models, are that they are rational and fully informed \cite{dieci2018heterogeneous}. This introduced two key scaling challenges: the size of the state space, and nonlinear state space dynamics. The state space can become so large that the agents can not reasonably behave as neither fully informed nor rational decision-makers. Artificially forcing bounded rationality onto the agents \cite{lebaron2000agent} can become an {\it ad hoc} assumption aimed at retaining some form of rationality in decision-making to accommodate feasible interactions with a static environment. In financial markets and many social systems, the state space can change due to adaption, interaction, and feedback between the agents and the changing environment they create and compete within. This implies that even with bounded rationality, the data sets extracted or bootstrapped from historical data are unlikely to be representative of either the current system configuration or future outcomes.

Adopting a bottom-up rule-based approach using simple online updating representations of reasonable trading constraints can be an effective alternative  \cite{farmer2005predictive}. Although these types of Agent-Based Models (ABMs) explicitly exclude top-down causation from higher complexity structures, they can still include the impact of a changing environment. Such ABMs provide a bottom-up approach to modelling a complex system where there can be multiple online heterogeneous interacting agents in a single level of the hierarchy but within a changing environment. Minimally intelligent agents with simple trading rules can be used to generate reasonably realistic dynamics \cite{leal2016rock}, which can be calibrated to the system's observations \cite{plattgebbie2016problem}. Financial ABMs that model the Limit Order Book (LOB) attribute LOB dynamics to the emergent phenomena caused by the interactions of trading agents programmed by these simple rules.

\subsection*{Agent-based models} 

When building ABMs, a good place to start is to understand the dynamics that can be realised from the interactions of minimally intelligent agents before adding on more complex forms of decision-making. Farmer {\it et al.} \cite{farmer2005predictive} used minimally intelligent agents within a continuous double auction and were able to explain 96\% of the variance of the spread and 76\% of the variance of the price diffusion rate. They also recovered the price impact function and thus demonstrated the simple laws relating price to order flows. The work by Farmer {\it et al.} also suggests institutions strongly shape agent behaviours so that some of the properties of markets may result from the constraints imposed by these institutions rather than the rationality of the individuals \cite{farmer2005predictive}. 

This is an attractive method for modelling the dynamics of the LOB because there has been an increase in the amount of high-frequency observational data available to the researcher, and this data is a direct measurement of how markets unfold through time. This level of detail can provide a picture of the behaviour of individual agents, which can then guide the development of new ABMs while providing methods for validating the models \cite{lebaron2000agent,AlexandruPeter2017,platt2020comparison}. ABMs are computationally expensive. This introduces a fundamental trade-off: balancing the computational cost of generating more sample paths against the additional statistical accuracy \cite{platt2020comparison}. The most common method for validating an ABM is the empirical {\it vs.} simulation comparison of {\it stylised facts} that are present in the generated time series, {\it e.g.} fat-tails in the return distributions, the presence of volatility clustering, the leverage effect, and an appropriate decay in the autocorrelations of price changes. 

This approach is computationally convenient and is essentially a form of exploratory data analysis. However, this approach has been criticised in the literature in the context of intraday ABMs because many parameters in ABMs are related to each other and share coupled impacts in the observed stylised facts. A given model's generated time series should reproduce the statistical properties of empirically measured transaction and order-book data through a pragmatic combination of sensitivity analysis, calibration, and stylised fact verification while accepting some level of parameter degeneracy. Given this, the empirical {\it vs.} simulation stylised fact comparison remains a good way to test the validity of the class of ABM of the limit-order book considered in this work \cite{fabretti2013problem,platt2020comparison,plattgebbie2018can}.

When market clearing takes place in event time rather than at preset times, the model needs to be able to update based on each individual event, and to be able to react to each individual event in a manner that is asynchronous for all the other agent realisations. In general, an open event-driven system model is incompatible with preset event times, whether randomly or uniformly sampled. However, if a particular agent-based model is the only model framework interacting via a given matching engine, then all the possible events will be generated by that given agent-based model. The model is then a closed system. 

This is the {\it hybrid} framework \cite{arxiv2021abm} where preset times can be used to streamline the computations even if the implementation is reactive. In such a closed system, new agents cannot be introduced as they may trigger events not accounted for by the preset event times. This means that when new agents are introduced in an open system without a method for agent synchronisation, {\it e.g} using batched market clearing, the system does not know {\it a-priori} when events will occur as they arise from interactions. In an open event-driven system, the model requires agents to be implemented in a reactive framework to make order matching realistic and event-driven.

\subsection*{Reactive agent-based models}

Reactive programming is a method of solving problems by manipulating or reacting to events in data streams generated by asynchronous observers using functional programming. Observables are the emitters of events in a data stream and can asynchronously emit events. There are a variety of reactive software implementations; we use Rocket.jl \cite{JuliaLang,Rocketjl} where actors \cite{hewitt2011actormodel} subscribe to an observable, and when a new event occurs in the data stream, the observable emits this event, and the actors consume this data and can act based on that event. In an event-driven ABM, the observable emits the current state of the LOB; the data stream consists of the states of the LOB through time. When the LOB is updated, the observable emits this as an event, and the actors - in this work, the minimal intelligence agents - can reactively trade based on the agent specification rules.

A reactive ABM lends itself to extending current ABM models in two novel ways: a fundamentally different model of time, and a fully decentralised control logic \cite{crafa2021agent}. In most of the current research, the ABMs that are used to model financial markets use a synchronous model of time where the model progresses through a sequence of global iterations where the agents are modelled as functions that are called in each of these iterations \cite{crafa2021agent}. This is not conducive to modelling financial agents as agents may respond to the same event at different times in reality. For example, one trader may be able to respond quickly to a large increase in the price of a particular stock price while another may have to unwind other positions to participate in the increase \cite{crafa2021agent}. 

Giles \cite{daniel2006asynchronous} demonstrated that the dynamics of the ABM can change depending on whether the agents trade synchronously or asynchronously. When agents trade synchronously, at each time point, all the agents are polled and then send in their orders and update their state. However, when agents trade asynchronously, they may place an order, hold, or be inactive. Aloud {\it et al.} \cite{aloud2017modeling} built an ABM of the foreign exchange (FX) market using agents that trade asynchronously and was able to closely approximate the stylised facts of the market. Asynchronous trading adds an important additional level of realism. However, the market clearing mechanism still operated synchronously since at each time step, all the agents' market orders were executed sequentially, and all the limit orders were cleared.

We decouple the matching engine and the model management system to ensure the agents are simulated in a continuous double auction with no synchronous market clearing. The system is now open. This has the benefit of allowing us to add additional agent classes that can interact with existing agents based on the state of the system. This can include agents learning to take actions based on system states. 

\subsection*{Reinforcement learning}

In a simple case of Reinforcement Learning (RL), an agent interacts with an environment at discrete time steps. At each time step, the agent observes the system's current state and chooses an action out of a set of finite actions. One time step later, the agent receives the reward for taking that particular action in that state and observes the new state of the environment. The goal of reinforcement learning is to find the mapping between states and actions to maximise the long-term reward of the agent. This framing of the reinforcement learning problem considers RL to solve a Markov Decision Problem (MDP) \cite{barto2003recent}. RL problems are not restricted to solving discrete-time finite state-action space MDPs. However, this is an ideal framework \cite{barto2003recent} for the simulation of simple RL algorithms and provides a simple baseline foundation for comparison with future extensions that is computationally inexpensive.

The advantages of using RL as compared to dynamic programming \cite{bellman1954theory} (for example) are twofold \cite{sutton2018reinforcement}. Firstly, the RL agents do not need full knowledge of the environment and learn an optimal policy. They can learn the optimal policy by interacting with the environment and recording the rewards for taking actions in certain states. Secondly, to increase the scalability of the learning process, the agent can use function approximations to represent their knowledge \cite{almgren2001optimal}. The RL algorithm that will be implemented in this paper is the \textit{Q-learning} algorithm \cite{watkins1989learning}. These simple RL agents are used to solve the optimal execution problem to represent a single learning agent with bounded rationality.

\subsection*{Optimal execution}

The optimal execution problem is defined as the problem of finding a way to trade a block of shares through time at the lowest possible cost \cite{almgren2001optimal, bertsimas1998optimal, obizhaeva2013optimal,cartea2015algorithmic}. Much of the existing research solves this problem using the tools of stochastic dynamic programming, which assumes that the prices will follow some combination of random walks in discrete time, or Brownian motions in continuous time, to approximate the data-generating processes. Data-driven representations of the data-generating processes have become popular due to the availability of market micro-structure data and improved computational resources. This has in turn made it possible to consider learning algorithms.

\subsection*{Problem Statement}

Our overarching aim is to investigate how learning affects, and is affected by, adaptation in markets. An additional objective is to set the scene for the calibration and simulation environment testing for a more sophisticated and detailed market microstructure analysis. In particular, to provide a market simulation framework that contains both bottom-up and top-down causation \cite{WilcoxGebbie2015}. We take initial steps towards this end, giving rise to the following research questions we consider in this work.  Firstly,  we ask {\it "Is learning feasible?"}. Secondly, we ask {\it "How do learning agent interactions change a minimally intelligent financial agent-based model?"}. 

Many agent-based models have some form of batch clearing \cite{plattgebbie2018can}, where either batches of market orders are executed against the limit-order book at preset calendar times, or rolling auctions are used. Batch clearing can introduce spurious correlations between parameters and elements of the model as the system is synchronised by an external feature that is not present in real financial markets. Consequently, realistic models should clear markets in event time so that the sampling time of events is emergent from agent interactions and not externally imposed. From a modelling perspective, one approach is to directly use a matching engine to avoid batch market clearing. 

A key technical issue then arises, the requirement that a reactive model is necessary to simulate the learning environment so that any number of additional agents can be included via direct interactions through the matching engine \cite{SoftwareXcointossx} without the need to specify the model update times. 

\subsection*{Our Approach}

We extend the ABM framework of Jericevich {\it et al.} \cite{arxiv2021abm} as the learning environment. Their model recovered several conventional stylised facts, including persistent auto-correlation in the trade signs, fat-tail return distributions, and volatility clustering, and recovered a realistic price impact function. In particular, recovers the expected power-law relationship between the log difference in the price changes and the size of the normalised volumes. Finally, and most importantly, the trades were simulated in event time, rather than the usual calendar time. 

The RL algorithm that will be implemented in this paper is the \textit{Q-learning} algorithm \cite{watkins1989learning}. Our implementation is inspired by the simple RL agent implemented by Hendricks and Wilcox \cite{hendricks2014reinforcement}. Their RL agent was trained on market data offline, with the assumption that the order book is resilient. This means that the trading of the RL agent does not affect future prices, and other agents in the market can't react to its trading. This was a necessary constraint, but not allowing other agents to react to the RL agent's trading may increase the observed profits above what will be realised in the real market. We aim to give some insight into the effects of training a similar RL agent in an online manner in an environment provided by a simple ABM. 

The choice of performance benchmark for the learning algorithm is important. This work uses a Time-Weighted-Average-Price (TWAP) benchmark because it is the optimal execution strategy in an efficient market. We measure performance by the Profit-and-Loss (PL) relative to immediate execution. This is discussed further in the Results section (See Figure \ref{rl-reward-convergnce}). It should be noted that in the real world setting, the AC benchmark can be a more appropriate and pragmatic choice \cite{hendricks2014reinforcement}.  

\subsection*{Contributions}

Our first contribution is the presentation of a novel ABM framework to facilitate the investigation of learning agents within a LOB. Our framework is novel for three reasons: i.) the change from modelling in calendar or stochastic time to event time is unusual in the current literature, ii.) the training of a learning trading agent within an environment with feedbacks is again unusual, with much of the machine learning literature moving towards surrogate data sets to capture the environment, and iii.) the investigation of the impact of a learning agent on a reasonably realistic minimally intelligent agent-based model to demonstrate the impact of strategic order-splitting on the price impact curves (See Figure \ref{rl-price-impact})  and on the trade-sign auto-correlations and log-micro-price fluctuations (See Figures \ref{rl-trade-sign-autocorrelation}). 

Our second and third contributions are our findings related to our research questions and objectives:
i) First, it shows that learning is feasible within an event-based ABM simulation of the market environment, even with a very simplistic reward function. 
ii) Second, this environment is relatively realistic but can be further enhanced with the inclusion of learning. 
iii) Thirdly, the inclusion of learning agents can be constructed within agent-based market models to include additional sources of top-down causation through adaption \cite{AulettaEllisJaeger2008}. iv) Lastly, a reactive framework is advisable if one aims to explore more realistic trading algorithm simulation when the order matching is event-based. The inclusion of many learning agents, agents with more complex cost functions, and alternative performance benchmarks has been explored in a preliminary manner elsewhere \cite{dicks2023learning}. 

\section*{The Model} \label{sec:AgentSpecifications}

A key feature of the combined model design is that we do not expect the underlying ABM model for the training environment to fully capture the required trade-sign Auto-Correlation Function (ACF). This is because the ABM model describing the trading environment has not been explicitly calibrated to describe the trade-sign ACF features, and the ABM model does not have agents explicitly engaging in strategic order-splitting. This is the same design choice made in prior calibration work \cite{plattgebbie2018can}. Only the learning agent is engaging in strategic order-splitting. This is important because both herding and order-splitting have been used to explain the order-flow auto-correlation, with order-splitting being given as the main reason \cite{toth2015equity}. So, in our foundational (minimally intelligent) ABM, the auto-correlations can only be explained by the herding of the chartists, and we show that this produces an effect, but one that is too weak to fully account for what is observed in the real market. 

The model calibration task is to both provide a method to estimate model parameters relative to the measured real-world data in a framework amenable to interactions with a learning agent and to also provide a method, with summary statistics, that can be used to measure the extent to which the learning agent, once introduced to the environment provided by the minimally intelligent ABM, is found to distort this environment. The method of simulated moments is convenient in this regard as it provides: i.) moments, ii.) an approach to managing the expected co-dependencies and correlations between these moments, and iii.) a method to bootstrap a reasonable spread of moments and their impact on parameters to coarsely measure model stability via indicative confidence intervals. However, the calibration is unsuitable for inference because of its weak link to the distributional properties of the moments themselves. Their purpose in this paper is not to statistically motivate the meaningfulness of particular parameter values relative to the real-world data, but to summarise various stylised facts for comparison once we introduce the learning agents into the market environment. 

Our model design has two key components. First, the underlying minimally intelligent ABM that will be used to model the learning environment. Second, the learning agent that interacts with and modifies the overall model behaviour, where the overall model is the minimally intelligent model ABM combined with an agent using RL. We now consider the learning agent specification, then the minimally intelligent foundation model of the environment that it will interact with. After this, we will discuss how the learning agent has changed the overall model. 

\subsection*{The Learning Agent} \label{ssec:AgentSpecifications-RLA}

The learning agent will learn how best to execute market orders using RL as part of an optimal liquidation problem. The optimal execution RL agent implemented here uses a simple tabular Q-learning approach with simple state attributes, actions, and reward functions \cite{sutton2018reinforcement}. The state space used and the agent's actions are based on the RL agent proposed by Hendricks and Wilcox \cite{hendricks2014reinforcement}. The state space parameters are given in Table \ref{rl-statespace-variables}. To specify the optimal execution agent, we need to specify the initial volume to be liquidated, the volume trajectory, the actions the agent can take, and the rewards the agent will attain after these actions. Let $X_{0}$ be the total initial volume to be liquidated, and is chosen so that the RL agents trade initial volumes of 21,500, 43,000, and 86,000 shares, and these are selected to be approximately 3\%, 6\%, and 12\% of the ABM's Average Daily Volume (ADV) traded.

\subsubsection*{Agent volume trajectory}

The agent is a selling agent and aims to maximise the profit received from liquidating shares \cite{hendricks2014reinforcement} where the agent aims to modify a given volume trajectory using only market orders. We do not use the AC volume trajectory \cite{hendricks2014reinforcement}. The volume trajectory to be modified is the event time analogue of the Time-Weighted-Average-Price (TWAP) volume trajectory. Since our model is implemented in event time, setting aside specific calendar times in the simulation for the agent to trade is inconvenient. 

Therefore, we have modified the TWAP strategy to work in the reactive event time approach without imposing update times on the learning agent. This volume trajectory will be referred to as the event-based TWAP volume trajectory to remind the reader that the formulation of time has changed. This is necessary because there is no external clock providing uniform, or event stochastic, time steps for a calendar time TWAP implementation {\it e.g.} at every 5 minutes in the market's calendar time. This is because no top-down causation process imposes an external calendar time in our implementation. However, in our model, it does not matter whether the TWAP schedule is implemented at random times or uniformly sampled times; at each event update, it triggers the update of the sequence of same-sized trading lots that make the trading schedule. This is computationally convenient as we could have triggered learning agent updates in machine time, and this has no impact on the results. 

The number of decision points is $N_{dp}$, which is the number of times an order of size $X_{0}/N_{dp}$ needs to be executed to fully liquidate. A decision point count is the number of times the LOB is updated, and the RL agent is asked to make a trading decision. $N_{dp}$ governs how long an RL agent will trade in the simulation if it follows the TWAP strategy. If $N_{dp}$ is small then the agent will trade for a short amount of time if it follows the TWAP volume trajectory and there is enough liquidity, and {\it vice versa} for large $N_{dp}$. In this implementation, we have set $N_{dp}$ so that if the typical agent trades exactly $x_{i}$ at each decision point and there is enough liquidity, the agent will trade on average for the whole simulation and fully liquidate. We have chosen $N_{dp}$ this way because it most closely resembles a TWAP strategy implemented over an entire 8-hour calendar time trading day.

\subsubsection*{Agent actions}

The RL agent is constrained to submitting only market orders \cite{hendricks2014reinforcement}. The volume of the market order is found from a given parent order $X_{0}$ and a volume trajectory $\{x_i\}_{i=1}^{_{N_{dp}}}$ such that the parent order is fully executed by the volume trajectory: $X_{0}=\sum \{ x_{i} \}_{i=1}^{_{N_{dp}}}$. Then at the $i^{\mathrm{th}}$ decision point the agent can increase, decrease, or trade $x_{i}$. At the $i^{th}$ decision point the agent chooses an action $a_{i} \in \left\{0,0.25,0.5,...,2\right\}$ such that the modified volume is given by $\tilde{x}_{i} = a_{i} x_{i}$. The goal is to have the agent learn in which states it is better to increase or decrease the volume traded to maximise the reward. If, at the end of the trading period, the agent has not fully liquidated their position, the agent will submit a maximum order of $\tilde{x}_{i} = 2 x_{i}$ to try to trade as much of the remaining inventory allowed by the actions. This does not guarantee that all the inventory will be traded. This ensures that there is a penalty for having inventory left over, as the trader will then not receive the additional profit. 

\subsubsection*{Agent rewards}

The goal for this agent is to maximise the total profit from selling the large initial parent order. Given this goal, the reward signal must be defined such that the return corresponds to the goal. Let there be $N$ market orders submitted by the RL agent. For the $n^{th}$ order, in an episode, let the total profit received be the reward for that trade:
$R_{n} = \sum_{j=1}^{J} p_{j} x_{j}$,
for an order that has walked $J$ levels down the order book. Given that each simulation is used as an episode, we then have that an episode's return is: $G = \sum_{n=1}^{N} R_{n}$,
where $R_{n}$ is the $n^{th}$ order in an episode. This corresponds to the total profit from selling inventory in that episode. Therefore, the agent will maximise the total profit, which is its goal.

For the Q-learning algorithm, the convergence guarantees are based on the fact that the underlying model generating the state, action, and reward samples evolve according to a Markov process, where the next state and the reward depend only on the previous state and the action taken in that state. Given the simplicity of the state space, it is unlikely that it encodes enough information about the LOB to ensure that the process is Markov. This means that the convergence guarantees for the Q-learning algorithm are most likely violated \cite{watkins1992q}. However, we show that this is sufficient to encode enough information for the agent to learn simple, intuitive rules for liquidating a position in this model and implementation. We use standard updating equations in an episodic setting because we do not use a finite horizon formulation \cite{garcia1998learning,hendricks2014reinforcement}.

\subsubsection*{Agent state space}

The state space consists of four variables: two variables that summarise the state of the individual agent's current execution and two variables that summarise the state of the LOB. \textit{Remaining time, t,} and \textit{Remaining inventory, i,} are used to summarise the state of execution. The \textit{Spread, s,} and \textit{Best bid volume, v,} are used as the variables to summarise the state of the LOB, with the idea being that the agent will learn to increase (resp. decrease) volume traded when the spreads are narrow (resp. wide), and the best bid volume is high (resp. low). To be compatible with tabular Q-learning, these variables must be discredited, and course-grained to ensure that the state space does not grow too large. The state space parameters are shown in Table \ref{rl-statespace-variables}. 

For the time and inventory variables, the total trading time is divided into $n_{_T}$ intervals, and the total inventory to be traded is divided into $n_{_I}$ blocks. As the time and inventory decrease through the simulation, the time interval and the inventory blocks that the state belongs to determine its state value. The state values are given by indices $i$ and $t$ as shown in Table \ref{rl-statespace-variables}. For the spread and volume states, we would like to know whether the spread and the best bid volume are ``high" or ``low". This is done by comparing the current values to historical distributions of the spread and volume \cite{hendricks2014reinforcement}. To generate the historical distributions, we ran the calibrated simulation model 365 times and recorded the spreads and the best bid volumes. The state value is determined by which quantiles the spread falls in between. The details are provided in Supplementary Table S6. For the volume distribution, we use the method of Hendricks and Wilcox \cite{hendricks2014reinforcement}, with the details provided in Supplementary Table S5. 

\begin{table*}[h]
\caption{The learning agent parameters: The RL agent training parameters are the session length in machine time $T_0$, the different values for the initial inventories for the parent order $X_0$, and state space variables representing the different parameter combinations. The volume trajectory lot sizes are given in the brackets. 
The number of decision points, the number of episodes, and the discounting and learning rates are fixed. The parameters used to create the state space are: the remaining time index is $t=$ $1,2,\ldots,n_{_T}$, the remaining inventory index is $i=$ $1,2,\ldots,n_{_I}$, the spread state index $s=$ $1,2,\ldots,n_{_S}$, and the volume state index $v = 1,2,\ldots,n_{_V}$. The spread and volume state indices map into values based on measured historic feature distributions.}
\label{rl-training-parameters}
\label{rl-statespace-variables}
\centering
\begin{tabular}{l l l l l}
\toprule
& Parameters & Description & Values & Interpretation \\
\toprule
\multirow{5}{*}{\rotatebox{90}{Trajectories}} &$T_0$ & Trading episode length & 24.5s & 8 hours of trading\\
&$X_{0}$ & Parent order and lot size & 21,500 (50) & 3\% of model ADV\\ 
&    & & 43,000 (100) & 6\% of model ADV\\
&    & & 86,000 (200) & 12\% of model ADV\\
&    $N_{dp}$ & Counts & 430 & Decisions per episode \\
    \hline
\multirow{4}{*}{\rotatebox{90}{States}} &$n_{_T}$ & Number of time states & 5, 10 & ``small", ``large"\\ 
&$n_{_I}$ & Number of inventory states & 5, 10 & ``small", ``large"\\ 
&$n_{_S}$ & Number of spread states & 5, 10 & ``small", ``large" \\ 
&$n_{_V}$ & Number of volume states & 5, 10 & ``small", ``large" \\ 
\hline
\multirow{3}{*}{\rotatebox{90}{Training}} &$N_{E}$ & Number of training episodes & 1,000 &  \\
&$\gamma$ & Discounting rate & 1 & \\
&$\alpha$ & Learning rate & 0.1 & \\
\bottomrule
\end{tabular}
\end{table*}

\subsubsection*{Training parameters and implementation}

The number of training episodes, the discounting factor and the learning rate used are given in Table \ref{rl-training-parameters}. In Q-learning, there are two policies: the target policy that we aim to optimise, and the behaviour policy, which generates the agent's behaviour used to optimise the target policy. In our implementation of Q-learning, the target policy is the greedy policy, and the behaviour policy is the $\epsilon$-greedy policy, where $\epsilon$ controls the amount of exploration versus exploitation in the optimisation. For a given state and $N_{A}$ actions, the $\epsilon$-greedy policy chooses the action with the highest value with probability $1-\epsilon + (\epsilon/N_{A})$ and the rest with probability $\epsilon/N_{A}$. Setting $\epsilon=1$ will mean an agent selects each action with equal probability regardless of its value, and setting $\epsilon=0$ will ensure the agent always selects the action with the highest value -- the greedy policy. To ensure a good balance between exploration and exploitation, the $\epsilon$ parameter was S-curve decayed throughout the simulation.

To update the value of the current state-action pair, the Q-learning update equation needs to observe the next state that the agent transitioned to. In a live implementation, this state is not observed until the message from the matching engine has been received, indicating if and then how the order was matched. To remedy this, the update to the state-action pair's value is done one step later when the next state is observed. This is shown in Supplementary Algorithm 4.

\subsection*{The Environment}

The ABM represents the environment in which a learning agent will be trained and then simulated. It is populated with minimally intelligent liquidity takers and liquidity providers interacting in a continuous-double auction market. The liquidity takers (LTs) submit only Market Orders (MOs), and the liquidity providers (LPs) submit only Limit Orders (LOs). Market orders are submitted for immediate execution, while limited orders have a reservation price. We do not know when agents will be activated because the model has no global time. To solve this activation problem, we add a spread correction of $\pm \frac{1}{2} s_t$ to the liquidity takers rules \cite{arxiv2021abm}. The liquidity takers would then act asynchronously in event time because they can now trade, hold, or do nothing as activated by the spread correction. In the case of High-Frequency (HF) agents, the environment's liquidity providers, a simplifying assumption allows them to trade in each event loop iteration. Concretely, this is because the LPs are modelling HF market makers, and we can assume that as the LOB is updated, there will be an HF trader attempting to use their latency advantage to exploit these updates to profit from the spread via round-trip trades -- either taking advantage of the strategic or competition effect. The volume of the orders is sampled from a power-law distribution density function  \cite{arxiv2021abm}, where the distribution parameters for the cut-off $x_{m}$ and the shape $\alpha$ are determined based on the state of the historical order book for both the liquidity takers and the liquidity providers. The environment is schematically represented in Figure \ref{fig:environment}.

\begin{figure}[!htb]
    \centering
    \begin{tikzpicture}[node distance=1.5cm, scale = 0.03]
        \node (ltf) [rectangle,rounded corners,minimum width=2.4cm,minimum height=1cm,text centered,align=center,yshift=-1.5cm,xshift=0.5cm,draw=black,thick] {Liquidity Taker: \\ Fundamentalist};
        \node (ltc) [rectangle,rounded corners,minimum width=2.4cm,minimum height=1cm,text centered,align=center,draw=black,thick,right of=ltf,xshift=1.2cm] {Liquidity Taker: \\ Chartist};
        \node (lp) [rectangle,rounded corners,minimum width=2.4cm,minimum height=1cm,text centered,align=center,draw=black,thick,right of=ltc,xshift=1.4cm] {Liquidity Provider: \\ HFT/ELP};
        \node (lob) [rectangle,rounded corners,minimum width=3cm,minimum height=1cm,text centered,draw=black,thick,above of=ltc,xshift=0.8cm] {LOB state};
        \node (ltfside) [rectangle,aspect=1.5,minimum width=1cm,minimum height=1cm,text centered,draw=black,thick,below of=ltf,align=center,yshift=0cm,xshift=-0.0cm] {MO Side (\ref{fund-update-rule}): \\$f_i-(m_k \mp \frac{1}{2}s_k)$  };
        \node (ltcside) [rectangle,aspect=1.5,minimum width=1cm,minimum height=1cm,text centered,draw=black,thick,below of=ltc,align=center,yshift=0cm,xshift=0.3cm] {MO Side (\ref{chart-update-rule}): \\$\bar m_{i,k} -(m_k \pm \frac{1}{2}s_k)$ };
        \node (lpside) [rectangle,aspect=1.5,minimum width=2cm,minimum height=1cm,text centered,draw=black,thick,below of=lp,align=center,xshift=0.2cm, yshift=0.0cm] {LO Side (\ref{bidprob}): \\$\theta = \frac{1}{2}(\rho_t + 1)$};
        \node (lpbs) [diamond,minimum width=3cm,minimum height=1cm,text centered,draw=black,thick,below of=lpside,align=center,aspect=2,yshift=-0.2cm,xshift=-0.6cm] {Buy/Sell($\pm$)};
        \node (lpdepth) [rectangle,minimum width=3.5cm,minimum height=1cm,text centered,draw=black,thick,below of=lpbs,align=center,yshift=-0.2cm] {Placement Depth:\\$\eta \sim f(x)$ (\ref{eta})};
        \node (lpprice) [rectangle,minimum width=3.5cm,minimum height=1cm,text centered,draw=black,thick,below of=lpdepth,align=center] {Limit Price (\ref{placementprice}): \\$p_k = b_k/a_k \pm 1 \pm \eta$};
        \node (lpalpha) [rectangle,minimum width=3.5cm,minimum height=1cm,text centered,draw=black,thick,below of=lpprice,align=center] {Volume Shape: \\$\alpha = 1 \pm \frac{\rho_k}{\nu}$};
        \node (ltbs) [diamond,minimum width=3.5cm,minimum height=1cm,text centered,draw=black,thick,left of=lpbs,align=center,xshift=-2.8cm,aspect=2] {Buy/Sell($\pm$)};
        \node (ltalpha) [rectangle,minimum width=3.5cm,minimum height=1cm,text centered,draw=black,thick,left of=lpalpha,align=center,xshift=-2.8cm] {Volume Shape: \\ $\alpha = 1 \pm \frac{\rho_k}{\nu}$};
        \node (ltminvol) [rectangle,minimum width=3.5cm,minimum height=1cm,text centered,draw=black,thick,below of=ltalpha,align=center] {Minimum Volume: \\  $x_m$ (\ref{fun-xm} or \ref{char-xm})};
        \node (ltvolume) [rectangle,minimum width=3.5cm,minimum height=1cm,text centered,draw=black,thick,below of=ltminvol,align=center] {Order Volume: \\$\omega_{i,k} \sim f(x ,x_m,\alpha)$ (\ref{eq:ordervolume}) };
        \node (lpvolume) [rectangle,minimum width=3.5cm,minimum height=1cm,text centered,draw=black,thick,right of=ltvolume,align=center,xshift=+2.8cm] {Order Volume:\\ $\omega_{i,k} \sim f(x, 5,\alpha)$ (\ref{eq:ordervolume})};
        \node (cointossx) [rectangle,rounded corners,minimum width=5cm,minimum height=1cm,text centered,draw=black,thick,below of=ltvolume,align=center,xshift=2cm] {Matching Engine (CoinTossX)};\node[align=center,xshift=-2.0cm,trapezium,trapezium left angle=70,trapezium right angle=110,minimum width=3cm,minimum height=1cm,text centered,draw=black,thick,left of=lob] (listener) {Market Data\\Listener};
        \draw [thick,->,>=stealth,dotted] (cointossx) -| (-40, -40) -- (listener);
        \draw [thick,->,>=stealth] (listener) -- (lob);
        \draw [thick,->,>=stealth] (lob) -- (lp); 
        \draw [thick,->,>=stealth] (lob) -- (ltf); 
        \draw [thick,->,>=stealth] (lob) -- (ltc);
        \draw [thick,->,>=stealth] (lp) -- (lpside); 
        \draw [thick,->,>=stealth] (ltf) -- (ltfside); 
        \draw [thick,->,>=stealth] (ltc) -- (ltcside);
        \draw [thick,->,>=stealth] (ltfside) -- (ltbs); 
        \draw [thick,->,>=stealth] (ltcside) -- (ltbs);
        \draw [thick,->,>=stealth] (lpside) -- (lpbs);
        \draw [thick,->,>=stealth] (lpbs) -- (lpdepth); 
        \draw [thick,->,>=stealth] (ltbs) -- (ltalpha);
        \draw [thick,->,>=stealth] (ltalpha) -- (ltminvol); 
        \draw [thick,->,>=stealth] (ltminvol) -- (ltvolume); 
        \draw [thick,->,>=stealth] (lpdepth) -- (lpprice);
        \draw [thick,->,>=stealth] (lpprice) -- (lpalpha); 
        \draw [thick,->,>=stealth] (lpalpha) -- (lpvolume);
        \draw [thick,->,>=stealth,dotted] (ltvolume) -- (cointossx);
        \draw [thick,->,>=stealth,dotted] (lpvolume) -- (cointossx);
    \end{tikzpicture}
    \caption{Learning environment state flow schematic as adapted from \citet{arxiv2021abm}. Starting with the state of the order book, three different classes of minimally intelligent agents can be activated by events: fundamentalists, chartists, and liquidity providers. The agents are index by $i$ and the events by $k$. Liquidity takers use Market Orders (MOs). The liquidity providers use only Limit Orders (LOs). Parameters are defined in the text following the equation numbers. Any resulting orders are submitted into the Matching Engine (ME) (CoinTossX  \cite{SoftwareXcointossx}). In a reactive formulation, any number of additional agents can inject orders into the matching engine (CointossX) when triggered by appropriate events from the Market Data Listener (MDL). See algorithms 1, 2 and 3 in the supplementary materials.} \label{fig:environment}
\end{figure}

\subsubsection*{Liquidity takers} \label{sssec:AgentSpecifications-LTs}

There are two types of liquidity takers: First, there are the fundamentalists who trade based on a perceived fundamental value of the asset. Second, there are the chartists, or trend followers, who trade based on the current asset price trend. There are $N^{f}_{_{\mathrm{LT}}}$ fundamentalists and $N^{c}_{_{\mathrm{LT}}}$ chartists that make up the population of liquidity takers. The direction and volumes of their market orders are determined by simple mechanistic decision rules. For the fundamentalists:
\begin{align}
    D^{f}_{i,k} = \begin{cases} 
      \text{sell} \;\;\; &\text{if} \;\; f_{i} < m_{k} - \frac{1}{2} s_{k} \\
      \text{buy} \;\;\; &\text{if} \;\; f_{i} > m_{k} + \frac{1}{2} s_{k}, 
      \end{cases}  \label{fund-update-rule}
\end{align}
where $m_{k}$ and $s_{k}$ are the mid-price and spread at event k, and $f_{i}$ is the fundamental price for the $i^{th}$ fundamental agent. Here the fundamental price is $f_{i} = m_{0} e^{x_{i}}$ where $x_{i} \sim \mathcal{N}(0, \sigma^{2}_{f})$ and
$m_{0}$ is the initial mid-price, and $\sigma^{2}_{f}$ is the variance of the normal distribution that generates the log-normal distribution. Each fundamentalist's fundamental price is determined at the start of the simulation and is kept constant throughout. The spread correction, $- \frac{1}{2} s_{k}$ for sells and $+ \frac{1}{2} s_{k}$ for buys, modifies the fundamentalists' trading rule \cite{arxiv2021abm} to ensure that the fundamentalists never buy (sell) a stock for more (less) than their fundamental value. It also ensures that the fundamentalists are trading asynchronously. 
For the chartist:
\begin{align}
    D^{c}_{i,k} = \begin{cases} 
      \text{sell} \;\;\; &\text{if} \;\; \Bar{m}_{i,k} > m_{k} + \frac{1}{2} s_{k} \\
      \text{buy} \;\;\; &\text{if} \;\; \Bar{m}_{i,k} < m_{k} - \frac{1}{2} s_{k}, 
      \end{cases}  \label{chart-update-rule}
\end{align}
where $\Bar{m}_{i}$ is the exponentially weighted moving average (EWMA) of the mid-price computed by the $i^{th}$ chartist agent. A similar spread correction was implemented for the chartists, as was implemented for the fundamentalists. The moving average is computed using: $\Bar{m}_{i,k+1} = \Bar{m}_{i,k} + \lambda_{i} (\Bar{m}_{i,k} - m_{k})$. The forgetting factor, $\lambda_{i}$, is not updated using the same method used in the hybrid ABM \cite{arxiv2021abm}. This is because, in event time, the chartists can converge and herd if they use the same moving average. The forgetting factor may not update fast enough and can become consistently close to one. This will cause the chartist agents to track the mid-price too closely, which may not generate trades. The forgetting factor is sampled for each agent at the beginning of the simulation from the Uniform distribution, $\lambda_{i} \sim U(\lambda_{min}, \lambda_{max})$. Sampling the uniform distribution gives each agent an equal probability of having a long or short trading horizon. As with the hybrid ABM's fundamentalists, the rule has been modified to include a spread correction. This ensures that the chartists' rules are not always activated and allows greater heterogeneity in the chartist agent behaviour.

Once the liquidity takers have determined the direction of the market order, the next step is to determine the volume of the order. The volume distribution's minimum volume $x_{m}$ is determined based on how far the fundamental value, or the moving average, is from the mid-price relative to a threshold $\delta$, which is given as a percentage of the current mid-price. The agent parameters are given in Table \ref{parameters-table}. For fundamentalists, the minimum volume is:
\begin{align}\label{fun-xm}
    x_{m} = \begin{cases} 
      20 \;\;\; &\text{if} \;\; |f_{i} - m_{k}| \leq \delta m_{k} \\
      50 \;\;\; &\text{if} \;\; |f_{i} - m_{k}| > \delta m_{k}.
      \end{cases}
\end{align}
For chartists, this is:
\begin{align}\label{char-xm}
    x_{m} = \begin{cases} 
      20 \;\;\; &\text{if} \;\; |\Bar{m}_{i,k} - m_{k}| \leq \delta m_{k} \\
      50 \;\;\; &\text{if} \;\; |\Bar{m}_{i,k} - m_{k}| > \delta m_{k}. 
      \end{cases}
\end{align}
This increases the aggression of the agents when the state of the system presents a favourable trading opportunity relative to their fundamental value or trading strategy.

The volume of each agent's order is sampled from a power law distribution function if $x \geq x_{\mathrm{m}}$ \cite{arxiv2021abm}:
\begin{equation}\label{eq:ordervolume}
    f(x)= \frac{\alpha x_{\mathrm{m}}^{\alpha}}{x^{\alpha+1}}, 
\end{equation}
 and is otherwise zero. The parameters $x_{\mathrm{m}}$ and $\alpha$ will depend on the state of the LOB and activation rules. The shape parameter of the volume distribution, $\alpha$, is determined with respect to the order imbalance $\rho_k$ and an additional model parameter $\nu$: $\alpha = 1\pm\rho_{k}/\nu$ with + (-) for buy (sell) MOs \cite{arxiv2021abm}. The order-imbalance is: $\rho_{k} = \sfrac{v_{k}^{a} - v_{k}^{b}}{v_{k}^{a} + v_{k}^{b}}$, where $v_{k}^{a}$ and $v_{k}^{b}$ is the volume of the asks and the bids respectively. The parameter $\nu > 1$ ensures that $\alpha$ never equals 0. Specifying $\alpha$ in this way achieves two effects. First, if the contra side of the order book is thicker, the agent will submit larger orders to take advantage of the extra liquidity because it will likely reduce the price impact of a large order. Second, if the same side is thicker, the orders will be smaller to reduce the price impact. 

\subsubsection*{Liquidity providers}\label{sssec:AgentSpecifications-LPs}

Each LP agent acts on every LOB modification in every event loop. Thus, an event loop can naively be thought to batch sets of limit order events for each agent. However, the orders are independently submitted to and matched by the matching engine. This approach is used for three reasons. First, it ensures that the simulation does not result in a state where no trade occurs because of the lack of liquidity provision. This would cause the simulation to halt prematurely and would, in turn, be problematic when calibrating the model. Second, the LP agents, in collective, are simulating high-frequency market-makers or electronic liquidity suppliers that are profiting from the spread. So it is natural that they want to submit as many orders as quickly as possible in the hope of generating lots of small profits. Moreover, given enough high-frequency agents in the market, it is reasonable to assume that at any moment LOs are being placed based on the state of the LOB. Third, they ensure that the order book is reasonably stable throughout the simulation and reduce the chances of liquidity crashes.

During each event loop, the LPs place an ask with a probability of $\theta$, and a bid with a probability of $1-\theta$, where 
\begin{align} \label{bidprob}
\theta = \frac{1}{2}(\rho_{k} + 1).
\end{align}
This setup has LPs provide liquidity, on average, to the side that has the least liquidity. After determining the direction of the order, the price placement and the volume have to be determined. 

The price placement of the limit order is given by:
\begin{align} \label{placementprice}
    p_{k} = \begin{cases} 
      b_{k} + 1 + \lfloor \eta \rfloor \;\;\; &\text{for asks}\\
      a_{k} - 1 - \lfloor \eta \rfloor \;\;\; &\text{for bids}, 
      \end{cases}
\end{align}
where $a_{k}$ and $b_{k}$ are the best offer and best bid prices at event k respectively. The threshold parameter $\eta$ is sampled from a Gamma distribution: 
\begin{align} \label{eta}
f(x) = \beta^{s_k} \Gamma^{-1}(s_k) x^{s_k - 1} e^{-\beta x}
\end{align}
found using the current spread $s_{k}$. Here $\beta$ is set using the order imbalance $\rho_k$: $\beta =  e^{\pm\rho_{k}/\kappa}$ with + (-) for bids (offers) where $\kappa > 0$  is a model parameter. The mean value for $\eta$ at event k is $\Bar{\eta} = s_{k}e^{\pm\rho_{k}/\kappa}$ with + (-) for bids (offers). This price placement achieves several effects. First, when the order book is relatively balanced, the orders are placed close to the best bid and offer prices. Second, when the order book is thicker on the contra side, the orders are placed further away from the best bid and offer prices, capturing the {\it strategic effect} \cite{Alexandru2015}. Third, when the same side of the order book is thicker, the orders are placed closer to the best bid and offer prices, capturing the {\it competition effect} \cite{Alexandru2015}. The larger (resp. smaller) the value of the model parameter $\kappa$, the lower (resp. higher) the intensity of the strategic and competition effects.

The volume of the limit orders is sampled from the same power-law distribution as the market orders given in Equation \ref{eq:ordervolume} but with different parameter values \cite{arxiv2021abm}. To ensure enough liquidity, an increase in the ratio of the number of LPs to LTs is needed, but this ensures that there are more small orders rather than more big orders that can saturate the best bid/offer. The minimum volume distribution cut-off parameter is 5; where it was 10 in the hybrid model \cite{arxiv2021abm}. The shape parameter for the volume distribution for the liquidity providers is: $\alpha = 1\pm\rho_{k}/\nu$ with +(-) for bids (asks). This achieves two effects. First, when the order book is thicker on the contra side, the volume will be larger. Second, when the order book is thicker on the same side, the volume will be smaller. The rationale behind these effects is that they will stabilise the order book. Frequently cancelled orders are a feature of high-frequency traders, and this ABM will model this behaviour using prior work \cite{arxiv2021abm} where orders are cancelled if they are not completely filled within a certain amount of time $\phi$.

\begin{table*}[h]
\centering
\caption{The environment parameters: The foundational ABM describing the environment that the learning agent is introduced into. Some parameters were kept fixed, while others were allowed to vary in the calibration. These choices were informed by a sensitivity analysis and prior work \cite{arxiv2021abm}. Additional detail is provided in the supporting materials. Time is measured in machine time, where 1 second of machine time is approximately equivalent to 20 minutes of calendar time trading. The calibrated parameters are shown in the last three columns. The moments generated by the calibrated model are provided in Table \ref{calibrated-params-moments-table}.}
\label{parameters-table}
\label{calibrated-params-table}
\begin{tabular}{rllllll}
\small \\
\hline
&&&& \multicolumn{3}{c}{Calibrated Values} \\
Parameters & Description & Type & Range & $\hat{\theta}_{0.025}$ & $\hat{\theta}$ & $\hat{\theta}_{0.975}$  \\
\hline
$T$ & LOB simulation (machine) time. & Fixed & 25s & & - &\\
$\phi$ & Time to cancellation of an order. & Fixed & 1s & & - & \\
$N_{_{\mathrm{LP}}}$ &  The number of liquidity providers & Fixed & 30 & & - &\\
$N^{c}_{_{\mathrm{LT}}}$ &  The number of chartist liquidity takers & Free & $\left[1,20 \right]$ & 7.256 & 8.000 & 8.744 \\
$N^{f}_{_{\mathrm{LT}}}$ & The number of fundamentalist liquidity takers & Free & $\left[1,20 \right]$ & 4.302 & 6.000 & 7.698  \\
$\delta$ & Minimum volume mid-price cut-off $\%$ & Free & $\left[0,10 \right]$ & 0.12 & 0.125 & 0.13 \\
$\kappa$ & Liquidity provider limit order placement intensity & Free & $\left[0,\infty \right)$ & 3.254 & 3.289 & 3.524  \\
$\nu$ &  Order-imbalance shape parameter. & Free & $\left[1.5,\infty \right)$ & 6.267 & 7.221 & 8.175 \\
$\sigma_{f}$ &  Fundamental value variance & Free & $\left[0,0.05 \right]$ & 0.0402 & 0.041 & 0.0418  \\ 
$\lambda_{min}$ & Minimum chartist forgetting factor & Fixed & 0.0005 & & - & \\
$\lambda_{max}$ & Maximum chartist forgetting factor & Fixed & 0.05 & & - &  \\
\bottomrule
\end{tabular}
\end{table*}

\subsubsection*{Agent-based model calibration}

The calibrated parameters $\hat \theta$ are shown in Table \ref{calibrated-params-table}. A single day of Naspers trade-and-quote (TAQ) from the Johannesburg Stock Exchange (JSE) was used as the empirical data to calibrate the model and compare the stylised facts. The day of data that was used was 08-07-2019, and it was extracted from a larger static trading research dataset \cite{Jericevich2020data,gebbiechangjericevich2020comparing, arxiv2021abm}. The data set used is 8 hours of trading in calendar time. For this day of trading, we found that the number of limit orders and trade events on the JSE was 15,104 and 2,510, respectively. The average number of limit order events in 100 simulations was 12,071 with a 95\% CI (10,988, 12,953), and similarly, the average number of trade events is 5,467 with a 95\% CI (4,566, 5, 798). The model's Average Daily Volume (ADV) is 745,362 with a 95\% CI of (593,536, 1,008,027) estimated over 100 simulations. This is comparable to the ADV of 498 981 for the trading week from which the calibration data was taken. A typical day of simulated trading using the calibrated parameters is shown in Figure \ref{sim-price-path}.

The data was first cleaned into a ``top-of-book" format for the calibrated price paths, simulated price paths, and measured real-world data. The mid/micro-prices obtained from this data are then used to estimate the moments and compare the stylised facts. This ensures a like-for-like data comparison between the simulated and measured data. It also means that the data is compared in event time on both markets, where an event is classified as an order (limit order or trade) that modifies the top of book and trade data. During calibration, we use the micro-prices from all updates to the LOB. This enables a like-for-like comparison between the hybrid ABM \cite{arxiv2021abm} and our reactive extension, and also reduces the large computational overhead of cleaning the data in each calibration iteration.

We calibrated the ABM representing the environment to data using the method-of-moments with simulated minimum distance (MM-SMD) \cite{gilli2003global}. This method was chosen because it is more computationally tractable than Bayesian inference, does not require a closed-form solution like the maximum-likelihood method, and provides a simple link between model parameters and moments to summarise important properties of the data. To optimise the MM-SMD objective function, we used the Nelder-Mead algorithm with Threshold Accepting (NMTA) using adaptive parameters \cite{gao2012implementing}. A description of MM-SMD and a more detailed discussion can be found in the supplementary materials. Briefly, the method of simulated moments starts with a moving block bootstrap of the measured data to generate a covariance between the estimated moments: $\Sigma_e = \Cov[{m}^e]$. This is then used to form a weighted distance function between the simulated moments $m^s(\theta)$ and the estimated moments $m^e$: $f(\theta)=\E[m^s(\theta) - m^e]^{_T} \Sigma_e^{-1} \E[m^s(\theta) - m^e]$. The expectation represents a Monte Carlo step using five random seeds. This distance is used as the objective function that is minimised to estimate the parameters: $\hat \theta =\argmin_{\theta} f(\theta)$. 

The trade and limit order volumes are highly dependent on the model's parameters. We chose to let the calibration procedure determine the parameters, which will produce the volume distributions. Even though there are some discrepancies between the counts, {\it i.e.} fewer limit orders and more trades in the ABM, when compared to the JSE, increasing or decreasing the length of time the simulation was run for will only increase the differences. Based on this, we are confident that the machine time period of 25s generates event counts that are comparable to that found in a day of real-world trading, as observed in the calibration data. 

As part of the calibration methodology, parameter convergence plots were investigated, and it was found that there was sufficient evidence that the NMTA algorithm was able to converge to a local optimum during calibration. It was found that the values of the objective function for the best vertex did not significantly improve in the last 32 iterations, and the variance in the objective function values across the simplex showed convergence. The calibrated parameters and associated indicative confidence intervals are shown in Table \ref{calibrated-params-table}. We could have estimated confidence intervals using a computationally expensive multi-day bootstrap that re-samples and then re-estimates over multiple trading days -- this was computationally infeasible, particularly when a similar method would be required for each learning agent experiment. However, from the sensitivity analysis, we can get insight into the relative stability of the parameters and the moments from the model perspective. 

The sensitivity analysis was used to find a feasible range of initial parameters and to check the objective function surface structure related to these to inform the choices summarised in Table \ref{parameters-table}. This is described in the supporting materials. The resulting parameter chains also provide combinations of parameters and moments that we can use to approximate the model-generated dependencies between the parameters, the parameters and the moments, and the moments. We can estimate $\Sigma_s = \Cov[m^s|\theta]$ to inform the expected model generated variability near the estimated moments, as well as the exposures between the six parameters and eight moments: $B=\Cov[\theta|m^s]/\Var[m^s]$. These can be used to approximate the model-generated parameter dependencies $\Sigma_{\theta} = B^{_T} \Sigma_e B$. These allow us to provide indicative model-generated confidence intervals for the parameters to provide guidance to the relative stability and influences across various parameters when the random number seeds are fixed.

The associated moments are presented in Table \ref{calibrated-params-moments-table}. The table also provides the simulated moments that capture the sample variations from random paths arising from many different random seeds in the simulation using the fixed calibrated parameters from Table \ref{parameters-table}. The calibrated moments represent the model moments from the calibration, but with a reduced fixed set of only five seed vectors for the random numbers used to simulate the environment. The same five seeds are reused for all training episodes. This ensures that comparable paths are recovered for each episode if no training agent is present. The training agent changes the overall model, and hence will change these paths, which changes the moments. These changes are not due to the variation in the random number sequences used nor parameters in the underlying ABM representing the training environment. Model training environment-specific variations can arise from model degeneracy and not the learning agent interactions. These are represented by the indicative confidence intervals for the calibration moments in Table \ref{calibrated-params-moments-table}. The table demonstrates that there is no significant difference between the simulated paths and the paths and resulting moments used in the calibration. The high computational costs of training and simulating learning agents motivate this approach to the simulation of paths.    

\begin{table*}[h]
\caption{The simulated, calibrated and empirical moments describing the learning environment and their respective variability. The simulated moments $\boldsymbol{m^{s*}}$ were obtained from the log-micro-price paths generated by the model parameters found in Table \ref{parameters-table} and are found from 100 replications using different seeds for each path. The calibrated moments $\boldsymbol{m^{s}}$ represent the combination of the same model parameters and the five random number seed vectors used to calibrate the environment. The same seeds and the calibrated parameters are used to simulate the training environment. The empirical moments $\boldsymbol{m^{e}}$ were estimated on a single day of transaction data using a moving window block bootstrap. These moments act as reference moments that summarise what is expected from the simulated environment. This is described in detail in the supporting materials.}
\label{calibrated-params-moments-table}
\label{estimated-simulated-moments-table}
\centering
\begin{tabular}{lrclrclrcl}
\toprule
& \multicolumn{3}{c}{Simulated moments} &  \multicolumn{3}{c}{Calibrated moments} & \multicolumn{3}{c}{Empirical moments} \\
Moment & $m^{s*}_{0.025}$ & $m^{s*}$ & $m^{s*}_{0.975}$ & $m^{s}_{0.025}$ & $m^{s}$ & $m^{s}_{0.975}$ & $m^{e}_{0.025}$ & $m^{e}$ & $m^{e}_{0.975}$\\
\hline
Mean & 0 & 0 & 0 & 0 & 0 & 0 & 0 & 0 & 0 \\
Std ($\times 10^{-4}$) & 0 & 4.479 & 10.79 & 1.638 & 2.342 & 3.046  & 1.190 & 1.389 & 1.588\\
KS & 0.103 & 0.217 & 0.332 & 0.117 & 0.175 & 0.234 & -0.013 & 0 & 0.013 \\
Hurst & 0.244 & 0.316 & 0.387 & 0.329 & 0.395 & 0.462 & 0.411 & 0.468 & 0.524 \\
GPH & 0.407 & 0.591 & 0.775 & 0.382 & 0.506 & 0.629 & 0.299 & 0.443 & 0.587\\
ADF & -165.6 & -151.7 & -138.0 & -152.2 & -148.9 & -145.6 & -139.8 & -136.2 & -132.6 \\ 
GARCH & 0.733 & 1.058 & 1.382 & 0.876 & 0.948 & 1.02 & 0.958 & 0.995 & 1.032 \\ 
Hill & 0.383 & 0.894 & 1.406 & 0.807 & 1.227 & 1.647 & 1.722 & 1.993 & 2.263 \\ 
\bottomrule
\end{tabular}
\end{table*}

To better understand how varying the parameters causes changes to the simulated price paths, we first perform a sensitivity analysis on the parameters of the ABM. To detect the changes in the price paths nine moments have been chosen to characterise the behaviour of a single price path: the means and standard deviations of the price paths, the Kolmogorov-Smirnov (KS) statistic \cite{Massey1951}, the Hurst (H) exponent \cite{Hurst1951,MandelbrotWallis1969}, the Geweke and Porter-Hudak (GPH) integration parameter \cite{geweke1983estimation}, the Augmented Dickey-Fuller (ADF) statistic \cite{DickeyFuller1979}, the sum of the parameters of a GARCH(1,1) model \cite{winker2007objective}, and the ``improved" Hill Estimator (HE) \cite{nuyts2010inference}. These moments were chosen because they were found to be robust enough to characterise a price path but flexible enough to discriminate between parameter settings \cite{winker2007objective,arxiv2021abm}. For most of the parameters, the distribution of moments across their values does not change considerably, and there is a large overlap of the distributions, indicating that changing a single parameter does not cause a large change to the overall model price paths. An important exception is that if we increase the number of fundamentalist agents {\it ceteris paribus}, the distribution of the Hurst exponent shifts to smaller values, showing that fundamentalists are engaging in mean-reverting trading strategies. This can also be seen in the correlation matrix between the parameters, the moments, and the objective function, which can be found in Supplementary Fig. S9.

There are only two large correlations: the standard deviation of the price paths with the objective function (0.918), and the Hurst exponent with the number of fundamentalists ( -0.699). Specific parameter values may tend to produce large outliers, {\it e.g.} smaller values of the limit order placement intensity  $\kappa$ increased the number of outliers. From surface plots of the objective functions, we find a decrease from areas of higher values to lower values, giving heuristic evidence that optimisation is feasible. However, the optimisation of the objective function is non-trivial because there is also evidence of local minima and parameter degeneracy. 

The Hurst exponent partially captures the mean reversion observed in the empirical data and is replicated by the model. However, despite an overlap in distributions, the ABM has overestimated the amount of mean reversion. The overlap in the GARCH estimates can be explained by the high variance in the simulated moment's distribution. Even with the high variance, the model was still able to recover the fact that the sum of the GARCH(1,1) parameters should be close to one. This indicates an integrated time series. Of the four moments that did not have enough overlap in confidence intervals, the ADF statistic is the least concerning because both time series exhibit little evidence of a unit root. The KS statistic is the moment with the largest difference in the confidence intervals, which gives evidence for the log-return distributions being different. 

The ``improved" Hill estimator measures the power-law behaviour in the right tail of the log-return distribution. The simulated returns have a lower estimate across the entire confidence interval, giving us that the distribution has fatter tails. This is confirmed by estimating the power law exponent for the probability of finding an increment larger than a given threshold \cite{gopikrishnanetal1998}. There is an asymptotic equivalence between the Hill estimator and such a tail exponent for the probabilities of exceedances. The exceedance probability is visualised in Figure \ref{fig:rl-tail-event-probabilities} for the learning agent configurations. These were estimated in event time, and no binning was used. This confirms the results from our preferred moment, the Hill estimator, that the learning agent introduces more extreme events but that this can be tuned to better conform with the observed real-world data.    

The simulated returns also have a higher variance when compared with the empirical returns distribution. For each moment, the simulated moments have a higher variance. This can result from fewer orders and higher variability in the liquidity in the simulation. It must be noted that these estimates are for a single day of JSE tick-by-tick data, and it is possible for moments to change across days. Overall, the calibration can be considered a partial success as the simulated moments were not considerably different from the empirical data while being reasonably stable. We are thus confident that this is a reasonably realistic market simulator in which to train a learning agent.  

\section*{Results} \label{sec:RLAgentSimulationResults}

\begin{figure*}[h]
    \centering
    \includegraphics[width=17cm]{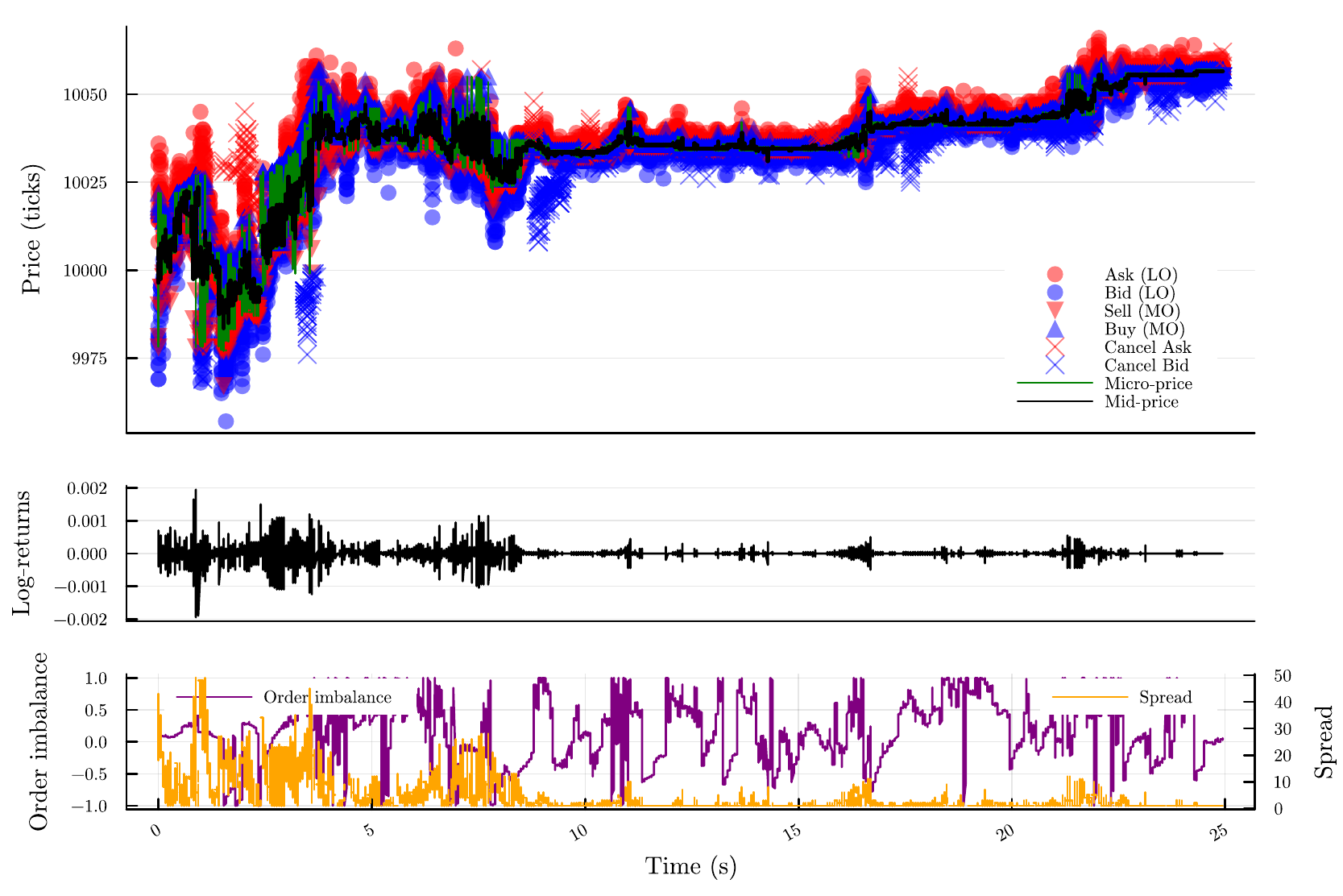}
    \caption{At the top of the figure is a 25 second (in machine time) simulated price path for the combined model configuration using a 6\% ADV learning agent with a small state space (see Table \ref{rl-training-parameters}). The environment is simulated using the calibrated parameters from Table \ref{calibrated-params-table}. This is equivalent to a single trading day, {\it i.e.} 8 hours of typical trading in calendar time. The different markers and colours correspond to the different types of orders and the different sides of the LOB, respectively. The fluctuations in the log-micro-price returns, as well as the spread and order imbalance, are plotted in the middle and the bottom, respectively. These results visually demonstrate how the ABM and learning agent have produced well-known stylised facts. This is captured in more detail in Table \ref{estimated-simulated-moments-table} with further details in Supplementary Figures S1-8.}
    \label{sim-price-path}
\end{figure*}

Figure \ref{sim-price-path} plots a simulated price path of the combined model using the calibrated ABM model parameters and the small state-space learning agent liquidating a medium-sized parent order that is 6\% of the Average Daily Volume (ADV). Table \ref{rl-statespace-variables} gives the remaining learning agent configurations. The fluctuations in the log-returns, as well as the spread and order imbalance, are shown. This is a single price path that is subject to variations due to changes in the fundamental prices, the forgetting factors, the randomness of the samplings, as well as the interactions with the learning agent.

\subsection*{Learning agents}

\begin{figure*}[h!]
    \centering
    \includegraphics[width=18cm, height=12cm]{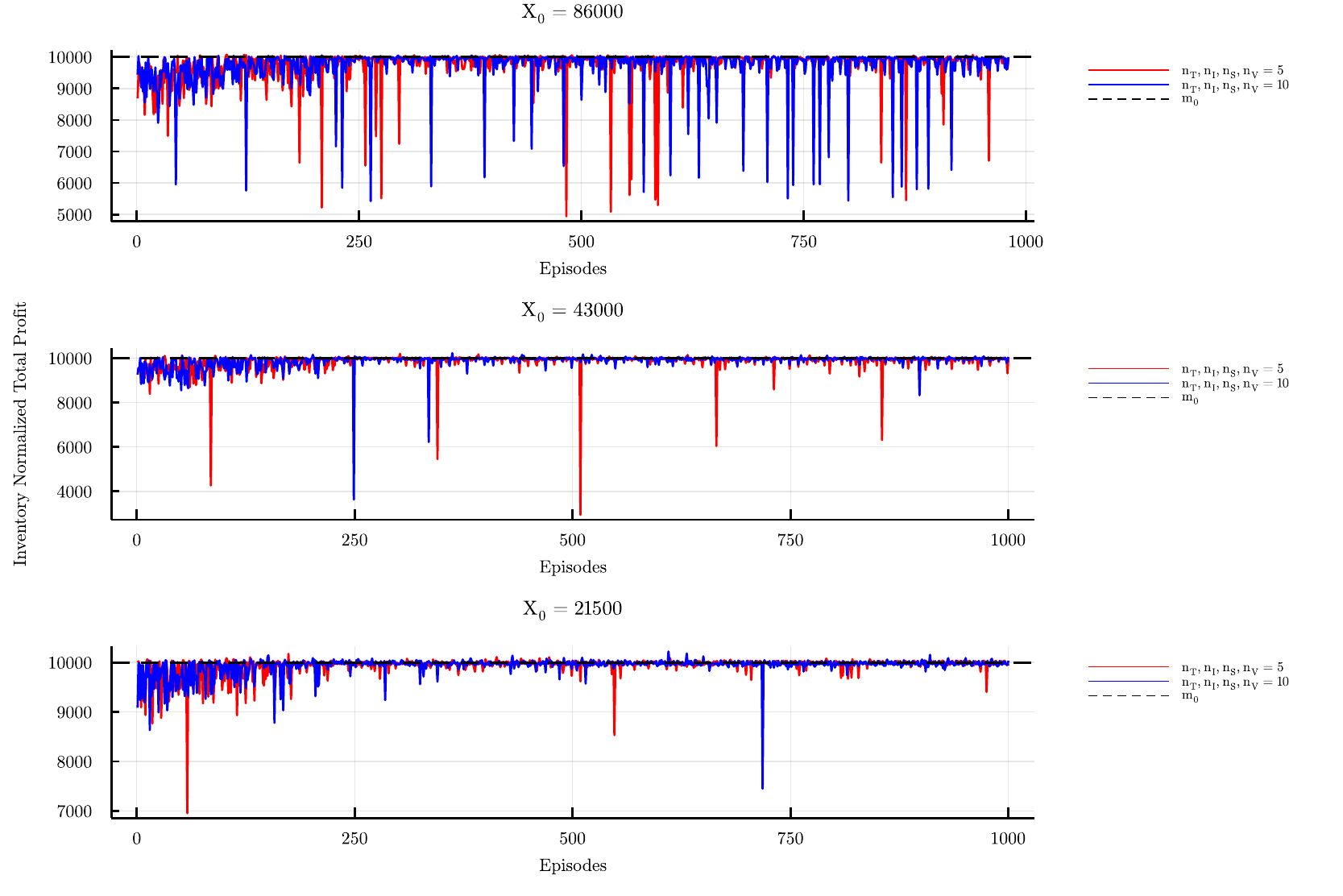}
    \caption{Plots the convergence in the profit of the RL agents. The amount of inventory the agent needed to trade increased from the bottom plot to the top plot. In each plot, we compare the profit convergence of the agents with smaller state space sizes to agents with larger state space sizes. $m_{0}$ is the initial mid-price, and if all agents were able to trade all shares at $m_{0}$ then the Inventory Normalised Total Profit (INTP) would be 10000, and the implementation shortfall would be 0. For the agents with the size of the parent order being 86000, the INTP was winsorised at the 1\%  and 99\% levels to allow for easier comparisons. All agents demonstrate improvement and convergence in the profit. As the inventory increases, we start to see liquidity effects, as there are more frequent liquidity crashes. The size of the state space does not seem to cause significant differences in the profit, and all the agents were unable to significantly improve the profit received over the implementation shortfall amount.}
    \label{rl-reward-convergnce}
\end{figure*}

The learning agent explicitly changes the environment, necessitating a heuristic approach to demonstrate policy convergence. We visually show that we have reasonable policy convergence in the presence of Q-matrix difference stability and, furthermore, that the actions learnt are explicable, scaling as expected with the size of the parent order over multiple training episodes. This is visually shown in the supporting materials and reported in this paper. Figure \ref{rl-reward-convergnce} demonstrates that all the agents showed an improvement and convergence in profit over the episodes. From episodes 0 to approximately 250, there is an increase in the average profit and a decrease in the variance of the profit. After this period, the agents receive roughly the same profit when we ignore liquidity shocks of the agents with larger initial parent orders: $X_{0}$. 

As $X_{0}$ increases, the frequency of large price decreases and liquidity crashes increase -- this is shown in Figure \ref{fig:rl-tail-event-probabilities}. Larger orders have a larger price impact and push the price further down. The impact of larger initial parent orders on the overall price impact is shown in Figure \ref{rl-price-impact}. Increasing the inventory traded will increase the chance of a large order being submitted, which can remove all the liquidity on the contra side of the order book before liquidity providers can replenish it, causing liquidity crashes. 

\begin{figure*}
   \centering
    \begin{subfigure}[b]{0.5\textwidth}
    \centering
    \includegraphics[width=\textwidth]{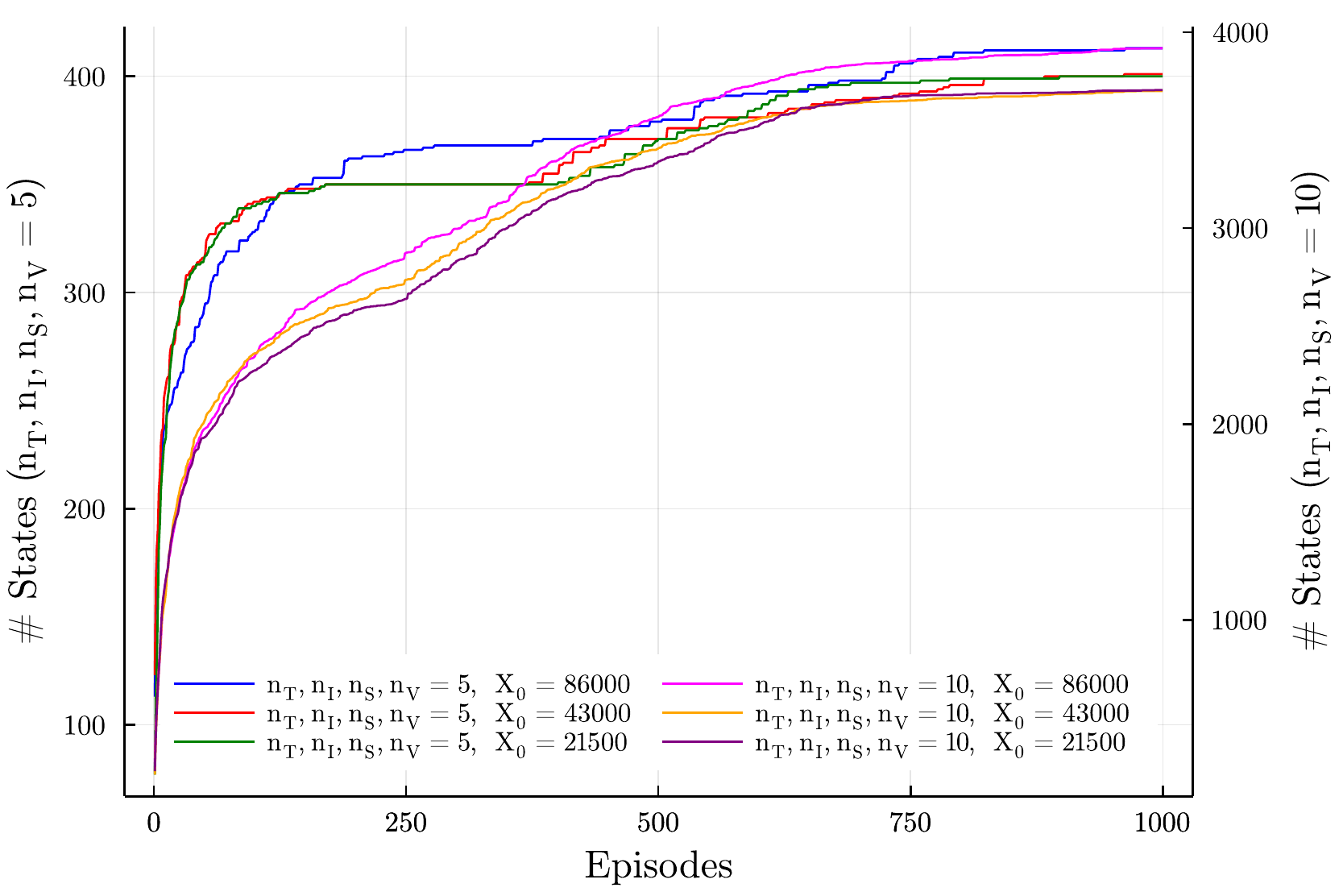}
    \caption{Total number of Q-matrix states found}%
    \label{rl-number-states-convergence}
    \end{subfigure}%
    \centering
    \begin{subfigure}[b]{0.5\textwidth}
    \centering
    \includegraphics[width=\textwidth]{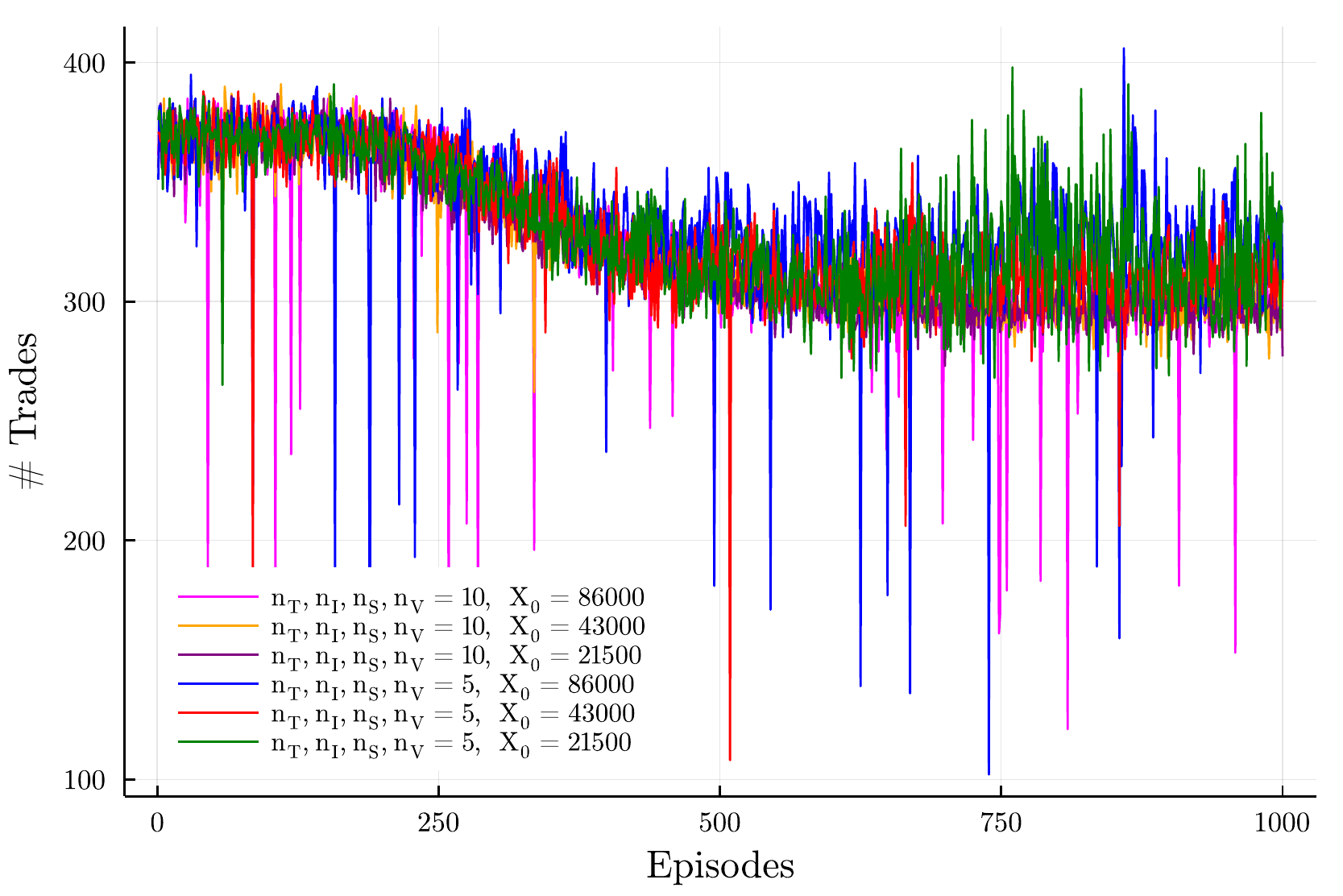}
    \caption{The number of agent trade events}%
    \label{rl-number-trades-convergence}
    \end{subfigure}%
       \caption{The total number of states found in the Q-matrix of each RL agent in each episode is shown on the left (Figure \ref{rl-number-states-convergence}). The smaller state space agents converge faster than the larger state space agents and have more jumps, whereas the larger state space agents exhibit a smoother convergence. The number of trade events an agent submits in each episode is plotted in Figure \ref{rl-number-trades-convergence} and shows the decrease in the number of trades over the episodes and the convergence of this decrease.}
    \label{rl-number-states-trades-convergence}
\end{figure*}

There is no significant difference between the profit obtained by the smaller state space agents and the larger state space agents when they are trading the same amount of initial inventory. The difference between the number of states visited by the smaller and the larger state space agents is shown in Figure \ref{rl-number-states-trades-convergence}. Agents cannot significantly improve the profit received compared to if they could trade all inventory at the initial mid-price. The agents have learned to only reduce the implementation shortfall. Unsurprisingly, the smaller state space converges faster than the larger state space. The smaller state space also exhibits more jumps, likely due to the clustering of states. In particular, the more coarse-grained space would cause the agent to sample from one cluster and then jump to another, while the larger state space would have fewer distinct clusters, and the jumps between states would be less significant, giving a smoother convergence. Having far more states means that there will be fewer returns to individual states over the same number of episodes, which means the agents with the larger state space will likely have greater difficulty in learning a good policy over the same number of episodes. 

The order volume trajectories learnt by the different agent configurations in the training environment are shown in Figure \ref{fig:rl-trading-schedules}. The six different configurations are from Table \ref{rl-statespace-variables}. All the agents learned to execute more volume early in the trading session. There is more erratic trading for larger parent orders $X_0$ as the trading algorithm responds to larger volatility in the market states that actions respond to. The trading agent liquidating the larger parent order generates relatively more extreme events, as shown in Figure \ref{fig:rl-tail-event-probabilities}.  

\subsection*{Model stylised facts}

It is difficult to assess the ``goodness" of one ABM relative to another as most assessments are qualitative comparisons of stylised facts, and often, a variety of quite different ABMs can produce similar stylised facts. However, demonstrating the extent to which a calibrated model can reproduce the stylised facts of a market, and then that this is with relative stability, is a good starting point in validating an ABM, and it is a necessary step if one is to demonstrate the impact of model changes. We use the calibrated model and the associated moments estimated from the simulated samples to compare the different model configurations as the moments are changed in the presence of the learning agent.

\begin{figure*}
    \centering
    \begin{subfigure}[b]{0.6\textwidth}
    \centering
    \includegraphics[width=\textwidth]{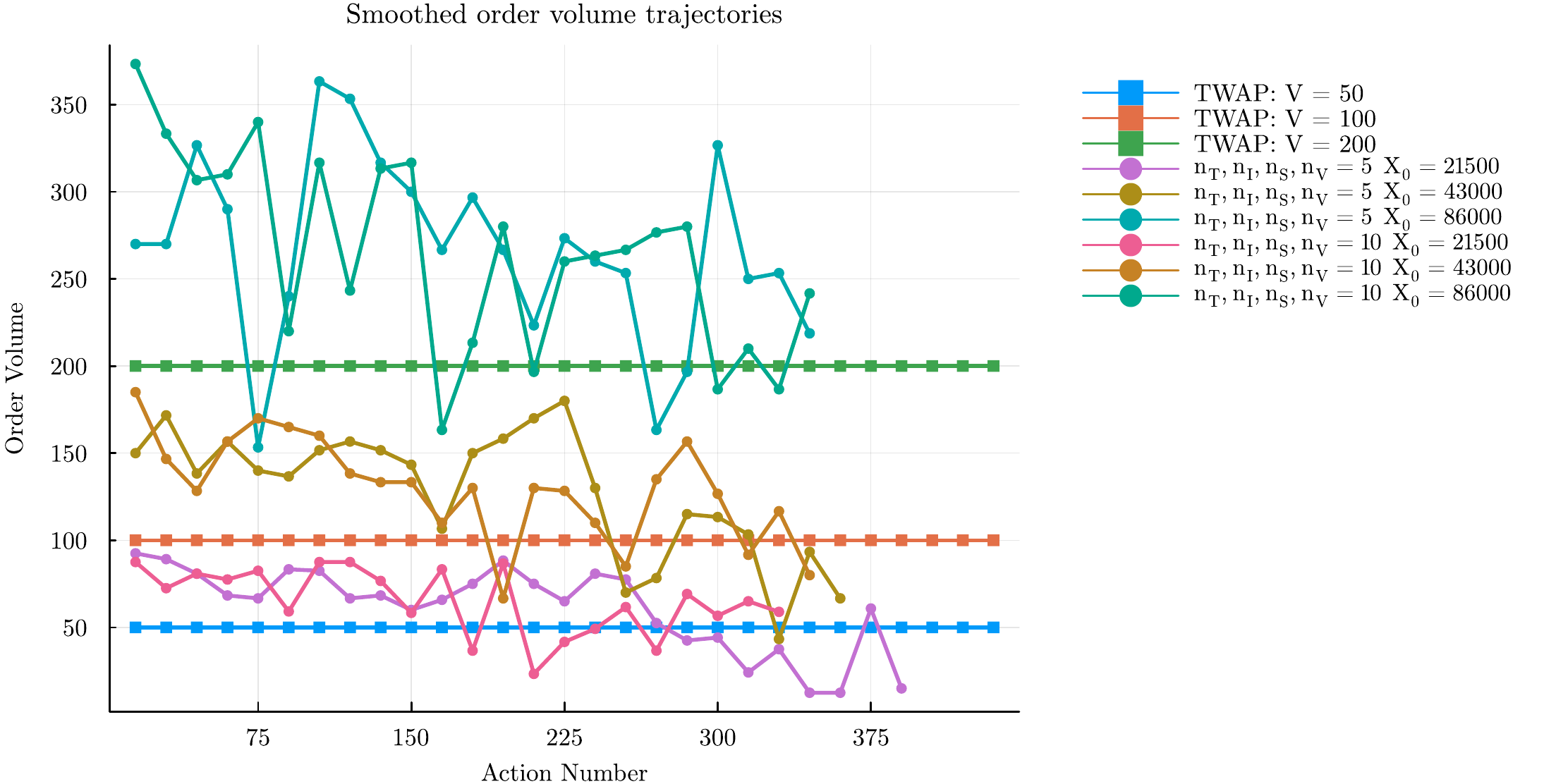}
    \caption{Learning agent order volume trajectories}%
    \label{fig:rl-trading-schedules}
    \end{subfigure}%
    \centering
    \begin{subfigure}[b]{0.4\textwidth}
    \centering
    \includegraphics[width=\textwidth]{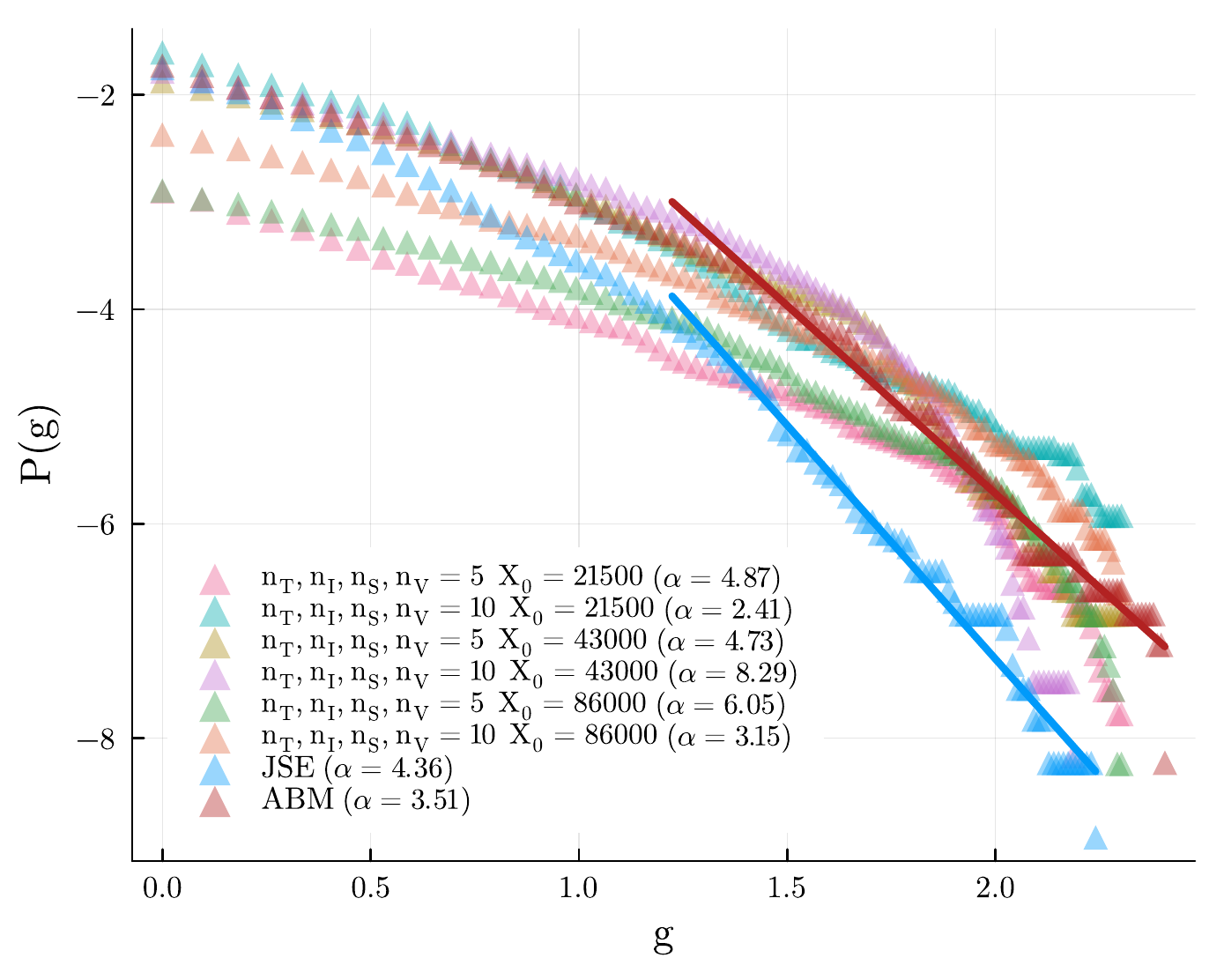}
    \caption{The probability of exceedances}%
    \label{fig:rl-tail-event-probabilities}
    \end{subfigure}%
    \caption{The order volume trading trajectories for the six different learning agent configurations are in Figure \ref{fig:rl-trading-schedules} (left). $X_0$ is the initial parent order. The TWAP for each of the three different initial order volumes is shown. V is the lot size. The trading agents have learnt to trade more early in the session and less later. The larger state spaces (10) are more erratic than the smaller ones (5). Agents with larger initial parent orders have more erratic trajectories. The associated changes in the exceedance probabilities are in Figure \ref{fig:rl-tail-event-probabilities} (right), where $\alpha$ is the slope in the tail \cite{gopikrishnanetal1998}. The exceedance probabilities for the ABM environment (red) are shown with a best-fit line in the linear region; similar plots are shown for the JSE (blue). The ABM has a higher number of extreme events due to liquidity shocks.}
    \label{rl-trading-schedules-events}
\end{figure*}

The first stylised fact that will be considered is the fat-tails of the micro-price log-returns in event time. The ABM and the empirical returns were found to have fatter tails than that of the normal distribution, and the ABM was able to reproduce the stylised fact. Table \ref{estimated-simulated-moments-table} shows that the variance for the ABM is higher than what is observed in the market. The tails of the return distributions exhibited power-law or Pareto-like distributions with a tail-index that is finite \cite{mandelbrot1963variation, gopikrishnanetal1998}. The ABM also has return distributions with near power-law behaviour in their tails. We found that the tail-index estimate for the simulated returns was less than the tail-index of the empirical returns, meaning there are fatter tails in the simulated return distribution; this is visualised across learning agent configurations in Figure \ref{fig:rl-tail-event-probabilities} and was also observed in the moments in Table \ref{estimated-simulated-moments-table}. Given that the distributions do not exactly follow a power-law, no conclusion about the existence of moments can be drawn from the tail-index parameter.

At high-frequency times scales, we have significant negative autocorrelations for the first lags, and the autocorrelations for the remaining lags are insignificant. This has been cited as evidence of a mean-reverting component of the time series that could potentially be attributed to market making \cite{cont2001empirical}. We found significant negative autocorrelations for the first few lags where the autocorrelation in the simulated returns is far higher than the empirical returns, which may result from the frequency of limit orders placed by the high-frequency agents in the ABM. This is consistent with the estimated moments as the Hurst exponent in Table \ref{estimated-simulated-moments-table} suggests more mean-reversion in the simulated returns. Volatility clustering is the feature where large (resp. small) price variations are followed by further large (resp. small) price variations. This feature was calibrated using the GPH statistic, which measures the long-range dependence in the absolute log-returns. However, we found that the decay in the autocorrelations is far quicker for the simulated returns. Trade-signs have been shown to have persistence in their autocorrelations \cite{lillo2004long,bouchaud2003fluctuations,toth2015equity}, meaning that buy (resp. sell) orders are followed by more buy (resp. sell) orders. To estimate the autocorrelation of the trade-signs each trade needs to be classified as a buy (+1) or a sell (-1). We use the Lee and Ready classification rules \cite{lee1991inferring}. We then find that the model's autocorrelation function (ACF) does not exhibit the same behaviour found in the empirical data. 
The ABM has captured the decay of the ACF for increasing lags. However, this decay is slower than the decay of the ACF for the empirical data. The model does reproduce the decay of the order-flow autocorrelation for increasing lags, but the volatility clustering effect is too strong, and it does not correctly describe the power-law decay seen in real markets. This only partially matches what is observed in the empirical data used to calibrate the ABM. This is not unexpected and was by design as made clear in the model discussion. 

The impact of the different approach to time can be seen by comparing the stylised facts generated by the event-based reactive model, where there is no globally imposed model time, with those of the hybrid model, where there is a global time; these are otherwise the same models. Both models were able to reproduce the fat-tailed return distributions. However, one limitation of the hybrid model that has been resolved in our model is that the simulated returns have data more evenly distributed from the mean and less concentrated near zero. This also means that there is a smaller discrepancy between the tail-index estimates for the extreme-value distributions of the log-returns. The event-based ABM also captures that only the first few lags should be significantly negative, whereas significant lags (both positive and negative) up to lags of fifty were reported \cite{arxiv2021abm} in the hybrid model. The hybrid model's decay in the absolute log-returns is faster than what is observed in our model. This means there is a weaker volatility clustering effect in the hybrid model. 

However, the hybrid model can more accurately depict the decay in the trade-sign autocorrelation. In both these models, the order-flow autocorrelation is due to the herding effect caused by the chartists. We have changed the chartist rule to better suit event-driven activations without knowledge of a global calendar time. In contrast, in the hybrid model, the agent specification rule may better capture this feature because it is defined in the stochastic time setting. The selection of a global time is a potential source of top-down causation in the hybrid model that is not present in the reactive model. The chartists are more likely to be trading in events that are very close together, while in the hybrid model, stochastic time provides an external feature that spreads events away from event clusters. This can increase the herding effect seen in the event-based model. It must be noted that, for the hybrid mode, the trade-sign autocorrelation was not tested for the long-memory property. This means that even if the hybrid models better capture the decay, it is unclear if it indicates a long-memory process. Both models could accurately recover the depth profiles and price impact curves.

\section*{Conclusions}  

A reactive event-based ABM representing a trading environment was built using two classes of agents: {\it liquidity takers} who only submit market orders, and {\it liquidity suppliers} who only submit limit orders. The liquidity takers were split in turn into two sub-classes: {\it fundamentalists} and {\it chartists}. The fundamentalist started each trading session, such as a trading day, with different notions of fair or fundamental value, and traded accordingly. At the same time, the chartists considered a moving average of past price changes to trigger trading decisions. The ABM was implemented in event time using a reactive framework, so we did not need to specify at what calendar times agent activations could occur. Agents responded to events, not times. This simulated model for the environment recovers an asymmetry between the bid and ask depth profile curves, and both curves exhibit a rapid decrease in liquidity as the price moves away from the best bid/ask, remaining consistent with empirical observation \cite{bouchaud2002statistical,gould2013limit,potters2003more}. The maximum volume at the best bid/ask is also consistent with previous research \cite{potters2003more}. The simulated trading environment was calibrated to real-world data. 

The real market observations were based on trade and order-book data extracted from the Johannesburg Stock Exchange (JSE) \cite{Jericevich2020data,arxiv2021abm} for a day of trading. The ABM representing the learning environment was calibrated using a simulated method of moments and calibrated parameters are provided in Table \ref{calibrated-params-table}. This method of calibration was chosen to balance computational speed with stability. We found that the most significant parameters are the variance in the fundamental value $\sigma_{f}$, the intensity of the HFT order placement $\kappa$ and the number of chartist liquidity takers $N^{c}_{_{\mathrm{LT}}}$. Of lesser importance is the shape of the order-imbalance $\nu$, the number of fundamental liquidity takers $N^{f}_{_{\mathrm{LT}}}$ and the mid-price cut-off that sets the minimum volume lot sizes $\delta$. The model can recover log-returns that exhibit fat-tails with extreme value distributions similar to those observed in the market, with volatility clustering, and replicating the log-returns' ACF. The ABM also captured the decay in the auto-correlation of the trade-signs, but with a slower decay rate when compared with the JSE. The JSE exhibited the expected long-memory property \cite{lillo2004long}. However, this was not fully replicated by the ABM. The decay of the auto-correlation in the absolute values of the log-returns was slower in the real market than in the ABM. 

After the calibration of the training environment, a single learning agent with bounded rationality was added to the market to interact with this ABM. The learning agent was represented by a simple tabular Q-learning RL algorithm that was learning to solve an optimal execution selling problem using only market orders. This RL agent learns how to modify an existing TWAP trading schedule to optimally execute its sell order. The algorithm considered two market states based on features representing the limit-order book: the spread and the best bid volume, and then two states representing the current execution agents states: the remaining time available in the trading session, and the agents remaining inventory. The agent was tested with different parent order sizes and different state space sizes to observe whether any agents' profits could be found to increase and converge and improve a random initial agent configuration. Our work shows that it can. 

\begin{figure*}[h]
        \centering
        \begin{subfigure}[b]{0.5\textwidth}
          \centering
         \includegraphics[width=9cm, height=6cm]{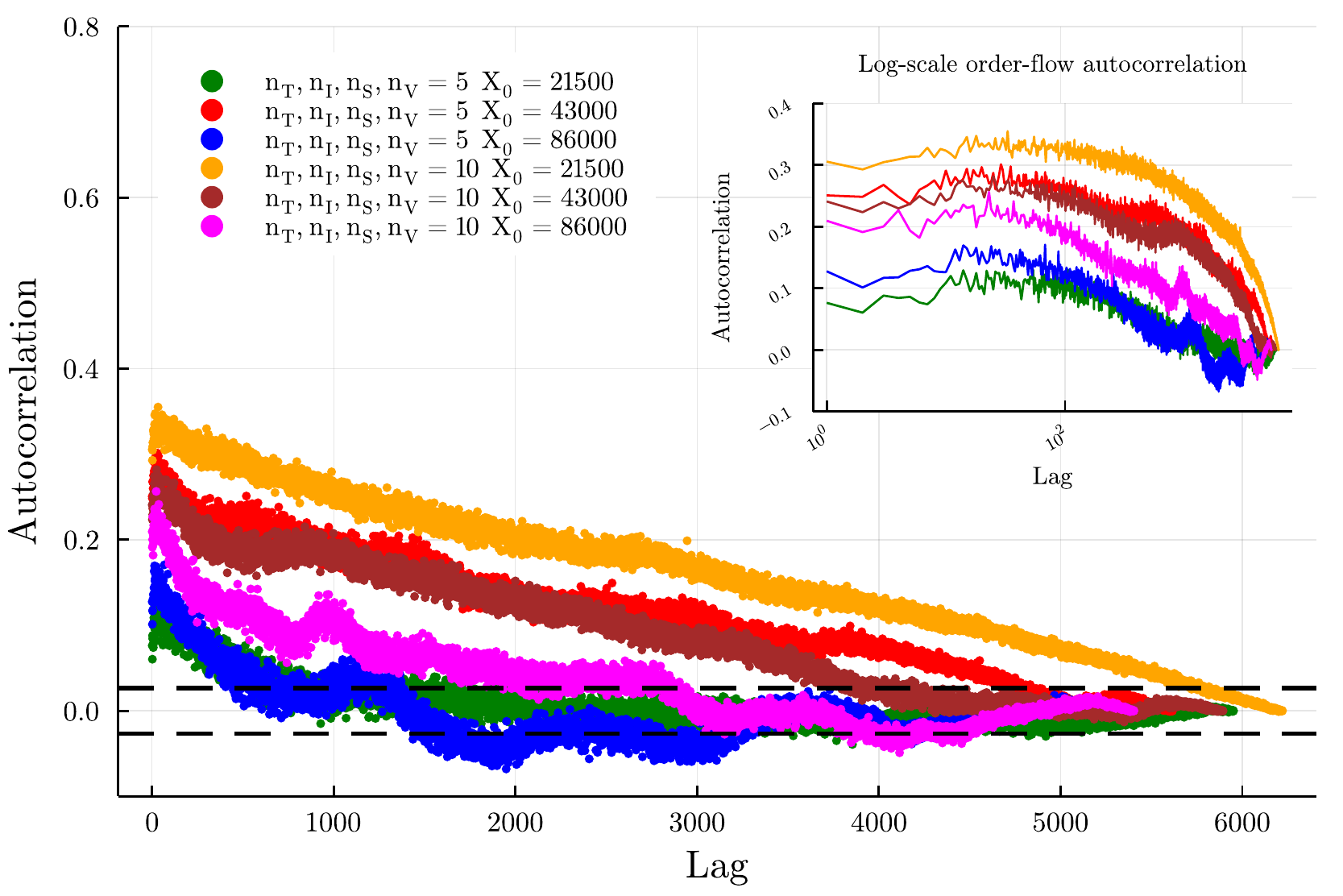}
         \caption{Trade-sign autocorrelations}%
         \label{rl-trade-sign-autocorrelation-(a)}
           \end{subfigure}%
           \begin{subfigure}[b]{0.5\textwidth}
          \centering
            \includegraphics[width=9cm]{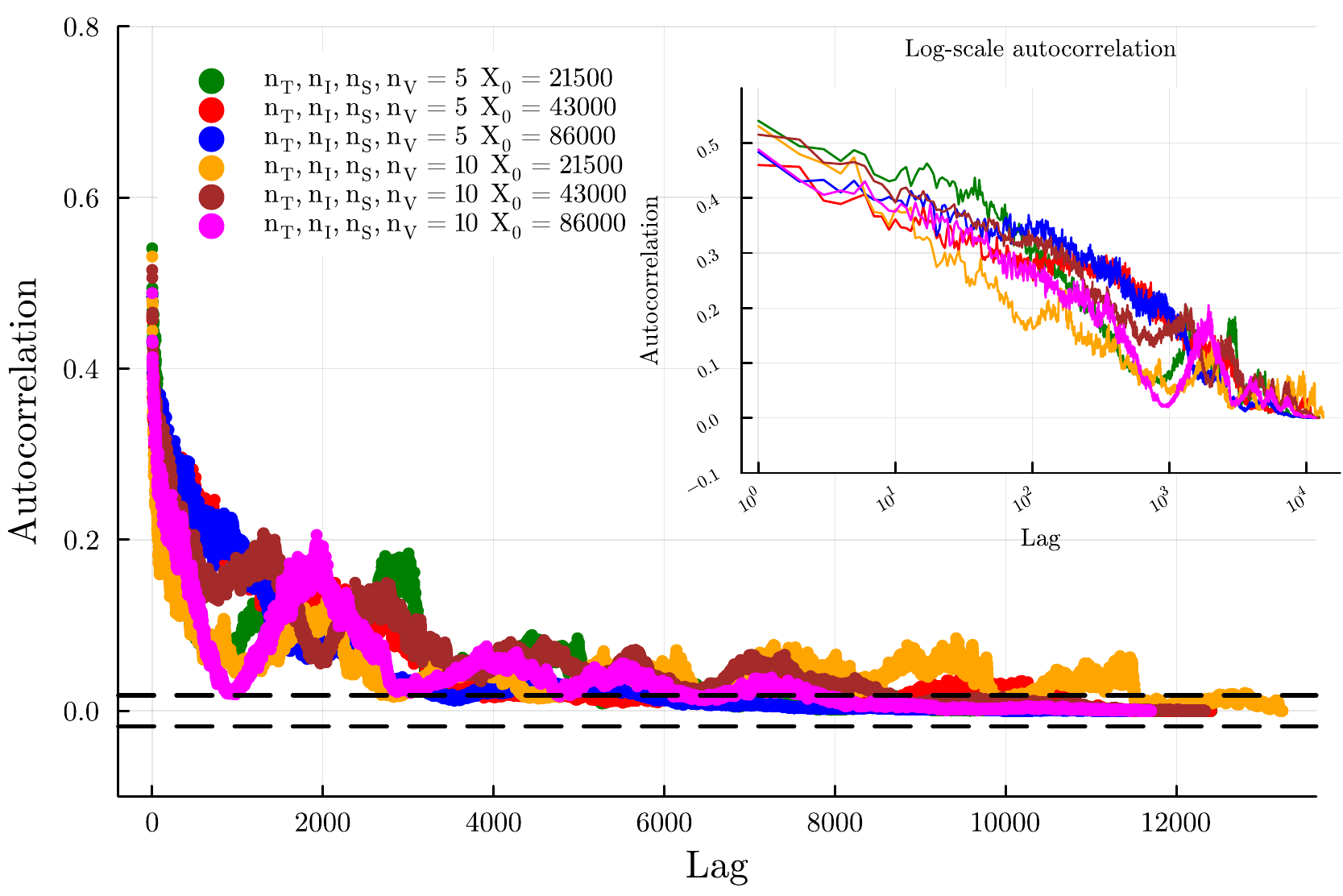}
              \caption{Absolute return autocorrelations}%
          \label{rl-abslog-return-autocorrelation-(c)}
        \end{subfigure}%
        \caption{The trade-sign autocorrelation plots for the ABM with the different RL agents are shown in Figure \ref{rl-trade-sign-autocorrelation-(a)} and the autocorrelations in the absolute log-returns are in Figure \ref{rl-abslog-return-autocorrelation-(c)}. The state space sizes and initial inventories are given in the legend. Adding a learning agent using order-splitting does not guarantee the correct decay in the trade-sign autocorrelation. Agents plotted in green, blue and pink in Figure \ref{rl-trade-sign-autocorrelation-(a)} show that these agents can reduce the trade-sign autocorrelations. These same agents can boost the autocorrelation in the absolute returns as seen for the agents plotted in green, blue and pink in Figure \ref{rl-abslog-return-autocorrelation-(c)}.}
        \label{rl-trade-sign-autocorrelation}
\end{figure*}

\begin{figure*}[ht!]
        \centering
        \begin{subfigure}[b]{.5\textwidth}
          \centering
            \includegraphics[width=8cm, height=6cm]{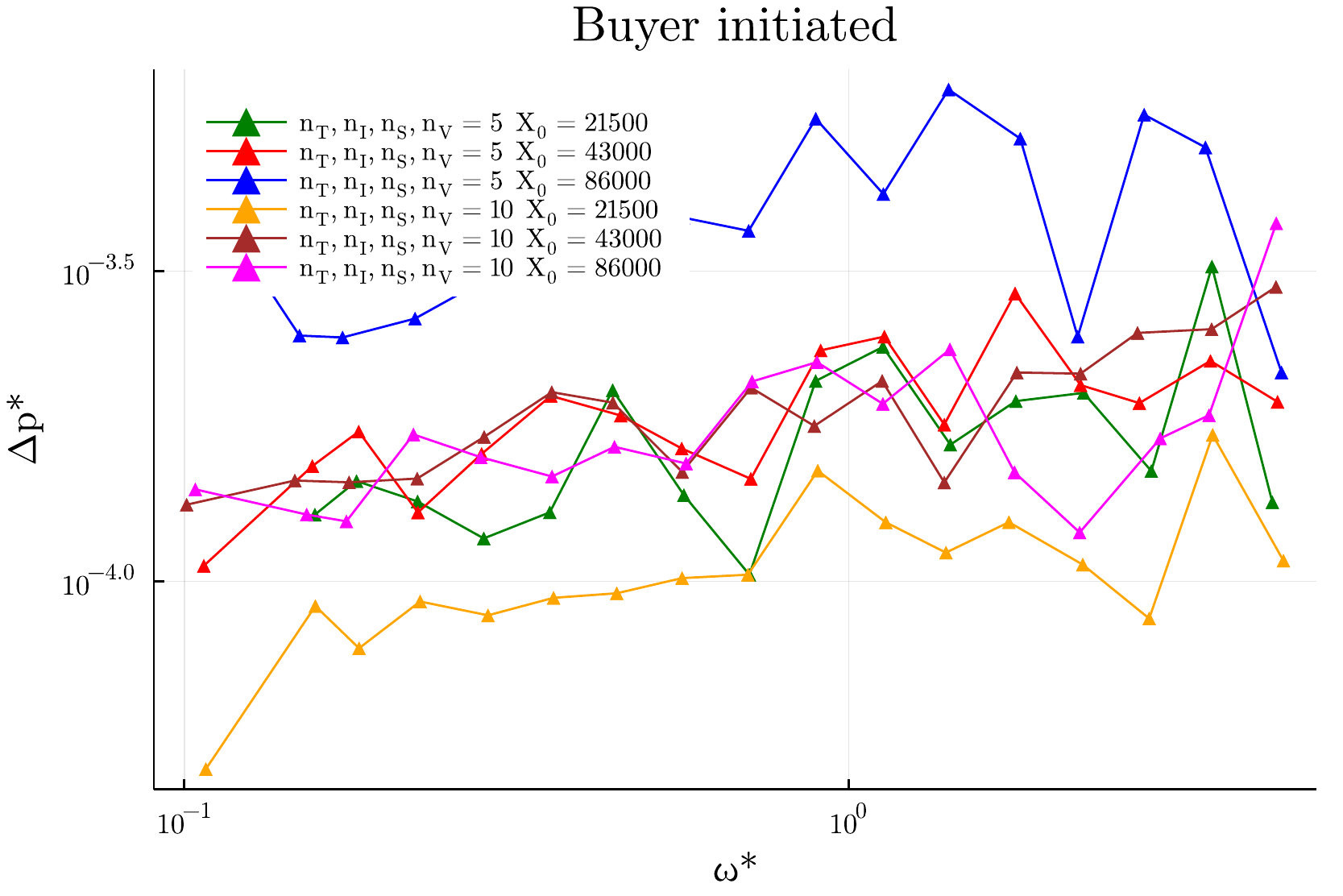}
            \caption{Buyer initiated price impact}
        \label{rl-price-impact-buyer-(a)}
        \end{subfigure}%
        \begin{subfigure}[b]{.5\textwidth}
          \centering
              \includegraphics[width=8cm, height=6cm]{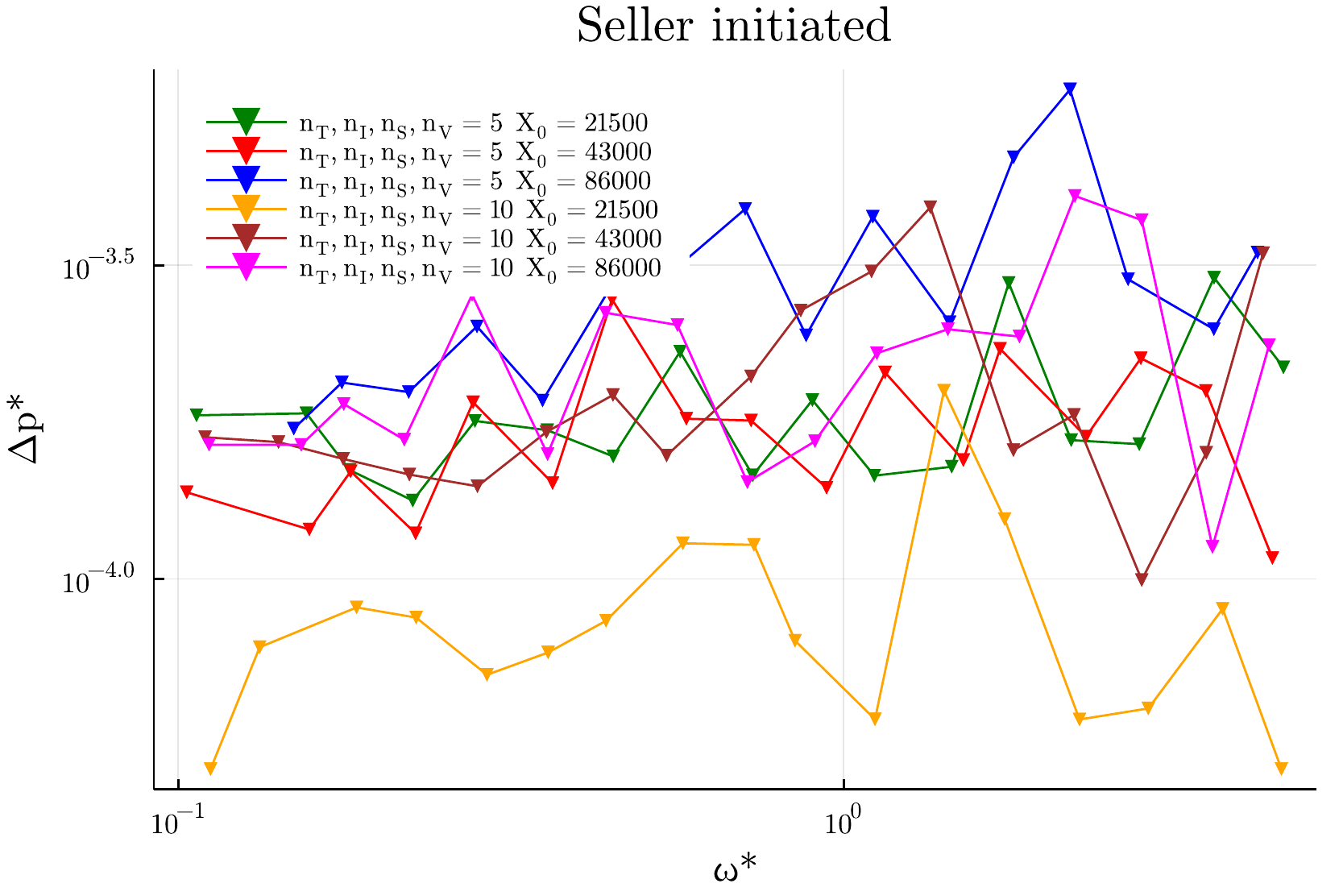}
              \caption{Seller initiated price impact}
              \label{rl-price-impact-seller-(b)}
        \end{subfigure}
        \caption{Plots the price impact curves for the ABMs that include the RL agents, for buyer initiated (Figure \ref{rl-price-impact-buyer-(a)}) and seller initiated (Figure \ref{rl-price-impact-seller-(b)}) trades. The linear upward trend in the price impact on the log-log scale indicates a power-law price impact function, where the price impact increases with the trade volume.  Adding in the RL agents does not change the slope of the impact function. The observed price impact increases as the inventory traded increases and these price impact curves remain consistent with previous research \cite{gebbie2017deviations,gebbiechangjericevich2020comparing,LFM2003,arxiv2021abm}.}
        \label{rl-price-impact}
\end{figure*}

The impact of the different learning agent configurations on the environment is visually demonstrated in Figure \ref{rl-trade-sign-autocorrelation} and Figure \ref{rl-trading-schedules-events}. In Figure \ref{rl-trade-sign-autocorrelation}, we see that the learning agents can suppress the trade-sign autocorrelations but boost the autocorrelations in the absolute micro-price value. This depends on the state space used in the learning algorithm, but learning agents with larger parent orders suppress the trade-sign autocorrelations and introduce more price volatility. Figure \ref{rl-price-impact} shows that the larger parent orders induce a larger overall price impact even when the learning agent only engages in selling. The learnt volume trajectories for the different agents are shown in Figure \ref{fig:rl-trading-schedules} and their associated exceedance probabilities in Figure \ref{fig:rl-tail-event-probabilities}.

We compared the ABM representing the environment using its calibrated moments against the combination of the same ABM with a single strategic learning agent. We found that, except for the Hurst exponent, the moments were robust to adding a single simple learning agent, even with increasing initial order sizes. The combined market stylised facts conform to those observed in the measured market data. In particular, the combined model is found to better replicate the trade-sign and return absolute value autocorrelations, and these are shown in Figure \ref{rl-trade-sign-autocorrelation}. The model can replicate price impact functions with the correct shape, and an intuitive relationship with initial order sizes, as evidenced in Figure \ref{rl-price-impact}. When we include learning agents in the market environment, we can replicate the observed stylised facts better.

It was found that all agents could increase the profits over their initial random agent, demonstrating that each agent has learned. This is shown in Figure \ref{rl-reward-convergnce}. However, an upper bound on the profits was observed. No agent could learn how to improve upon the profit of immediate execution (assuming that the execution volume was available). This shows that although the agents could reduce the implementation shortfall, they could not increase profits enough to observe a consistent and significant negative implementation shortfall. This is considered to be the result of the very simple reward function used. It was found that adding RL agents with increasing order sizes caused the ABM to have significant liquidity effects, as expected. 

There were clear differences in the dynamics for agents with different state space sizes. The smaller state space agents had the number of states it visited converge much faster than the larger ones. They quickly finished the state discovery process while the larger state space agents were still rapidly discovering new states, limiting their learning ability. The larger state space agents showed less switching of actions in each state when compared with the small state space agents due to the agents not returning to individual states as often to update their state-action values. 

This showed that the larger state space agents had difficulty learning in such few episodes. This was also evident in the behaviour learned. If a large state space agent had a large amount of inventory at the beginning of the simulation versus a smaller amount at the end, the agent did not learn to change behaviour while the smaller state space agent learned to trade more initially and then reduce the rate of trading at the end. There was also evidence that the smaller state space agent could learn to trade intuitively using the spread and volume states. Figure \ref{rl-number-trades-convergence} shows that all agents learned to reduce the number of trades and guarantee the execution of the entire inventory. They also learned some structure during the training process, as evidenced by observed changes in the action volume trajectories over the training period.

This work promotes the idea there is value in combining minimally intelligent ABMs of the environment with the training of learning agents.  We demonstrate that learning is feasible with trading agents comparable to those seen in real markets and that the combined model can generate relatively realistic market dynamics. From a market modelling perspective, we show that with strategic order-splitting, when combined with learning, we can more easily recover the observed auto-correlations of trade-signs and the auto-correlations of the absolute of the micro-prices while retaining realistic volume profiles, price impact profiles, and exceedance probabilities. Learning agents engaging in strategic order-splitting can suppress trade-sign auto-correlations, boost volatility, and increase mean-reversion effects. 

\section*{Code and Data Availability}

The JSE data used in this study is available at the \href{https://zivahub.uct.ac.za/}{ZivaHub} repository, \cite{Jericevich2020data} \href{https://doi.org/10.25375/uct.13187591.v1}{10.25375/uct.13187591.v1}. The code used to implement the model management system is available 
at the \href{https://zivahub.uct.ac.za/}{ZivaHub} repository, \cite{Dicks2022Julia} \href{https://doi.org/10.25375/uct.21163723.v1}{10.25375/uct.21163723.v1}, and the matching engine with the UDP socket is available at the \href{https://zivahub.uct.ac.za/}{ZivaHub} repository, \cite{Jericevich2022CoinTossXUDP} \href{https://doi.org/10.25375/uct.21163993.v1}{10.25375/uct.21163993.v1}  

\section*{Reproducibility of the Research}
The Julia code can be found in our GitHub site  \cite{Dicks2022Julia}.

\section*{Acknowledgements}

We thank Ivan Jericevich for support and advice with respect to the matching engine and hybrid agent-based model implementation. We thank the journal reviewers for their thoughtful feedback and critique.

\section*{Author contributions statement}

T.G. and M.D. conceived the experiments and models,  M.D. implemented and conducted the experiments, T.G., A.P.  and M.D. analysed the results and reformulated the experiments as required. All authors reviewed the manuscript. 


\balance
\bibliographystyle{elsarticle-harv}
\bibliography{MDAPTG-PhysA-003.bib}

\begin{thebibliography}{57}
\expandafter\ifx\csname natexlab\endcsname\relax\def\natexlab#1{#1}\fi
\providecommand{\url}[1]{\texttt{#1}}
\providecommand{\href}[2]{#2}
\providecommand{\path}[1]{#1}
\providecommand{\DOIprefix}{doi:}
\providecommand{\ArXivprefix}{arXiv:}
\providecommand{\URLprefix}{URL: }
\providecommand{\Pubmedprefix}{pmid:}
\providecommand{\doi}[1]{\href{http://dx.doi.org/#1}{\path{#1}}}
\providecommand{\Pubmed}[1]{\href{pmid:#1}{\path{#1}}}
\providecommand{\bibinfo}[2]{#2}
\ifx\xfnm\relax \def\xfnm[#1]{\unskip,\space#1}\fi
\bibitem[{Almgren and Chriss(2001)}]{almgren2001optimal}
\bibinfo{author}{Almgren, R.}, \bibinfo{author}{Chriss, N.},
  \bibinfo{year}{2001}.
\newblock \bibinfo{title}{Optimal execution of portfolio transactions}.
\newblock \bibinfo{journal}{Journal of Risk} \bibinfo{volume}{3},
  \bibinfo{pages}{5--40}.
\newblock \DOIprefix\doi{10.21314/JOR.2001.041}.
\bibitem[{Aloud et~al.(2017)Aloud, Fasli, Tsang, Dupuis and
  Olsen}]{aloud2017modeling}
\bibinfo{author}{Aloud, M.}, \bibinfo{author}{Fasli, M.},
  \bibinfo{author}{Tsang, E.}, \bibinfo{author}{Dupuis, A.},
  \bibinfo{author}{Olsen, R.}, \bibinfo{year}{2017}.
\newblock \bibinfo{title}{Modeling the high-frequency fx market: An agent-based
  approach}.
\newblock \bibinfo{journal}{Computational Intelligence} \bibinfo{volume}{33},
  \bibinfo{pages}{771--825}.
\newblock \DOIprefix\doi{10.1111/coin.12114}.
\bibitem[{Auletta et~al.(2008)Auletta, Ellis and
  Jaeger}]{AulettaEllisJaeger2008}
\bibinfo{author}{Auletta, G.}, \bibinfo{author}{Ellis, G.F.R.},
  \bibinfo{author}{Jaeger, L.}, \bibinfo{year}{2008}.
\newblock \bibinfo{title}{Top-down causation by information control: from a
  philosophical problem to a scientific research program}.
\newblock \bibinfo{journal}{J. R. Soc. Interface} ,
  \bibinfo{pages}{1159–1172}\URLprefix
  \url{https://royalsocietypublishing.org/doi/10.1098/rsif.2008.0018},
  \DOIprefix\doi{https://doi.org/10.1098/rsif.2008.0018}.
\bibitem[{Barto and Mahadevan(2003)}]{barto2003recent}
\bibinfo{author}{Barto, A.G.}, \bibinfo{author}{Mahadevan, S.},
  \bibinfo{year}{2003}.
\newblock \bibinfo{title}{Recent advances in hierarchical reinforcement
  learning}.
\newblock \bibinfo{journal}{Discrete event dynamic systems}
  \bibinfo{volume}{13}, \bibinfo{pages}{41--77}.
\newblock \DOIprefix\doi{10.1023/A:1022140919877}.
\bibitem[{Bellman(1954)}]{bellman1954theory}
\bibinfo{author}{Bellman, R.}, \bibinfo{year}{1954}.
\newblock \bibinfo{title}{The theory of dynamic programming}.
\newblock \bibinfo{journal}{Bulletin of the American Mathematical Society}
  \bibinfo{volume}{60}, \bibinfo{pages}{503--515}.
\newblock \DOIprefix\doi{10.1090/S0002-9904-1954-09848-8}.
\bibitem[{Bertsimas and Lo(1998)}]{bertsimas1998optimal}
\bibinfo{author}{Bertsimas, D.}, \bibinfo{author}{Lo, A.W.},
  \bibinfo{year}{1998}.
\newblock \bibinfo{title}{Optimal control of execution costs}.
\newblock \bibinfo{journal}{Journal of Financial Markets} \bibinfo{volume}{1},
  \bibinfo{pages}{1--50}.
\newblock \DOIprefix\doi{10.1016/S1386-4181(97)00012-8}.
\bibitem[{Bezanson et~al.(2021)Bezanson, Edelman, Karpinski and
  Shah}]{JuliaLang}
\bibinfo{author}{Bezanson, J.}, \bibinfo{author}{Edelman, A.},
  \bibinfo{author}{Karpinski, S.}, \bibinfo{author}{Shah, V.B.},
  \bibinfo{year}{2021}.
\newblock \bibinfo{title}{{The Julia Programming Language}}.
\newblock \URLprefix \url{https://julialang.org/}. \bibinfo{note}{accessed:
  2021-07-27}.
\bibitem[{biasLab(2022)}]{Rocketjl}
\bibinfo{author}{biasLab}, \bibinfo{year}{2022}.
\newblock \bibinfo{title}{{Rocket.jl Documentation}}.
\newblock \URLprefix \url{https://biaslab.github.io/Rocket.jl/stable/}.
  \bibinfo{note}{biasLab/Rocket.jl v1.4.0. Accessed: 2022-09-09}.
\bibitem[{Bouchaud et~al.(2003)Bouchaud, Gefen, Potters and
  Wyart}]{bouchaud2003fluctuations}
\bibinfo{author}{Bouchaud, J.P.}, \bibinfo{author}{Gefen, Y.},
  \bibinfo{author}{Potters, M.}, \bibinfo{author}{Wyart, M.},
  \bibinfo{year}{2003}.
\newblock \bibinfo{title}{Fluctuations and response in financial markets: the
  subtle nature ofrandom'price changes}.
\newblock \bibinfo{journal}{Quantitative finance} \bibinfo{volume}{4},
  \bibinfo{pages}{176}.
\newblock \DOIprefix\doi{10.1080/14697680400000022}.
\bibitem[{Bouchaud et~al.(2002)Bouchaud, M{\'e}zard and
  Potters}]{bouchaud2002statistical}
\bibinfo{author}{Bouchaud, J.P.}, \bibinfo{author}{M{\'e}zard, M.},
  \bibinfo{author}{Potters, M.}, \bibinfo{year}{2002}.
\newblock \bibinfo{title}{Statistical properties of stock order books:
  empirical results and models}.
\newblock \bibinfo{journal}{Quantitative finance} \bibinfo{volume}{2},
  \bibinfo{pages}{251--256}.
\newblock \DOIprefix\doi{10.1088/1469-7688/2/4/301}.
\bibitem[{Cartea et~al.(2015)Cartea, Jaimungal and
  Penalva}]{cartea2015algorithmic}
\bibinfo{author}{Cartea, {\'A}.}, \bibinfo{author}{Jaimungal, S.},
  \bibinfo{author}{Penalva, J.}, \bibinfo{year}{2015}.
\newblock \bibinfo{title}{Algorithmic and high-frequency trading}.
\newblock \bibinfo{publisher}{Cambridge University Press}.
\bibitem[{Cont(2001)}]{cont2001empirical}
\bibinfo{author}{Cont, R.}, \bibinfo{year}{2001}.
\newblock \bibinfo{title}{Empirical properties of asset returns: stylized facts
  and statistical issues}.
\newblock \bibinfo{journal}{Quantitative Finance} \bibinfo{volume}{1},
  \bibinfo{pages}{223--236}.
\newblock \DOIprefix\doi{10.1080/713665670}.
\bibitem[{Crafa(2021)}]{crafa2021agent}
\bibinfo{author}{Crafa, S.}, \bibinfo{year}{2021}.
\newblock \bibinfo{title}{From agent-based modeling to actor-based reactive
  systems in the analysis of financial networks}.
\newblock \bibinfo{journal}{Journal of Economic Interaction and Coordination}
  \bibinfo{volume}{16}, \bibinfo{pages}{649--673}.
\newblock \URLprefix \url{https://doi.org/10.1007/s11403-021-00323-8},
  \DOIprefix\doi{10.1007/s11403-021-00323-8}.
\bibitem[{Dickey and Fuller(1979)}]{DickeyFuller1979}
\bibinfo{author}{Dickey, D.A.}, \bibinfo{author}{Fuller, W.A.},
  \bibinfo{year}{1979}.
\newblock \bibinfo{title}{Distribution of the estimators for autoregressive
  time series with a unit root}.
\newblock \bibinfo{journal}{Journal of the American Statistical Association}
  \bibinfo{volume}{74}, \bibinfo{pages}{427--431}.
\newblock \DOIprefix\doi{10.1080/01621459.1979.10482531}.
\bibitem[{Dicks and Gebbie(2022)}]{Dicks2022Julia}
\bibinfo{author}{Dicks, M.}, \bibinfo{author}{Gebbie, T.},
  \bibinfo{year}{2022}.
\newblock \bibinfo{title}{{A simple learning agent interacting with an
  agent-based market model: Julia code}}.
\newblock \bibinfo{journal}{figshare} \URLprefix
  \url{https://zivahub.uct.ac.za/articles/software/A_simple_learning_agent_interacting_with_an_agent-based_market_model_Julia_code/21163723},
  \DOIprefix\doi{{10.25375/uct.21163723.v1}}.
\bibitem[{Dicks et~al.(2023)Dicks, Paskaramoothy and
  Gebbie}]{dicks2023learning}
\bibinfo{author}{Dicks, M.}, \bibinfo{author}{Paskaramoothy, A.},
  \bibinfo{author}{Gebbie, T.}, \bibinfo{year}{2023}.
\newblock \bibinfo{title}{Many learning agents interacting with an agent-based
  market model}.
\newblock \href{http://arxiv.org/abs/2303.07393}{{\tt arXiv:2303.07393}}.
\bibitem[{Dieci and He(2018)}]{dieci2018heterogeneous}
\bibinfo{author}{Dieci, R.}, \bibinfo{author}{He, X.Z.}, \bibinfo{year}{2018}.
\newblock \bibinfo{title}{Heterogeneous agent models in finance}.
\newblock \bibinfo{journal}{Handbook of computational economics}
  \bibinfo{volume}{4}, \bibinfo{pages}{257--328}.
\newblock \URLprefix
  \url{https://www.sciencedirect.com/science/article/pii/S157400211830008X},
  \DOIprefix\doi{10.1016/bs.hescom.2018.03.002}.
\bibitem[{Fabretti(2013)}]{fabretti2013problem}
\bibinfo{author}{Fabretti, A.}, \bibinfo{year}{2013}.
\newblock \bibinfo{title}{On the problem of calibrating an agent based model
  for financial markets}.
\newblock \bibinfo{journal}{Journal of Economic Interaction and Coordination}
  \bibinfo{volume}{8}, \bibinfo{pages}{277--293}.
\newblock \DOIprefix\doi{10.1007/s11403-012-0096-3}.
\bibitem[{Farmer et~al.(2005)Farmer, Patelli and Zovko}]{farmer2005predictive}
\bibinfo{author}{Farmer, J.D.}, \bibinfo{author}{Patelli, P.},
  \bibinfo{author}{Zovko, I.I.}, \bibinfo{year}{2005}.
\newblock \bibinfo{title}{The predictive power of zero intelligence in
  financial markets}.
\newblock \bibinfo{journal}{Proceedings of the National Academy of Sciences}
  \bibinfo{volume}{102}, \bibinfo{pages}{2254--2259}.
\newblock \URLprefix \url{https://www.pnas.org/content/102/6/2254},
  \DOIprefix\doi{10.1073/pnas.0409157102}.
\bibitem[{Gao and Han(2012)}]{gao2012implementing}
\bibinfo{author}{Gao, F.}, \bibinfo{author}{Han, L.}, \bibinfo{year}{2012}.
\newblock \bibinfo{title}{Implementing the nelder-mead simplex algorithm with
  adaptive parameters}.
\newblock \bibinfo{journal}{Computational Optimization and Applications}
  \bibinfo{volume}{51}, \bibinfo{pages}{259--277}.
\newblock \DOIprefix\doi{10.1007/s10589-010-9329-3}.
\bibitem[{Garcia and Ndiaye(1998)}]{garcia1998learning}
\bibinfo{author}{Garcia, F.}, \bibinfo{author}{Ndiaye, S.M.},
  \bibinfo{year}{1998}.
\newblock \bibinfo{title}{A learning rate analysis of reinforcement learning
  algorithms in finite-horizon}, in: \bibinfo{booktitle}{ICML '98: Proceedings
  of the Fifteenth International Conference on Machine Learning},
  \bibinfo{publisher}{Morgan Kaufmann Publishers Inc.}, \bibinfo{address}{San
  Francisco, CA, USA}. pp. \bibinfo{pages}{215--223}.
\bibitem[{Geweke and Porter-Hudak(1983)}]{geweke1983estimation}
\bibinfo{author}{Geweke, J.}, \bibinfo{author}{Porter-Hudak, S.},
  \bibinfo{year}{1983}.
\newblock \bibinfo{title}{The estimation and application of long memory time
  series models}.
\newblock \bibinfo{journal}{Journal of time series analysis}
  \bibinfo{volume}{4}, \bibinfo{pages}{221--238}.
\newblock \DOIprefix\doi{10.1111/j.1467-9892.1983.tb00371.x}.
\bibitem[{Gilles(2006)}]{daniel2006asynchronous}
\bibinfo{author}{Gilles, D.}, \bibinfo{year}{2006}.
\newblock \bibinfo{title}{Asynchronous simulations of a limit order book}.
\newblock Master's thesis. University of Manchester.
\bibitem[{Gilli and Winker(2003)}]{gilli2003global}
\bibinfo{author}{Gilli, M.}, \bibinfo{author}{Winker, P.},
  \bibinfo{year}{2003}.
\newblock \bibinfo{title}{A global optimization heuristic for estimating agent
  based models}.
\newblock \bibinfo{journal}{Computational Statistics \& Data Analysis}
  \bibinfo{volume}{42}, \bibinfo{pages}{299--312}.
\newblock \DOIprefix\doi{10.1016/S0167-9473(02)00214-1}.
\bibitem[{Gopikrishnan et~al.(1998)Gopikrishnan, Meyer, Amaral and
  Stanley}]{gopikrishnanetal1998}
\bibinfo{author}{Gopikrishnan, P.}, \bibinfo{author}{Meyer, M.},
  \bibinfo{author}{Amaral, L.A.N.}, \bibinfo{author}{Stanley, H.},
  \bibinfo{year}{1998}.
\newblock \bibinfo{title}{Inverse cubic law for the distribution of stock price
  variations}.
\newblock \bibinfo{journal}{Eur. Phys. J. B} \bibinfo{volume}{3},
  \bibinfo{pages}{139--140}.
\newblock \DOIprefix\doi{https://doi.org/10.1007/s100510050292}.
\bibitem[{Gould et~al.(2013)Gould, Porter, Williams, McDonald, Fenn and
  Howison}]{gould2013limit}
\bibinfo{author}{Gould, M.D.}, \bibinfo{author}{Porter, M.A.},
  \bibinfo{author}{Williams, S.}, \bibinfo{author}{McDonald, M.},
  \bibinfo{author}{Fenn, D.J.}, \bibinfo{author}{Howison, S.D.},
  \bibinfo{year}{2013}.
\newblock \bibinfo{title}{Limit order books}.
\newblock \bibinfo{journal}{Quantitative Finance} \bibinfo{volume}{13},
  \bibinfo{pages}{1709--1742}.
\newblock \DOIprefix\doi{10.1080/14697688.2013.803148}.
\bibitem[{Harvey et~al.(2016)Harvey, Hendricks, Gebbie and
  Wilcox}]{gebbie2017deviations}
\bibinfo{author}{Harvey, M.}, \bibinfo{author}{Hendricks, D.},
  \bibinfo{author}{Gebbie, T.}, \bibinfo{author}{Wilcox, D.},
  \bibinfo{year}{2016}.
\newblock \bibinfo{title}{Deviations in expected price impact for small
  transaction volumes under fee restructuring}.
\newblock \bibinfo{journal}{Physica A: Statistical Mechanics and its
  Applications} \bibinfo{volume}{471}, \bibinfo{pages}{416--426}.
\newblock \DOIprefix\doi{10.1016/j.physa.2016.11.042}.
\bibitem[{Hendricks and Wilcox(2014)}]{hendricks2014reinforcement}
\bibinfo{author}{Hendricks, D.}, \bibinfo{author}{Wilcox, D.},
  \bibinfo{year}{2014}.
\newblock \bibinfo{title}{A reinforcement learning extension to the
  almgren-chriss framework for optimal trade execution}, in:
  \bibinfo{booktitle}{2014 IEEE Conference on Computational Intelligence for
  Financial Engineering \& Economics (CIFEr)}, \bibinfo{organization}{IEEE}.
  pp. \bibinfo{pages}{457--464}.
\newblock \DOIprefix\doi{10.1109/CIFEr.2014.6924109}.
\bibitem[{Hewitt(2011)}]{hewitt2011actormodel}
\bibinfo{author}{Hewitt, C.}, \bibinfo{year}{2011}.
\newblock \bibinfo{title}{Actor model of computation: Scalable robust
  information systems}.
\newblock \bibinfo{journal}{Proceedings of Inconsistency Robustness, Stanford}
  .
\bibitem[{Hurst(1951)}]{Hurst1951}
\bibinfo{author}{Hurst, H.E.}, \bibinfo{year}{1951}.
\newblock \bibinfo{title}{Long-term storage capacity of reservoirs}.
\newblock \bibinfo{journal}{Transactions of the American Society of Civil
  Engineers} \bibinfo{volume}{116}, \bibinfo{pages}{770--799}.
\newblock \DOIprefix\doi{10.1061/TACEAT.0006518}.
\bibitem[{Jericevich et~al.(2020a)Jericevich, Chang and
  Gebbie}]{gebbiechangjericevich2020comparing}
\bibinfo{author}{Jericevich, I.}, \bibinfo{author}{Chang, P.},
  \bibinfo{author}{Gebbie, T.}, \bibinfo{year}{2020}a.
\newblock \bibinfo{title}{{Comparing the market microstructure between two
  South African exchanges}}.
\newblock \href{http://arxiv.org/abs/2011.04367}{{\tt arXiv:2011.04367}}.
\bibitem[{Jericevich et~al.(2021)Jericevich, Chang and Gebbie}]{arxiv2021abm}
\bibinfo{author}{Jericevich, I.}, \bibinfo{author}{Chang, P.},
  \bibinfo{author}{Gebbie, T.}, \bibinfo{year}{2021}.
\newblock \bibinfo{title}{Simulation and estimation of an agent-based
  market-model with a matching engine}.
\newblock \href{http://arxiv.org/abs/2108.07806}{{\tt arXiv:2108.07806}}.
\bibitem[{Jericevich et~al.(2020b)Jericevich, Chang, Pillay and
  Gebbie}]{Jericevich2020data}
\bibinfo{author}{Jericevich, I.}, \bibinfo{author}{Chang, P.},
  \bibinfo{author}{Pillay, A.}, \bibinfo{author}{Gebbie, T.},
  \bibinfo{year}{2020}b.
\newblock \bibinfo{title}{Supporting test data: Comparing the market
  microstructure between two south african exchanges}.
\newblock \URLprefix
  \url{https://zivahub.uct.ac.za/articles/dataset/Supporting_Test_Data_Comparing_the_Market_Microstructure_Between_two_South_African_Exchanges/13187591},
  \DOIprefix\doi{10.25375/uct.13187591.v1}.
\bibitem[{Jericevich et~al.(2022a)Jericevich, Sing and
  Gebbie}]{SoftwareXcointossx}
\bibinfo{author}{Jericevich, I.}, \bibinfo{author}{Sing, D.},
  \bibinfo{author}{Gebbie, T.}, \bibinfo{year}{2022}a.
\newblock \bibinfo{title}{Cointossx: An open-source low-latency high-throughput
  matching engine}.
\newblock \bibinfo{journal}{SoftwareX} \bibinfo{volume}{19},
  \bibinfo{pages}{101136}.
\newblock \URLprefix
  \url{https://www.softxjournal.com/article/S2352-7110(22)00087-5/pdf},
  \DOIprefix\doi{10.1016/j.softx.2022.101136}.
\bibitem[{Jericevich et~al.(2022b)Jericevich, Sing, Gebbie and
  Dicks}]{Jericevich2022CoinTossXUDP}
\bibinfo{author}{Jericevich, I.}, \bibinfo{author}{Sing, D.},
  \bibinfo{author}{Gebbie, T.}, \bibinfo{author}{Dicks, M.},
  \bibinfo{year}{2022}b.
\newblock \bibinfo{title}{{CoinTossX with market data feed}}.
\newblock \bibinfo{journal}{figshare} \URLprefix
  \url{https://zivahub.uct.ac.za/articles/software/CoinTossX_with_market_data_feed/21163993},
  \DOIprefix\doi{10.25375/uct.21163993.v1}.
\bibitem[{Leal et~al.(2016)Leal, Napoletano, Roventini and
  Fagiolo}]{leal2016rock}
\bibinfo{author}{Leal, S.J.}, \bibinfo{author}{Napoletano, M.},
  \bibinfo{author}{Roventini, A.}, \bibinfo{author}{Fagiolo, G.},
  \bibinfo{year}{2016}.
\newblock \bibinfo{title}{Rock around the clock: An agent-based model of
  low-and high-frequency trading}.
\newblock \bibinfo{journal}{Journal of Evolutionary Economics}
  \bibinfo{volume}{26}, \bibinfo{pages}{49--76}.
\bibitem[{LeBaron(2000)}]{lebaron2000agent}
\bibinfo{author}{LeBaron, B.}, \bibinfo{year}{2000}.
\newblock \bibinfo{title}{Agent-based computational finance: Suggested readings
  and early research}.
\newblock \bibinfo{journal}{Journal of Economic Dynamics and Control}
  \bibinfo{volume}{24}, \bibinfo{pages}{679--702}.
\newblock \URLprefix
  \url{https://www.sciencedirect.com/science/article/pii/S0165188999000226},
  \DOIprefix\doi{10.1016/S0165-1889(99)00022-6}.
\bibitem[{Lee and Ready(1991)}]{lee1991inferring}
\bibinfo{author}{Lee, C.M.C.}, \bibinfo{author}{Ready, M.J.},
  \bibinfo{year}{1991}.
\newblock \bibinfo{title}{Inferring trade direction from intraday data}.
\newblock \bibinfo{journal}{The Journal of Finance} \bibinfo{volume}{46},
  \bibinfo{pages}{733--746}.
\newblock \DOIprefix\doi{10.1111/j.1540-6261.1991.tb02683.x}.
\bibitem[{Lillo and Farmer(2007)}]{lillo2004long}
\bibinfo{author}{Lillo, F.}, \bibinfo{author}{Farmer, J.D.},
  \bibinfo{year}{2007}.
\newblock \bibinfo{title}{The long memory of the efficient market}.
\newblock \bibinfo{journal}{Studies in Nonlinear Dynamics \& Econometrics}
  \bibinfo{volume}{8}, \bibinfo{pages}{1--1}.
\newblock \DOIprefix\doi{10.2202/1558-3708.1226}.
\bibitem[{Lillo et~al.(2003)Lillo, Farmer and Mantegna}]{LFM2003}
\bibinfo{author}{Lillo, F.}, \bibinfo{author}{Farmer, J.D.},
  \bibinfo{author}{Mantegna, R.}, \bibinfo{year}{2003}.
\newblock \bibinfo{title}{Master curve for price-impact function}.
\newblock \bibinfo{journal}{Nature} \bibinfo{volume}{421},
  \bibinfo{pages}{129--30}.
\newblock \DOIprefix\doi{10.1038/421129a}.
\bibitem[{Mandelbrot(1963)}]{mandelbrot1963variation}
\bibinfo{author}{Mandelbrot, B.B.}, \bibinfo{year}{1963}.
\newblock \bibinfo{title}{The variation of certain speculative prices}.
\newblock \bibinfo{journal}{The Journal of Business} \bibinfo{volume}{36},
  \bibinfo{pages}{394--419}.
\newblock \URLprefix \url{http://www.jstor.org/stable/2350970}.
\bibitem[{Mandelbrot and Wallis(1969)}]{MandelbrotWallis1969}
\bibinfo{author}{Mandelbrot, B.B.}, \bibinfo{author}{Wallis, J.R.},
  \bibinfo{year}{1969}.
\newblock \bibinfo{title}{Robustness of the rescaled range r/s in the
  measurement of noncyclic long run statistical dependence}.
\newblock \bibinfo{journal}{Water Resources Research} \bibinfo{volume}{5},
  \bibinfo{pages}{967--988}.
\newblock \DOIprefix\doi{https://doi.org/10.1029/WR005i005p00967}.
\bibitem[{Mandeş(2015)}]{Alexandru2015}
\bibinfo{author}{Mandeş, A.}, \bibinfo{year}{2015}.
\newblock \bibinfo{title}{Microstructure-based order placement in a continuous
  double auction agent based model}.
\newblock \bibinfo{journal}{Algorithmic Finance} \bibinfo{volume}{4},
  \bibinfo{pages}{105--125}.
\newblock \DOIprefix\doi{10.3233/AF-150049}.
\bibitem[{Mandeş and Winker(2017)}]{AlexandruPeter2017}
\bibinfo{author}{Mandeş, A.}, \bibinfo{author}{Winker, P.},
  \bibinfo{year}{2017}.
\newblock \bibinfo{title}{Complexity and model comparison in agent based
  modeling of financial markets}.
\newblock \bibinfo{journal}{Journal of Economic Interaction and Coordination}
  \bibinfo{volume}{12}, \bibinfo{pages}{469--506}.
\newblock \DOIprefix\doi{10.1007/s11403-016-0173-0}.
\bibitem[{Massey(1951)}]{Massey1951}
\bibinfo{author}{Massey, F.J.J.}, \bibinfo{year}{1951}.
\newblock \bibinfo{title}{The kolmogorov-smirnov test for goodness of fit}.
\newblock \bibinfo{journal}{Journal of the American Statistical Association}
  \bibinfo{volume}{46}, \bibinfo{pages}{68--78}.
\newblock \DOIprefix\doi{10.1080/01621459.1951.10500769}.
\bibitem[{Nuyts(2010)}]{nuyts2010inference}
\bibinfo{author}{Nuyts, J.}, \bibinfo{year}{2010}.
\newblock \bibinfo{title}{Inference about the tail of a distribution:
  Improvement on the hill estimator}.
\newblock \bibinfo{journal}{International Journal of mathematics and
  mathematical sciences} \bibinfo{volume}{2010}.
\newblock \DOIprefix\doi{10.1155/2010/924013}.
\bibitem[{Obizhaeva and Wang(2013)}]{obizhaeva2013optimal}
\bibinfo{author}{Obizhaeva, A.A.}, \bibinfo{author}{Wang, J.},
  \bibinfo{year}{2013}.
\newblock \bibinfo{title}{Optimal trading strategy and supply/demand dynamics}.
\newblock \bibinfo{journal}{Journal of Financial Markets} \bibinfo{volume}{16},
  \bibinfo{pages}{1--32}.
\newblock \DOIprefix\doi{10.1016/j.finmar.2012.09.001}.
\bibitem[{Platt(2020)}]{platt2020comparison}
\bibinfo{author}{Platt, D.}, \bibinfo{year}{2020}.
\newblock \bibinfo{title}{A comparison of economic agent-based model
  calibration methods}.
\newblock \bibinfo{journal}{Journal of Economic Dynamics and Control}
  \bibinfo{volume}{113}, \bibinfo{pages}{103859}.
\newblock \URLprefix
  \url{https://www.sciencedirect.com/science/article/pii/S0165188920300294},
  \DOIprefix\doi{10.1016/j.jedc.2020.103859}.
\bibitem[{Platt and Gebbie(2016)}]{plattgebbie2016problem}
\bibinfo{author}{Platt, D.}, \bibinfo{author}{Gebbie, T.},
  \bibinfo{year}{2016}.
\newblock \bibinfo{title}{The problem of calibrating an agent-based model of
  high-frequency trading}.
\newblock \href{http://arxiv.org/abs/1606.01495}{{\tt arXiv:1606.01495}}.
\bibitem[{Platt and Gebbie(2018)}]{plattgebbie2018can}
\bibinfo{author}{Platt, D.}, \bibinfo{author}{Gebbie, T.},
  \bibinfo{year}{2018}.
\newblock \bibinfo{title}{Can agent-based models probe market microstructure?}
\newblock \bibinfo{journal}{Physica A: Statistical Mechanics and its
  Applications} \bibinfo{volume}{503}, \bibinfo{pages}{1092--1106}.
\newblock \URLprefix
  \url{https://www.sciencedirect.com/science/article/pii/S0378437118309956},
  \DOIprefix\doi{10.1016/j.physa.2018.08.055}.
\bibitem[{Potters and Bouchaud(2003)}]{potters2003more}
\bibinfo{author}{Potters, M.}, \bibinfo{author}{Bouchaud, J.P.},
  \bibinfo{year}{2003}.
\newblock \bibinfo{title}{More statistical properties of order books and price
  impact}.
\newblock \bibinfo{journal}{Physica A: Statistical Mechanics and its
  Applications} \bibinfo{volume}{324}, \bibinfo{pages}{133--140}.
\newblock \DOIprefix\doi{10.1016/S0378-4371(02)01896-4}.
\bibitem[{Sutton and Barto(2018)}]{sutton2018reinforcement}
\bibinfo{author}{Sutton, R.S.}, \bibinfo{author}{Barto, A.G.},
  \bibinfo{year}{2018}.
\newblock \bibinfo{title}{Reinforcement learning: An introduction}.
\newblock \bibinfo{publisher}{MIT press}.
\bibitem[{Toth et~al.(2015)Toth, Palit, Lillo and Farmer}]{toth2015equity}
\bibinfo{author}{Toth, B.}, \bibinfo{author}{Palit, I.},
  \bibinfo{author}{Lillo, F.}, \bibinfo{author}{Farmer, J.D.},
  \bibinfo{year}{2015}.
\newblock \bibinfo{title}{Why is equity order flow so persistent?}
\newblock \bibinfo{journal}{Journal of Economic Dynamics and Control}
  \bibinfo{volume}{51}, \bibinfo{pages}{218--239}.
\newblock \DOIprefix\doi{10.1016/j.jedc.2014.10.007}.
\bibitem[{Watkins and Dayan(1992)}]{watkins1992q}
\bibinfo{author}{Watkins, C.J.}, \bibinfo{author}{Dayan, P.},
  \bibinfo{year}{1992}.
\newblock \bibinfo{title}{Q-learning}.
\newblock \bibinfo{journal}{Machine learning} \bibinfo{volume}{8},
  \bibinfo{pages}{279--292}.
\newblock \DOIprefix\doi{10.1007/BF00992698}.
\bibitem[{Watkins(1989)}]{watkins1989learning}
\bibinfo{author}{Watkins, C.J.C.H.}, \bibinfo{year}{1989}.
\newblock \bibinfo{title}{Learning from delayed rewards}.
\newblock Ph.D. thesis. King's College, Cambridge United Kingdom.
\bibitem[{Wilcox and Gebbie(2014)}]{WilcoxGebbie2015}
\bibinfo{author}{Wilcox, D.}, \bibinfo{author}{Gebbie, T.},
  \bibinfo{year}{2014}.
\newblock \bibinfo{title}{Hierarchical causality in financial economics}.
\newblock \bibinfo{journal}{Available at SSRN:} \URLprefix
  \url{https://ssrn.com/abstract=2544327},
  \DOIprefix\doi{http://dx.doi.org/10.2139/ssrn.2544327}.
\bibitem[{Winker et~al.(2007)Winker, Gilli and
  Jeleskovic}]{winker2007objective}
\bibinfo{author}{Winker, P.}, \bibinfo{author}{Gilli, M.},
  \bibinfo{author}{Jeleskovic, V.}, \bibinfo{year}{2007}.
\newblock \bibinfo{title}{An objective function for simulation based inference
  on exchange rate data}.
\newblock \bibinfo{journal}{Journal of Economic Interaction and Coordination}
  \bibinfo{volume}{2}, \bibinfo{pages}{125--145}.
\newblock \DOIprefix\doi{10.1007/s11403-007-0020-4}.

\end{thebibliography}


\end{document}


\flushbottom
\maketitle
	\tableofcontents
\thispagestyle{empty}

\section{System implementation} \label{app:SystemImplementation}

\captionsetup[table]{labelsep=space, singlelinecheck=off}

\begin{table}[htbp]
    \caption{The hardware specifications on which the simulations were run and shows indicative runtimes for a single simulation of the ABM, the calibrations and sensitivity analysis, as well as the training time for a single RL agent and the total amount of time need to train all the RL agents. The sensitivity analysis and calibration were run on Virtual Machine 1 and the RL experiments were run on the upgraded Virtual Machine 2. The approximate runtimes for a single run of the simulation as well as the sensitivity analysis, calibration, training time for a single RL agent and the time it took to train all RL agents.}
        \label{table-hardware-runtimes}
        \begin{tabular}{lllp{4cm}}
            \toprule
            Hardware & Specifications & Primary Task & Runtime \\
            \hline
            Virtual Machine 1 & Standard A4m V2 &  Sensitivity Analysis & 30 hours\\
            Operating system & 64-bit Ubuntu 20.04.4 LTS  &  Calibration & 8.3 hours\\
            CPU & Intel(R) Xeon(R) Platinum 8171M &&\\
                & @ 2.60GHz with 4 vCPUs &&\\
            RAM & 32GiB \\
            \hline
            Virtual Machine 2 & Standard F16s V2 &  1 simulation run & 26 secs (includes JIT o.h.)  \\
            Operating system & 64-bit Ubuntu 20.04.4 LTS & RL agent training time & 8.3 hours \\
            CPU & Intel(R) Xeon(R) Platinum 8272CL & Total RL training time & 50 hours \\
                & @ 2.60GHz with 16 vCPUs & &\\
            RAM & 32GiB & &\\
            \bottomrule
            \end{tabular}
\end{table}

The simulations are run on a system that consists of two parts: the Matching Engine (ME) and the Model Management System (MMS). The matching engine is its own piece of software that receives orders from the MMS, processes them, and sends messages back to the MMS notifying it that an event has occurred. The MMS consists of two parts, the first is the market-data listener that listens for update messages coming from the matching engine \cite{arxiv2021abm}, and the second is where the LOB, maintained by the MMS, gets updated and the actors submit orders to the matching engine based on the current state of the LOB \cite{Dicks2022Julia}.  

\subsection{Matching Engine (ME)} \label{ssec:SystemImplementation-ME}

The matching engine implementation is CoinTossX \cite{sing2017cointossx}. The first release was not designed to support a continuous market-data feed, which is needed to run an event-based simulation. Previous work that implemented CoinTossX submitted orders to the matching engine and requested snapshots of the LOB maintained by CoinTossX \cite{arxiv2021hawkes}. This is a high-coupling and low-cohesion design, as the matching engine is doing work that should be implemented by the MMS. Requesting a snapshot of the LOB is a time-consuming task that affects the low latency requirement of the matching engine and it was included in the first release as a convenience to be used by developers as a debugging tool, and not as part of a simulation workflow \cite{arxiv2021abm}. For this reason, a UDP socket was added to the matching engine\cite{arxiv2021abm,Jericevich2022CoinTossXUDP} to send event messages (Event messages refer to the messages detailing that a trade, cancellation or limit order has occurred.) to the MMS and the internal debugging tools depreciated. This then achieves the desired low-coupling and high-cohesion design pattern required for simulation work.

Logging in and out, starting and stopping the simulation, and the submission of orders are submitted to the matching engine by calling functions using a java wrapper (The java wrapper package used is called JavaCall.) implemented in \textit{Julia}. These orders are processed by CoinTossX and messages are sent to the MMS over a network using a UDP socket. These messages are captured by an asynchronous task in the MMS which allows the MMS to build its own LOB. Messages received from the matching engine represent events in the event-based simulation that the agents will trade-off. 

A key requirement of the event-based ABM is that the matching engine is low latency. To ensure this the messages in the continuous data feed are encoded and decoded in binary format to increase the speed of the transmission of the messages over the network \cite{arxiv2021abm}.

\subsection{Model Management System (MMS)} \label{ssec:SystemImplementation-MMS}

The MMS is split into 2 tasks: the first being the market-data listener, and the second being the event processor/generator. At the time of the project, \textit{Julia}'s multi-threading capabilities were relatively new and there were still issues in its implementation. Therefore the tasks could not be reliably implemented in a multi-threaded environment. Thus an asynchronous programming approach became necessary to split the two tasks in the MMS\cite{Dicks2022Julia}. Now because the code is run on the same thread the program is thread safe. The \textit{@async} macro was used to implement the asynchronous programming \cite{juliaLangAsync}. The main task is the event processor/generator where the messages received by the market-data listener are used to update the LOB in the MMS and then generate trades from the agents based on the updated state of the LOB. The secondary task consists of the market-data listener that receives messages from the matching engine. The market-data listener was implemented in the secondary task because the call to read from the UDP socket blocks when there are no messages in the UDP buffer. 

To switch between tasks the main task is slept for a brief moment to allow the market-data listener to read the new messages from the UDP buffer. The amount of time the main thread is slept for depends on the number of agents submitting orders to the matching engine and the speed of the hardware the simulation is running on. If the main thread is slept for too short a time and there are a large number of messages that need to be processed, messages get dropped. Therefore, as the number of agents increases the time that the main thread is slept for will need to increase.

\subsubsection{Market-data listener} \label{sssec:SystemImplementation-MMS-MDL}

The relationship between the two tasks can be thought of in the producer-consumer paradigm \cite{dijkstra1968cooperating}. The market-data listener produces the messages, which it receives from the matching engine, and the event processor/generator consumes these messages and uses them to update the LOB. Getting data from the market-data listener is not as simple as a function call because there may be more than one message it has received. \textit{Julia} provides a channel (A channel is a first-in-first-out (FIFO) queue that can have multiple tasks reading from it.) mechanism for solving this problem. The market-data listener reads messages from the UDP buffer and places them into a channel that is used by the event processor/generator to read the messages.

\subsubsection{Event processor/generator} \label{ssec:SystemImplementation-MMS-EPG}

The event processor/generator task is the main task. Since this is an event-based simulation the driver of the simulation is the event loop. In each event loop, all the new messages that were received by the market-data listener and placed in the channel are taken from the channel and used to update the LOB. The updated LOB is then sent to all the agents and they trade based on the state of the LOB, their current state, and the agent rules. See the pseudocode for the event loop in Algorithm \ref{Pseudocode-simulation}.

An event in the simulation is classified as an update to the LOB generated by all the order messages generated in the previous event loop. For example, if three orders were generated in the previous event loop the market-data listener would read the three new events from the UDP buffer and place them into the channel. Then in the next iteration of the event loop, those three events would be removed from the channel and they would be used to update the LOB. Classifying a single message updating the LOB as an event the agents would trade-off was tested, but it created a large lag between the matching engine's LOB used to match the orders and the MMS's LOB, which meant that agents were trading based on a stale LOB. The stale LOB lag effect was then exacerbated as more agents were submitting additional orders. 

A simple Q-learning RL agent needs to observe the state of the environment after taking an action and the lag would mean that the effect of their action would only be observed after a large number of iterations, and in the worst case, the effect of their actions may not be observed by the agent if the simulation has ended before the update is received. Additionally, the software is designed in such a way that adding more RL agents to the simulation is straightforward in order to support future work that will use this software to analyse what effects there may be when adding additional learning agents to the model.

The current setup means that the market-data listener and the updating of the LOB are not in the same task. In an actor-based model, the market-data listener and the updating of the LOB would likely be placed in the same thread because you would want the messages to directly update the LOB and not have to go through another process. One limitation of the model is that it could not be multi-threaded and asynchronous programming had to be used. Updating the LOB in a separate task was done to ensure that the market-data listener had the fastest execution time possible so that the main thread could sleep for the least amount of time. This increases the speed of the program as well as allowed more messages to be processed, which helped the simulation scale better as the number of agents increased.

\subsubsection{Ordering of LOB updates} \label{ssec:SystemImplementation-MMS-OLU}

The \texttt{Reactive.jl} framework delivers messages to actors in an ordered manner. The order is defined by the order in which the actors are initialised.  For example, if actor A was defined before actor B then a message passed to actors A and B, who subscribe to the same source, would cause actor A to receive and process the message before B. The processing of a message can generate a trade for the given agent. This means that the order of the agents receiving the messages from the source also imposes an ordering on the trades. For example, if agent A is defined before agent B, agent A will always receive a message before agent B, and if they both trade then agent A will always submit an order before agent B. This ordering affects the dynamics of the LOB. Using another example, if liquidity providers always submit orders before the liquidity takers there will always be additional liquidity provided in the LOB before liquidity is removed. This may make the order book appear more resilient than is defined by the model rules. To ensure that an arbitrary ordering of agents does not affect the ABM it was decided to randomise the order in which agents receive messages (LOB updates), which will cause the order of the trades submitted to also be randomised. A uniformly random ordering ensured that the results of the ABM do not depend on an arbitrarily defined ordering of agents.

\subsubsection{Simulation initialisation} \label{ssec:SystemImplementation-MMS-Initialisation}

The hybrid ABM\cite{arxiv2021abm} simulation was able to start with an empty limit order book. This did not work for the event-based ABM as it causes frequent liquidity crashes because the LTs can remove liquidity faster than it could be provided. Therefore, it was decided that it was necessary to start with a certain amount of liquidity in the LOB before the simulation starts. With the limit-order book initialised with some limited liquidity the frequency of crashes was reduced, and could be viewed as the outcome of an open auction.  

The order book is initialised by having the LPs submit limit orders before the simulation starts. The number of orders submitted is set at 1001. This value was chosen as it provided a good balance between reducing the number of liquidity crashes and reducing the likelihood of having the LOB lock up due to too much liquidity being provided at the top-of-book. Given the agent specifications for the LPs, we have that the number of bids and asks will be roughly balanced and therefore the initial order book will start with only a small order imbalance.

During the initialising of the LOB, the LPs are not allowed to place orders inside the initial spread. This maintains the initial spread and does not allow the LPs to close the spread due to the competition effect without it being pushed back by the liquidity takers. The initial mid-price and the initial spread are set arbitrarily at 10000 and 40 respectively. 

\subsubsection{Miscellaneous considerations} \label{ssec:SystemImplementation-MMS-MC}

In the Johannesburg Stock Exchange (JSE) a volatility auction is triggered when the execution price is larger than a 10\% deviation from the dynamic reference price \cite{JSEvolauction}. As per \cite{sing2017jse} the execution price is the price of the best bid or offer that is crossed when the market order hits the LOB, while the dynamic reference price is the deepest price level executed from the last transaction. 

The agent specifications defined by \cite{arxiv2021abm} do not provide rules for how to trade in a volatility auction. Therefore, they imposed a restriction on the LTs to ensure that a market order is not submitted if it will trigger a volatility auction. The event-based model also does not account for behaviour in volatility auctions and will follow the same method as \cite{arxiv2021abm} to prevent them from occurring. This also applies to the RL agents.

Finally, liquidity takers and RL agents will not submit market orders if the contra side of the LOB is empty. A recommendation from \cite{arxiv2021abm} is to not submit an order at all as this will help in the data cleaning process, because when a market order is submitted to an empty LOB the data feed returns an empty string. So, submitting a market order or not will lead to the same result. In the event-based model, this will also reduce the number of unnecessary orders submitted, which will allow the main task to sleep for a shorter period and therefore speed up the simulation. 

\subsubsection{Hardware specifications and run times} \label{ssec:SystemImplementation-MMS-HSRT}

The experiments run were all implemented using Microsoft Azure's Virtual machines \cite{MSAzureVM}. The hardware specifications for the virtual machines used are detailed in Table \ref{table-hardware-runtimes}. The sensitivity analysis and the calibration were run on Virtual Machine 1. Due to the reduced number of virtual CPUs (cores), there were instances of some UDP packet loss, which could cause errors in the MMS. Error handling was implemented for the sensitivity analysis and calibration work so that the simulations could resume at the place where errors occurred, in subsequent runs, but without incorrect data being recorded or actual data loss. 

For the RL experiments, the server was upgraded to provide better message load handling. The training of the RL agent was therefore done on the compute-optimised Virtual Machine 2. The increase in the number of vCPUs from 4 to 16 was enough to ensure that during the training of all the RL agents no packets were dropped and no errors occurred.

Table \ref{table-hardware-runtimes} shows indicative run times for each of the experiments. Note that the increase in computing power, from sensitivity analysis and calibration to the RL experiments, was not used to increase the speed of the simulation but was used to reduce the number of packets dropped. The time it takes to run a single simulation takes the same time on both sets of hardware because the simulation is run for a set length of time.

\newpage
\section{Stylised facts} \label{app:stylisedfacts}

\captionsetup[table]{labelsep=space, 
         justification=raggedright, singlelinecheck=off}
\begin{table}[h]
\centering
\caption{The key stylised facts that we aim to reproduce with our ABM.}
\label{stylised-facts-table}
\begin{tabular}{p{0.3\textwidth}p{0.7\textwidth}}
\toprule
Stylised Fact & Description\\
\toprule
Heavy tails & The unconditional distribution of the log-returns has been shown to have power-law behaviour in the tails \cite{mandelbrot1963variation}, with a finite tail-index between 2 and 5 \cite{gopikrishnanetal1998,cont2001empirical}.\\
\hline
Return autocorrelation & The log-returns exhibit an absence of significant autocorrelations \cite{cont2001empirical}. In high-frequency time scales the return series observes negative significant autocorrelation for the first few lags which are usually given as evidence for a fast mean-reverting component\cite{cont2001empirical}.  \\
\hline
Trade-sign autocorrelation & The trade sign time-series has been shown to have persistence in their autocorrelations \cite{lillo2004long,bouchaud2003fluctuations,toth2015equity}; implying that buy (resp. sell) orders are followed by more buy (resp. sell) orders. Herding and order-splitting have been used to explain this observed phenomena; with order-splitting being the main reason for the persistence in order flow \cite{toth2015equity}.  \\
\hline
Volatility Clustering & Here large (resp. small) price variations are often followed by large (resp. small) price variations. The persistence of the autocorrelation of the absolute log-returns are signatures of volatility clustering and long-range dependence in the volatility \cite{cont2001empirical}. \\
\hline
Depth profile & The Paris Bourse and the NASDAQ have an observed hump in their shape with a slight increase in volume as the price moves away from the best bid/offer, with it reaching a maximum a few ticks away from the best bid/offer before a steep decline \cite{bouchaud2002statistical,gould2013limit}. The Standard and Poor's Depository Receipts (SPY) showed a maximum at the best bid/offer \cite{potters2003more}. \\
\hline
Price impact & Price impact can be measured as the change in bid/ask price caused by a trade of a given volume. As the volume increases the observed price changes increase linearly on a log-log scale indicating a power-law relationship \cite{gebbie2017deviations,gebbiechangjericevich2020comparing,LFM2003}. \\ 
\bottomrule
\end{tabular}
\end{table}

\begin{figure}[h!]
        \centering
        \begin{subfigure}{.45\textwidth}
          \centering
              \includegraphics[width=8cm, height=5cm]{Figures/Stylised Facts/Log-ReturnDistribution4.png}
              \caption{ABM log-return distributions}
        \end{subfigure}%
        \begin{subfigure}{.45\textwidth}
          \centering
          \includegraphics[width=8cm, height=5cm]{Figures/Stylised Facts/Log-ReturnDistributionJSE.png}
          \caption{JSE log-return distribution}
        \end{subfigure}
        \caption{Plots the log-return distributions for the ABM (green) and the JSE (purple). Using QQ-plots the insets compare the distributions to the normal distribution fitted to the data using MLE and reveal that the log-return distributions have fatter tails than the normal distribution, demonstrating the ABM's ability to reproduce the required stylised fact.}
        \label{logreturn-style}
\end{figure}

\begin{figure}[h!]
        \centering
        \begin{subfigure}{.45\textwidth}
          \centering
              \includegraphics[width=8cm, height=5cm]{Figures/Stylised Facts/ExtremeLog-ReturnPercentilesDistribution4.png}
              \caption{ABM extreme value distributions}
        \end{subfigure}%
        \begin{subfigure}{.45\textwidth}
          \centering
          \includegraphics[width=8cm, height=5cm]{Figures/Stylised Facts/ExtremeLog-ReturnPercentilesDistributionJSE.png}
          \caption{JSE extreme value distributions}
        \end{subfigure}
        \caption{The extreme distributions taken from the upper 95th percentile as well as the lower 5th percentile for the ABM (green) and JSE (purple) are plotted. This figure demonstrates that both return distributions have near power-law behaviour in their tails. If their distributions were exactly a power-law we would expect that the data would be linear on a log-log scale. However, we observe a slight downward curve which indicates only near power-law behaviour. Comparing the empirical-versus-simulated returns we see that both distributions observe this behaviour, giving credibility to the returns generated by the ABM. We also note that the ABM's distribution is fatter tailed when compared with the JSE's distribution.}
        \label{extremelogreturn-style}
\end{figure}

\begin{figure}[h!]
        \centering
        \begin{subfigure}{.45\textwidth}
          \centering
              \includegraphics[width=8cm, height=5cm]{Figures/Stylised Facts/CDF_Inverse_Cube4.png}
              \caption{ABM extreme value distributions}
        \end{subfigure}%
        \begin{subfigure}{.45\textwidth}
          \centering
          \includegraphics[width=8cm, height=5cm]{Figures/Stylised Facts/CDF_Inverse_Cube_JSE.png}
          \caption{JSE extreme value distributions}
        \end{subfigure}
        \caption{The exceedance probabilities are shown for the left and right tail of the log-return distributions, these are plotted for the ABM (left) and the JSE (right). The plot is on the log-log scale with the x-axis and y-axis denoting the exponents\cite{gopikrishnanetal1998}. The JSE has extreme events of similar sizes for both the left and the right tail, while the ABM has larger extreme events in the right tail and this is a simulation artifact. We see that $\alpha$ is close to three for the ABM showing that the distribution for this price path follows the inverse cubic law \cite{gopikrishnanetal1998}. This value was determined by fitting a linear regression to the linear region of the exceedances in a log-log scale. The JSE shows a larger $\alpha$ when compared with the ABM.}
        \label{InverseCubice-style}
\end{figure}

\begin{figure}[h!]
        \centering
        \begin{subfigure}{.45\textwidth}
          \centering
              \includegraphics[width=8cm, height=5cm]{Figures/Stylised Facts/Log-ReturnAutocorrelation4.png}
              \caption{ABM log-return autocorrelations}
        \end{subfigure}%
        \begin{subfigure}{.45\textwidth}
          \centering
          \includegraphics[width=8cm, height=5cm]{Figures/Stylised Facts/Log-ReturnAutocorrelationJSE.png}
          \caption{JSE log-return autocorrelations}
        \end{subfigure}
        \caption{Plots the autocorrelations in the log-returns and the inset plots the autocorrelations in the absolute log-returns, for the ABM (green) and the JSE (purple). The significant negative autocorrelation in the log-returns for the first few lags is present in both the plots. The absolute log-returns for both the simulated and empirical both have persistence in their autocorrelations, meaning that both time series exhibit evidence for volatility clustering. However, the decay in the autocorrelations is far quicker for the simulated returns.}
        \label{logreturn-auto-style}
\end{figure}

\begin{figure}[h!]
        \centering
        \begin{subfigure}[b]{0.45\textwidth}
          \centering
            \includegraphics[width=8cm]{Figures/Stylised Facts/AbsLog-ReturnAutocorrelation4_inset.png}
            \caption{ABM absolute return autocorrelations}
        \label{abslog-return-autocorrelation-ctx-(a)}
        \end{subfigure}%
        \begin{subfigure}[b]{0.45\textwidth}
          \centering
              \includegraphics[width=8cm]{Figures/Stylised Facts/AbsLog-ReturnAutocorrelationJSE_inset.png}
              \caption{JSE absolute return autocorrelations}
              \label{abslog-return-autocorrelation-jse-(b)}
        \end{subfigure}
        \caption{Plots the autocorrelation in the absolute log-returns for the calibrated ABM in Figure \ref{abslog-return-autocorrelation-ctx-(a)} and the JSE in Figure \ref{abslog-return-autocorrelation-jse-(b)}. Those for the the models that include the RL agents are shown in the paper.}
        \label{abslog-return-autocorrelations}
\end{figure}

\begin{figure}[h!]
        \centering
        \begin{subfigure}{.45\textwidth}
          \centering
              \includegraphics[width=8cm, height=5cm]{Figures/Stylised Facts/Trade-SignAutocorrelation4.png}
              \caption{ABM trade-sign autocorrelations}
        \end{subfigure}%
        \begin{subfigure}{.45\textwidth}
          \centering
          \includegraphics[width=8cm, height=5cm]{Figures/Stylised Facts/Trade-SignAutocorrelationJSE.png}
          \caption{JSE trade-sign autocorrelations}
        \end{subfigure}
        \caption{The autocorrelations in the trade-signs are plotted for the ABM (green) and the JSE (purple). The model's autocorrelation function (ACF) does not exhibit the same behaviour as the JSE. The ABM does capture the stylised fact of persistent decay of the trade-signs. However, the decay is far slower than that found in the JSE data. The insets plot the trade-sign decay on a log-linear scale where the difference in the rate of decay is also clearly demonstrated.}
        \label{tradesign-auto-style}
\end{figure}

\begin{figure}[h!]
    \centering
    \begin{minipage}[t]{0.497\textwidth}
        \centering
        \includegraphics[width=8cm, height=5cm]{Figures/Stylised Facts/DepthProfile4.png}
        \caption{The depth profile is computed as the average volume observed at seven price levels throughout the simulation \cite{arxiv2021abm}. The maximum volume occurring at the best bid/ask and the rapid decrease in volume as the price increases is consistent with empirical data \cite{potters2003more} as well as previous ABM results \cite{arxiv2021abm}.}
        \label{depth-profile}
    \end{minipage}
    \begin{minipage}[t]{0.497\textwidth}
        \centering
        \includegraphics[width=8cm, height=5cm]{Figures/Stylised Facts/Trade-SignAutocorrelationlog-logJSE.png}
        \caption{The power-law decay in the order-flow autocorrelation shows the long-memory property \cite{lillo2004long}. Here we plot the trade-sign autocorrelations for the JSE on a log-log scale and observe an approximately linear decrease to recover similar results. Using the MLE of the power-law density \cite{clauset2009power}, we show that $\alpha = 1.286$, which means that the distribution is not integrable and exhibits the long-memory property.}
        \label{tradesign-auto-log-log-style}
    \end{minipage}\hfill

\end{figure}

\begin{figure*}[h!]
        \centering
        \begin{subfigure}{.45\textwidth}
          \centering
              \includegraphics[width=\textwidth]{Figures/Stylised Facts/PriceImpact4.png}
              \caption{Calibrated ABM price impact curves}
              \label{fig:ABMpriceimpact}
        \end{subfigure}%
        \begin{subfigure}{.45\textwidth}
          \centering
          \includegraphics[width=\textwidth]{Figures/Stylised Facts/PriceImpactJSE.png}
          \caption{Measured market price impact curves}
          \label{fig:JSEpriceimpact}
        \end{subfigure}
        \caption{The price impact curves are shown for the ABM (Figure \ref{fig:ABMpriceimpact}) and the real world observations from the JSE (Figure \ref{fig:JSEpriceimpact}). They are both plotted on a log-log scale and the simulated and empirical price impact curves both capture the linear upward trend, meaning that there is a power-law relationship between the volume traded and the price impact caused by the trade. These findings are consistent with previous empirical research \cite{gebbie2017deviations,gebbiechangjericevich2020comparing,LFM2003} as well as previous ABM results \cite{arxiv2021abm}. The key differences relate to the impact of many small trades, which became particularly noticeable after the removal of the price floor in the JSE pricing model\cite{gebbie2017deviations}}
        \label{price-impact-style}
\end{figure*}

\section{Sensitivity Analysis} \label{app:sensitivity}

Due to computational constraints, six parameters were chosen to be varied in the sensitivity analysis. These are the free parameters defined in Table 1 in the paper and were informed by prior work \cite{arxiv2021abm}. The parameters that were held constant during the sensitivity analysis as well as their values are also shown. The $\lambda_{min}$ and $\lambda_{max}$ parameters were chosen in such a way that a chartist agent has an equal probability of having a long or short trading horizon, and the time to cancellation, of the LPs limit-orders, was arbitrarily tuned to ensure that liquidity is not being removed too quickly from the order book, which would cause frequent liquidity crashes, and having too slow a cancellation time, which would cause almost no cancellations. This would cause behaviour that does not represent that of a high-frequency agent, which the LPs are trying to model. The total simulation time, after initialisation, $T$ was set at 25 seconds in machine time, because it was the time that produced order counts most similar to the day of trading on the JSE that we will use to calibrate the model. On an Average Daily Volume (ADV) basis 25 seconds of machine time simulation is considered equivalent to both 8 hours of typical calendar time trading, and on an ADV count, as a measure of the typical number of volume time events, the volume count is equivalent. Here simulation time is measured in machine time.  

Comparing the free parameters in our model with that of \cite{arxiv2021abm}, it can be seen that We are varying the number of chartists and the number of fundamentalists while keeping the number of high-frequency agents constant. However, \cite{arxiv2021abm} kept the number of chartists and fundamentalists constant, at one agent each, and varied the number of high-frequency agents to determine how the ratio of liquidity providers to liquidity takers, $N = \frac{N_{_{\mathrm{LP}}}}{N_{_{\mathrm{LT}}}}$ , changed the simulated price paths. During initial testing of the model, it was observed that having different values for $N^{f}_{_{\mathrm{LT}}}$ and $N^{c}_{_{\mathrm{LT}}}$ changed the behaviour of the simulated price paths. For example, when there are more chartists ($N^{c}_{_{\mathrm{LT}}}=12$) than fundamentalists ($N^{f}_{_{\mathrm{LT}}} = 5$) the log-return distribution has a smaller variance and but has a lower tail-index when compared with more fundamentalists and less chartists. A possible reason for this is there are fewer large orders placed by fundamentalists and smaller orders placed by the chartists on account of equations 3 and 4 in the paper. The initial lags in the trade-sign autocorrelation are larger when there are more chartists and fundamentalists, which is possibly attributed to greater herding behaviour. Furthermore, as the number of fundamentalists increases more large orders are placed, and this caused wider spreads, which would give rise to more extreme price moves.

Due to these observed changes, it was decided that a better parameter set would be to keep the number of high-frequency agents constant and vary both the number of chartists and fundamentalists. Since we are still changing the ratio of liquidity takers to liquidity providers we will also be changing N and therefore, we will still be modelling how a changing N affects the moments as was done in \cite{arxiv2021abm}.

When implementing the sensitivity analysis we considered 4 different values for each parameter, and for each value, we varied all other parameters and estimated the moments of all the simulated price paths, which leads to a total of 4096 parameter combinations. Therefore, for each parameter value and each moment, there were 1024 values to be aggregated over. Another benefit of this method is that it allowed us to create parameter surfaces for the 2 way interactions for each of the parameters. These surfaces allow us to visualise how the interactions of two parameters affect the statistical properties of the simulated price paths and the objective function. These effects can be seen in Figures \ref{app:SensitivityAnalysis-ObjectiveSurfacePlot} and \ref{app:SensitivityAnalysis-MomentSurfacesPlot}. 

It must be noted that when implementing the sensitivity analysis, for each of the parameter combinations the moments were estimated for a single run of the simulation with a single seed, which ensured the same fundamental prices and the same chartist forgetting factors. This was done due to time and computational cost constraints. Therefore, these results should be considered indicative rather than presenting the exact surfaces and moment values as there will be uncertainty around these values that have not been reduced through Monte Carlo replications over multiple seeds.

To ensure that the simulations for each of the parameter combinations remain independent of each other we had to start and stop trading sessions with CoinTossX before and after each simulation was run. This means, the LOB was cleared before the start of the next session and there were no orders from the previous session that could be matched with orders from the current session.

Due to a large number of parameter combinations, after each simulation the \textit{Julia} garbage collector was run. Helping to free up memory from the \textit{Julia} program and free up memory in the JVM created by the \textit{Julia} program. This was recommended in the JavaCall documentation. They note that the pressure on the \textit{Julia} and java heaps are different and the \textit{Julia} garbage collector may not be run even if the java heap is full and memory could be freed with garbage collection. Therefore, they recommend periodically manually running the \textit{Julia} garbage collector to ensure this issue does not cause an error.

\begin{figure}[h!]
    \centering
    \includegraphics[width=15cm, height=12cm]{Figures/Sensitivity Analysis/CorrelationMatrixMicroPrice.png}
    \caption{The correlation matrix shown depicts the correlations between the parameters (Table 1 in the paper), the moments (Table \ref{moment-table}), and the optimisation objective function. There are only two large correlations namely the standard deviation of the price paths with the objective function (0.918), and the price path Hurst exponent with the number of fundamentalists trading agents (-0.699). A mild negative correlation (-0.505) is found between the number of fundamentalist traders and the KS statistic is also present.}
    \label{app:SensitivityAnalysis-CorrelationMatrixPlot}
\end{figure}

\subsection{Marginals} \label{app:SensitivityAnalysis-Marginals}

Figure \ref{app:SA-micro-price-marginals}, shows the distributions of each of the moments for different values of each of the parameters as well as the objective function. These distributions consist of holding one parameter value fixed and varying all other parameters giving us 1024 samples for each of the distributions. For almost all of the parameters, the distributions of the moments across their values do not change considerably, and there is a large overlap of the distributions. Indicating that a change to a single parameter value is not going to cause significant changes to the model. One parameter-moment combination where we do see a change in distribution when changing the parameter values is between the number of fundamentalist traders and the Hurst exponent. As we increase the number of fundamentalists the distribution of the Hurst exponent shifts to smaller values where most of the distribution lies further below 0.5. As described in Table \ref{moment-table}, this represents more mean-reversion in the time series. This is consistent with the fact that the fundamentalists follow a mean-reverting trading strategy.

Due to large outliers and large variations in the objective function, for each of the parameters, it is difficult to see a significant difference in the distributions of the objective function across the range of parameter values. The only observation we can make is that certain parameter values may be subject to producing large outliers. For example, smaller values of the $\kappa$ parameter produce larger outliers.

\begin{sidewaysfigure}
        \centering
        \begin{minipage}{.15\textwidth}
          \centering
          \includegraphics[width=3cm, height=2.5cm]{Figures/Sensitivity Analysis/DeltaObjective.png}
        \end{minipage}%
        \begin{minipage}{.15\textwidth}
          \centering
          \includegraphics[width=3cm, height=2.5cm]{Figures/Sensitivity Analysis/KappaObjective.png}
        \end{minipage}%
        \begin{minipage}{.15\textwidth}
          \centering
          \includegraphics[width=3cm, height=2.5cm]{Figures/Sensitivity Analysis/NtObjective.png}
        \end{minipage}%
        \begin{minipage}{.15\textwidth}
          \centering
          \includegraphics[width=3cm, height=2.5cm]{Figures/Sensitivity Analysis/NuObjective.png}
        \end{minipage}%
        \begin{minipage}{.15\textwidth}
          \centering
          \includegraphics[width=3cm, height=2.5cm]{Figures/Sensitivity Analysis/NvObjective.png}
        \end{minipage}%
        \begin{minipage}{.15\textwidth}
          \centering
          \includegraphics[width=3cm, height=2.5cm]{Figures/Sensitivity Analysis/SigmaVObjective.png}
        \end{minipage}
        
        \begin{minipage}{.15\textwidth}
          \centering
              \includegraphics[width=3cm, height=2.5cm]{Figures/Sensitivity Analysis/DeltaADF.png}
        \end{minipage}%
        \begin{minipage}{.15\textwidth}
          \centering
          \includegraphics[width=3cm, height=2.5cm]{Figures/Sensitivity Analysis/KappaADF.png}
        \end{minipage}%
        \begin{minipage}{.15\textwidth}
          \centering
          \includegraphics[width=3cm, height=2.5cm]{Figures/Sensitivity Analysis/NtADF.png}
        \end{minipage}%
        \begin{minipage}{.15\textwidth}
          \centering
          \includegraphics[width=3cm, height=2.5cm]{Figures/Sensitivity Analysis/NuADF.png}
        \end{minipage}%
        \begin{minipage}{.15\textwidth}
          \centering
          \includegraphics[width=3cm, height=2.5cm]{Figures/Sensitivity Analysis/NvADF.png}
        \end{minipage}%
        \begin{minipage}{.15\textwidth}
          \centering
          \includegraphics[width=3cm, height=2.5cm]{Figures/Sensitivity Analysis/SigmaVADF.png}
        \end{minipage}

        \begin{minipage}{.15\textwidth}
          \centering
          \includegraphics[width=3cm, height=2.5cm]{Figures/Sensitivity Analysis/DeltaGARCH.png}
        \end{minipage}%
        \begin{minipage}{.15\textwidth}
          \centering
          \includegraphics[width=3cm, height=2.5cm]{Figures/Sensitivity Analysis/KappaGARCH.png}
        \end{minipage}%
        \begin{minipage}{.15\textwidth}
          \centering
          \includegraphics[width=3cm, height=2.5cm]{Figures/Sensitivity Analysis/NtGARCH.png}
        \end{minipage}%
        \begin{minipage}{.15\textwidth}
          \centering
          \includegraphics[width=3cm, height=2.5cm]{Figures/Sensitivity Analysis/NuGARCH.png}
        \end{minipage}%
        \begin{minipage}{.15\textwidth}
          \centering
          \includegraphics[width=3cm, height=2.5cm]{Figures/Sensitivity Analysis/NvGARCH.png}
        \end{minipage}%
        \begin{minipage}{.15\textwidth}
          \centering
          \includegraphics[width=3cm, height=2.5cm]{Figures/Sensitivity Analysis/SigmaVGARCH.png}
        \end{minipage}
        
        \begin{minipage}{.15\textwidth}
          \centering
          \includegraphics[width=3cm, height=2.5cm]{Figures/Sensitivity Analysis/DeltaGPH.png}
        \end{minipage}%
        \begin{minipage}{.15\textwidth}
          \centering
          \includegraphics[width=3cm, height=2.5cm]{Figures/Sensitivity Analysis/KappaGPH.png}
        \end{minipage}%
        \begin{minipage}{.15\textwidth}
          \centering
          \includegraphics[width=3cm, height=2.5cm]{Figures/Sensitivity Analysis/NtGPH.png}
        \end{minipage}%
        \begin{minipage}{.15\textwidth}
          \centering
          \includegraphics[width=3cm, height=2.5cm]{Figures/Sensitivity Analysis/NuGPH.png}
        \end{minipage}%
        \begin{minipage}{.15\textwidth}
          \centering
          \includegraphics[width=3cm, height=2.5cm]{Figures/Sensitivity Analysis/NvGPH.png}
        \end{minipage}%
        \begin{minipage}{.15\textwidth}
          \centering
          \includegraphics[width=3cm, height=2.5cm]{Figures/Sensitivity Analysis/SigmaVGPH.png}
        \end{minipage}
        
        \begin{minipage}{.15\textwidth}
          \centering
          \includegraphics[width=3cm, height=2.5cm]{Figures/Sensitivity Analysis/DeltaHill.png}
        \end{minipage}%
        \begin{minipage}{.15\textwidth}
          \centering
          \includegraphics[width=3cm, height=2.5cm]{Figures/Sensitivity Analysis/KappaHill.png}
        \end{minipage}%
        \begin{minipage}{.15\textwidth}
          \centering
          \includegraphics[width=3cm, height=2.5cm]{Figures/Sensitivity Analysis/NtHill.png}
        \end{minipage}%
        \begin{minipage}{.15\textwidth}
          \centering
          \includegraphics[width=3cm, height=2.5cm]{Figures/Sensitivity Analysis/NuHill.png}
        \end{minipage}%
        \begin{minipage}{.15\textwidth}
          \centering
          \includegraphics[width=3cm, height=2.5cm]{Figures/Sensitivity Analysis/NvHill.png}
        \end{minipage}%
        \begin{minipage}{.15\textwidth}
          \centering
          \includegraphics[width=3cm, height=2.5cm]{Figures/Sensitivity Analysis/SigmaVHill.png}
        \end{minipage}
        
        \begin{minipage}{.15\textwidth}
          \centering
          \includegraphics[width=3cm, height=2.5cm]{Figures/Sensitivity Analysis/DeltaHurst.png}
        \end{minipage}%
        \begin{minipage}{.15\textwidth}
          \centering
          \includegraphics[width=3cm, height=2.5cm]{Figures/Sensitivity Analysis/KappaHurst.png}
        \end{minipage}%
        \begin{minipage}{.15\textwidth}
          \centering
          \includegraphics[width=3cm, height=2.5cm]{Figures/Sensitivity Analysis/NtHurst.png}
        \end{minipage}%
        \begin{minipage}{.15\textwidth}
          \centering
          \includegraphics[width=3cm, height=2.5cm]{Figures/Sensitivity Analysis/NuHurst.png}
        \end{minipage}%
        \begin{minipage}{.15\textwidth}
          \centering
          \includegraphics[width=3cm, height=2.5cm]{Figures/Sensitivity Analysis/NvHurst.png}
        \end{minipage}%
        \begin{minipage}{.15\textwidth}
          \centering
          \includegraphics[width=3cm, height=2.5cm]{Figures/Sensitivity Analysis/SigmaVHurst.png}
        \end{minipage}
        
        \begin{minipage}{.15\textwidth}
          \centering
          \includegraphics[width=3cm, height=2.5cm]{Figures/Sensitivity Analysis/DeltaKS.png}
        \end{minipage}%
        \begin{minipage}{.15\textwidth}
          \centering
          \includegraphics[width=3cm, height=2.5cm]{Figures/Sensitivity Analysis/KappaKS.png}
        \end{minipage}%
        \begin{minipage}{.15\textwidth}
          \centering
          \includegraphics[width=3cm, height=2.5cm]{Figures/Sensitivity Analysis/NtKS.png}
        \end{minipage}%
        \begin{minipage}{.15\textwidth}
          \centering
          \includegraphics[width=3cm, height=2.5cm]{Figures/Sensitivity Analysis/NuKS.png}
        \end{minipage}%
        \begin{minipage}{.15\textwidth}
          \centering
          \includegraphics[width=3cm, height=2.5cm]{Figures/Sensitivity Analysis/NvKS.png}
        \end{minipage}%
        \begin{minipage}{.15\textwidth}
          \centering
          \includegraphics[width=3cm, height=2.5cm]{Figures/Sensitivity Analysis/SigmaVKS.png}
        \end{minipage}
        \caption{Plots the distributions of each of the moments and the objective function for different values of each of the parameters. The columns and rows represent the values of the moments and parameters respectively. These distributions consist of holding one parameter value fixed and varying all other parameters giving us 1024 samples for each of the distributions.}
        \label{app:SA-micro-price-marginals}
\end{sidewaysfigure}

\subsection{Objective surfaces} \label{app:SensitivityAnalysis-ObjectiveSurfaceSection}

A better way to visualise the changes in the objective function as the parameters change is to look at the behaviour of the objective function for an interaction of two variables. Figure \ref{app:SensitivityAnalysis-ObjectiveSurfacePlot} plots the micro-price objective function when holding two variables fixed and averaging over all other combinations of the other parameters. Recall that the sensitivity analysis was run once for a single seed so there were no Monte Carlo replications. The surfaces are also course grained and made up of a 4x4 grid. Therefore, these surfaces are just an indicative representation of the objective function and are not to be interpreted as the true objective function. Despite this these surfaces let us see some indicative patterns.

In almost all the plots we have that the objective function decreases from areas of extremely high values to much lower values, giving evidence that optimisation is possible. However, optimisation of the objective function will not be trivial since there is evidence for local minima where the optimisation can get stuck. For example, consider the surface for parameters $N^{f}_{_{\mathrm{LT}}}$ and $\nu$. The optimisation algorithm could possible get stuck in the local minima at $N^{f}_{_{\mathrm{LT}}}=9$ and $\nu=7$ rather than reaching the actual minimum at $N^{f}_{_{\mathrm{LT}}}=3$ and $\nu=2$. Another reason that the objective function will be difficult to minimise is that the minima that the optimiser converges to may depend heavily on the initial state of the optimiser. For an extreme example, consider the interaction surface between $N^{f}_{_{\mathrm{LT}}}$ and $\delta$. Depending on which side of the saddle you are on will cause you to find different minima. These two issues make it extremely hard to calibrate the ABM. Calibration difficulties are not new and for detailed discussions on the difficulty of calibrating ABMs, we refer the reader to \cite{plattgebbie2016problem, platt2020comparison}.
\begin{figure}[h!]
        \centering
        \begin{minipage}{.3\textwidth}
          \centering
              \includegraphics[width=6cm, height=4cm]{Figures/Sensitivity Analysis/DeltaKappaObjective.png}
        \end{minipage}%
        \begin{minipage}{.3\textwidth}
          \centering
          \includegraphics[width=6cm, height=4cm]{Figures/Sensitivity Analysis/DeltaNuObjective.png}
        \end{minipage}
        \begin{minipage}{.3\textwidth}
          \centering
          \includegraphics[width=6cm, height=4cm]{Figures/Sensitivity Analysis/DeltaSigmaVObjective.png}
        \end{minipage}
        \begin{minipage}{.3\textwidth}
          \centering
          \includegraphics[width=6cm, height=4cm]{Figures/Sensitivity Analysis/KappaNuObjective.png}
        \end{minipage}%
        \begin{minipage}{.3\textwidth}
          \centering
          \includegraphics[width=6cm, height=4cm]{Figures/Sensitivity Analysis/KappaSigmaVObjective.png}
        \end{minipage}%
        \begin{minipage}{.3\textwidth}
          \centering
          \includegraphics[width=6cm, height=4cm]{Figures/Sensitivity Analysis/NtDeltaObjective.png}
        \end{minipage}
        \begin{minipage}{.3\textwidth}
          \centering
          \includegraphics[width=6cm, height=4cm]{Figures/Sensitivity Analysis/NtKappaObjective.png}
        \end{minipage}%
        \begin{minipage}{.3\textwidth}
          \centering
          \includegraphics[width=6cm, height=4cm]{Figures/Sensitivity Analysis/NtNuObjective.png}
        \end{minipage}%
        \begin{minipage}{.3\textwidth}
          \centering
          \includegraphics[width=6cm, height=4cm]{Figures/Sensitivity Analysis/NtNvObjective.png}
        \end{minipage}
         \begin{minipage}{.3\textwidth}
          \centering
          \includegraphics[width=6cm, height=4cm]{Figures/Sensitivity Analysis/NtSigmaVObjective.png}
        \end{minipage}%
        \begin{minipage}{.3\textwidth}
          \centering
          \includegraphics[width=6cm, height=4cm]{Figures/Sensitivity Analysis/NuSigmaVObjective.png}
        \end{minipage}%
        \begin{minipage}{.3\textwidth}
          \centering
          \includegraphics[width=6cm, height=4cm]{Figures/Sensitivity Analysis/NvDeltaObjective.png}
        \end{minipage}
        \begin{minipage}{.3\textwidth}
          \centering
          \includegraphics[width=6cm, height=4cm]{Figures/Sensitivity Analysis/NvKappaObjective.png}
        \end{minipage}%
        \begin{minipage}{.3\textwidth}
          \centering
          \includegraphics[width=6cm, height=4cm]{Figures/Sensitivity Analysis/NvNuObjective.png}
        \end{minipage}%
        \begin{minipage}{.3\textwidth}
          \centering
          \includegraphics[width=6cm, height=4cm]{Figures/Sensitivity Analysis/NvSigmaVObjective.png}
        \end{minipage}
        \caption{Plots the micro-price objective surfaces for all the two-way interactions of the parameters. For each parameter combination, two of the parameters are held fixed and we average over the micro-price objective function for all other parameters. The figure demonstrates that there are regions in the objective function that have higher and lower values and therefore parameters can be differentiated by their objective function value and the function can be optimised. It also shows that there are local minima in the objective surface that will have to be taken into account when performing optimisation.}
        \label{app:SensitivityAnalysis-ObjectiveSurfacePlot}
\end{figure}

\subsection{Moment surfaces} \label{app:SensitivityAnalysis-MomentSurfacesSection}


To show how the interaction of parameters affects the moments of the price paths, moment surfaces were produced in the same way as was done for the objective function. If the mean and standard deviation are not considered we have six moments and six parameters, which means we will have 90 surfaces. This is far too many surfaces to present here so we have considered just the surfaces for parameter combinations of $N^{c}_{_{\mathrm{LT}}}$ and $N^{f}_{_{\mathrm{LT}}}$ as well as $\kappa$ and $\nu$. These can be found in Figure \ref{app:SensitivityAnalysis-MomentSurfacesPlot}. Considering the Hurst exponent surface for the $N^{c}_{_{\mathrm{LT}}}$ and $N^{f}_{_{\mathrm{LT}}}$ parameters we can see that the Hurst exponent decrease drastically as the number of fundamentalists increase. There is also a small decrease in the Hurst exponent as the number of chartists increases but this is a much smaller effect and should not be interpreted as a trend. From this plot, there is evidence that the mean reversion of the fundamentalists is dominating the momentum from chartists. To illustrate some of the non-linear effects not captured by the correlations, consider the Hill estimator surface for the $N^{c}_{_{\mathrm{LT}}}$ and $N^{f}_{_{\mathrm{LT}}}$ parameters as well as the $\kappa$ and $\nu$ parameters. 

\begin{figure}[h!]
        \centering
        \begin{minipage}{.3\textwidth}
          \centering
              \includegraphics[width=6cm, height=4cm]{Figures/Sensitivity Analysis/NtNvADF.png}
        \end{minipage}%
        \begin{minipage}{.3\textwidth}
          \centering
          \includegraphics[width=6cm, height=4cm]{Figures/Sensitivity Analysis/NtNvGARCH.png}
        \end{minipage}%
        \begin{minipage}{.3\textwidth}
          \centering
          \includegraphics[width=6cm, height=4cm]{Figures/Sensitivity Analysis/NtNvGPH.png}
        \end{minipage}
        \begin{minipage}{.3\textwidth}
          \centering
          \includegraphics[width=6cm, height=4cm]{Figures/Sensitivity Analysis/NtNvHill.png}
        \end{minipage}%
        \begin{minipage}{.3\textwidth}
          \centering
          \includegraphics[width=6cm, height=4cm]{Figures/Sensitivity Analysis/NtNvHurst.png}
        \end{minipage}%
        \begin{minipage}{.3\textwidth}
          \centering
          \includegraphics[width=6cm, height=4cm]{Figures/Sensitivity Analysis/NtNvKS.png}
        \end{minipage}
        \begin{minipage}{.3\textwidth}
          \centering
          \includegraphics[width=6cm, height=4cm]{Figures/Sensitivity Analysis/KappaNuADF.png}
        \end{minipage}%
        \begin{minipage}{.3\textwidth}
          \centering
          \includegraphics[width=6cm, height=4cm]{Figures/Sensitivity Analysis/KappaNuGARCH.png}
        \end{minipage}%
         \begin{minipage}{.3\textwidth}
          \centering
          \includegraphics[width=6cm, height=4cm]{Figures/Sensitivity Analysis/KappaNuGPH.png}
        \end{minipage}
        \begin{minipage}{.3\textwidth}
          \centering
          \includegraphics[width=6cm, height=4cm]{Figures/Sensitivity Analysis/KappaNuHill.png}
        \end{minipage}%
        \begin{minipage}{.3\textwidth}
          \centering
          \includegraphics[width=6cm, height=4cm]{Figures/Sensitivity Analysis/KappaNuHurst.png}
        \end{minipage}%
        \begin{minipage}{.3\textwidth}
          \centering
          \includegraphics[width=6cm, height=4cm]{Figures/Sensitivity Analysis/KappaNuKS.png}
        \end{minipage}
        \caption{Micro-price moment surfaces show how the estimated moments change on average when varying two of the parameters. Due to a large number of moment surfaces we have just considered the surfaces for parameter combinations $N^{c}_{_{\mathrm{LT}}}$ and $N^{f}_{_{\mathrm{LT}}}$ as well as $\kappa$ and $\nu$. Looking at the Hurst exponent surface for the $N^{c}_{_{\mathrm{LT}}}$ and $N^{f}_{_{\mathrm{LT}}}$ parameters we can see that the Hurst exponent decreases drastically as the number of fundamentalists increases, which is consistent with the fact that the fundamentalists employ a mean reverting strategy.}
        \label{app:SensitivityAnalysis-MomentSurfacesPlot}
\end{figure}

\section{Calibration} \label{app:Calibration}

\begin{table}[h!]
\centering
\caption{The moments used to characterise the price paths. These are  used in the sensitivity analysis to measure how changes in the parameters change the price paths, and they are used to calibrate the ABM to measured market data.}
\label{moment-table}
\begin{tabular}{lp{0.8\textwidth}}
\toprule
Moments & Description \\
\toprule
Mean, Std. & The mean, $\mu$, and standard deviation, $\sigma$, measure the shape of the log-return distribution. \\
\hline
KS & The Kolmogorov-Smirnov \cite{Massey1951}(KS) statistic is used to compare the maximum absolute difference between the CDFs of the empirical and simulated log-returns and tests if the samples are drawn from the same distribution. \\
\hline
GPH & Geweke and Porter-Hudak \cite{geweke1983estimation} (GPH) provide an estimate for the fractional integration parameter $d \in [-0.5,0.5]$ in absolute log-returns time series by estimating the memory parameter in the log-periodogram. This measures the long-range dependence in log-returns.\\
\hline
ADF & The Augmented Dickey-Fuller \cite{DickeyFuller1979} (ADF) statistic tests for a unit root in the time series of the log-returns and the more negative the test statistic the less evidence there is for the presence of a unit root. \\
\hline
GARCH & The sum of the parameters of the GARCH(1,1) model provide a useful summary for the short-range dependence in the squared returns\cite{winker2007objective}.  \\
\hline
Hill & The ``improved" Hill Estimator (HE) is used to estimate the tail-index of a power-law distribution which can be used to infer the power-law behaviour in the tails of experimental distributions\cite{nuyts2010inference}.\\
\hline
Hurst (H) & The Hurst (H) exponent \cite{Hurst1951,MandelbrotWallis1969} provides a measure of the long-memory of the log-returns. If $H \in [0,1]$ is significantly far from 0.5 we have evidence that the log-returns do not follow a random walk. Furthermore, if $H \in [0,0.5)$ then the process is said to be mean-reverting and if $H \in (0.5,1]$ the process is said to be trending. \\
\bottomrule
\end{tabular}
\end{table}

Initial ABM research manually selected parameters for the ABM such that these parameters were able to produce the required log-return stylised facts \cite{fabretti2013problem}. There were no checks to see if the distribution of the simulated data came from the same distribution as that of the empirical data \cite{fabretti2013problem}. This presents a calibration problem in which the model must be calibrated to the empirical data \cite{plattgebbie2016problem}.

The three most common methods used to calibrate financial ABMs are (1) maximum likelihood estimation, (2) Bayesian inference and, (3) method-of-moment using simulated minimum distance (MM-SMD). Maximum likelihood is reserved for those models involving closed-form solutions usually for daily sampled data \cite{plattgebbie2016problem}, approximating closing auctions and not intraday trading. Therefore, this method was deemed unsuitable for calibrating our model. The second method, Bayesian inference, constructs a posterior distribution of the parameters using either the kernel density estimation (KDE) or parametric approaches. Parametric methods have been said to often not be flexible enough to accurately characterise the distribution \cite{plattgebbie2016problem}. However, Bayesian estimation has been shown to outperform MM-SMD in a wide range of circumstances using the KDE method \cite{platt2020comparison}, but this method is extremely computationally intensive for large ABMs such as the one used in this paper, and therefore was ruled out as a calibration method. The method chosen is MM-SMD, because it is more computationally tractable than Bayesian inference, it does not require a closed-form solution, and it means that we are consistent with the methods used in \cite{arxiv2021abm}, whose model we try to replicate in event time.

MM-SMD does have some pitfalls that need to be noted. The MM-SMD method has been criticised for producing parameter degeneracies and causing haphazard effects on the objective function \cite{plattgebbie2016problem}. These haphazard effects limit the ability of the optimiser to find identifiable parameters due to the optimiser converging to local optima. From the indicative objective surfaces in the sensitivity analysis, we see that there are likely local optima in our objective function, which could present the same calibration problems found in \cite{plattgebbie2016problem}. The method has also been criticised for the arbitrary selection of moments that will only represent a limited number of aggregate properties of the data, and therefore may not capture important dynamics \cite{plattgebbie2016problem}. Despite these shortcomings, MM-SMD presents a fast, simple and reliable (when there are more moments than parameters) method for calibrating an ABM. However, due to these shortcomings, these parameters should be considered indicative rather than robust \cite{arxiv2021abm}.   

\subsection{Calibration data} \label{app:ABMCalibration-CalData}

Since we are simulating a continuous-double-auction (CDA) all data relating to the opening and closing auctions needed to be discarded. Therefore, the only activity that occurred between 9:00 and 16:50 was considered. Moreover, it was noticed that the first minute of trading produced erroneous data and it was decided to remove the first minute of trading since it would make the model more difficult to calibrate. It is also vital to remove any events related to intraday volatility auctions and the impact of various futures close-outs. Given that the future close-outs occur only in March, June, September and December we do not need to consider this case. To remove most of the unwanted trades such as after-hour trades (LT), correction of previous day's trades (LC) and auction uncrossing price trades (IP) we only considered automated trades (AT). 

Often there are larger trades that get executed against a set of smaller trades. This means that single trade, at a given timestamp, gets split up into smaller trades that change the best bid and best ask as it moves up and down the order book. This event is represented in the data as multiple trades when it is a single trade. Also, some timestamps have multiple quotes. To deal with these events we performed trade and quote compacting. This is the process of changing the data to better represent the fact that a single trade has occurred, or to ensure that there is only a single quote per timestamp. For quotes with the same timestamp, the most recent quote in the sequence of quotes is kept and the rest are removed. For trades with the same timestamp, and the same order type, the volume of the trade is calculated to be the aggregated volume for that timestamp and the price of the trade is the volume weighted average price. 

\subsection{Selection of moments} \label{app:Calibration-MomentSelection}

The first step in the construction of MM-SMD is the selection of the moments. Winker \cite{winker2007objective} recommend selecting a large number of moments taking into account a large number of characteristics of the empirical data, which will increase the probability of identification of the model parameters. The moments that have been chosen are described in Table \ref{moment-table}. These moments have been chosen as they measure several stylised facts that we aim to reproduce. A description of these stylised facts can be found in Table \ref{stylised-facts-table}. We differ from \cite{arxiv2021abm} as we do not include kurtosis as a moment. The fat-tails of the log-return distribution call into question the existence of such higher order moments and kurtosis has been shown to not be a robust moment \cite{winker2007objective}. Prior work \cite{arxiv2021abm} found that it was the least reliable moment as large kurtosis values were present when there were extreme price changes initially but the price remained unchanged for the rest of the simulation. To measure the fat tails of the return distribution the ``improved" Hill estimator was used to measure the tail-index of the upper 5\% of the right tail of the log-return distribution \cite{winker2007objective}. Removing the kurtosis moment is also valid as it still ensures that we have more moments than parameters. The standard deviation of the log-returns was kept as it is a robust statistic \cite{winker2007objective} and is far more likely to exist when compared with the kurtosis. During the initial testing of the model, the maximum likelihood estimate of the tail-index of the Pareto distribution using the upper and lower 5\% of the tails of the log-return distribution was above 2 for all simulations, however, the estimate never went above 4 again giving evidence for the existence of the second moment and the nonexistence of the fourth.

The moments were estimated on the micro-price log-returns because prior work found that using the mid-prices was unreliable when trying to identify the distributional properties\cite{arxiv2021abm}.

\subsection{Simulated minimum distance} \label{app:Calibration-MM-SMD}

The calibration problem can be defined as the optimisation of a stochastic approximation of an objective function $\min_{\boldsymbol{\theta} \in \boldsymbol{\Theta}} f(\boldsymbol{\theta})$
where $\boldsymbol{\theta}$ is a parameter vector, $\boldsymbol{\Theta}$ is the space of all possible parameters and f is the objective function. 

The parameters that we will be optimising are described in Table 1 in the paper. We have that $\boldsymbol{\theta} = \{N^{c}, N^{f}, \delta, \kappa, \nu, \sigma_{f}\}$ are the free parameters of the model. There are several parameters in the model that were fixed to ensure that the fewest and most influential parameters were the only ones that varied. Keeping the number of free parameters as low as possible reduces the dimensionality of the surface, making it easier to optimise. The maximum and minimum forgetting factors of the uniform distribution, $\lambda_{min}$ and $\lambda_{max}$, that were sampled were kept fixed. These parameters aimed to establish a range of forgetting factors where each chartist, when sampling from the uniform distribution, had an equal chance of having a long or short trading horizon. The smallest forgetting factor was set low enough such that the moving average tracked the mid-price almost perfectly and the largest forgetting factor kept the moving average reasonably constant across the simulation. Since these values have already fulfilled their role there is no need to calibrate them. The time to trade, after initialisation, was set at 25 seconds because it was deemed to give the most similar order counts when compared with the day of data we calibrated to. The time to cancellation parameter for the high-frequency agent limit orders was set at 1 second as this was the fastest time to cancellation that did not remove liquidity too quickly to cause consistent liquidity crashes. The initial mid-price was arbitrarily set at 10000 and the minimum volume for the limit orders, $x_{m}$, was reduced by half when compared to \cite{arxiv2021abm} to account for the increase in the number of limit orders due to the high-frequency agents trading in every event loop. 

The MM-SMD method considers: the true moments, estimated moments and the moments estimated from the simulated paths. $\boldsymbol{m} = (m_{1},...,m_{d})$ represent the true moments of the $d$ selected moments of the log-returns. These values can't be known and they must be estimated from a particular realisation of a price path. The estimated moments $\boldsymbol{m^{e}} = (m_{1}^{e},...,m_{d}^{e})$ are estimated from the empirical intraday price path collected from the data source. Finally, moments $\left[ \boldsymbol{m^{s}|\theta} \right]= (m_{1}^{s},...,m_{d}^{s})$ can be estimated from the simulated price path of the model given a particular set of parameters $\boldsymbol{\theta}$. Using the standard method of moments condition $\mathbb{E}\left[ \left(\boldsymbol{m^{s}|\theta}\right)\right] = \boldsymbol{m}$. If we know $\boldsymbol{m}$ then
$\mathbb{E}\left[ \left(\boldsymbol{m^{s}|\theta}\right) - \boldsymbol{m}\right] = 0$ so that from estimating the expectation over $I$ simulation runs we have that $\frac{1}{I} \sum^{I}_{i = 1} \left[\left(\boldsymbol{m_{i}^{s}|\theta}\right) - \boldsymbol{m} \right] = 0$. However, we don't have the true moments $\boldsymbol{m}$, so instead, we use the estimated moments $\boldsymbol{m^{e}}$. We define an error measure in terms of the distance between the simulated moments and the estimated moments:
\begin{align}\label{mm-g-definition}
    \hat{G}(\boldsymbol{\theta}) = \frac{1}{I} \sum^{I}_{i = 1} \left[\left(\boldsymbol{m_{i}^{s}|\theta}\right) - \boldsymbol{m^{e}} \right]
\end{align}
Here $I=5$ for the entire calibration process following \cite{plattgebbie2016problem}. The Monte Carlo replications are key to reducing the variance of the stochastic approximation of the objective function. Moreover, for each simulation run, we have that the seed was also set to $i \in \{1,...,I\}$. The seed controls the fundamental prices and the forgetting factors for each of the fundamentalists and chartists respectively. Averaging over multiple seeds is key as we want to try to average out the effects of the random fundamental prices and forgetting factors. Five Monte Carlo replications are low and this value was used due to computational constraints. Please see Table \ref{table-hardware-runtimes} that shows the runtimes to get an idea of the computational burden. In the ideal setting, each seed should be averaged over far more than 5 times to reduce the variance caused by the randomness in each simulation as well as the random fundamental prices and forgetting factors.

We can construct the full objective function as a weighted sum-of-squares of the errors defined in $G$:
\begin{align}\label{full-obj-function}
f(\boldsymbol{\theta}) =  \hat{G}(\boldsymbol{\theta})^{\mathsmaller T} \boldsymbol{W} \hat{G}(\boldsymbol{\theta}).
\end{align}
Here $\boldsymbol{W}$ is a matrix of weights. Using the method of Winker \cite{winker2007objective} we choose $\boldsymbol{W}=\Cov\left[\boldsymbol{m^{e}}\right]^{-1}$. The intuition behind the choice is that it takes into account the uncertainty of the empirically estimated moments, $\boldsymbol{m^{e}}$, by assigning larger weights to moments with greater uncertainty \cite{plattgebbie2016problem}. The matrix $\boldsymbol{W}$ is estimated using a moving block bootstrap with a window of size 2000 \cite{arxiv2021abm}. The moving block bootstrap is used to preserve the autocorrelations in the time series. The total number of moving block bootstrap samples that were obtained was 1000. Therefore, we have 1000 samples in which the empirical moments could be estimated, and 1000 samples of each moment to estimate $\boldsymbol{W}$. Since we are inverting a covariance matrix the condition number needs to be checked to determine the invertibility of the matrix. The condition number of $\Cov\left[\boldsymbol{m^{e}}\right]$ is $1.9536096786628117 \times 10^{13}$. This is concerning however it is far less than \cite{arxiv2021abm} who's condition number was $8.375811152908104 \times 10^{19}$. Therefore the matrix may only just be invertible but it is at least comparable and better than that of \cite{arxiv2021abm}.

\subsection{Optimisation} \label{app:Calibration-NMTA} 

The objective function described in Section \ref{app:Calibration-MM-SMD} is computationally expensive to evaluate. A single run of the simulation lasts 25 seconds and because the number of Monte Carlo replications was set at 5, this means that it takes approximately 125 seconds to evaluate the objective function. Therefore, the Nelder-Mead (NM) optimisation algorithm \cite{nelder1965simplex} was chosen as it reduces the number of objective function evaluations by choosing efficient search steps \cite{gilli2003global}. The roughness of the stochastic approximation of the objective function will cause lots of local minima. To avoid the rapid descent into local minima the NM optimisation algorithm is combined with the threshold accepting algorithm (TA) to give us Nelder-Mead with threshold accepting (NMTA) used in \cite{gilli2003global, arxiv2021abm}. The NM algorithm has been shown to become inefficient in higher dimensions and therefore the NM algorithm implementation with adaptive parameters was chosen to increase the efficiency of the optimisation \cite{gao2012implementing}.

Given $n$ free parameters that need to be optimised the NM algorithm creates a simplex of $n+1$ vertices, where each vertex represents a potential solution to the optimisation. At each iteration, the simplex is transformed by a reflection and either an expansion, contraction or shrinkage operator. This is repeated until either the convergence criteria are met or until the optimisation is halted by a time limit or a fixed number of iterations. As noted previously this algorithm will converge quickly to local optima if the objective surface is rough.

The TA algorithm selects a single point on the objective function and perturbs it to create a new point. This point is then evaluated and if it has a better objective function value then it is accepted as the new solution. Furthermore, if it has a worse objective function value by no more than some threshold $\tau$ of the current solution then the new solution replaces the current solution \cite{winker2007objective}. The acceptance of worse solutions allows this algorithm to escape from local optima. The threshold, $\tau$, is decreased throughout the optimisation so that convergence can occur. Each value for $\tau$ is used for a set number of steps to allow the algorithm to search the parameter space using each threshold several times \cite{arxiv2021abm}.

Combining these two algorithms gives the NMTA algorithm. At each iteration either a single step in the TA algorithm is executed or a single step in the NM algorithm is executed. A step of either the TA or NM algorithm is performed with probability $\xi$ and $1-\xi$ respectively with NM being chosen with a higher probability. The TA algorithm is applied to the vertex with the lowest objective function value and the size of the perturbation is dependent on the mean value, computed over the simplex, of the parameter being chosen \cite{arxiv2021abm}. Moreover, for each of the operations in the simplex search the objective function values are also compared using the same TA thresholding to ensure that worse solutions are selected in the NM algorithm. 

Initial calibration attempts caused parameters of the model to fall outside reasonable bounds. For, example some vertices of the simplex would contain negative values for the number of liquidity takers. This is unrealistic and bounds on the parameters would need to be placed so that the simplex does not search areas of the objective function that have no grounds in reality. Table \ref{parameters-bounds-table} describes the parameter bounds and the justification for the bounds.

\begin{table}[h!]
\centering
\caption{Describes the bounds on the parameters of the model and gives a justification as to why the particular bounds were chosen.}
\label{parameters-bounds-table}
\begin{tabular}{p{0.2\linewidth}p{0.75\linewidth}}
\toprule
Parameter Bounds & Justification \\
\toprule
$N^{c}_{_{\mathrm{LT}}}$, $N^{f}_{_{\mathrm{LT}}} \in \left[1,20 \right]$ & The lower bound is used to ensure that there are always liquidity takers. The upper bound was set such that there is a case for more liquidity takers than providers but it was not set too high such that the program would drop messages from the matching engine. \\
\hline
$\delta \in \left[0,10 \right]$ & Assumes the fundamental prices and the moving averages would never be more than 10 times that of the mid-price. This is a reasonable assumption due to the bounds on the variance on the fundamental prices and the method used to sample the forgetting factors. During calibration, this parameter never went outside this range.\\
\hline
$\kappa \in \left[0,\infty \right)$ & Was allowed to vary on the positive real line.\\
\hline
$\nu \in \left[1.5,\infty \right)$ & The lower bound was adjusted from 1 to 1.5 as values too close to 1 would cause volumes so high that there were errors when converting the sampled volume from a float to an integer. These volumes were deemed unrealistic.\\
\hline
$\sigma_{f} \in \left[0,0.05 \right]$ & Ensures that the fundamental prices are within approximately 10\% of the mid-price. It was found that increasing this further did not make a difference as this value was rarely truncated and the prices did not move more than 10\% over the course of the calibrated simulations.\\ 
\bottomrule
\end{tabular}
\end{table}
Due to the bounds on the parameters the initial vertex was set at the mean values for each of the parameters and an affine transformation was applied to this vertex to give the initial simplex. The simulation was run for 100 iterations as it was determined that the simplex vertices converged to a state where the parameters in the vertices had stopped changing in a meaningful way at this point. The objective function was also shown to have converged to an optima in this time.

\subsection{Calibration convergence results} \label{app:Calibration-Convergence-Results}

\begin{figure}[h!]
        \centering
        \begin{subfigure}{.5\textwidth}
          \centering
              \includegraphics[width=9cm, height=6cm]{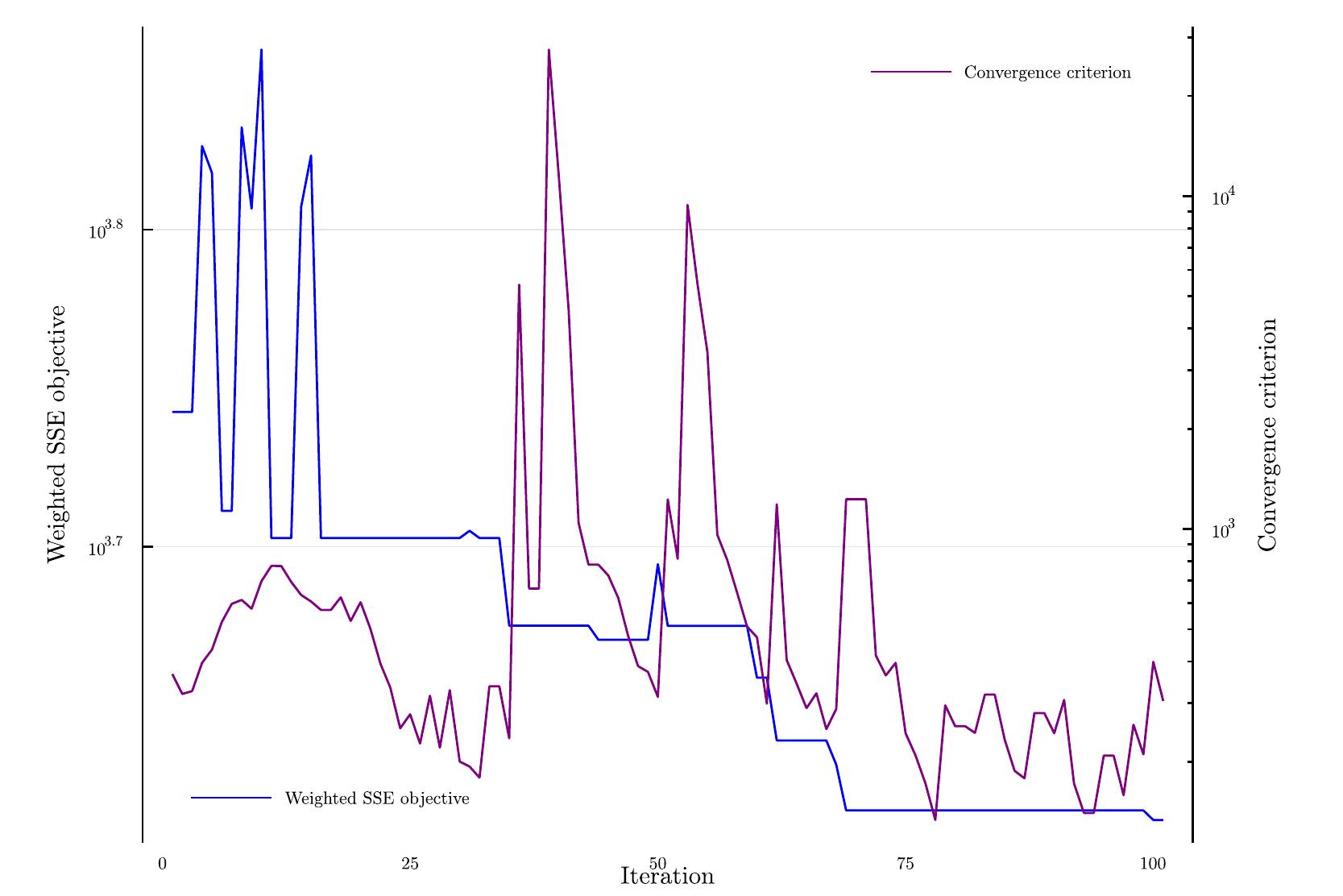}
              \caption{Best vertex convergence}
        \end{subfigure}%
        \begin{subfigure}{.4\textwidth}
          \centering
          \includegraphics[width=8cm, height=6cm]{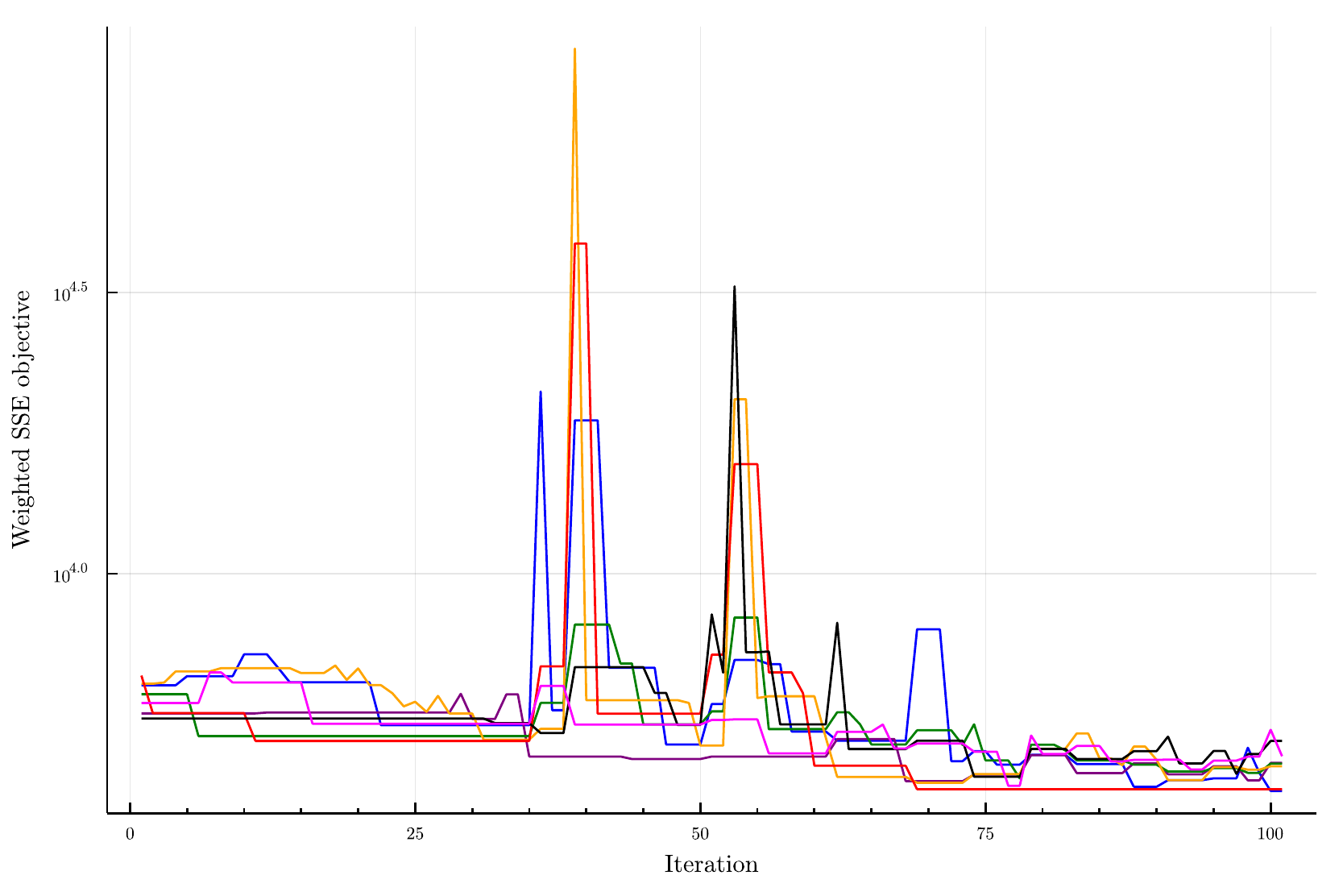}
          \caption{Vertex objective function convergence}
        \end{subfigure}
        \caption{The left plot shows the convergence for the best vertex over the number of iterations (blue) as well as the estimate for the scaled standard deviation of the objective function values of the entire simplex (purple). The purple line can be thought of as the convergence criteria of the objective function because as this decreases this means that the simplex vertices' objective function values are becoming more similar and are therefore likely closer on the objective surface. The plot on the right shows the convergence of the objective function for each of the vertices on the simplex. The objective function value for the best vertex does seem to have converged to an optima because there has only been a very small improvement in the last 32 iterations. The last 35 iterations had no thresholding applied. The convergence criteria have also decreased as the simplex values seem to have more similar objective function values. This is also shown in the right-hand side plot of the figure.}
        \label{obj-convergence}
\end{figure}

The plot on the left of Figure \ref{obj-convergence} depicts the convergence of the objective function for the best vertex over the number of iterations (blue) as well as the estimate for the scaled standard deviation of the objective function values over the entire simplex (purple). The purple line can be thought of as the convergence criteria of the objective function because as this decreases this means that the simplex values are becoming more similar in their objective function values and are therefore likely closer together on the objective function. The plot on the right shows the convergence of the objective function for each of the vertices on the simplex. The objective function value for the best vertex does seem to have converged to an optima because there has only been a very small improvement in the last 32 iterations. The last 35 iterations had no thresholding applied. The convergence criteria has also decreased as the simplex vertices seem to have more similar objective function values. This is also shown in the right-hand side plot of the figure. Since the objective surface is stochastic and most likely rough, there will always be noise in the objective function values even if they are very close in the parameter space. Therefore, just looking at the objective function convergence will not provide the full picture of convergence. The key question is whether the vertices in the simplex are similar enough where any of the simplex transformations or threshold accepting perturbations will cause the simplex to continue to move around the parameter space. 

\begin{figure}[h!]
        \centering
        \begin{subfigure}{.3\textwidth}
          \centering
              \includegraphics[width=6cm, height=4cm]{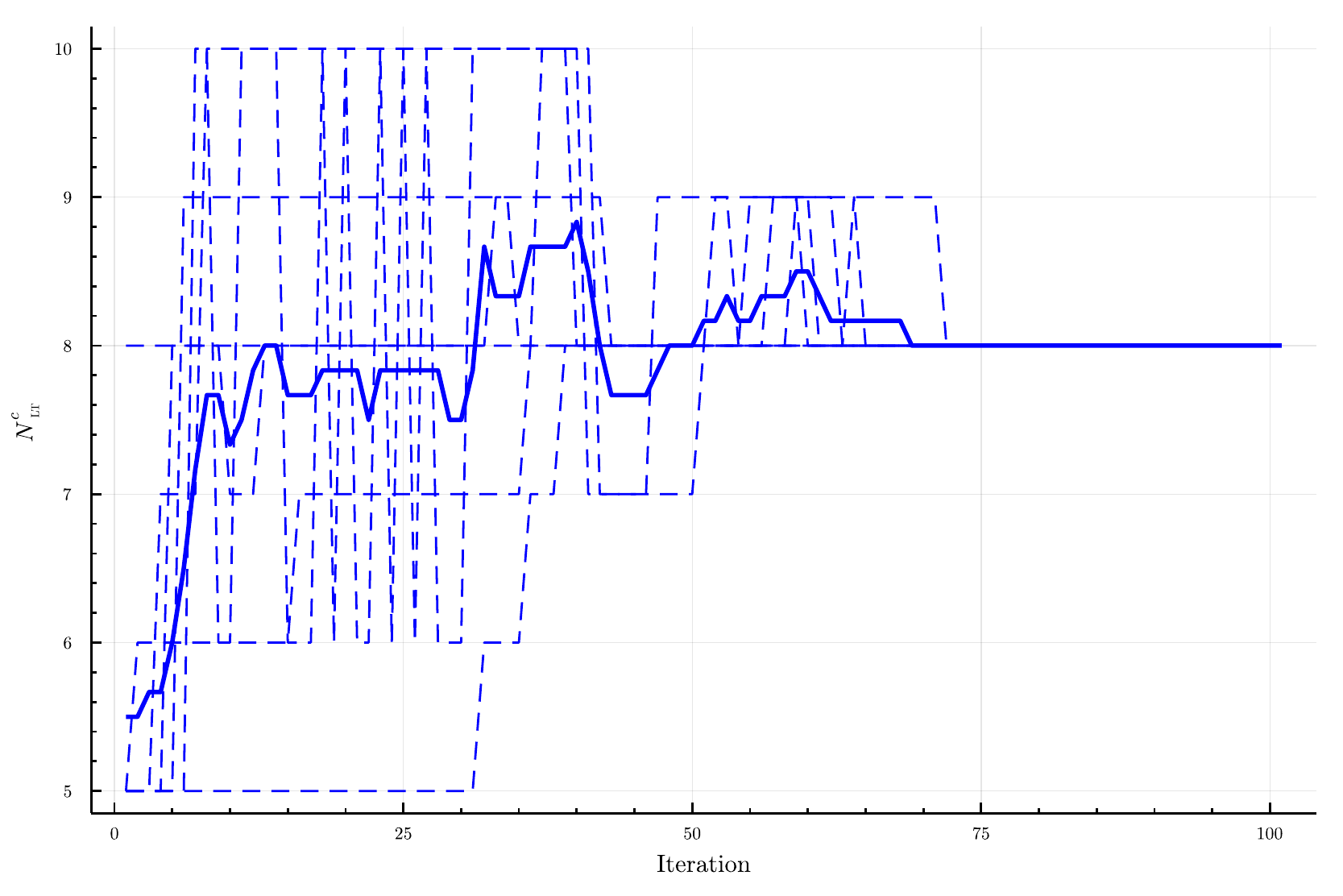}
              \caption{$N^{c}_{_{\mathrm{LT}}}$}
        \end{subfigure}%
        \begin{subfigure}{.3\textwidth}
          \centering
          \includegraphics[width=6cm, height=4cm]{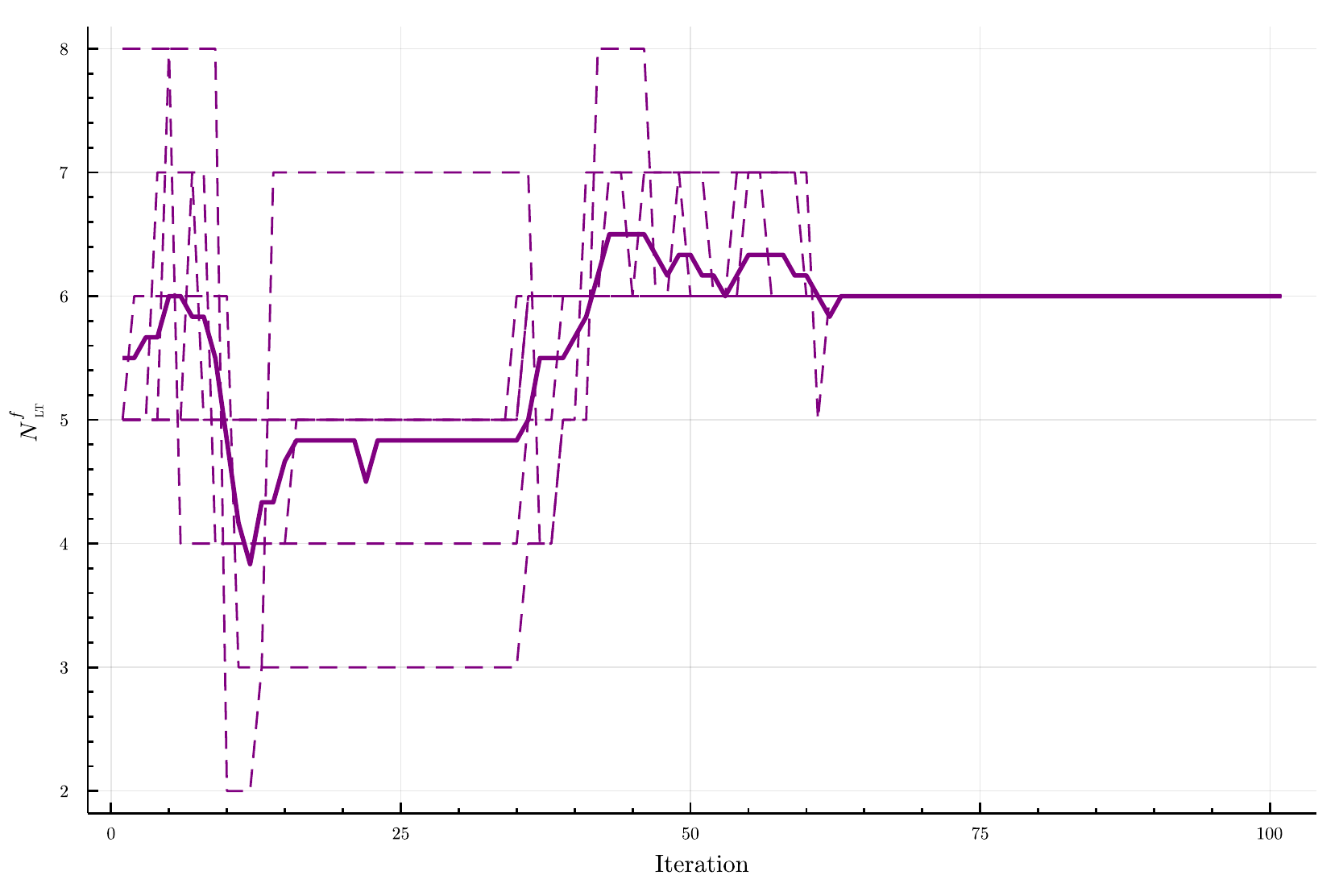}
          \caption{$N^{f}_{_{\mathrm{LT}}}$}
        \end{subfigure}%
        \begin{subfigure}{.3\textwidth}
          \centering
          \includegraphics[width=6cm, height=4cm]{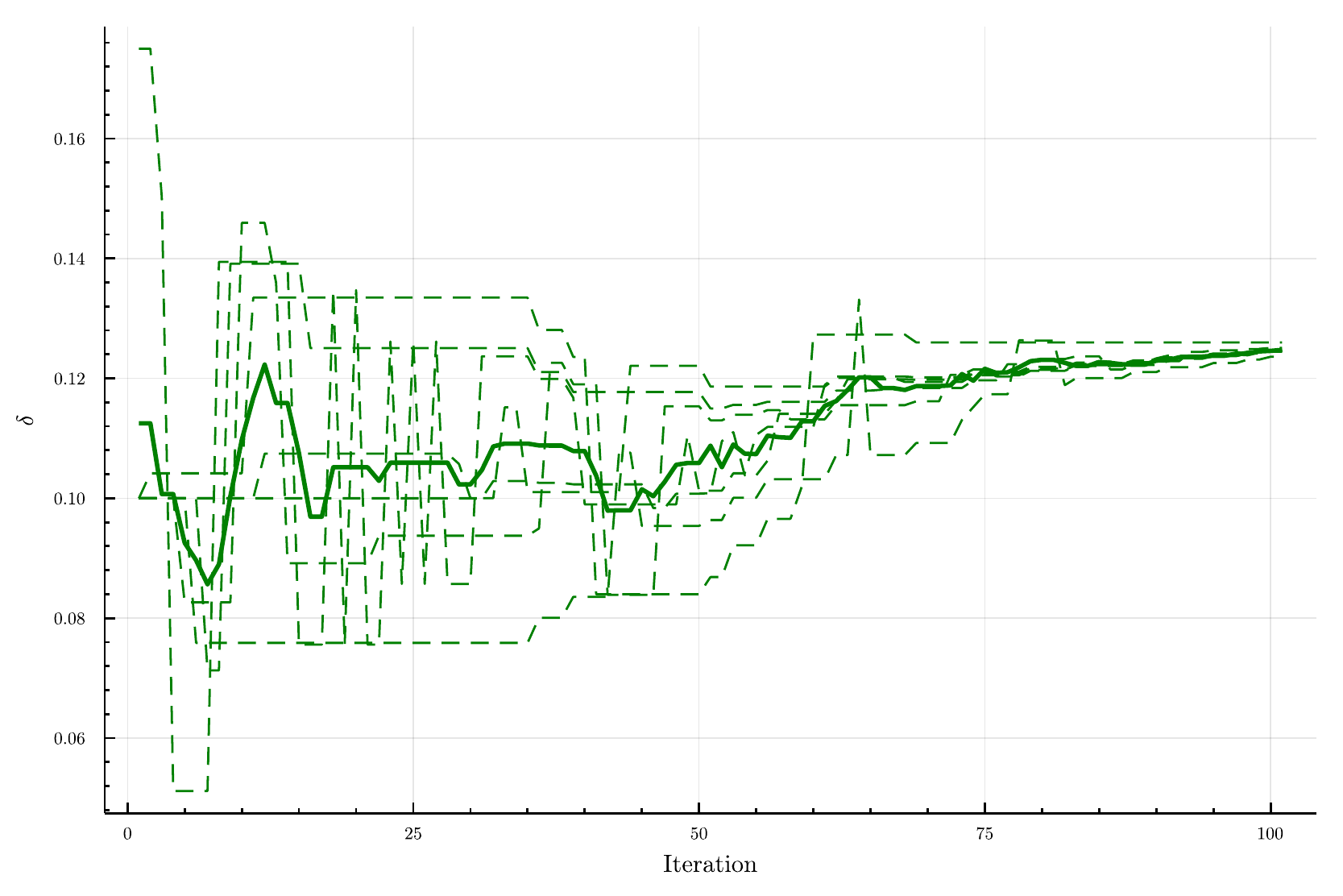}
          \caption{$\delta$}
        \end{subfigure}
        \begin{subfigure}{.3\textwidth}
          \centering
          \includegraphics[width=6cm, height=4cm]{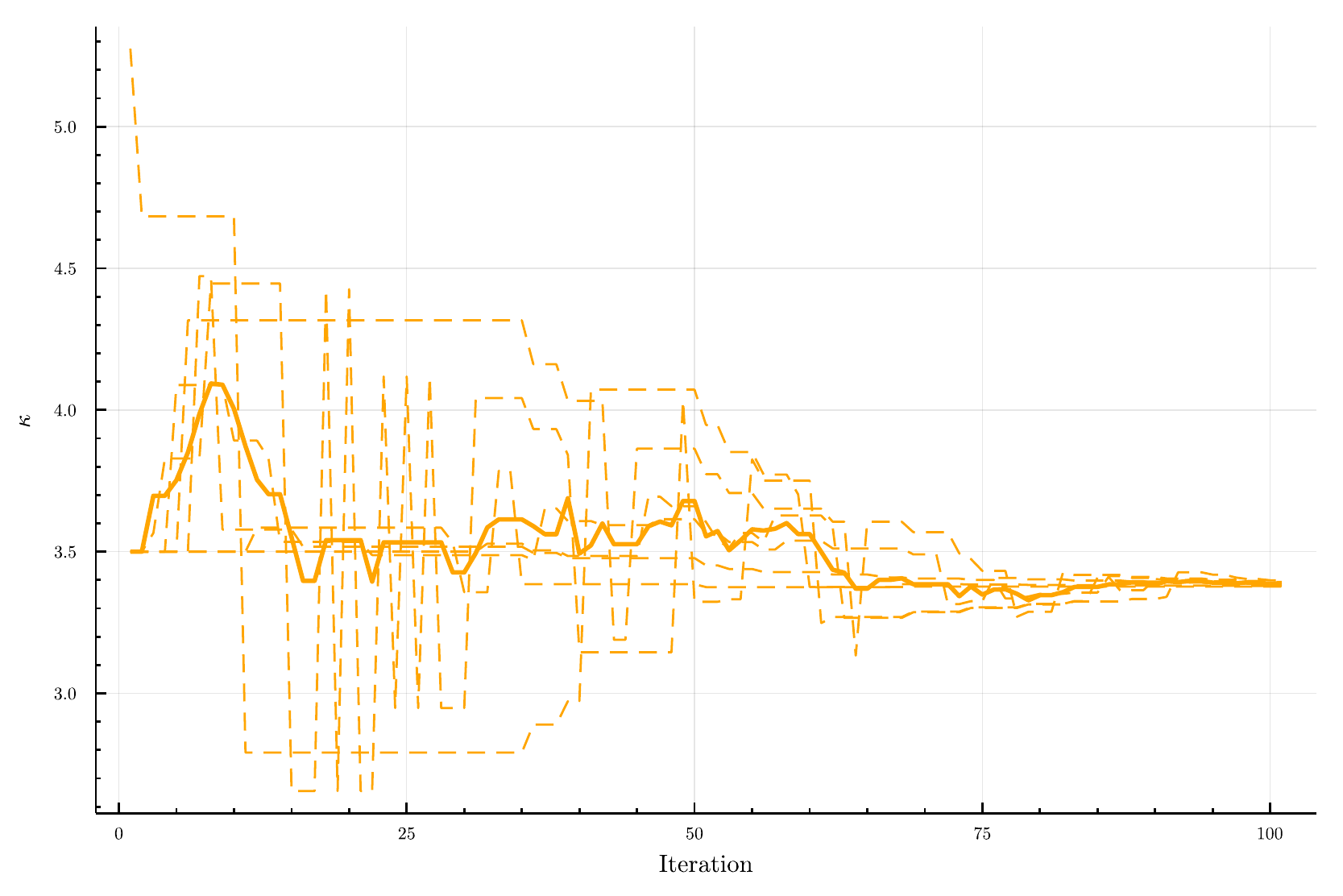}
          \caption{$\kappa$}
        \end{subfigure}%
        \begin{subfigure}{.3\textwidth}
          \centering
          \includegraphics[width=6cm, height=4cm]{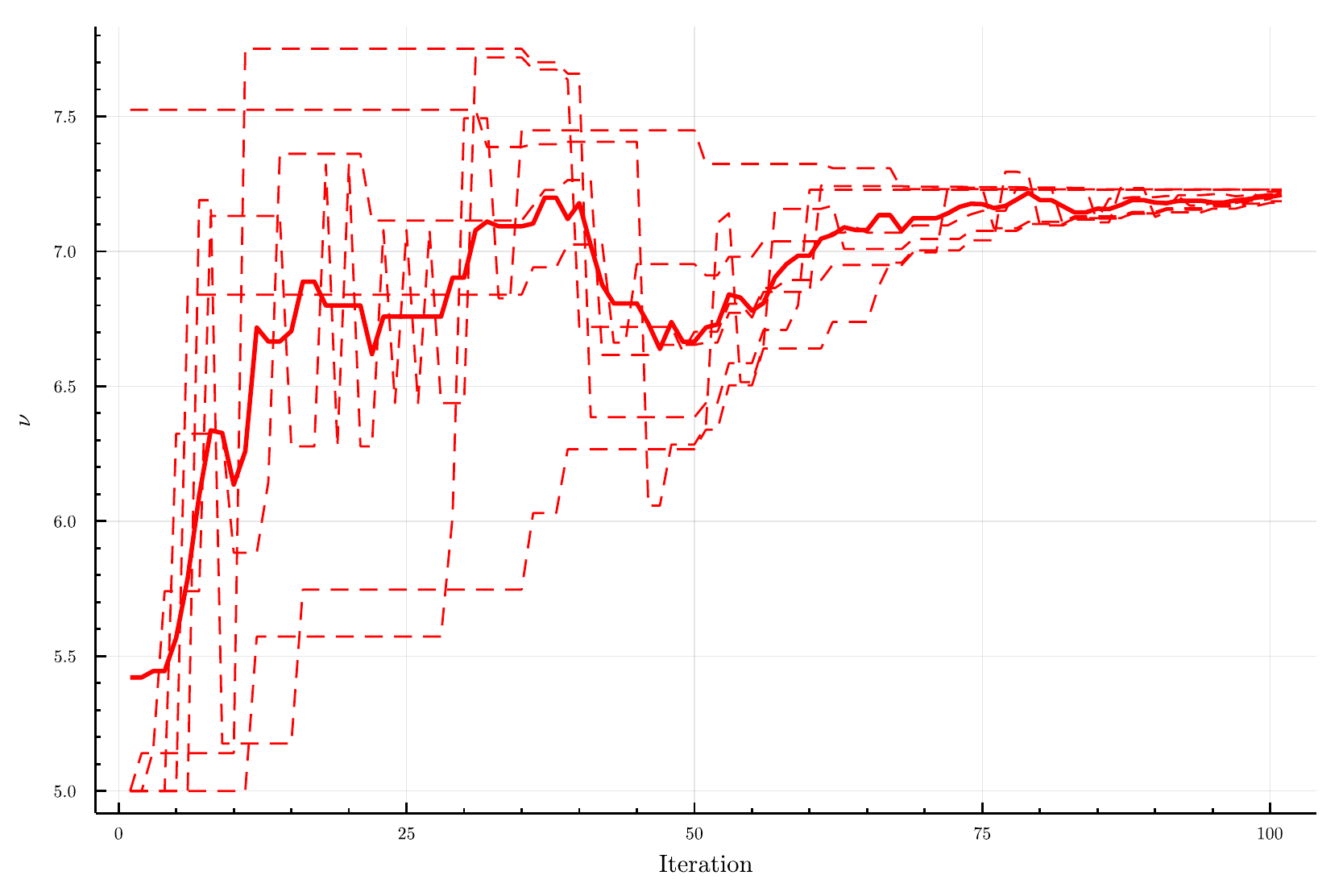}
          \caption{$\nu$}
        \end{subfigure}%
        \begin{subfigure}{.3\textwidth}
          \centering
          \includegraphics[width=6cm, height=4cm]{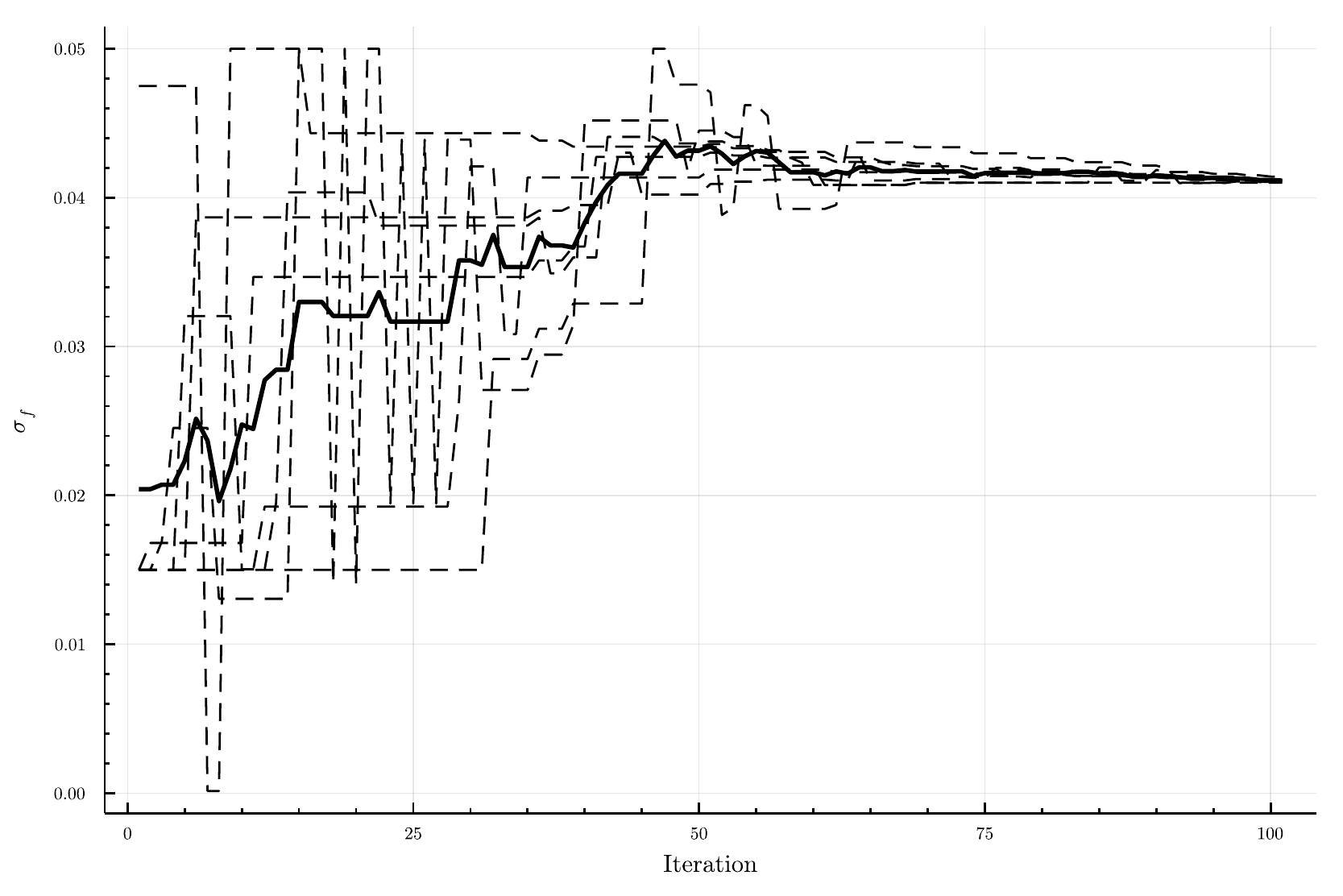}
          \caption{$\sigma_{f}$}
        \end{subfigure}
        \caption{To determine if the parameters converged we plot trace plots for the parameters. The parameter value for each of the vertices in the simplex (dashed lines) and the parameter value for the centroid of the simplex (solid lines) are plotted in these trace plots. Throughout the optimisation, we have that the simplex has converged on the objective surface because by the end of the calibration all the simplex vertices have essentially a single parameter value. This means that if the optimisation process was run for further iterations there would be no significant changes in the vertices of the simplex.}
        \label{param-convergence}
\end{figure}

Figure \ref{param-convergence} shows the parameter value for each of the vertices in the simplex (dashed lines) and the parameter value for the centroid of the simplex (solid line). These plots show that throughout the optimisation we have that each parameter value on each vertex has converged to the same place on the objective surface because by the end of the calibration all the simplex vertices have essentially the same parameter values. This means that if the optimisation process was run for further iterations there would be no significant changes in the vertices of the simplex. The simplex search transformations are based on the difference in the parameter values of the simplex and the centroid and because they are so similar there would not be significant changes to the vertices and the algorithm would become a threshold accepting algorithm. The threshold accepting steps are run with probability $\xi$ which is much smaller than $(1-\xi)$ and therefore will only be run very few times and the optimisation will become very inefficient to continue the search as the simplex will change very slowly if at all for the rest of the algorithm. For these reasons, the optimisation was halted at 100 iterations. It can also be noted that there is variance in the objective function values of each vertex in the simplex, in the last few iterations, even though the parameter values are not significantly different. Highlighting the stochastic and rough nature of the objective surface and its difficulty to optimise. 

The right-hand plot in Figure \ref{obj-convergence} has large spikes in the value of the objective function. These spikes are due to the shrink operator of the simplex algorithm. For a deterministic smooth function, the shrinkage operator will shrink the simplex closer to the minimum as it assumes it is closer to an optima. However, for a stochastic approximation of a function shrinking the simplex can produce vertices that have worse objective function values due to the variance of the objective function and its rough surface, especially if the simplex vertices still have significant differences.

\newpage
\section{Learning agent results} \label{app:RLAgentSimulationResults}
\subsection{Reinforcement learning} \label{app:ReinforcementLearning}

A Reinforcement Learning (RL) agent interacts with the environment at discrete time steps, $t = 1,2,3,...$ (We restrict ourselves to considering only the discrete-time case as this is what our agent will observe. However, we note that continuous-time extensions to this formulation are available (see \cite{bertsekas1996neuro,doya1996efficient}).). At each time step the agent observes the state of the environment $S_{t} \in \mathcal{S}$ where $\mathcal{S}$ is the set of possible states, and takes an action $A_{t} \in \mathcal{A}(S_{t})$ where $\mathcal{A}(S_{t})$ is the set of actions the agent can take in state $S_{t}$. At the following time step the agent observes the reward $R_{t+1} \in \mathcal{R} \subset \mathbb{R}$ (We use $R_{t+1}$ instead of $R_{t}$ to denote the reward received from taking action $A_{t}$ in state $S_{t}$ to indicate that $R_{t+1}$ and $S_{t+1}$ are observed at the same time. This convention was taken from \cite{sutton2018reinforcement}.), for taking action $A_{t}$ in state $S_{t}$, and it observes the new state $S_{t+1}$.

To select the actions at each time step the agent uses a \textit{policy}, which maps the state to the probabilities of selecting each possible action \cite{sutton2018reinforcement}. The policy is denoted by $\pi_{t}$ where $\pi_{t}(a|s)$ is the probability that $A_{t} = a$ when $S_{t} = s$ \cite{sutton2018reinforcement}.

The goal of the agent is to find the policy such that it maximises the total reward received. Or more formally, at each time step $t$ the agent wishes to maximise the \textit{expected return}, $G_{t}$, which is a function of the reward sequence. In the simplest case this is just the sum of the rewards: $G_{t} = R_{t+1} + R_{t+2} + R_{t+3} + \cdots + R_{T}$, where T is the final time step \cite{sutton2018reinforcement}. This is the expected return function that our agent will be using.

This formulation of the return maps nicely to problems that have a final time step, where the agent-environment interaction can be broken up into sub-sequences, such as repeated plays of a game \cite{sutton2018reinforcement}. For example, blackjack, or in our case repeated executions of a large parent order. These repeated tasks are called \textit{episodes} and tasks that can be broken up into episodes are called \textit{episodic tasks}. At the final time step, of an episode, the agent moves into a \textit{terminal state}, after which the state of the environment gets reset to some starting state.

Some tasks can't be broken up into episodes, where $T = \infty$. A trading agent that just aims to maximise its profit over its lifetime would be an example of such an agent. These tasks are called \textit{continuing tasks} \cite{sutton2018reinforcement}. Since $T = \infty$, we have that the expected return could be infinite. This is remedied by adding \textit{discounting}, which gives us the expected \textit{discounted return}: $G_{t} = \sum_{k = 0}^{\infty} \gamma^{k} R_{t + k + 1}$.
Here $0 \leq \gamma \leq 1$ is the \textit{discounting rate}, and it controls how much the agent values immediate vs. farsighted returns. As $\gamma$ approaches zero the agent becomes more myopic and values immediate rewards more heavily, while as $\gamma$ approaches one the agent becomes more farsighted and starts to value future returns more heavily. Now, an issue occurs because we have two tasks and two reward functions. To recover a single notation that describes both tasks we use the convention given to us by \cite{sutton2018reinforcement}, where we consider episode termination to be the case of entering an \textit{absorbing state} that transitions only to itself and has a reward of zero. This allows us to describe both tasks' return functions. 

Given action $a$ and state $s$, the probability of observing a reward $r$ and moving to state $s'$ is given by:
\begin{align} \label{app:rl-reward-state-probability}
    p(s',r|s,a) = \Pp(S_{t+1} = s', R_{t+1} = r | S_{t} = s, A_{t} = a). 
\end{align}
This defines the one-step dynamics of the model for the MDP. From this we can find the \textit{state-transition probabilities}:
\begin{align} \label{app:rl-state-transition-probability}
    p(s'|s,a) = \sum_{r \in \mathcal{R}} p(s',r|s,a), 
\end{align}
as well as the expected reward for taking action $a$ in state $s$:
\begin{align}\label{app:rl-expected-reward-mdp}
    r(s,a) = \sum_{r \in \mathcal{R}} r \sum_{s' \in \mathcal{S}} p(s',r|s,a).
\end{align}
Using these definitions we can derive equations for the \textit{state-value function} $v_{\pi}(s)$ and the \textit{action-value function} $q_{\pi}(s,a)$ for policy $\pi$, where $v_{\pi}(s)$ is the value of state $s$ under policy $\pi$, which is defined as:
\begin{align}\label{app:rl-state-value-function}
    v_{\pi}(s) = \mathbb{E}_{\pi|S_{t} = s}[G_{t}] 
    = \mathbb{E}_{\pi| S_t=s}\left[\sum_{k = 0}^{\infty} \gamma^{k} R_{t + k + 1}\right] 
    = \sum_{a} \pi(a|s) \sum_{s',r}  p(s',r|s,a) \left[r + \gamma v_{\pi}(s')\right].
\end{align}
Here the action-value function, which is the value of taking action $a$ in state $s$, and following $\pi$ thereafter, is:
\begin{align}\label{app:rl-action-value-function}
    q_{\pi}(s,a) &= \mathbb{E}_{\pi|\substack{ S_{t} = s \\ A_{t} = a}}[G_{t}] 
    = \mathbb{E}_{\pi|\substack{S_{t} = s \\ A_{t} = a }}\left[\sum_{k = 0}^{\infty} \gamma^{k} R_{t + k + 1}\right] 
    = \sum_{r,s'} p(s',r|s,a) \left[r + \gamma \sum_{a'} \pi(a'|s') q_{\pi}(s',a') \right]. 
\end{align}
Value functions give us a partial ordering of policies. We say that policy $\pi$ is better than or equal to a policy $\pi'$ if its expected return is greater than or equal to the expected return of $\pi'$ for all states \cite{sutton2018reinforcement}. The \textit{optimal policy} is the policy such that no other policy has a greater expected return for any of the states. There could be more than one of these policies, however, we denote the optimal policy as $\pi_{*}$. Using this ordering of policies we can define the \textit{optimal state-value function} as: $ v_{*}(s) = \max_{\pi} v_{\pi}(s)~ \forall s \in \mathcal{S}$, and the \textit{optimal action-value function} as: $q_{*}(s,a) = \max_{\pi}  q_{\pi}(s,a)$ $\forall s \in \mathcal{S}$ and $\forall a \in \mathcal{A}$. The optimal policy can then be found by maximising over $q_{*}(s,a)$ using that if $a = \argmax_{a \in \mathcal{A}} q_{*}(s,a)$ then $\pi^{*}(s,a)=1$, otherwise $\pi^{*}(s,a) =0$. Therefore if we know $q_{*}(s,a)$ we know the optimal policy. This is key since most RL algorithms, and the Q-learning algorithm that we will implement will aim to optimise the action-value function directly. Therefore, we will need to recover the optimal policy from the optimal action-value function.

We do not have the full model of the environment therefore, we can't use equations \ref{app:rl-state-value-function} and \ref{app:rl-action-value-function}. However, we can learn the optimal state-value and action-value functions from samples of sequences of states, actions, and rewards, that can be generated through actual or simulated interaction with the environment. The two main methods for learning these functions through experience are Monte Carlo and Temporal Difference (TD) learning. When using Monte Carlo methods to estimate the value functions, the value functions are only updated at the end of every episode. This can make learning inefficient. Moreover, Monte Carlo methods can't be used to learn in a step-by-step online way. Given the learning inefficiencies and the long runtimes of a simulation, it is likely that we won't be able to have enough episodes for the RL agent to accurately estimate the value functions, using Monte Carlo methods. Therefore, we will only consider TD learning. 

TD methods allow the agent to learn in each episode and does not require waiting for the episode to complete like Monte Carlo. The TD algorithm that we will be using is the Q-learning algorithm first proposed by \cite{watkins1989learning}. In this particular case, the update to the value function can be done after a single step in an episode.

Q-learning is an off-policy TD control algorithm which is in contrast to the SARSA on-policy control algorithm. This means that the policy that generates the actions is not the same as the target policy. Using this method simplifies the analysis of the algorithm and reduces the conditions needed for guaranteed convergence to $q_{*}$.

At a particular time t in an arbitrary episode, we give the simplest form of Q-learning, the one-step Q-learning update equation is as follows:

\begin{align}\label{app:rl-qlearning-update}
    Q(S_{t},A_{t}) =& Q(S_{t},A_{t}) + \alpha R_{t+1} + \alpha \left[ {\gamma \max_{a} Q(S_{t+1},a) - Q(S_{t},A_{t})}\right].
\end{align}
Here $\alpha$ is the learning rate. Provided each state is visited infinitely often and each action is taken in each state infinitely often we have that $Q$ converges to $q_{*}$ with probability 1, for appropriately decaying step sizes (The conditions on the step sizes are $\sum_{t} \alpha_{t} = \infty$ and $\sum_{t} \alpha_{t}^{2} < \infty$.) \cite{watkins1992q}.

\begin{figure}[h]
    \centering
    \includegraphics[width=9cm, height=6cm]{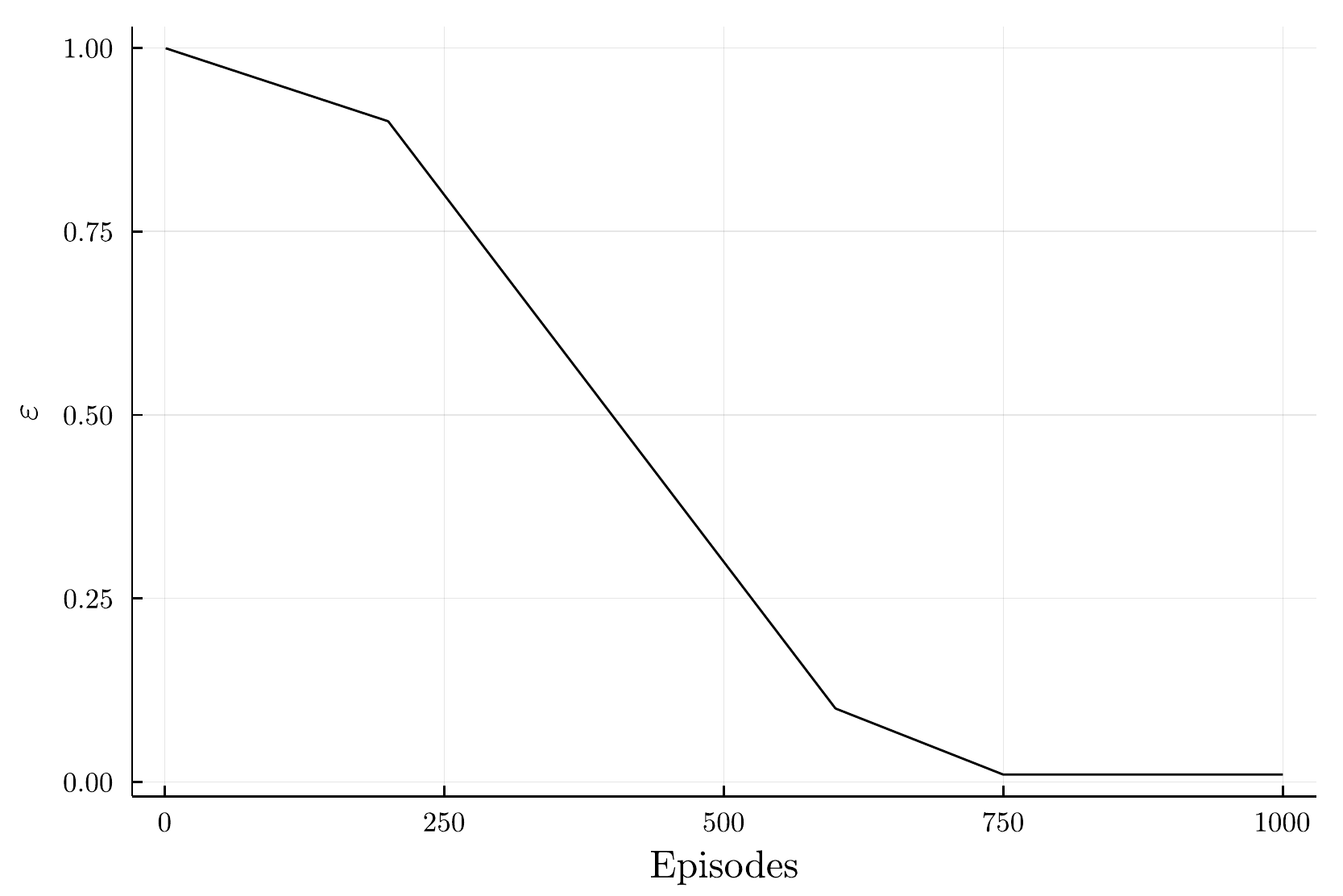}
    \caption{Plots the $\epsilon$ decay for $\epsilon$-greedy policy used to generate the actions of the RL agents. For a given state and $N_{A}$ actions, the $\epsilon$-greedy policy chooses the action with the highest value with probability $1-\epsilon + (\epsilon/N_{A})$ and the rest with probability $\epsilon/N_{A}$. Setting $\epsilon=1$ will mean an agent selects each action with equal probability regardless of its value and setting $\epsilon=0$ will ensure the agent always selects the action with the highest value (greedy policy).}
    \label{rl-epsilon-decay}
\end{figure}

In Q-learning there are two policies: the first being the target policy which we aim to optimise, and the second is the behaviour policy that generates the agents' behaviour used to optimise the target policy. In our implementation of Q-learning the target policy is the greedy policy (see equation \ref{app:rl-qlearning-update}), and the behaviour policy is the $\epsilon$-greedy policy. $\epsilon$ controls the amount of exploration versus exploitation in the optimisation. For a given state and $N_{A}$ actions, the $\epsilon$-greedy policy chooses the action with the highest value with probability $1-\epsilon + (\epsilon/N_{A})$ and the rest with probability $\epsilon/N_{A}$. Setting $\epsilon=1$ will mean an agent that selects each action with equal probability regardless of its value and setting $\epsilon=0$ will ensure the agent always selects the action with the highest value -- the greedy policy. To ensure a good balance between exploration and exploitation the $\epsilon$ parameter was decayed throughout the simulation. Figure \ref{rl-epsilon-decay} shows how the $\epsilon$ parameter is decayed over the episodes. For the first 200 episodes, the $\epsilon$ decays from 1 to 0.9, for the next 400 episodes $\epsilon$ decays from 0.9 to 0.1, and for the following 150 episodes, it decays from 0.1 to 0.01 where it remains for the rest of the episode.

\subsection{Best bid volume state spaces} \label{app:RLAgentSimulationResults-VolumeStateSpace}

The volume states are assigned based on the observed best bid after some event $k$. The values and probabilities of falling into each of the states are from historic distributions. This provides the probabilities of breaking through the best bid {\it e.g.} an order of size 50 is likely to break through the best bid with a probability of a little over 20\%, whereas an order of size 200 has a probability of a little less than 40\%. Agents trading with less inventory are far less likely to cause excessive price impact. Based on the depth profiles observed in the ABM most of the liquidity is at the best bid/offer, therefore if you break through the best bid you have a high chance of clearing that side of the order book. Based on this, it follows intuitively why agents with larger inventories are causing more frequent liquidity shocks.

\begin{table}[h!]
    \caption{Describes how the states, for the best bid volume ($v_{k}^{b}$), are determined as well as the approximate probability of being in each of those states, based on the observed best bid volume at some event $k$ and the simulated historical distributions. Table \ref{table-volume-states-s5} and \ref{table-volume-states-S10} shows how the volume states are assigned for the agents with the smaller and larger state spaces respectively.}
    \label{table-volume-states}
    \begin{subtable}[t]{.5\linewidth}
      \centering
       \caption{Smaller volume state space assignment: ($n_{T},n_{I},n_{S},n_{V} = 5$).}
        \label{table-volume-states-s5}
        \begin{tabular}{crlc}
            \toprule
            State & \multicolumn{2}{c}{Volume Range} & Prob. \\
             $v$   & \multicolumn{2}{c}{$ \cdot < v_{k}^{b} \leq \cdot$} & \\
            \hline
            1 & 0  & 31 & 0.2\\
            2 & 31 & 266 & 0.2\\
            3 & 266 & 1453 & 0.2\\
            4 & 1453 & 5209 & 0.2\\
            5 & 5209 & $\cdot$ & 0.2\\
            \bottomrule
            \end{tabular}
              \centering
    \end{subtable}%
    \begin{subtable}[t]{.5\linewidth}
      \centering
       \caption{Larger volume state space assignment:($n_{T},n_{I},n_{S},n_{V} = 10$).}
        \label{table-volume-states-S10}
        \begin{tabular}{crlc}
            \toprule
            State & \multicolumn{2}{c}{Volume Range} & Prob. \\
            $v$ & \multicolumn{2}{c}{$ \cdot < v_{k}^{b} \leq \cdot$} &  \\
            \hline
            1 & 0 & 11 & 0.1\\
            2 & 11 & 31 & 0.1\\
            3 & 31 & 93 & 0.1\\
            4 & 93 & 266 & 0.1\\
            5 & 266 & 636 & 0.1\\
            6 & 636 & 1453 & 0.1\\
            7 & 1453 & 2930 & 0.1\\
            8 & 2930 & 5209 & 0.1\\
            9 & 5209 & 2322 & 0.1\\
            10 & 12322 & $\cdot$ & 0.1\\
            \bottomrule
            \end{tabular}
              \centering
    \end{subtable} 
\end{table}

\subsection{Spread state spaces} \label{app:RLAgentSimulationResults-SpreadStateSpace}

\begin{table}[h!]
    \caption{Describes how the states, for the spread ($s_{k}$), are determined as well as the approximate probability of being in each of those states, based on the observed spread at some event $k$ and the simulated historical distributions. Table \ref{table-spread-states-s5} and \ref{table-spread-states-S10} shows how the spread states are assigned for the agents with the smaller and larger state spaces respectively.}
    \label{table-spread-states}
    \begin{subtable}[t]{.5\linewidth}
      \centering
        \caption{Smaller spread state space assignment: ($n_{T},n_{I},n_{S},n_{V} = 5$).}
        \label{table-spread-states-s5}
        \begin{tabular}{crlc}
            \toprule
           State & \multicolumn{2}{c}{Spread Range} & Prob. \\
            $s$ & \multicolumn{2}{c}{$ \cdot < s_{k}^{b} \leq \cdot$} &  \\
            \hline
            1 & 0 & 1 & 0.6 \\
            2 & 1 & 2 & 0.123 \\
            3 & 2 & 3 & 0.062 \\
            4 & 3 & 7 & 0.091 \\
            5 & ~~~~7 & $\cdot$~~~~ & 0.124 \\
            \bottomrule
            \end{tabular}
    \end{subtable}%
    \begin{subtable}[t]{.5\linewidth}
      \centering
        \caption{Larger spread state space assignment:($n_{T},n_{I},n_{S},n_{V} = 10$).}
        \label{table-spread-states-S10}
        \begin{tabular}{crlc}
            \toprule
                State & \multicolumn{2}{c}{Spread Range} & Prob. \\
            $s$ & \multicolumn{2}{c}{$ \cdot < s_{k}^{b} \leq \cdot$} &  \\
            \hline
            1 & 0 & 1 & 0.6 \\
            2 & 1 & 2 & 0.123 \\
            3 & 2 & 3 & 0.062 \\
            4 & 3 & 4 & 0.031 \\
            5 & 4 & 5 & 0.031 \\
            6 & 5 & 6 & 0.031 \\
            7 & 6 & 8 & 0.031 \\
            8 & 8 & 12 & 0.031 \\
            9 & 12& 19 & 0.031 \\
            10 & ~~~~19 & $\cdot$~~~~ & 0.031 \\
            \bottomrule
            \end{tabular}
    \end{subtable} 
\end{table}

\subsection{Convergence results} \label{app:RLAgentSimulationResults-Convergence}

\begin{figure*}[ht]
        \centering
        \begin{subfigure}{.5\textwidth}
          \centering
              \includegraphics[width=9cm, height=6cm]{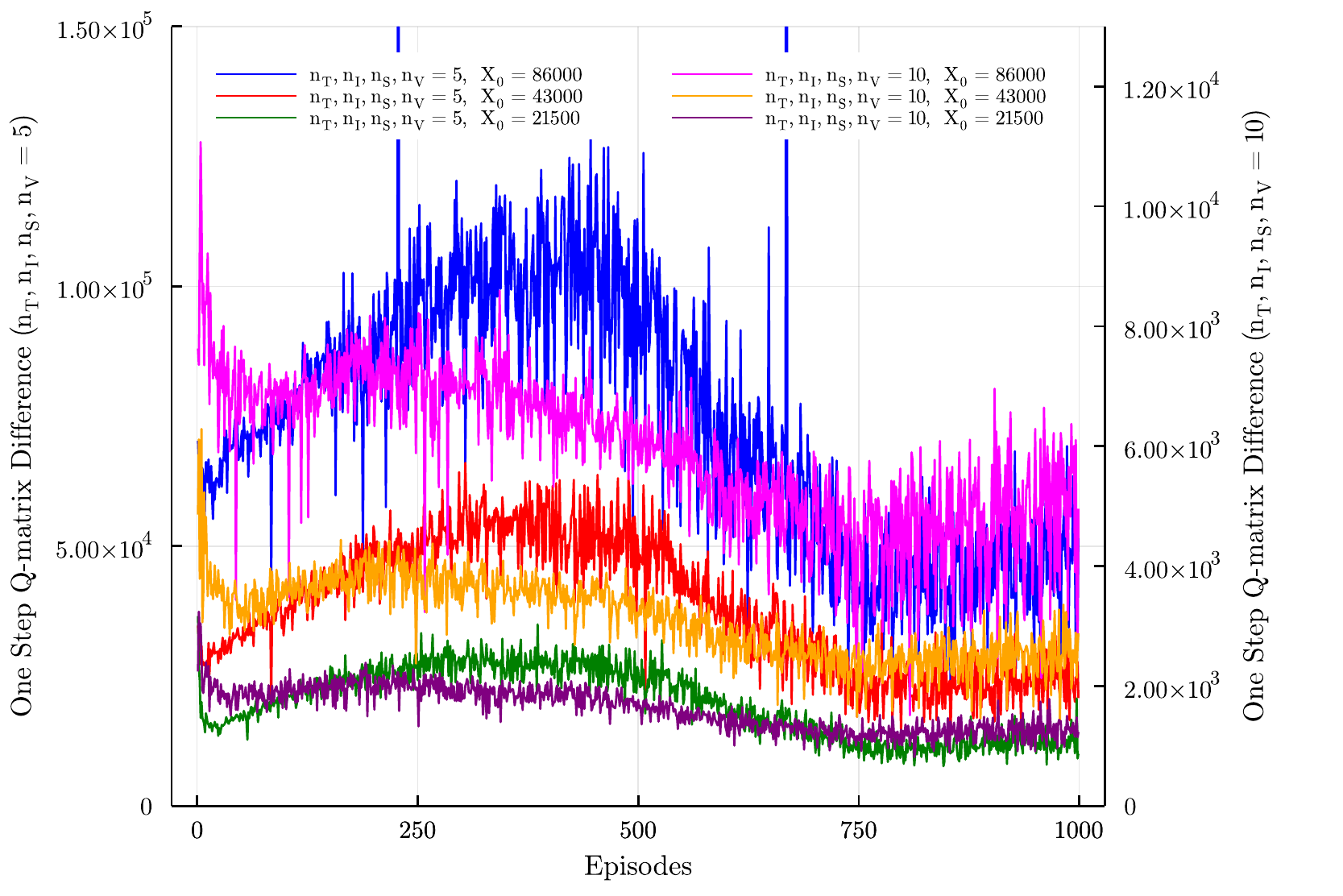}
              \caption{Q-matrix difference}
              \label{rl-QandPolicyConvergence-(a)}
        \end{subfigure}%
        \begin{subfigure}{.5\textwidth}
          \centering
          \includegraphics[width=7cm, height=6cm]{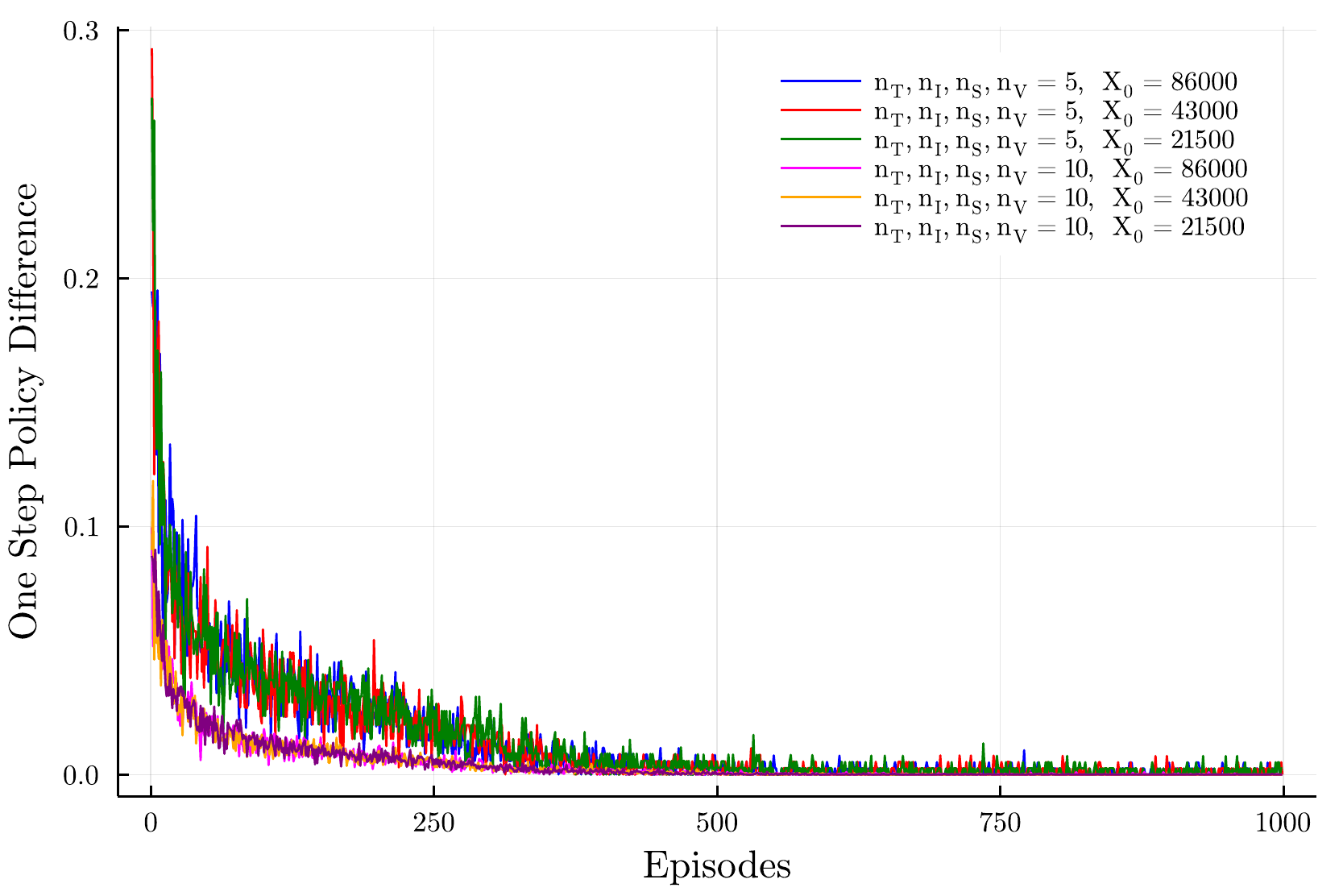}
          \caption{Policy difference}
          \label{rl-QandPolicyConvergence-(b)}
        \end{subfigure}
        \caption{One-step absolute differences in common states between Q-matrices (Figure \ref{rl-QandPolicyConvergence-(a)}) and the number of one-step action changes for the common states in the greedy policy (Figure \ref{rl-QandPolicyConvergence-(b)}). The Q-matrix differences will not converge to zero as the learning rate was not decayed, but we see convergence in all agents' Q-matrices. The one-step action changes for common states in the greedy policy demonstrate a rapid decrease in the number of action changes as the episodes increase. The larger state space converges faster than the smaller state space indicating fewer action changes.}
        \label{rl-QandPolicyConvergence}
\end{figure*}

The convergence in the behaviour of the agents is demonstrated by measuring the convergence of the Q-matrix and the convergence of the greedy policy. The left-hand side plot of Figure \ref{rl-QandPolicyConvergence} plots the one-step average absolute difference in the Q-matrix, for common states. Here let $\mathcal{A}_k$ and $\mathcal{S}_k$ be the set of actions to choose and states observed at time step $t$, and let $|\mathcal{A}_k|$ and $|\mathcal{S}_k|$ denote the cardinality of these sets. The one-step average absolute difference in the Q-matrix at event $k$ for common states is:
\begin{align}\label{rl-onestep-Q-convergence}
    \frac{1}{|\mathcal{S}_{k-1}|} \frac{1}{|\mathcal{A}_{k-1}|} \sum_{\substack{s \in \mathcal{S}_{k-1} \\ a \in \mathcal{A}_{k-1}}} \left|{ Q_{k}^{(s,a)} - Q_{k-1}^{(s,a)}}\right|.
\end{align}
It follows from $\mathcal{A}_{k-1} \subset \mathcal{A}_k$ and $\mathcal{S}_{k-1} \subset \mathcal{S}_k$ that we can average over the actions and states of the time step $k-1$ because at event $k$ we have that only more states could have been discovered. If Q converges to the optimal policy then the one-step difference should converge to zero when there are appropriately decayed learning rates. Here the learning rates are not decayed but are constant. This means that the differences will not converge to zero. However, it is still a good measure of the convergence of the Q-matrix values. 

From Figure \ref{rl-QandPolicyConvergence} we see that all the differences have decreased and converged around some constant, with some variance. The agents trading less inventory have smaller differences due to the smaller profit they will obtain, and they have a lower variance which could be partly due to the lower total profit and the smaller price impact. Comparing the smaller state space to the larger, we see that the smaller state space agents all exhibit a hump shape instead of a smoother more monotone convergence of the larger state space agents. 

Changes in the Q-matrix may not cause changes in the actions the agent selects. For this reason, in a one-step update, we consider the convergence of the greedy policy over the training of the agent. To measure this we use the one-step difference at event $k$:
\begin{align}\label{rl-onestep-policy-convergence}
    \frac{1}{|\mathcal{S}_{k-1}|} \sum_{s \in \mathcal{S}_{k-1}} \left|{
    \argmax_{a} Q_{k}^{(s,a)} \neq \argmax_{a} Q_{k-1}^{(s,a)}}\right|.
\end{align}
For a given state, this measures the difference between the actions chosen by the greedy policy, from event $k-1$ to event $k$. The idea behind this measure is that as the agent learns the optimal policy the agent should learn to select the same action for the same state. This measure is plotted on the right-hand side of Figure \ref{rl-QandPolicyConvergence}. There is clear convergence in the actions selected by the agents as the training progressed. With the agents selecting the same actions for the same states using the greedy policy. A surprising result is that the agents with the larger state space have the greedy policy converge faster and with less variance when compared with the agents with the smaller state space. Thus due to the size of the state space, individual states don't get visited as often and therefore updates to the policy are much slower demonstrating the need for more episodes needed to train the larger state space agent. This plot does not claim that the agent has found the optimal greedy policy, but it does show that the greedy policy is changing at a very slow rate. Moreover, we have that unless the number of episodes is drastically increased, the policies will not change significantly and hence the profit obtained will be unlikely to significantly improve.  

Trace plots for the actions taken in each state in each episode by the greedy policy for an agent with a smaller state space (left) and larger state space (right) with the same volume. The -1 action indicates a state that has not yet been visited and a jump from -1 to the other actions means the state has just been discovered. Looking at the smaller state space agent we can see that initially there is a large amount of changing of actions for each state, however, over the training period, the number of changes decreases and the agent starts to repeatedly select the same actions. Due to most of the states already being discovered the action changes are largely due to the action-values of previous states being modified. The larger state space agent seems to initially have lots of jumps which decrease over the episodes. However, it was found that this pattern was mainly due to state discovery rather than learning of actions. Trace plots for the remaining agents are similar and found in the supporting materials.

\begin{figure}[h!]
        \centering
        \begin{subfigure}{.5\textwidth}
          \centering
              \includegraphics[width=9cm, height=6cm]{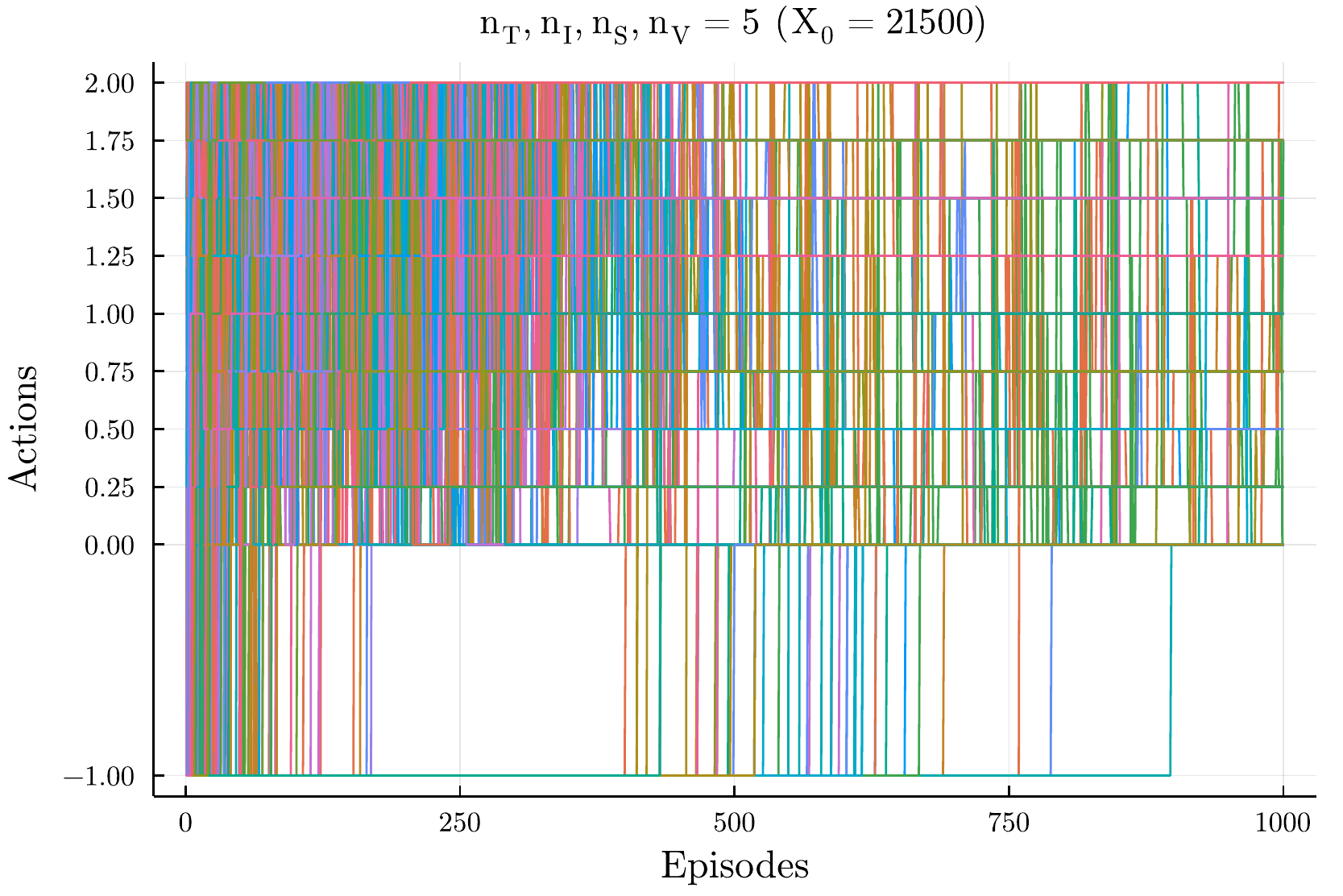}
              \caption{Small state space trace plots, small order.}
        \end{subfigure}%
        \begin{subfigure}{.5\textwidth}
          \centering
          \includegraphics[width=9cm, height=6cm]{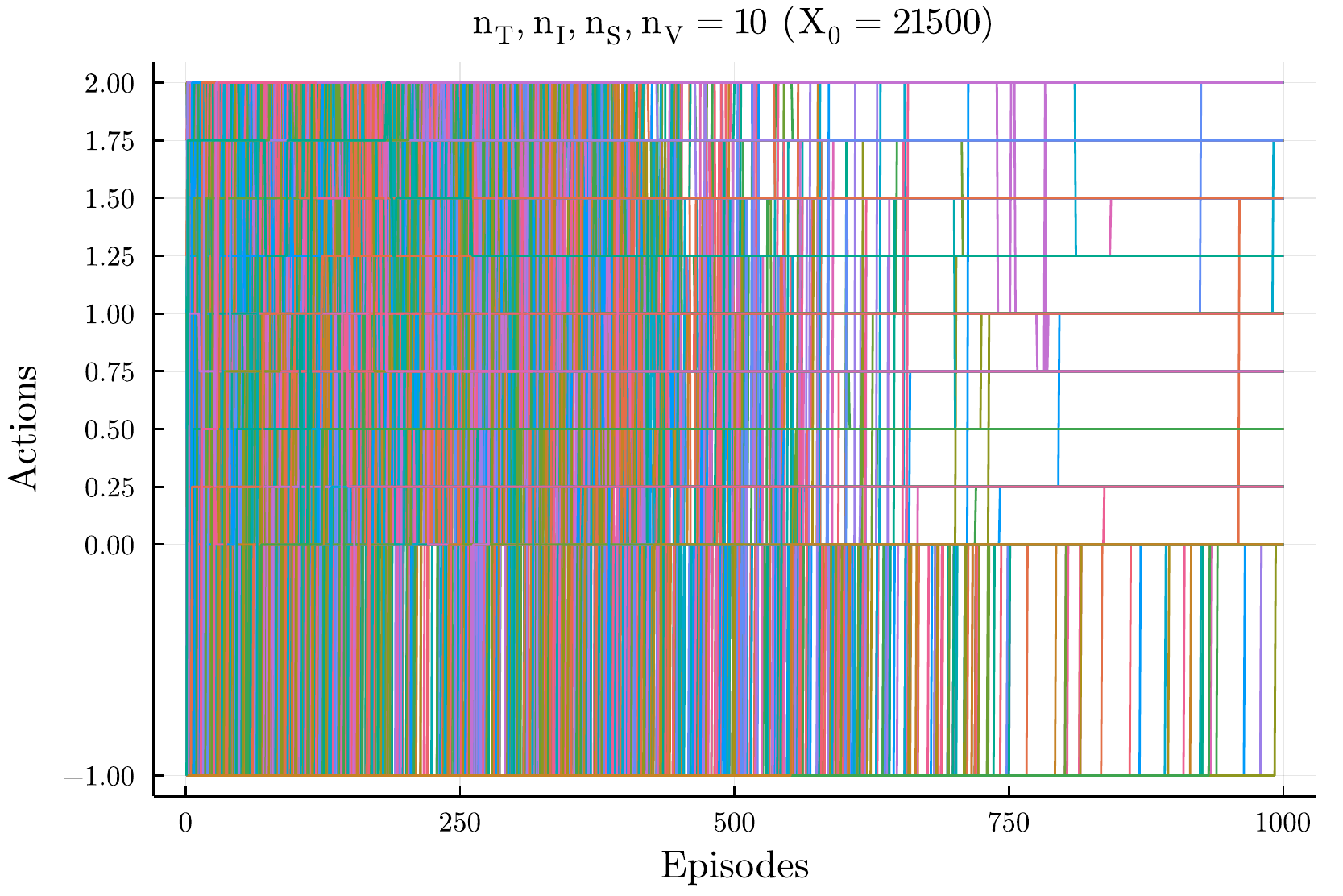}
          \caption{Large state space trace plots, small order.}
        \end{subfigure}
        \caption{Training agent action trace plots for $X_{0} = 21500$ and $n_{T},n_{I},n_{S},n_{V}=5$ (left) and $n_{T},n_{I},n_{S},n_{V}=10$ (right).}
        \label{app:rl-StateActionConvergence-V50}
\end{figure}

\begin{figure*}[h!]
        \centering
        \begin{subfigure}{.5\textwidth}
          \centering
              \includegraphics[width=9cm, height=6cm]{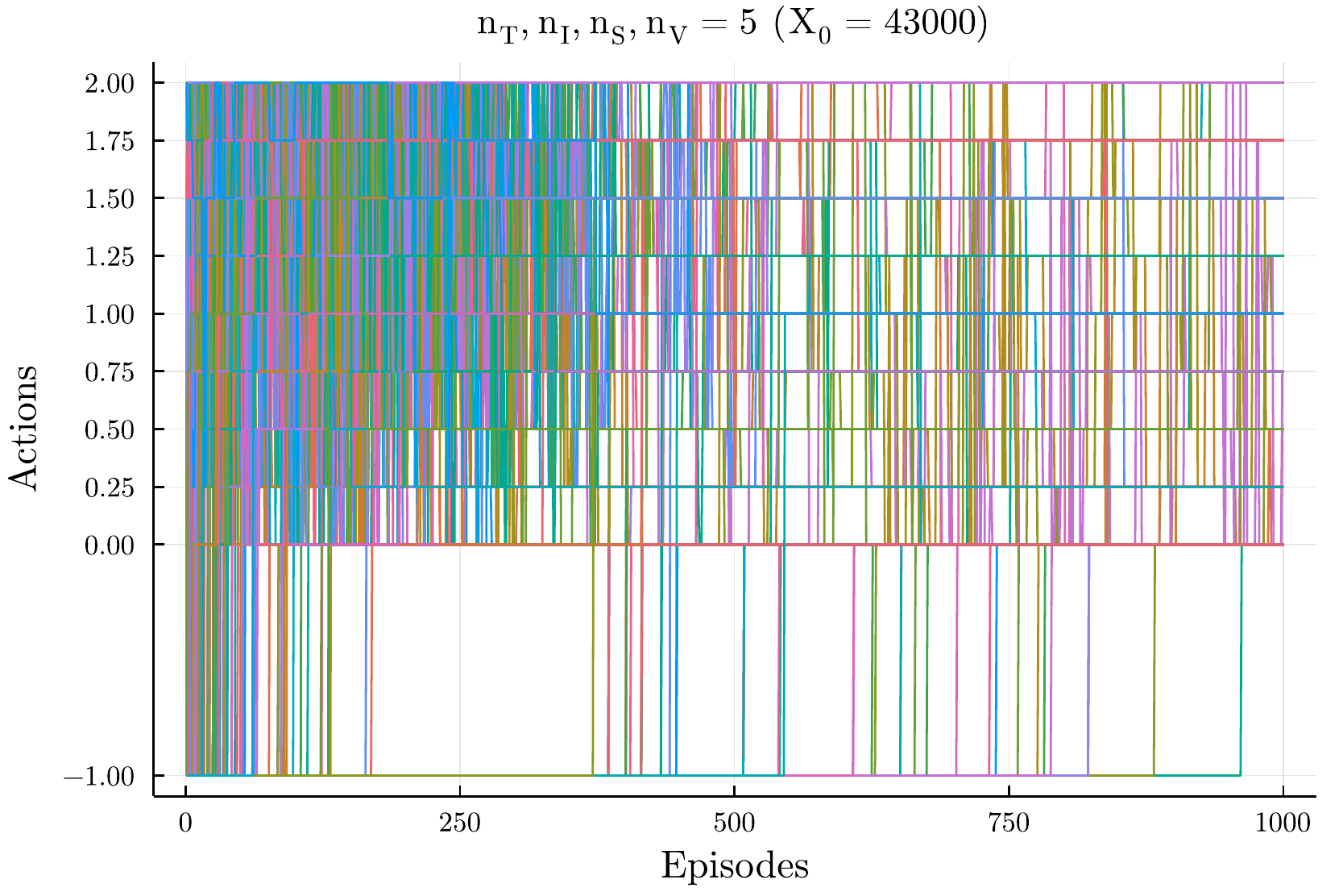}
              \caption{Small state space trace plots, medium order}
              \label{rl-StateActionConvergence-a}
        \end{subfigure}%
        \begin{subfigure}{.5\textwidth}
          \centering
          \includegraphics[width=9cm, height=6cm]{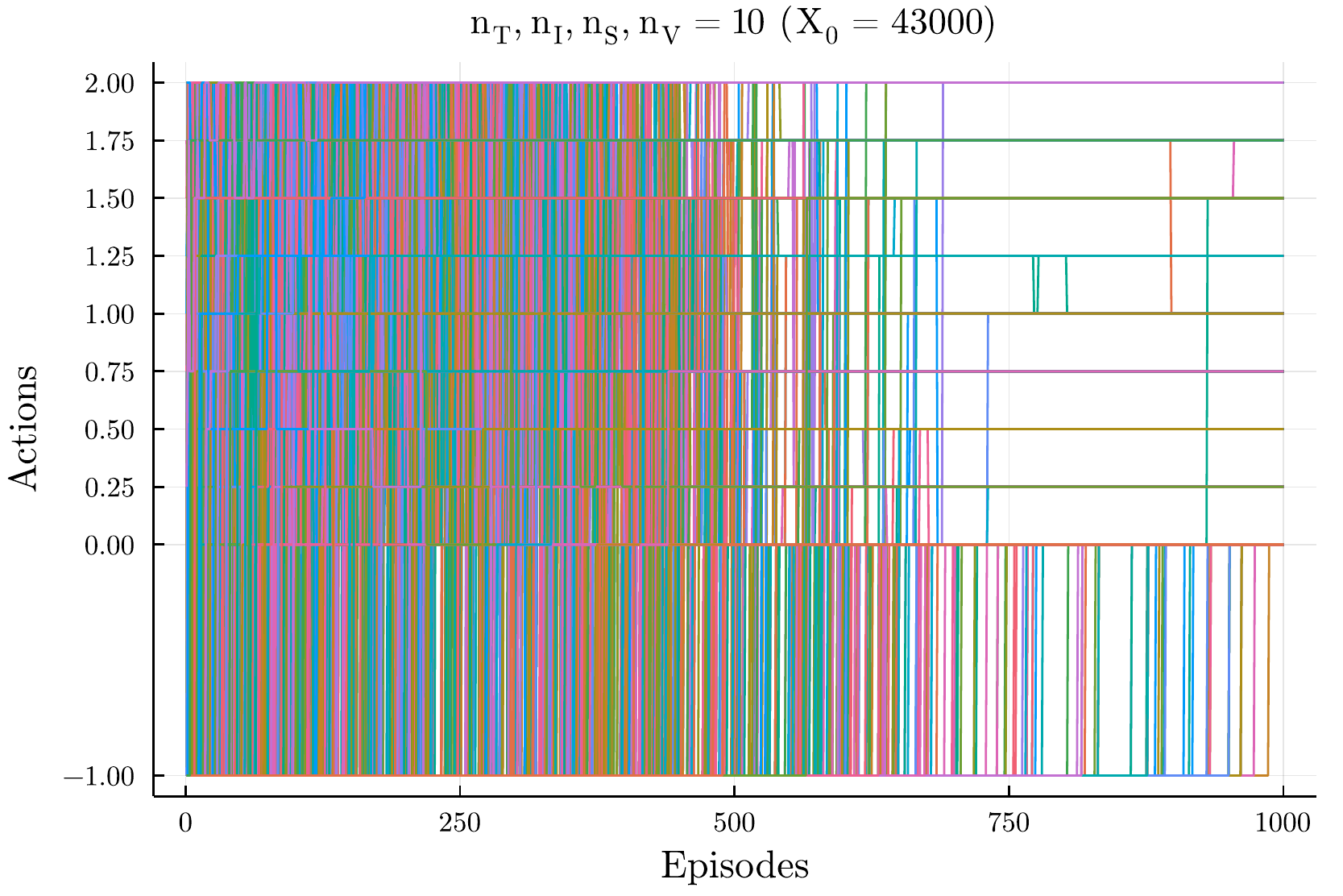}
          \caption{Large state space trace plots, medium order.}
          \label{rl-StateActionConvergence-b}
        \end{subfigure}
        \caption{Training agent action trace plots for $X_{0} = 43000$ and $n_{T},n_{I},n_{S},n_{V}=5$ (left) and $n_{T},n_{I},n_{S},n_{V}=10$ (right).}
        \label{rl-StateActionConvergence}
\end{figure*}

\begin{figure}[h!]
        \centering
        \begin{subfigure}{.5\textwidth}
          \centering
              \includegraphics[width=9cm, height=6cm]{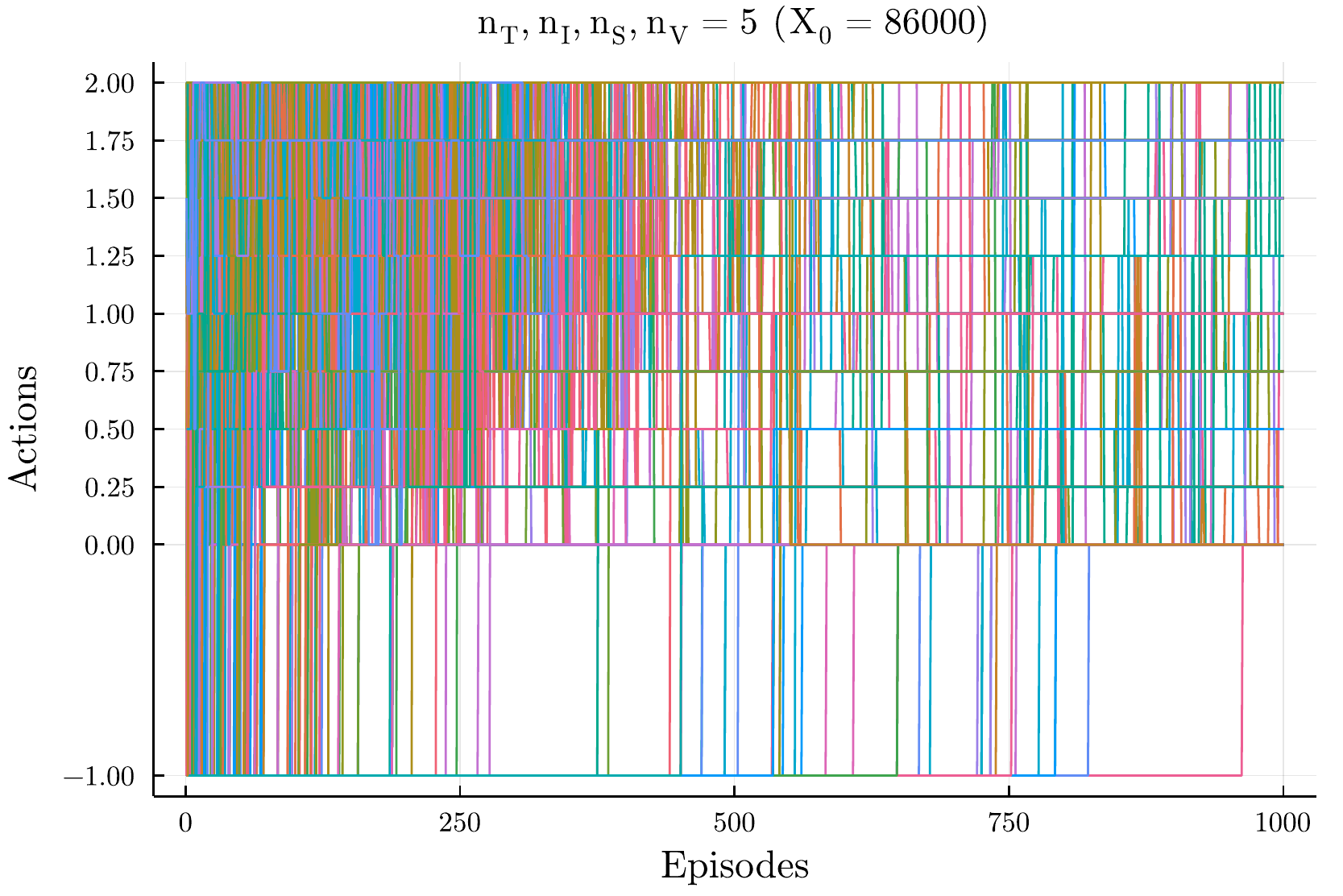}
              \caption{Small state space trace plots, large order}
        \end{subfigure}%
        \begin{subfigure}{.5\textwidth}
          \centering
          \includegraphics[width=9cm, height=6cm]{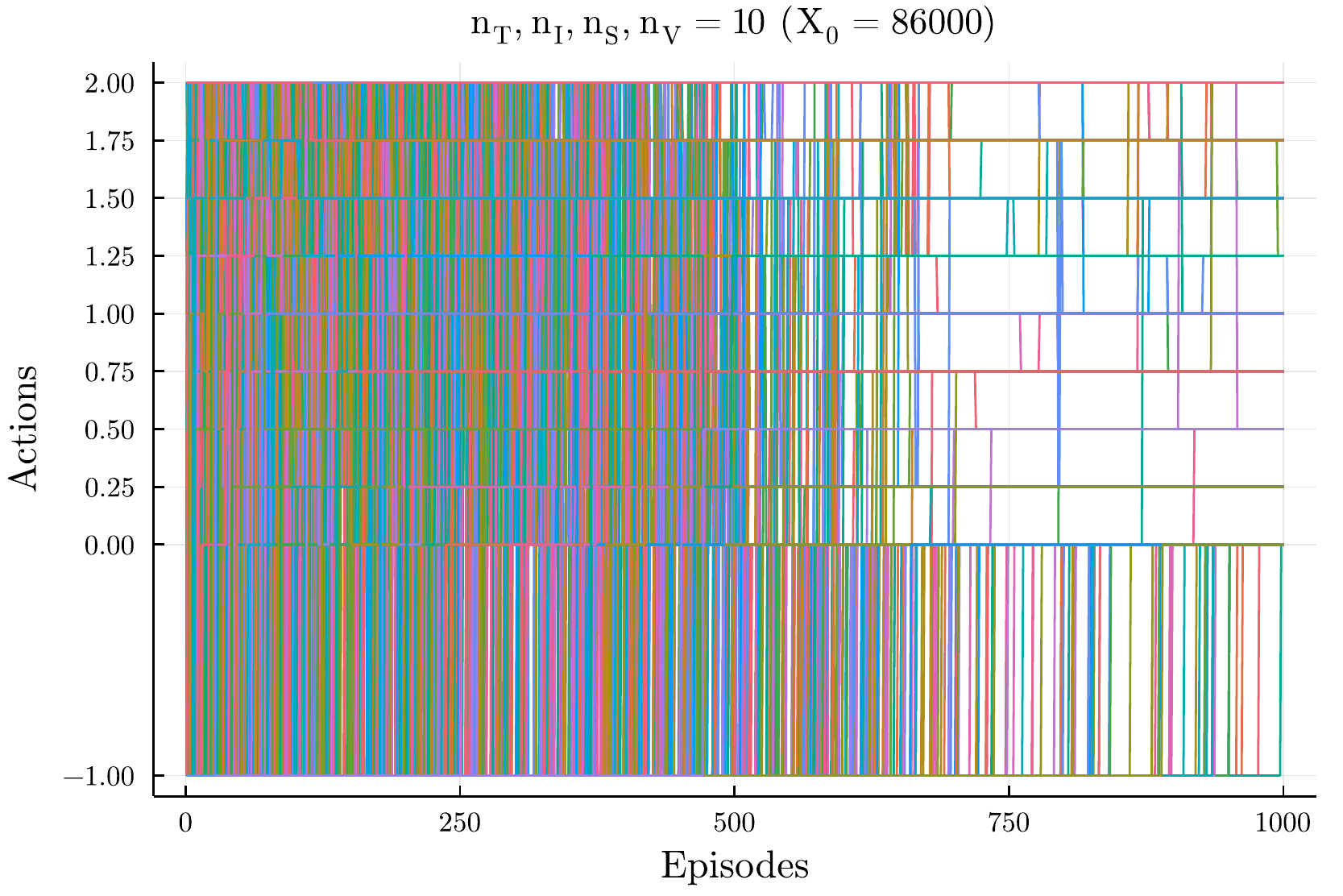}
          \caption{Large state space trace plots, large order}
        \end{subfigure}
        \caption{Training agent action trace plots for $X_{0} = 86000$ and $n_{T},n_{I},n_{S},n_{V}=5$ (left) and $n_{T},n_{I},n_{S},n_{V}=10$ (right).}
        \label{app:rl-StateActionConvergence-V200}
\end{figure}

\subsection{Behaviour analysis} \label{app:RLAgentSimulationResults-BehaviourAnalysis}

It was found that the number of trades decreased over the episodes for all the agents, this is shown in Figure 3a in the paper. If the agent were to trade exactly $x_{i}$ shares corresponding to action 1, and there is enough liquidity, the agent will be asked to make a trading decision approximately 430 times. There may not be liquidity on the contra side of the order book. The agent may then trigger a volatility auction if it trades, and because of the market rules of how this is implemented in this ABM, it will not be able to submit an order until the liquidity is replenished. In part, this explains why the number of trades is less than 430 because the illiquid nature of the ABM means that if the agent submits an order it may not be fully matched. This means that the agent is at some risk of not fully executing the parent order. Therefore, it can be seen that the agent has to learn to trade a higher volume, which means that it reduces the number of times it trades. It has correctly learnt to do this to mitigate the lack of liquidity and ensure that the parent order is fully executed. 

Using the reward function it could be expected that the agent learns to only trade the maximum volume because this could potentially provide the agent with the highest reward based on the volume traded. Figure 3a in the paper has shown that the agent has learned that this is not a feasible tactic as it has converged to a higher number of trades than would be allowed by always trading the maximum volume. The final greedy policy for the a single RL agent with initial inventory of 43000 and $n_{T},n_{I},n_{S},n_{V} = 5$ is shown in Figure \ref{rl-policy-V100-S5}. An action of -1 indicates that the state has not yet been reached. The states that are not visited correspond to situations the agent is highly unlikely to be in. For example, having a large (resp. small) amount of inventory when there is little (resp. lots) of time left. These states will seldom occur as the agent is unlikely to have traded most of the inventory when in the initial stages of the simulation nor will it have traded almost nothing for most of the simulation. 

If you fix a specific time state and move upwards in the figure you see that the agent has learned to trade more, averaged across the spread and volume states, as the inventory increases. It can also be seen that if you fix a specific inventory state and increase the amount of time that the simulation has been running for (by moving left) you can see that, on average, the agent is learning to trade more. The agent has clearly learned to trade at different rates depending on the time and inventory states. From this figure there are no obvious changes in actions as the spread and volume states change, however, these effects will be investigated more thoroughly in the next section. This policy is for a single agent but the results are consistent across all the agents and these are shown in the supporting materials. To see the change in behaviour of the agents across the episodes, the action volume trajectories for the first and last episodes are plotted in Figures \ref{rl-action-volume-V100-S5} and \ref{rl-action-volume-V100-S10}.

\begin{figure}[h!]
    \centering
        \begin{subfigure}[b]{0.45\textwidth}
          \centering
            \includegraphics[width=8cm, height=6cm]{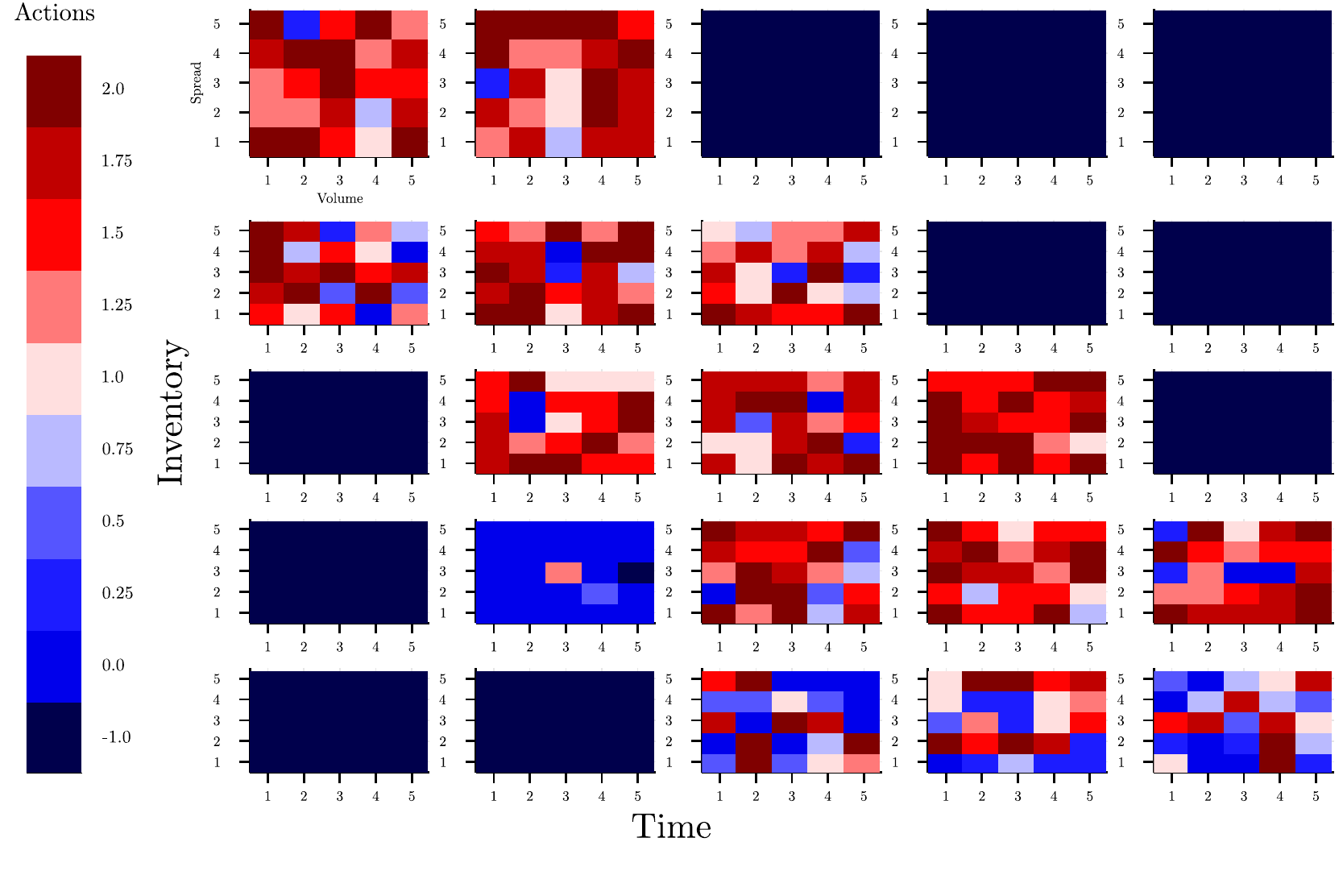}
            \caption{Plots the policy for the agent}
            \label{app:rl-policy-V50-S5}
        \end{subfigure}\hfill
        \begin{subfigure}[b]{0.45\textwidth}
          \centering
            \includegraphics[width=8cm, height=6cm]{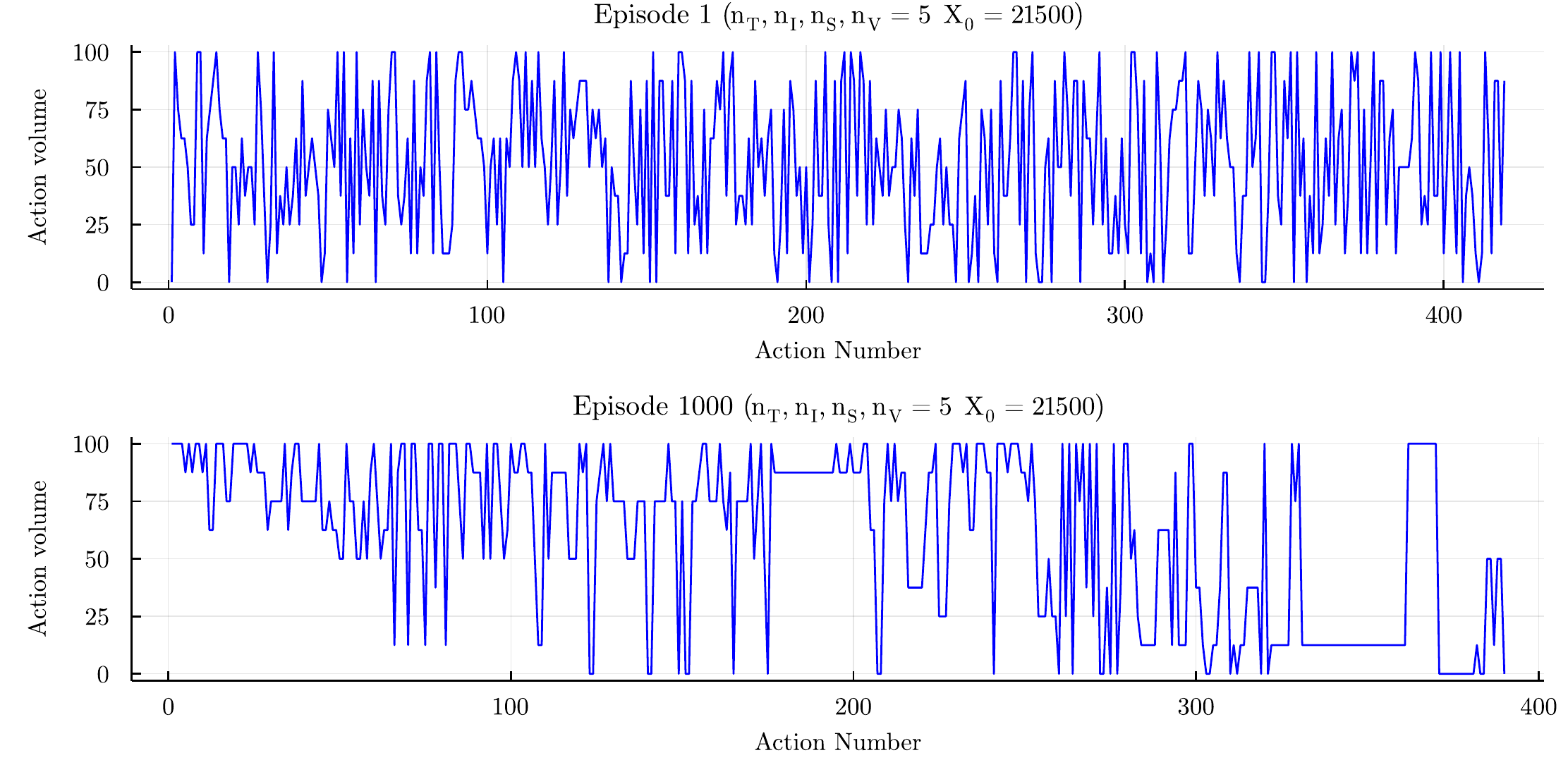}
            \caption{Plots the action volume trajectory}
            \label{app:rl-action-volume-V50-S5}
        \end{subfigure}
        \label{rl-policy-volume-trajectory-V50-S5}
        \caption{Plots the policy and the action volume trajectory for agent with $X_{0} = 21500$ and $n_{T},n_{I},n_{S},n_{V}=5$.}
\end{figure}

\begin{figure}[h!]
  \begin{subfigure}[b]{0.45\textwidth}
          \centering
             \includegraphics[width=\textwidth]{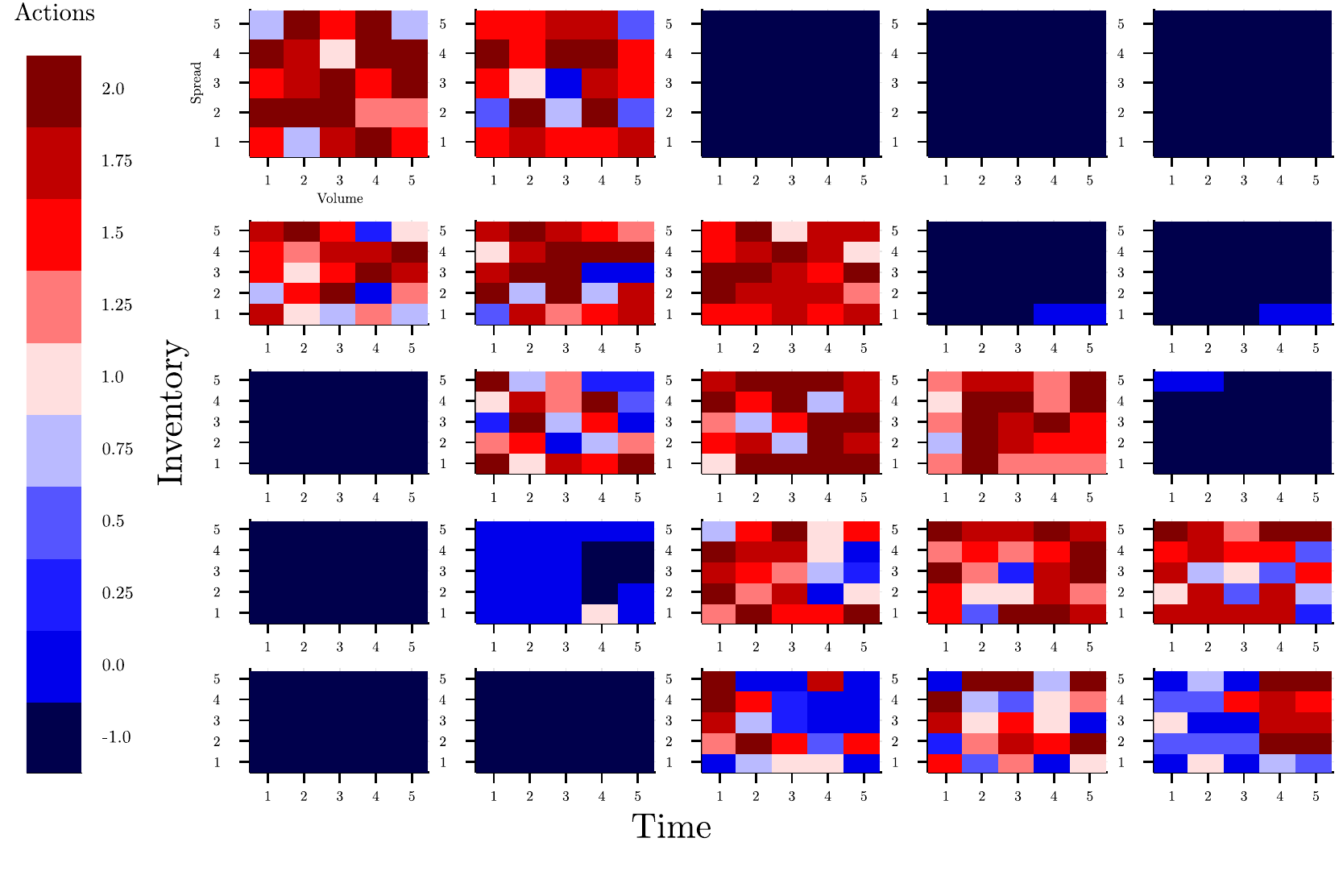}
            \caption{Plots the policy for the agent}
             \label{rl-policy-V100-S5}
        \end{subfigure}\hfill
        \begin{subfigure}[b]{0.45\textwidth}
          \centering
             \includegraphics[width=\textwidth]{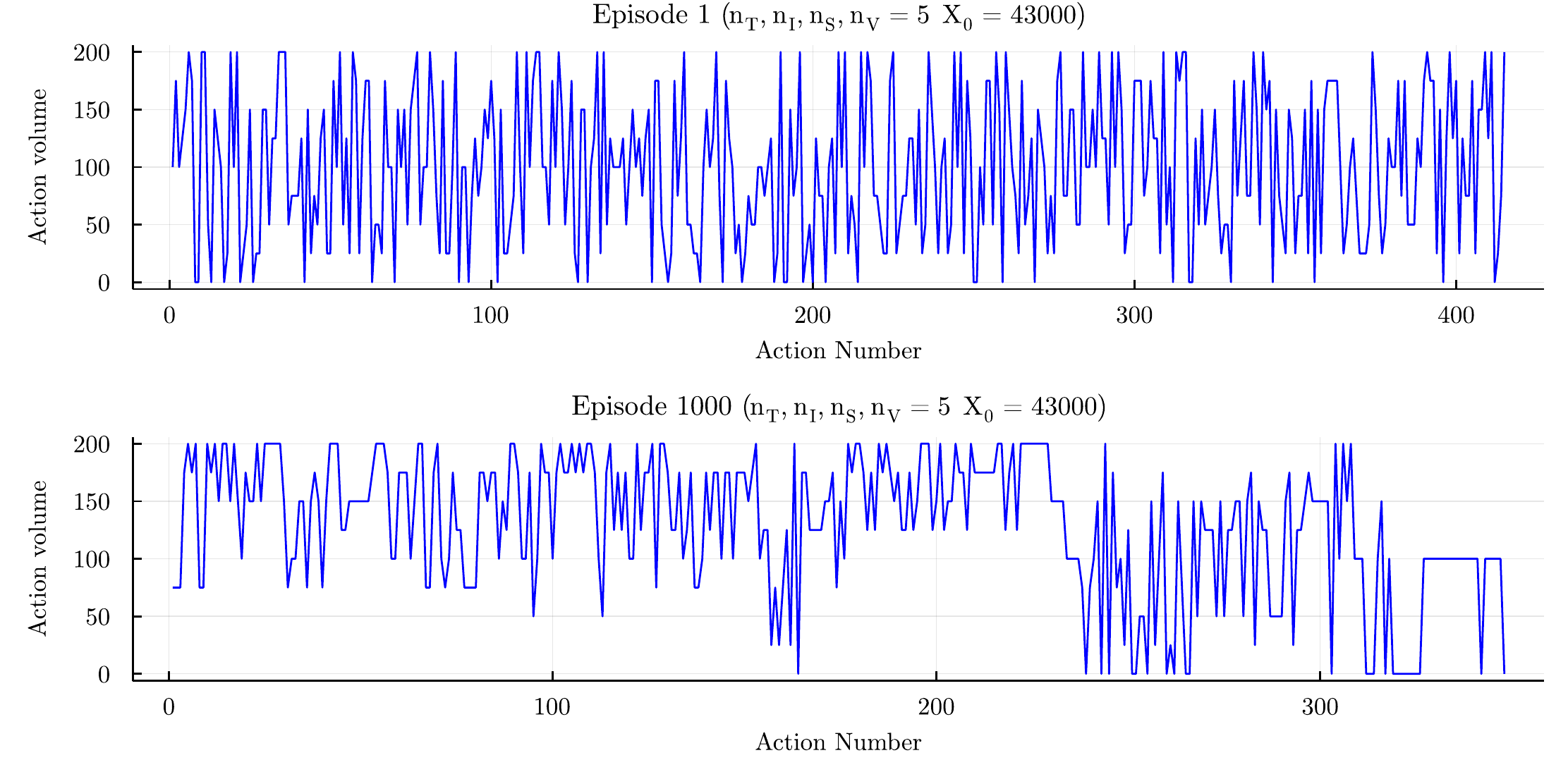}
            \caption{Action volume trajectories}%
            \label{rl-action-volume-V100-S5}
        \end{subfigure}\hfill
    \centering
    \caption{Plots the policy and the action volume trajectory for agents with $X_{0} = 43000$ and $n_{T},n_{I},n_{S},n_{V}=5$.}
\end{figure}

\begin{figure}[h!]
    \centering
        \begin{subfigure}[b]{0.45\textwidth}
          \centering
            \includegraphics[width=8cm, height=6cm]{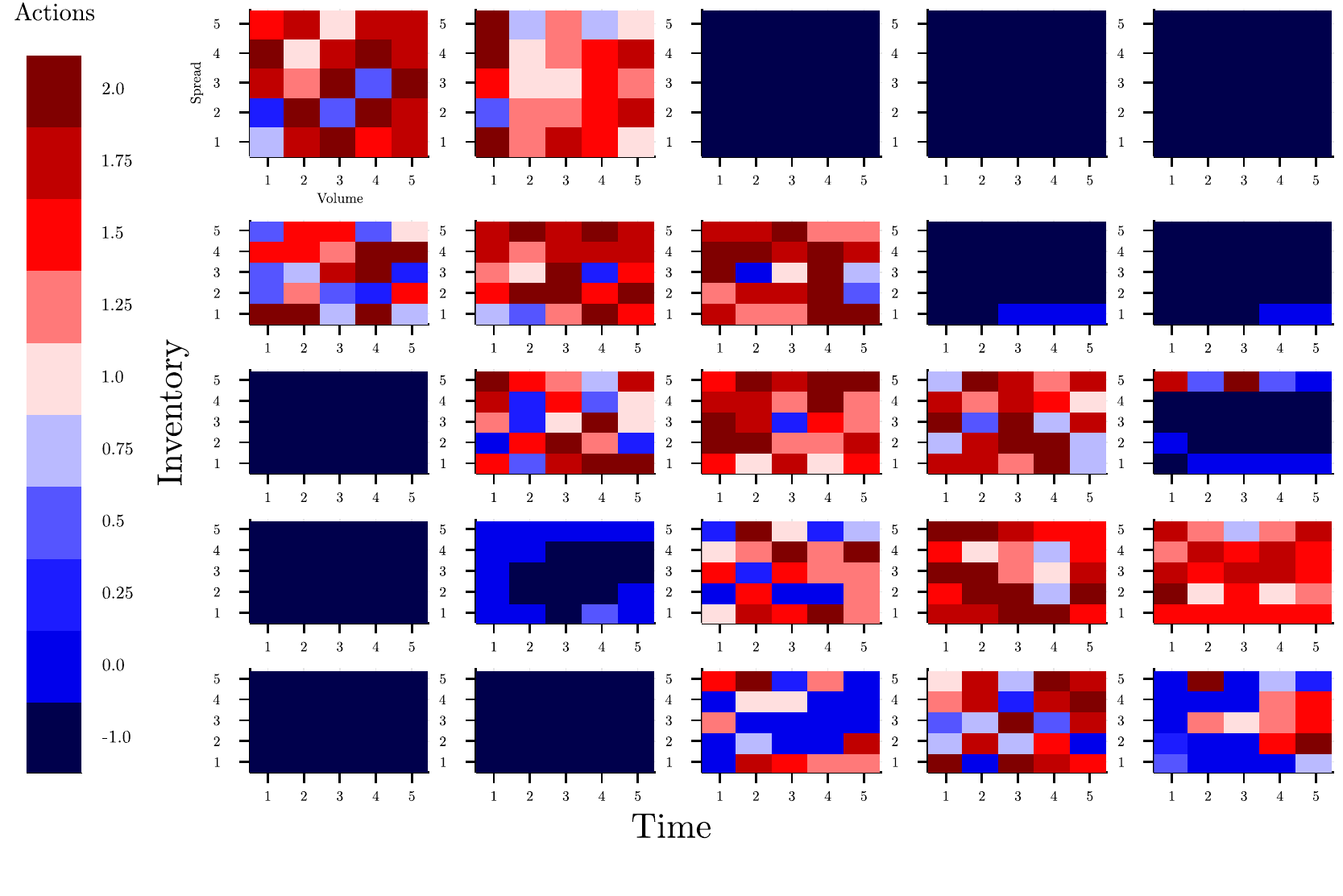}
            \caption{Plots the policy for the agent}
            \label{app:rl-policy-V200-S5}
        \end{subfigure}\hfill
        \begin{subfigure}[b]{0.45\textwidth}
          \centering
            \includegraphics[width=8cm, height=6cm]{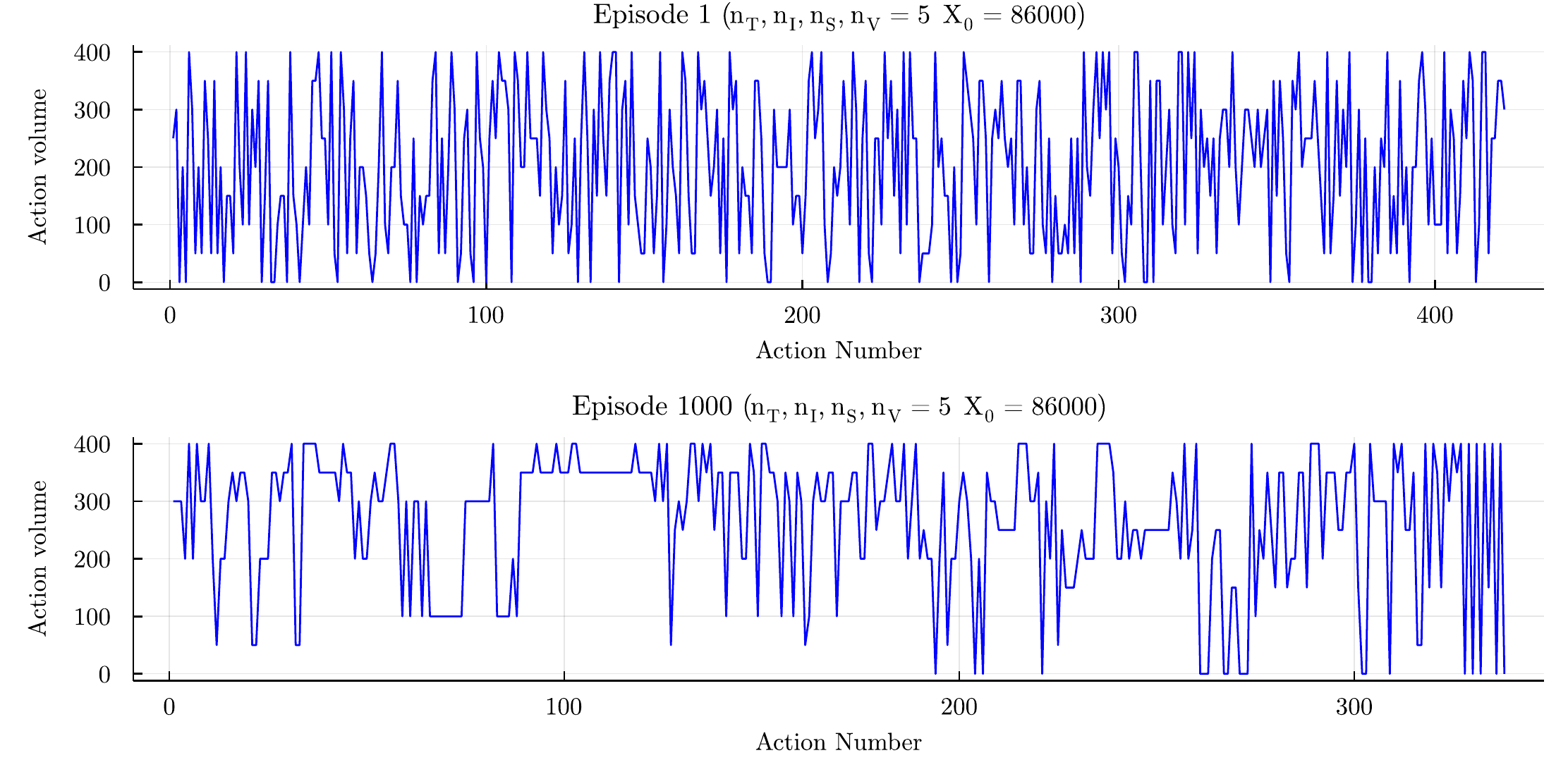}
            \caption{Plots the action volume trajectory for the agent}
            \label{app:rl-action-volume-V200-S5}
        \end{subfigure}
        \label{rl-policy-volume-trajectory-V200-S5}
        \caption{Plots the policy and the action volume trajectory for agents with $X_{0} = 86000$ and $n_{T},n_{I},n_{S},n_{V}=5$.}
\end{figure}

The action volume trajectory for the RL agent with an initial inventory of 43000 for the small and large state space are plotted with the action volume trajectories for the first episode (1) at the top, and the last episode (1000) at the bottom. In the first episode, the agent is randomly selecting actions and there is no pattern in the volume trajectory. For the last episode, the agent has learned to trade higher volumes initially and at the end of the simulation, when most of the inventory has been traded, the agent submits smaller orders. The large agent has also learned to trade more initially, however, there is more variation in the actions when compared with the smaller state space agent seen in Figures \ref{rl-action-volume-V100-S5}.

The states that have not yet been visited are given the action value of -1. The inventory states increase from top to bottom and the time passed increases left to right. For each time and inventory state combination, the actions taken across all volume and spread states are plotted. Fixing a particular inventory state and increasing the time past (moving left to right), you can see that the agent has learned to trade higher volume as the time increases when averaged across the spread and volume states. If you fix a time state and increase the inventory remaining (move bottom to top), you see that the agent has learned to trade at a higher rate when there is more inventory remaining.

\newpage
\begin{figure}[ht!]
    \centering
        \begin{subfigure}{\textwidth}
            \centering
            \begin{minipage}{.5\textwidth}
            \centering
            \includegraphics[width=9cm, height=6cm]{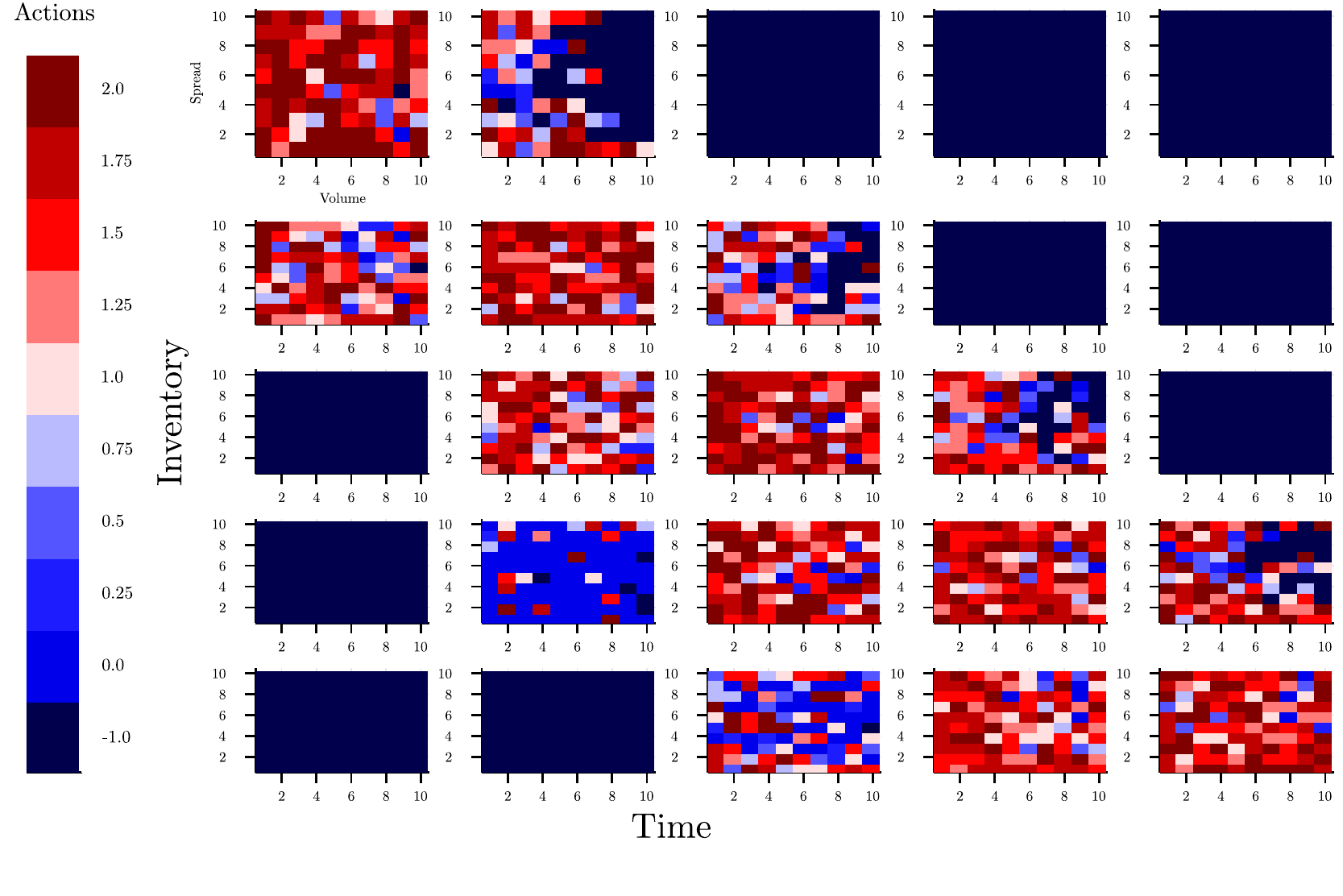}
            \end{minipage}%
            \begin{minipage}{.5\textwidth}
            \centering
            \includegraphics[width=9cm, height=6cm]{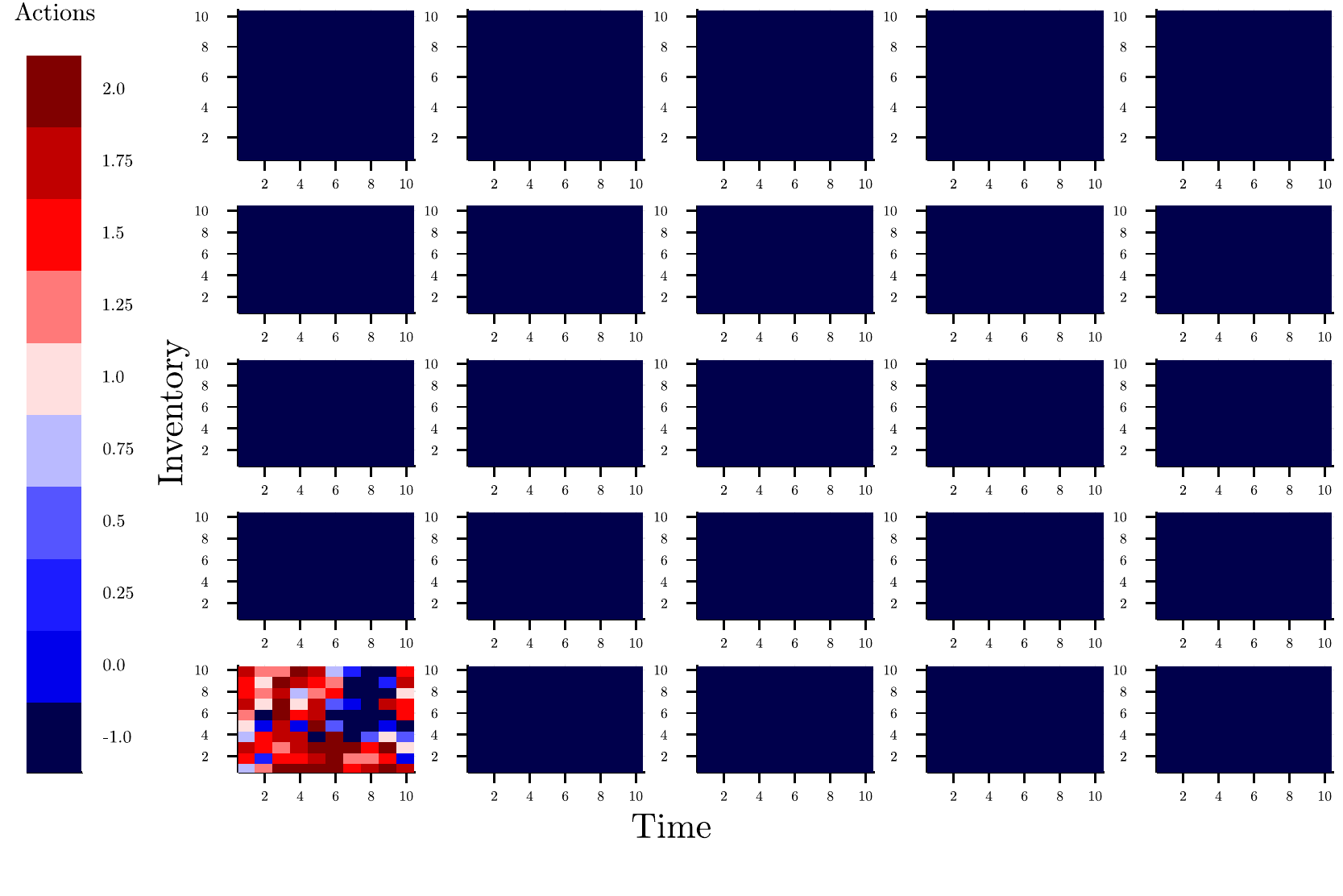}
            \end{minipage}
            \begin{minipage}{.5\textwidth}
            \centering
            \includegraphics[width=9cm, height=6cm]{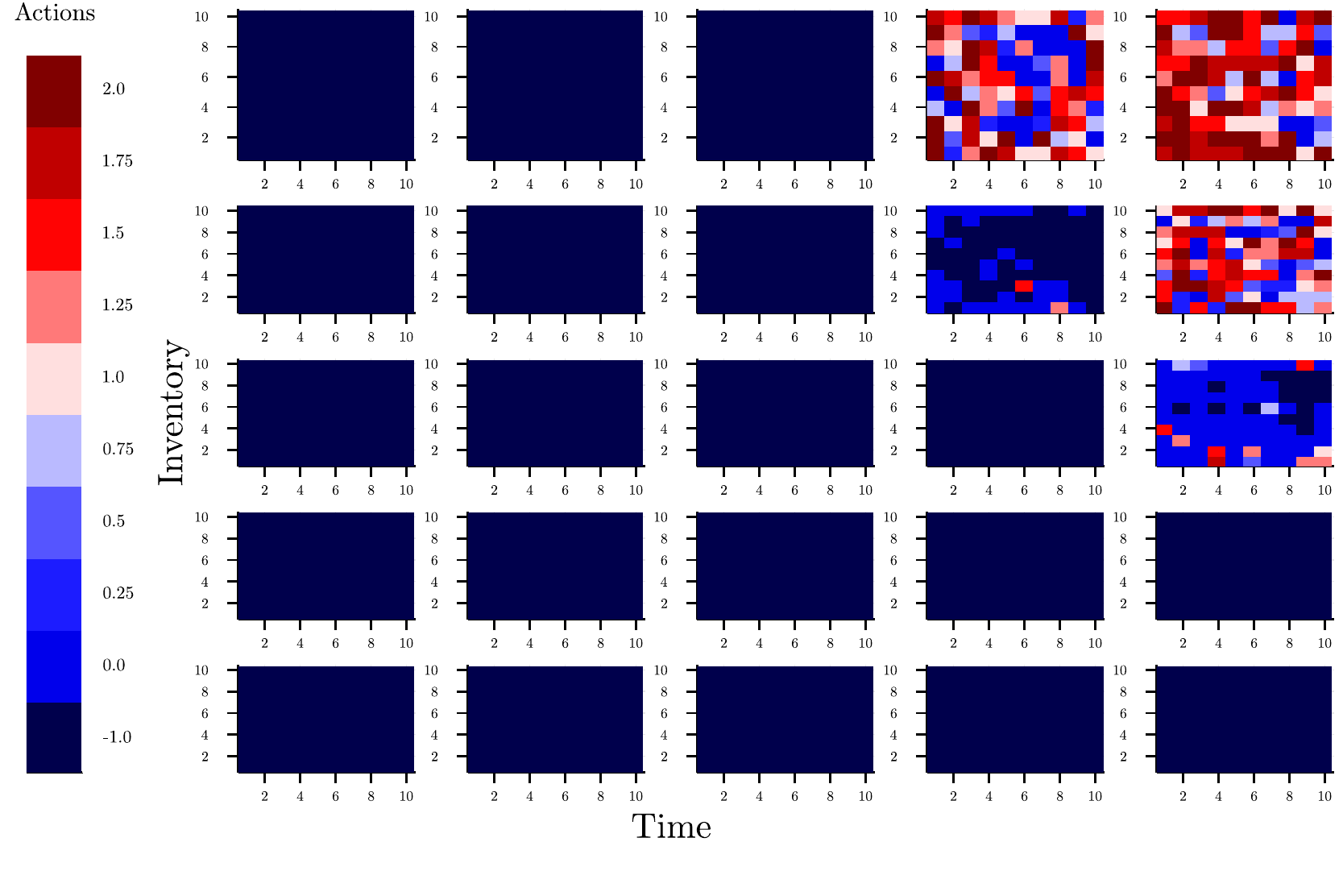}
            \end{minipage}%
            \begin{minipage}{.5\textwidth}
            \centering
            \includegraphics[width=9cm, height=6cm]{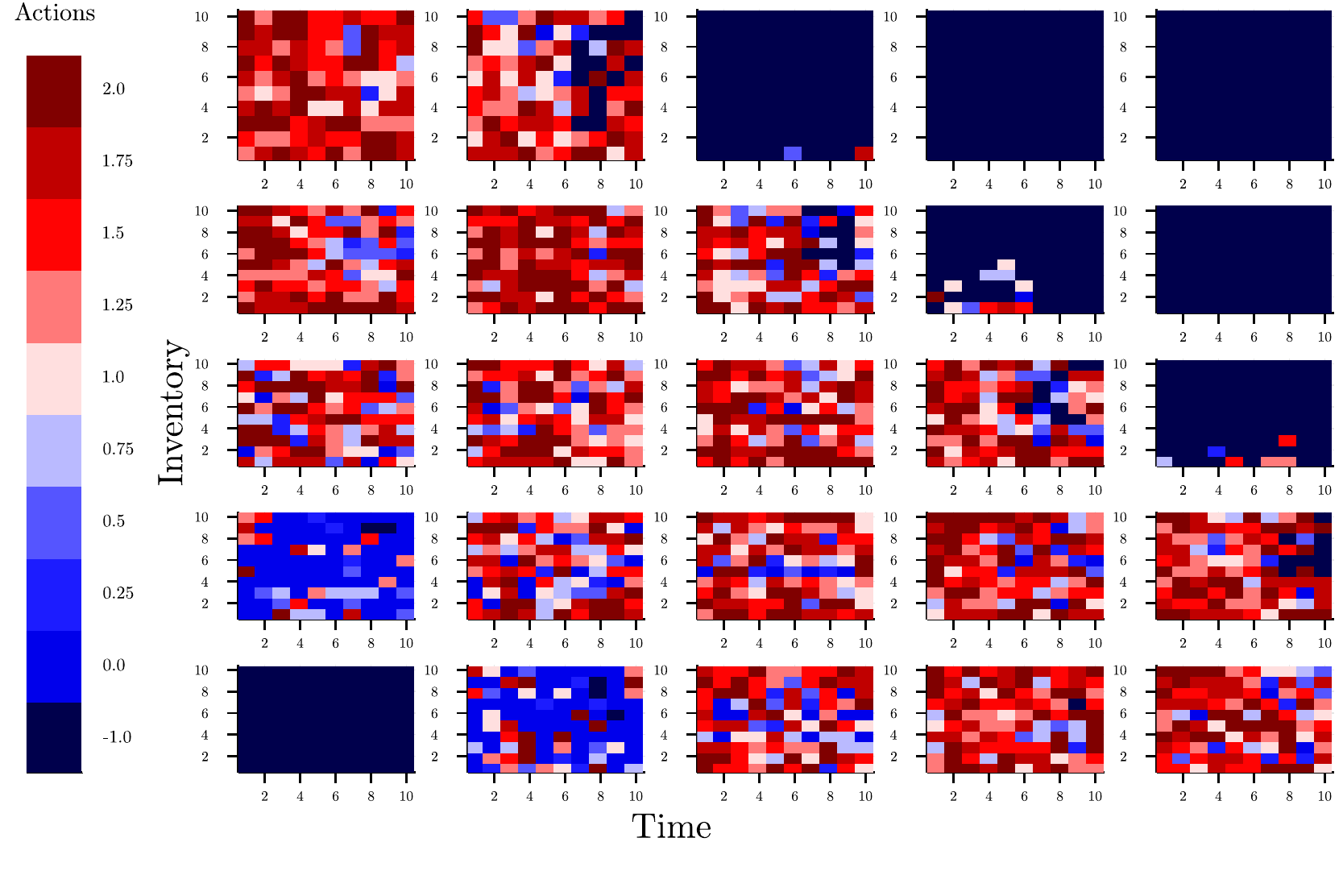}
            \end{minipage}
            \caption{Plots the policy for the agent with $X_{0} = 21500$ and $n_{T},n_{I},n_{S},n_{V}=10$}
            \label{app:rl-policy-V50-S10}
        \end{subfigure}
        \begin{subfigure}[b]{\textwidth}
            \includegraphics[width=\textwidth,
            height=8cm]{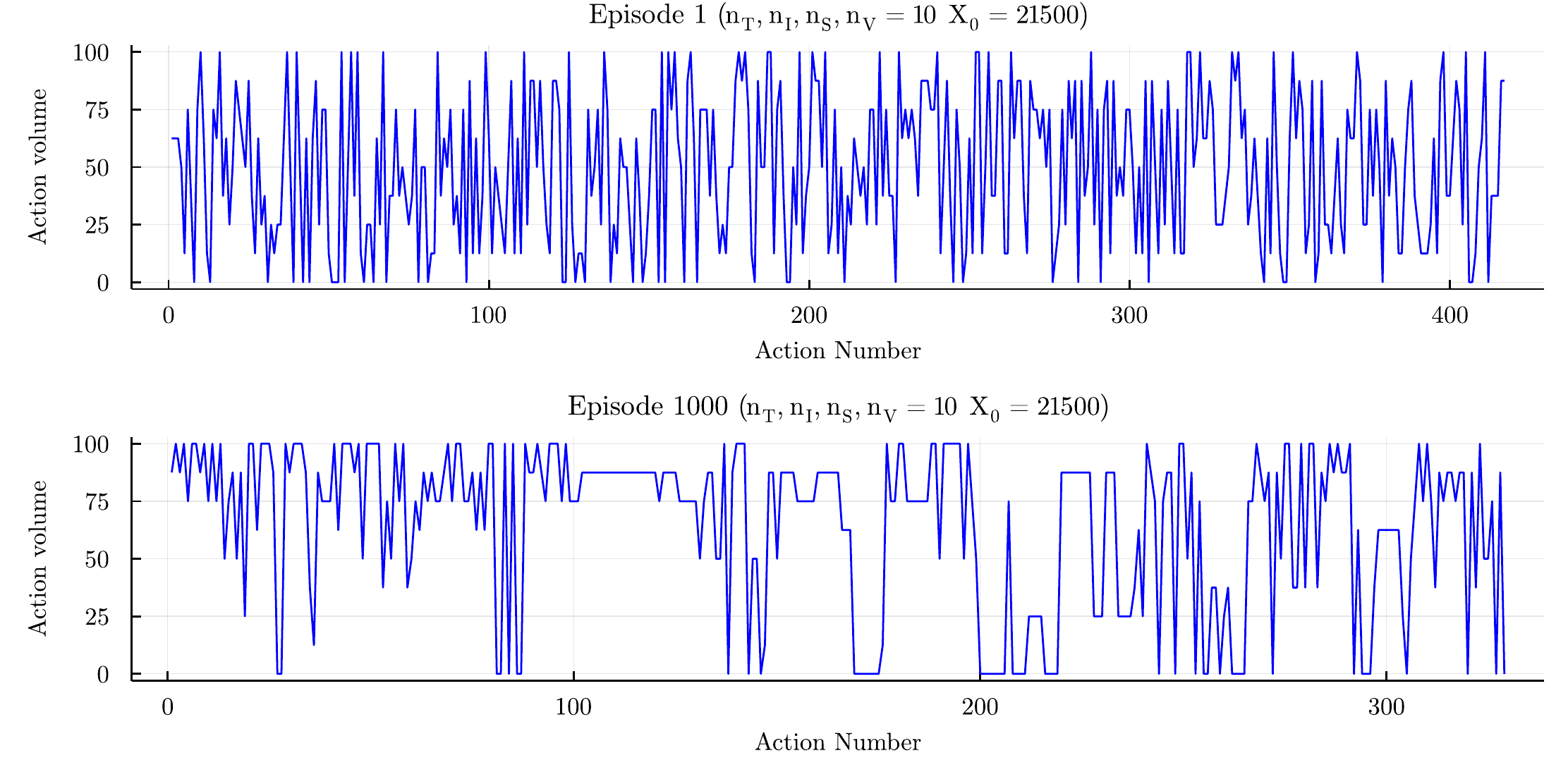}
            \caption{Plots the action volume trajectory for the agent with $X_{0} = 21500$ and $n_{T},n_{I},n_{S},n_{V}=10$.}
            \label{app:rl-action-volume-V50-S10}
        \end{subfigure}
        \caption{$X_{0} = 21500$ and $n_{T},n_{I},n_{S},n_{V}=10$}
\end{figure}
\newpage
\begin{figure}[ht!]
    \centering
        \begin{subfigure}{\textwidth}
            \begin{minipage}{.5\textwidth}
            \centering
            \includegraphics[width=9cm, height=6cm]{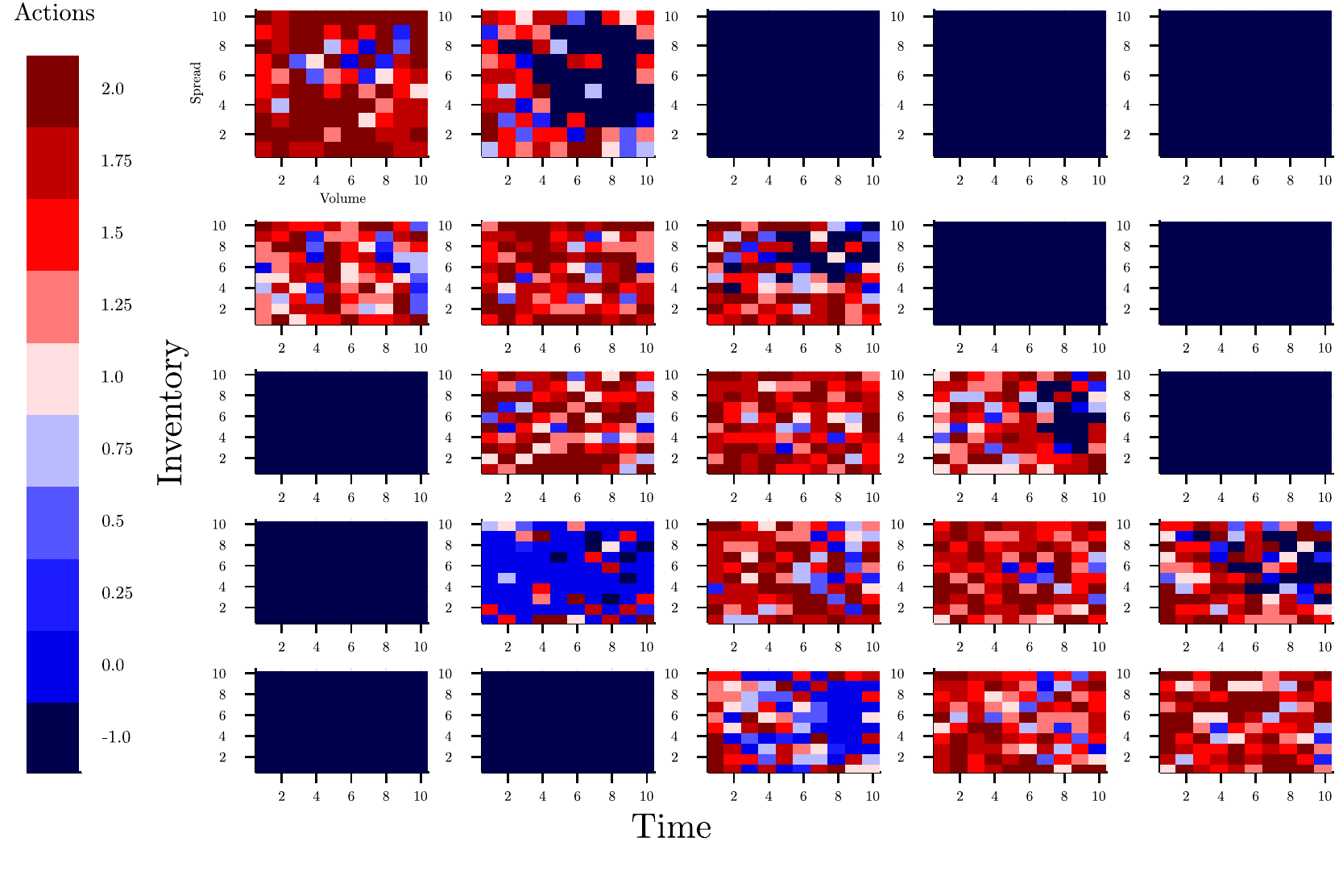}
            \end{minipage}%
            \begin{minipage}{.5\textwidth}
            \centering
            \includegraphics[width=9cm, height=6cm]{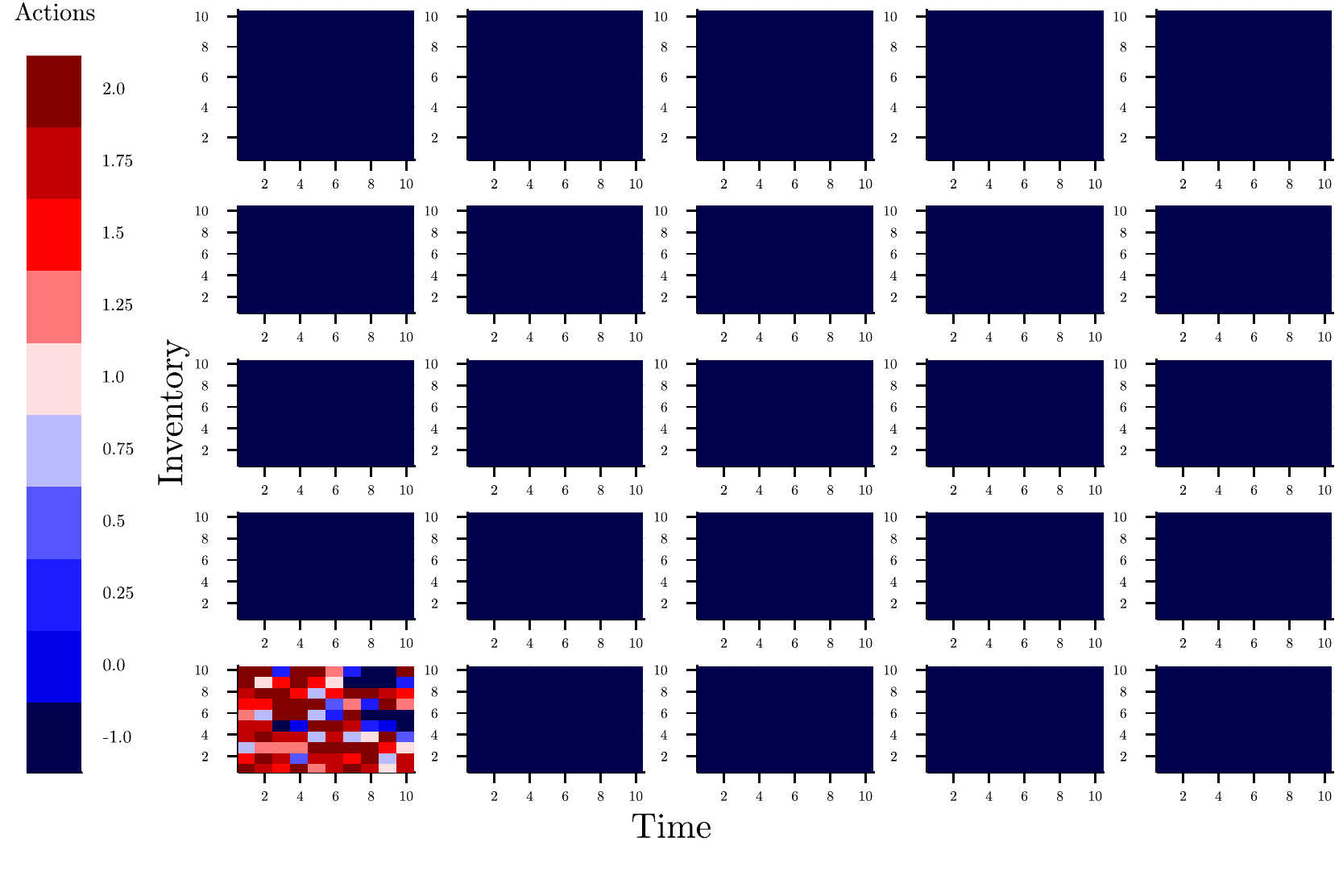}
            \end{minipage}
            \begin{minipage}{.5\textwidth}
            \centering
            \includegraphics[width=9cm, height=6cm]{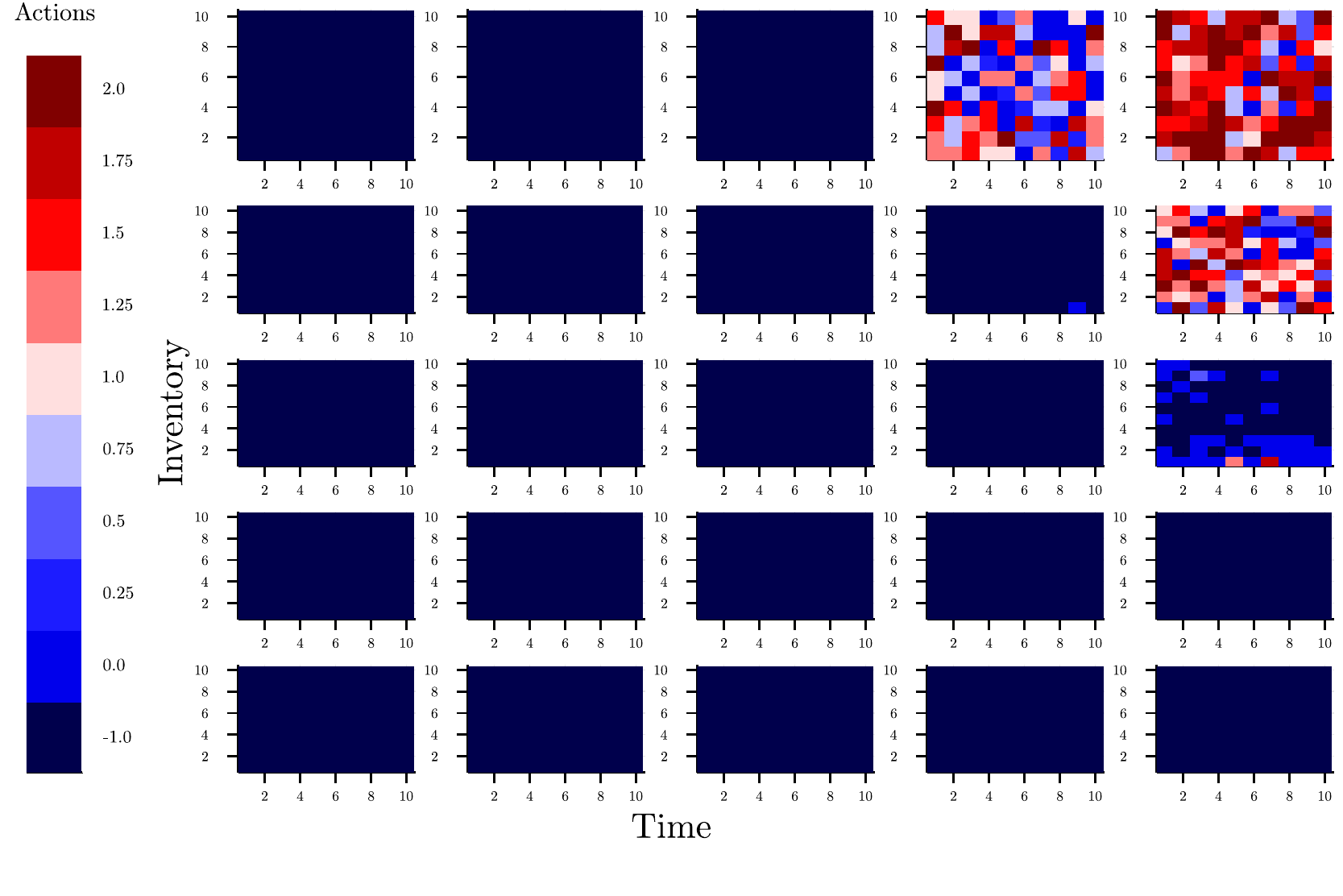}
            \end{minipage}%
            \begin{minipage}{.5\textwidth}
            \centering
            \includegraphics[width=9cm, height=6cm]{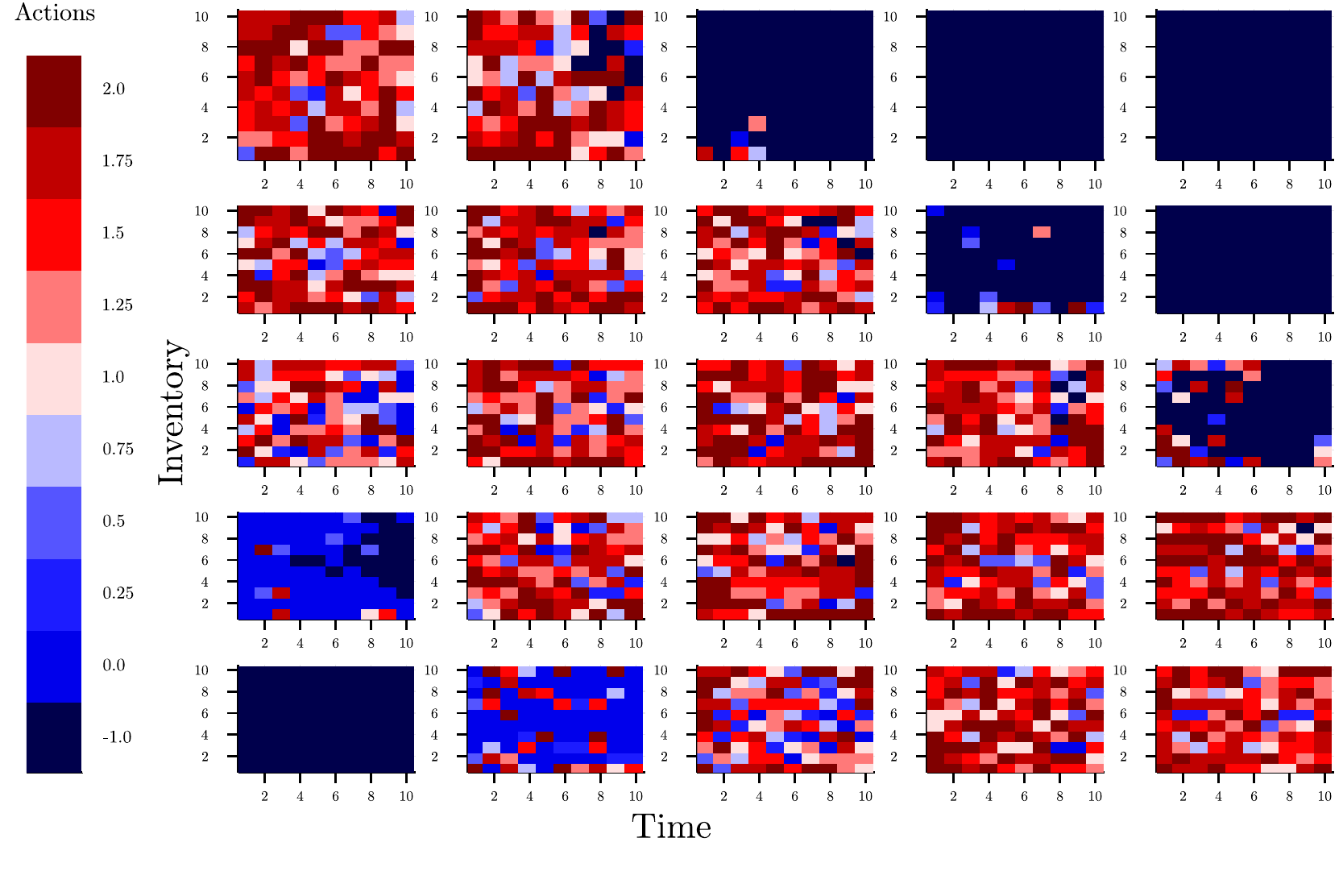}
            \end{minipage}
            \caption{Plots the policy for the agent with $X_{0} = 43000$ and $n_{T},n_{I},n_{S},n_{V}=10$.}
            \label{app:rl-policy-V100-S10}
        \end{subfigure}
        \begin{subfigure}[b]{\textwidth}
             \includegraphics[width=\textwidth, height=8cm]{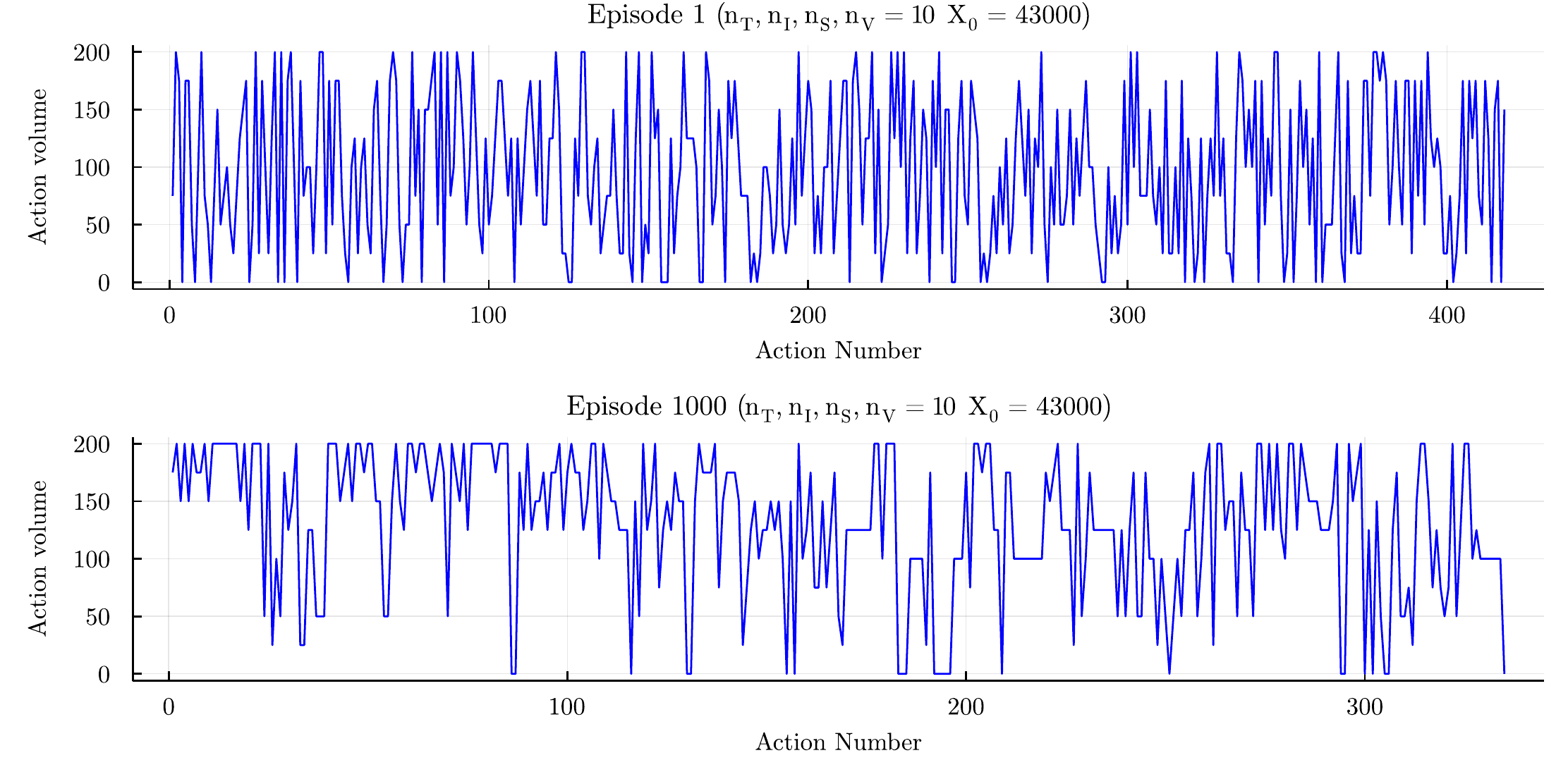}
            \caption{Large state space first and last volume trajectories}%
            \label{rl-action-volume-V100-S10}
        \end{subfigure}
        \caption{$X_{0} = 43000$ and $n_{T},n_{I},n_{S},n_{V}=10$}
        \label{rl-action-volume-V100}
\end{figure}

\newpage
\begin{figure}[ht!]
\begin{subfigure}{\textwidth}
        \centering
        \begin{minipage}{.5\textwidth}
          \centering
              \includegraphics[width=9cm, height=6cm]{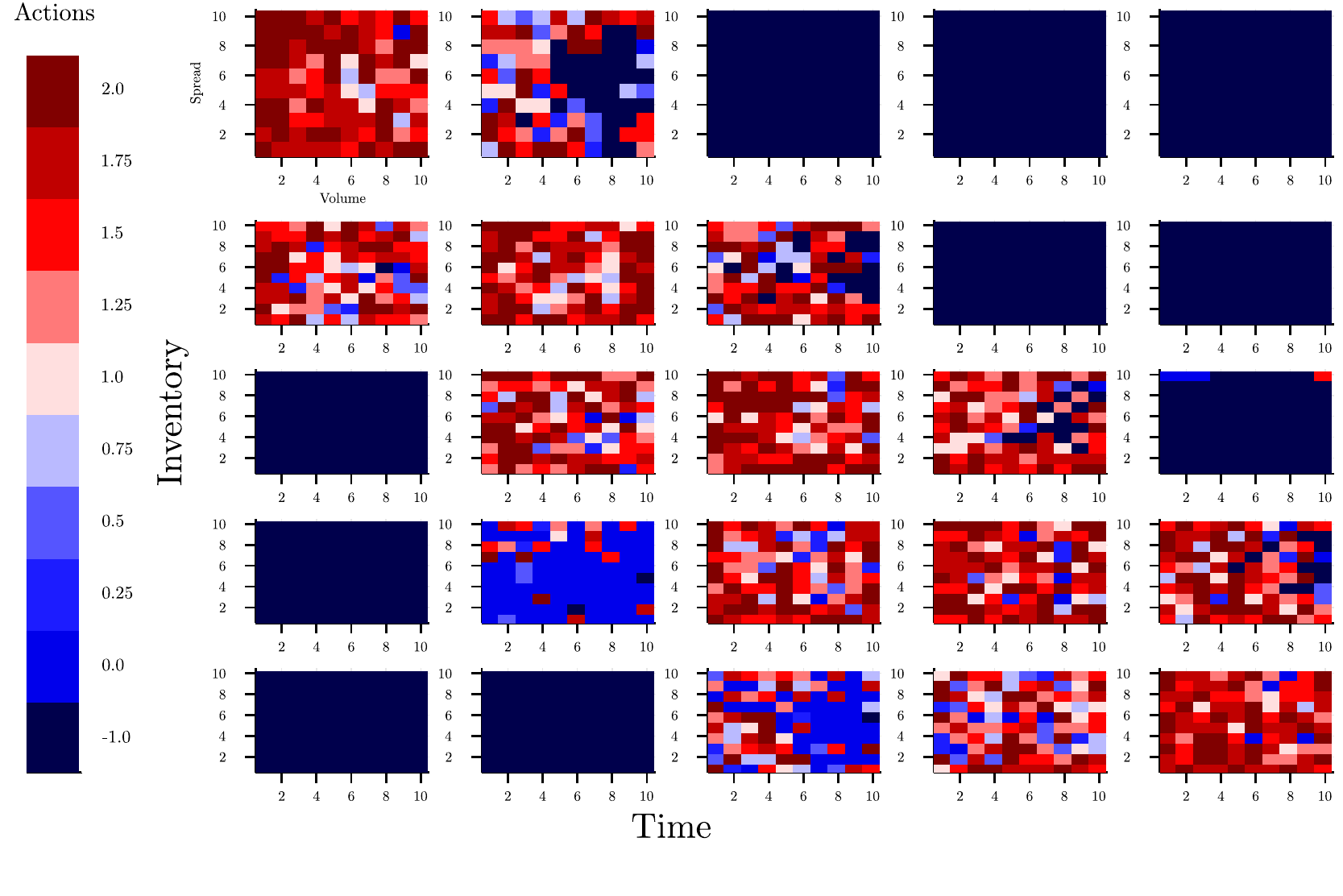}
        \end{minipage}%
        \begin{minipage}{.5\textwidth}
          \centering
          \includegraphics[width=9cm, height=6cm]{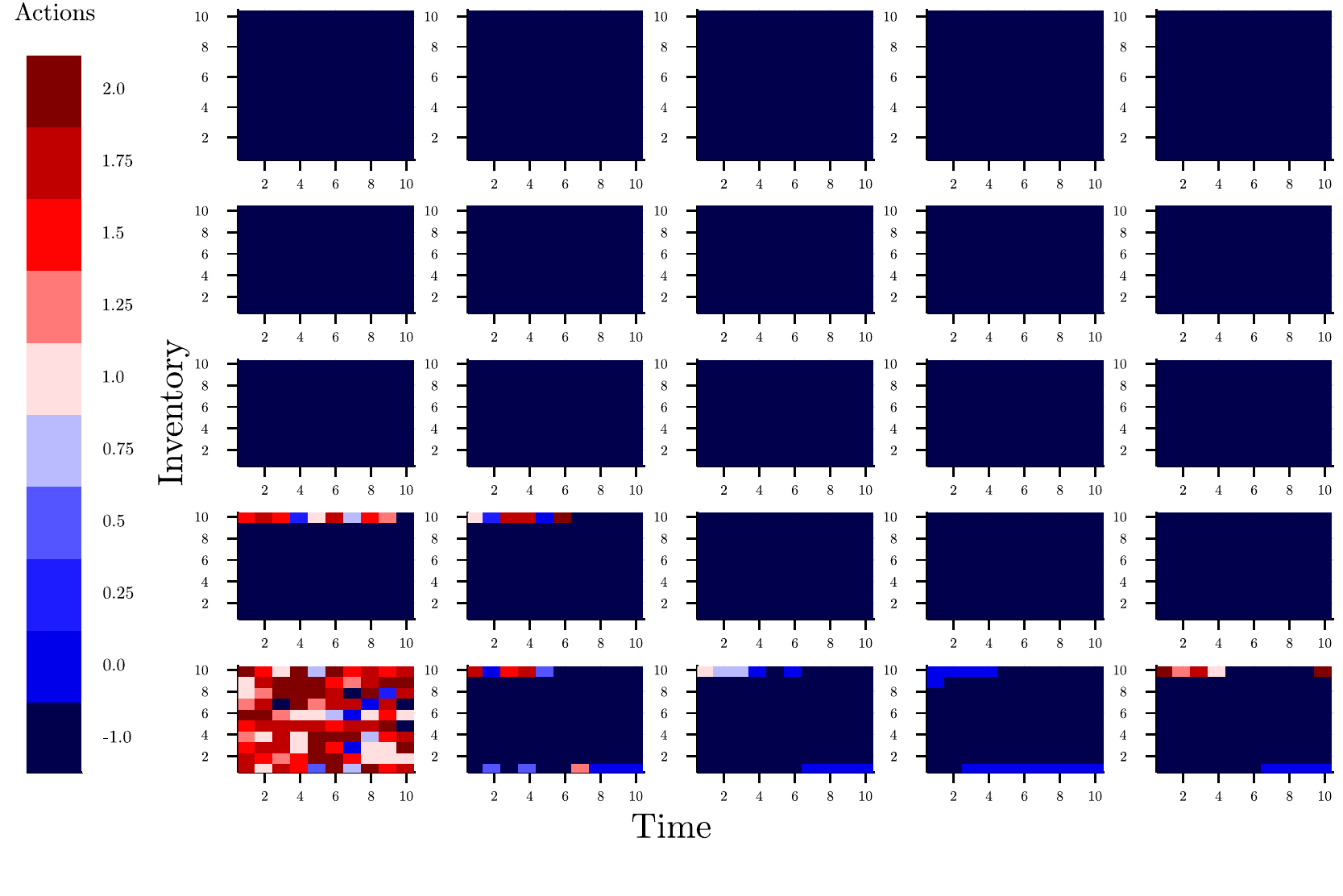}
        \end{minipage}
        \begin{minipage}{.5\textwidth}
          \centering
          \includegraphics[width=9cm, height=6cm]{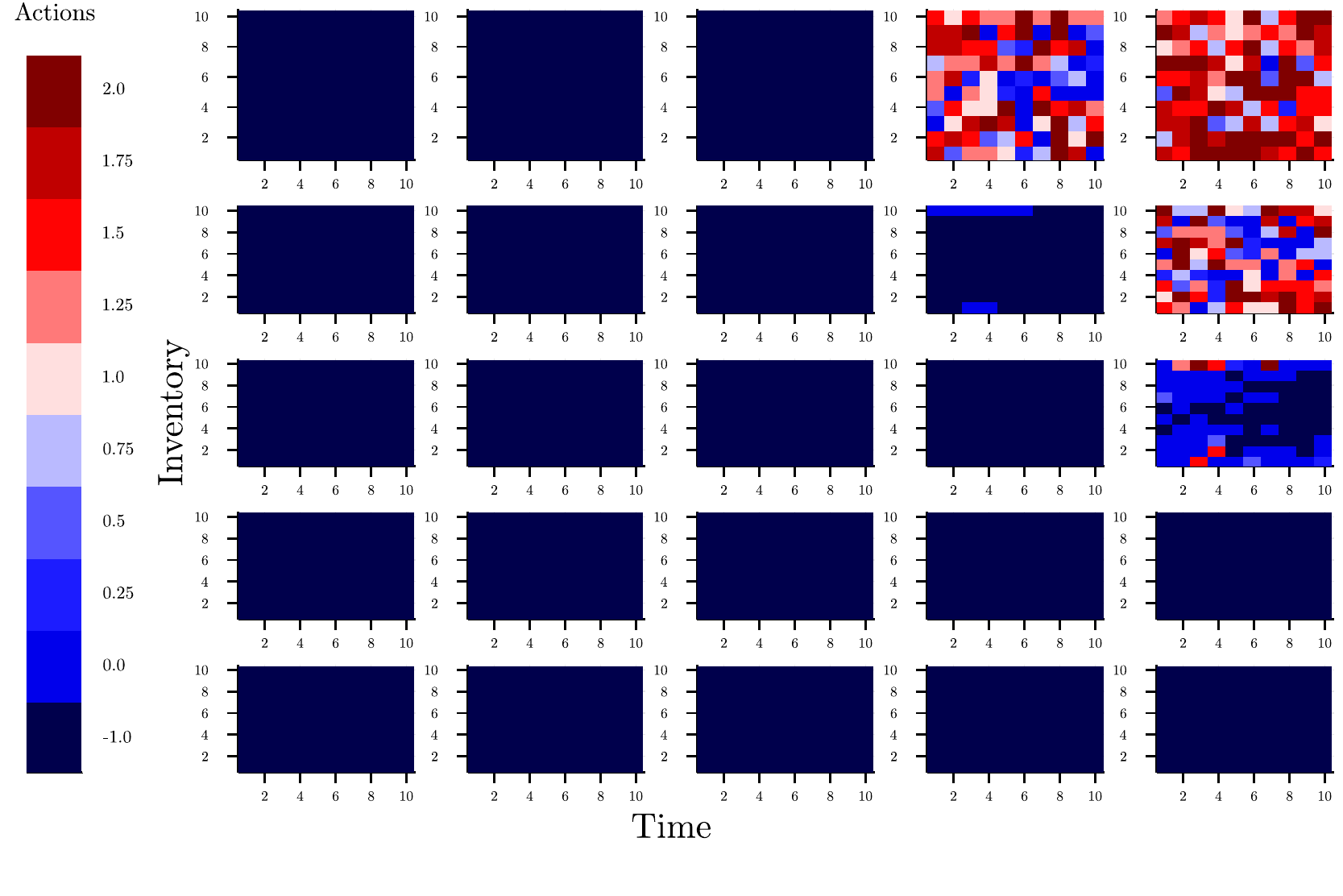}
        \end{minipage}%
        \begin{minipage}{.5\textwidth}
          \centering
          \includegraphics[width=9cm, height=6cm]{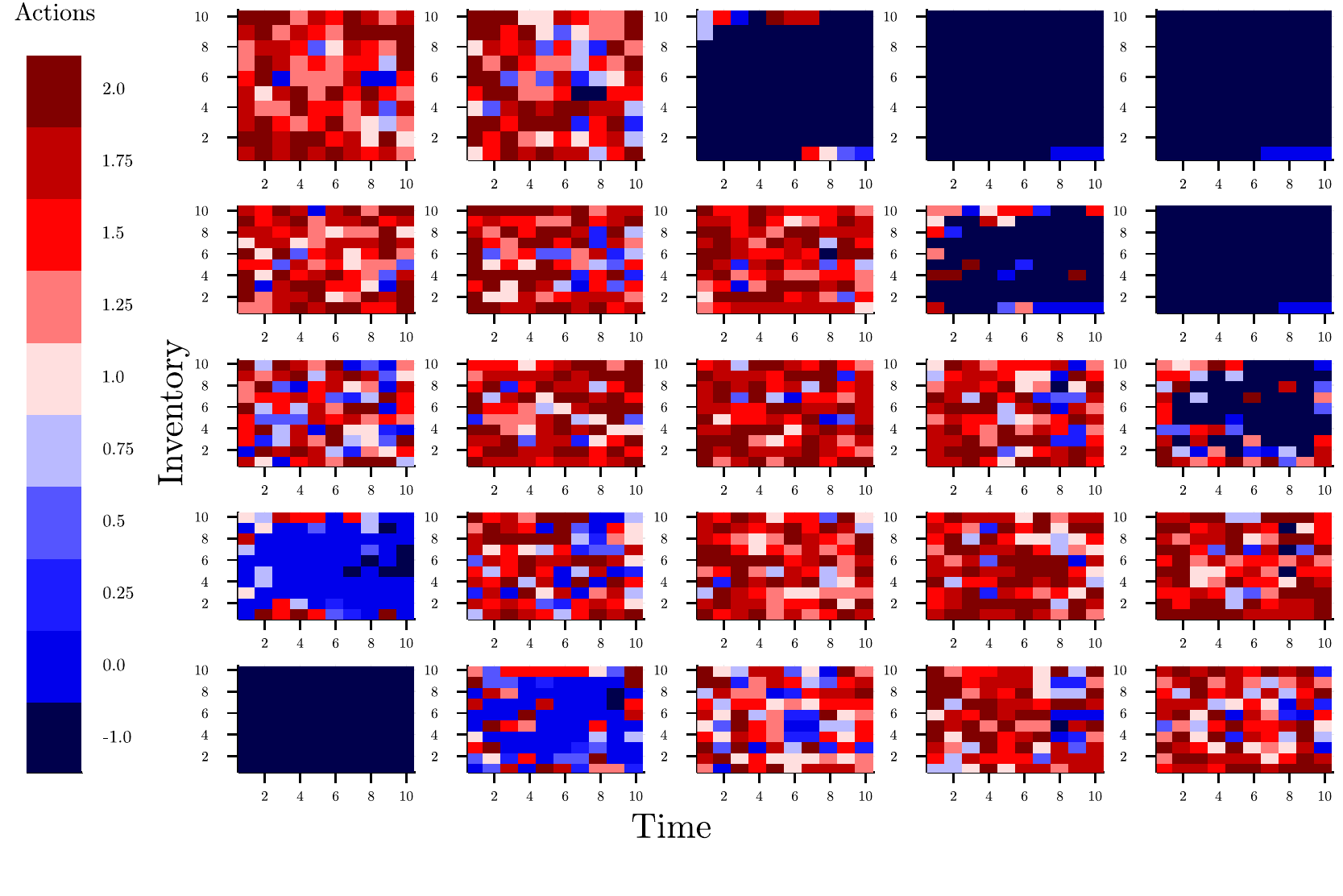}
        \end{minipage}
        \caption{Plots the policy for the agent with $X_{0} = 86000$ and $n_{T},n_{I},n_{S},n_{V}=10$.}
        \label{app:rl-policy-V200-S10}
\end{subfigure}
\begin{subfigure}{\textwidth}
    \centering
    \includegraphics[width=18cm, height=6cm]{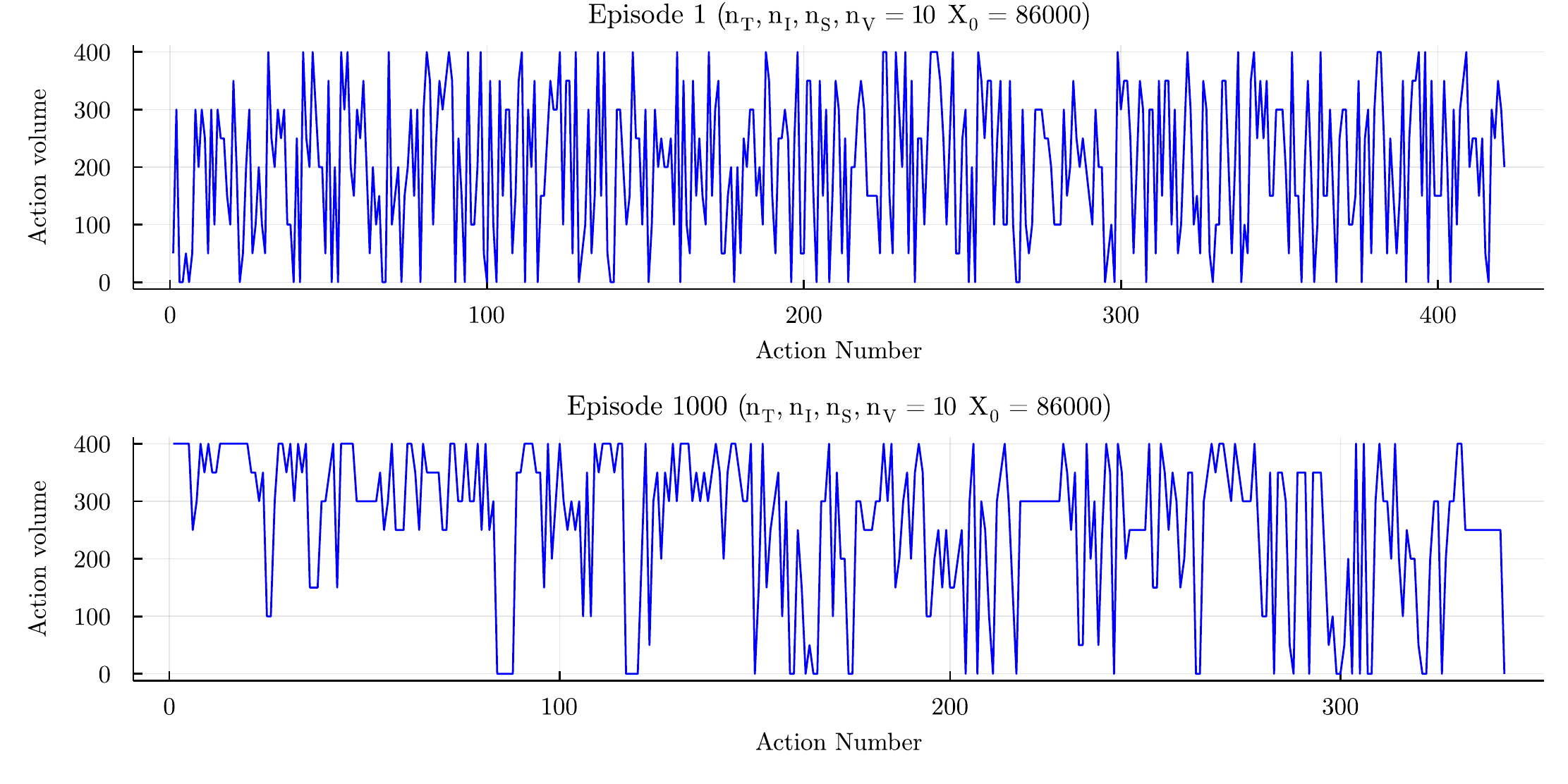}
    \caption{Plots the action volume trajectory for the agent with $X_{0} = 86000$ and $n_{T},n_{I},n_{S},n_{V}=10$.}
    \label{app:rl-action-volume-V200-S10}
\end{subfigure}
    \caption{$X_{0} = 86000$ and $n_{T},n_{I},n_{S},n_{V}=10$.}
\end{figure}

\newpage
\subsection{Learning dynamics} \label{app:RLAgentSimulationResults-LearningDynamics}
\label{ssec:RLAgentSimulationResults-BehaviourAnalysis}

Figure \ref{rl-action-volume-V100-S5} shows that in the first episode the agent is randomly selecting actions and there is no pattern in the volume trajectory. For the last episode the agent has learned to trade higher volumes initially, and at the end of the simulation, when most of the inventory has been traded, the agent starts to submit smaller orders. The higher inventory traded early in the simulation causes the agent to trade less when compared to the first episode. Figure \ref{rl-action-volume-V100-S5} shows the action volume trajectory for the RL agent with the larger state space. Here the agent has learned to initially trade more; however, there are more variations in the actions when compared with the smaller state space agent. These figures only visualise the results for two out of the six agents. The remainder of the agents' volume trajectories can be found in the supporting materials and give similar results.

\begin{figure}[h!]
        \centering
        \begin{subfigure}[b]{0.475\textwidth}
            \centering
            \includegraphics[width=\textwidth, height=6cm]{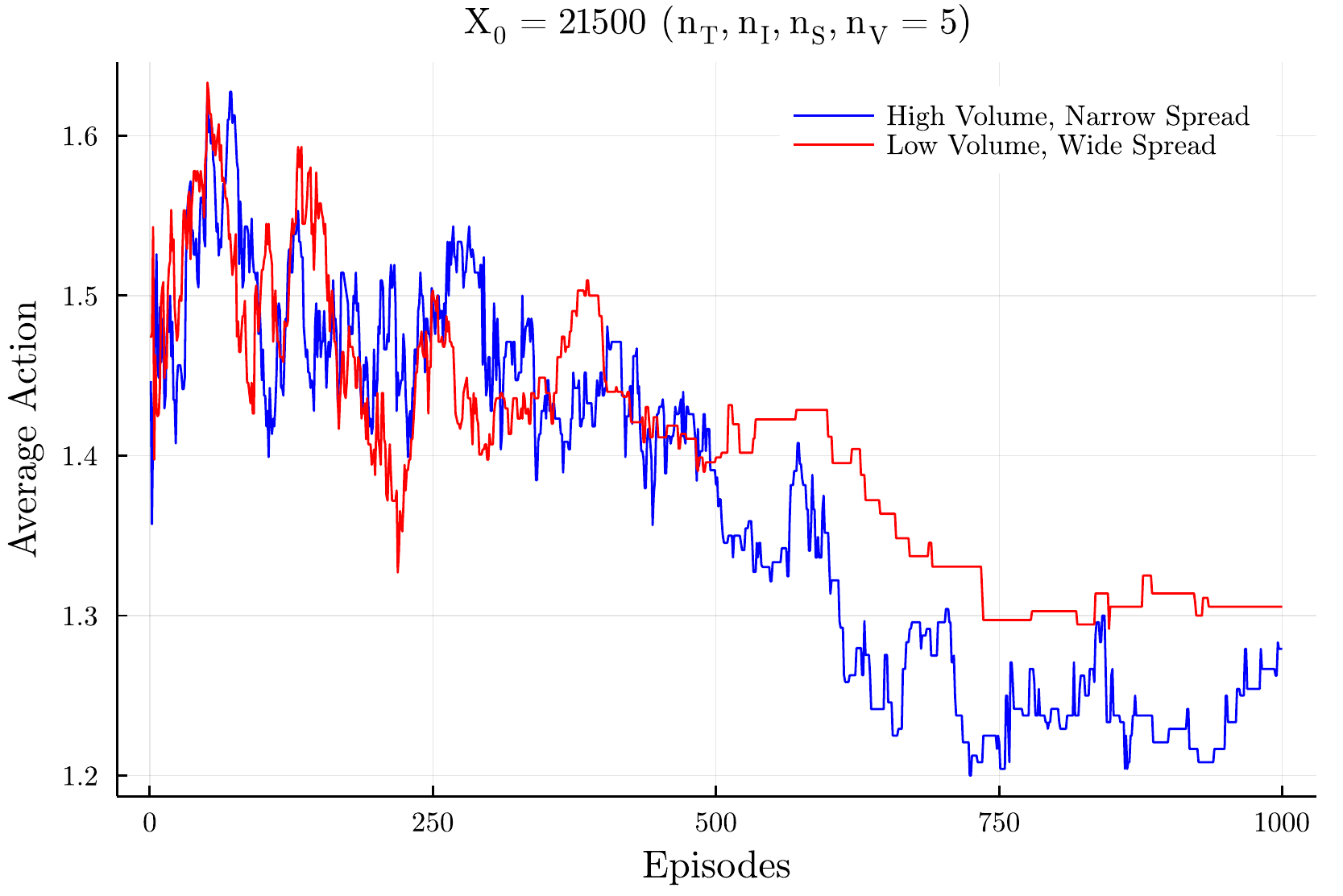}
            \caption[]{Small order, small state space}    
            \label{fig:average-action-SV-V50-S5}
        \end{subfigure}
        \hfill
        \begin{subfigure}[b]{0.475\textwidth}  
            \centering 
            \includegraphics[width=\textwidth, height=6cm]{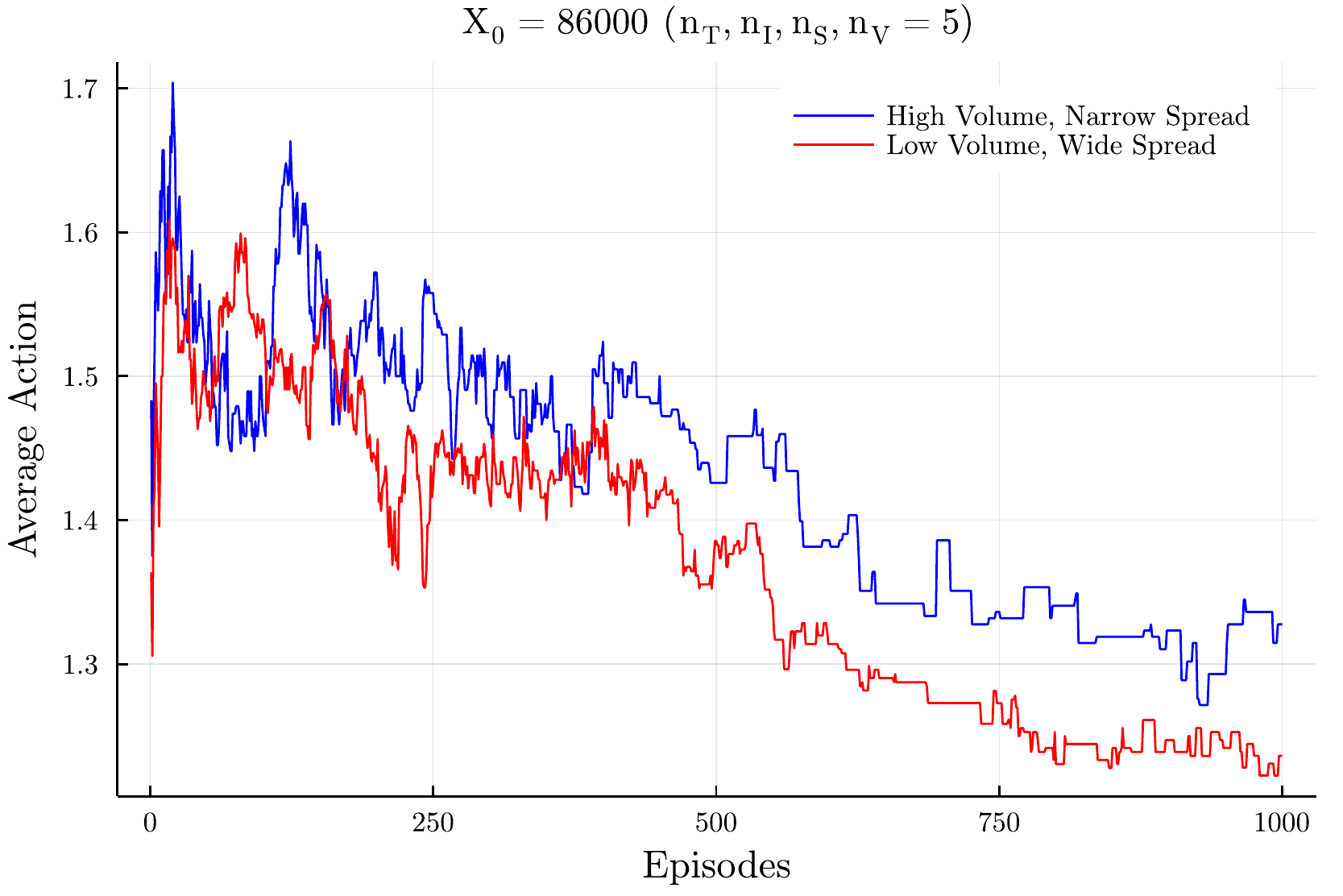}
            \caption[]{Large order, small state space}
            \label{fig:average-action-SV-V200-S5}
        \end{subfigure}
        \vskip\baselineskip
        \begin{subfigure}[b]{0.475\textwidth}
            \centering
            \includegraphics[width=\textwidth, height=6cm]{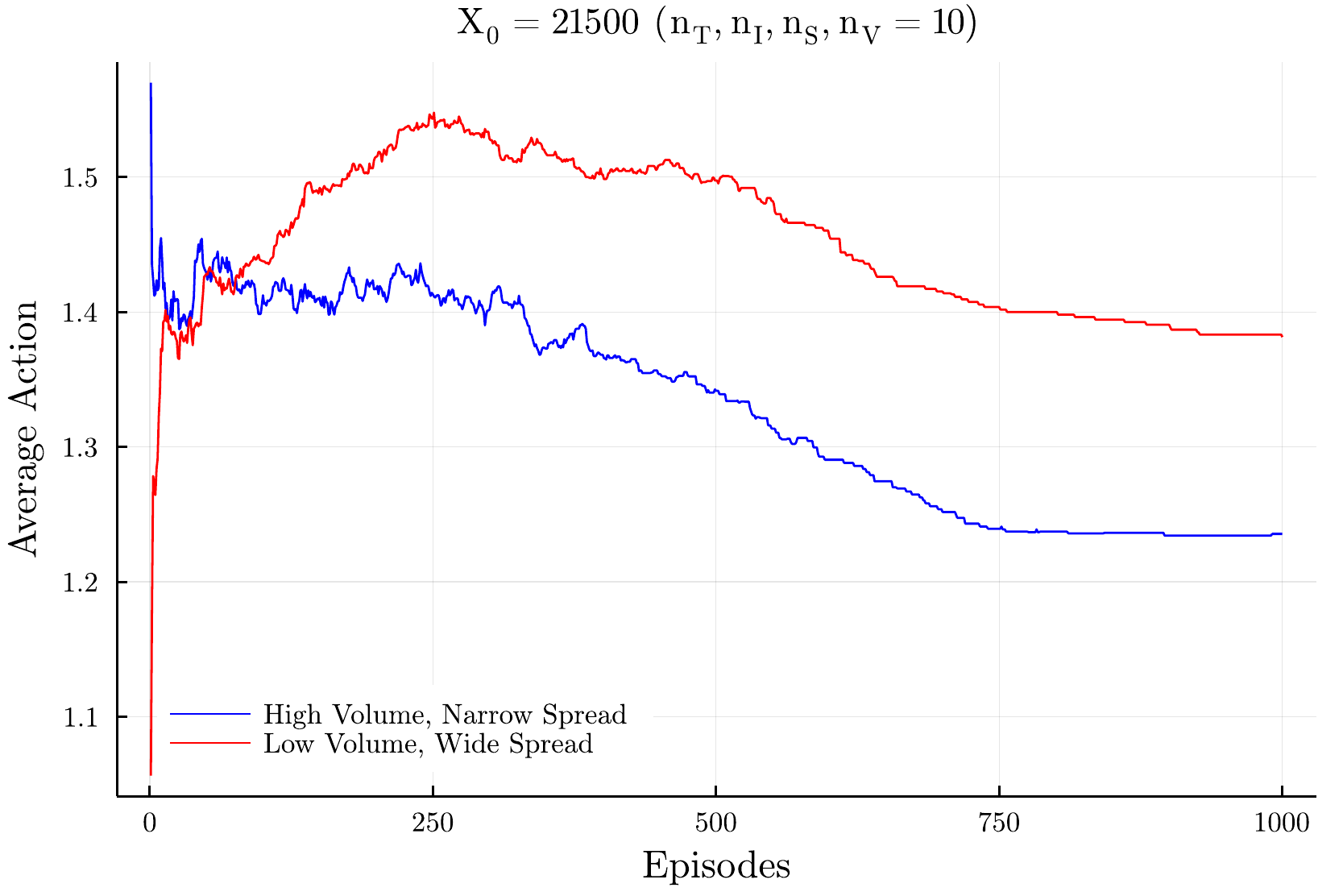}
            \caption[]{Small order, large state space}    
            \label{fig:average-action-V50-S10}
        \end{subfigure}
        \hfill
        \begin{subfigure}[b]{0.475\textwidth}  
            \centering 
              \includegraphics[width=\textwidth, height=6cm]{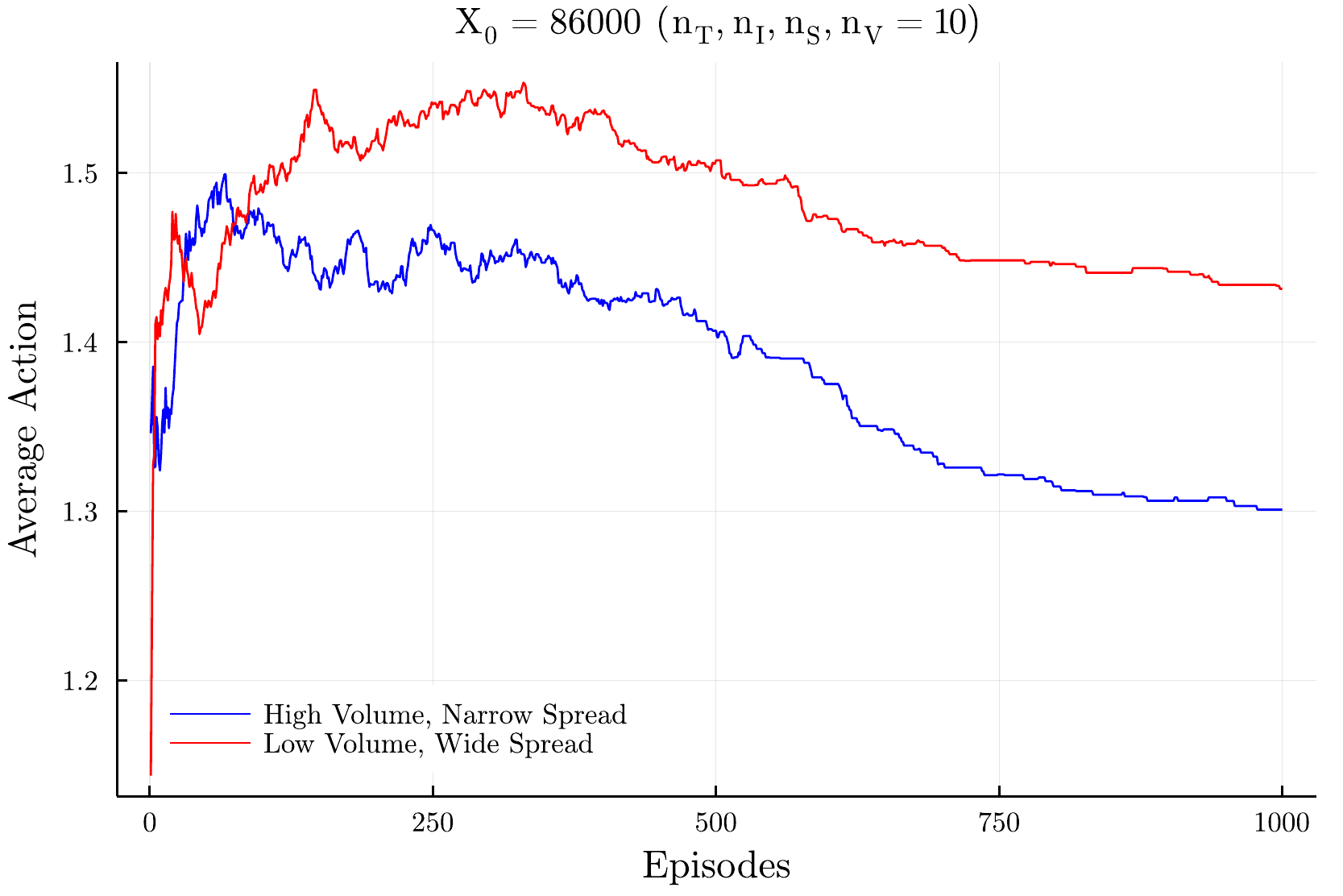}
            \caption[]{Large order, large state space}    
            \label{fig:average-action-V200-S10}
        \end{subfigure}

 \caption{We select the states that correspond to high (resp. low) volume and narrow (resp. wide) spread and take an average over the actions selected by the greedy policy to determine if the agent has learned to trade more (resp. less) when the volume is high (resp. low) and when the spreads are narrow (resp. wide). Figure (\ref{fig:average-action-SV-V50-S5}) shows that at lower initial inventory levels the agent was not able to learn this behaviour. However, for larger initial inventory levels (Figure \ref{fig:average-action-SV-V200-S5}) the agent was able to learn the correct intuitive actions. In both plots the agent has learned, on average, to decrease the amount of inventory traded. 
 The agents with the larger state space have not been able to learn the intuitive actions as they trade significantly more when the volumes are low and the spreads are wide and are shown in Figure \ref{fig:average-action-V50-S10} and Figure \ref{fig:average-action-V200-S10} in the lower row.}
        \label{rl-average-action-spread-volume-S10}
         \label{rl-average-action-spread-volume-S5}
    \end{figure}

Figure \ref{rl-average-action-spread-volume-S5} shows the average action for high (resp. low) volume and narrow (resp. wide) spreads for two of the RL agents with the smaller state space, but with different initial inventories. In both plots, we can see that the agent has learned to decrease the amount of inventory that is traded as the agent was trained for longer. This could be a sign that the agent was initially trading lots of inventory due to the reward signal but subsequently learned that this causes a larger price impact and hence reduced the inventory traded. For the lower initial inventory case (left), the agent has not learned to trade more when there is high volume and narrow spreads. However, when the initial inventory is increased (right) we see that it has learned the intuitive behaviour. 

To compare the learning dynamics of the agents with the larger state space to the agents with the smaller state space we plot the average action for the high (resp. low) and narrow (resp. wide) spread states. Figure \ref{rl-average-action-spread-volume-S10} shows these results. Low and high volume states are given by $v \leq 4$ and $v \geq 7$ respectively, and narrow and wide spread states are given by $s \leq 4$ and $s \geq 6$ respectively. Once again we see that the agents are learning to trade less on average as the agents get trained. The larger state space agents trade more when the spreads are wide and the volume is low instead of trading more when the spreads are narrow and the volume is high. These agents have not learned intuitive behaviour. 

\begin{figure}[h]
        \centering
        \begin{subfigure}[b]{0.475\textwidth}   
            \centering 
            \includegraphics[width=\textwidth, height=6cm]{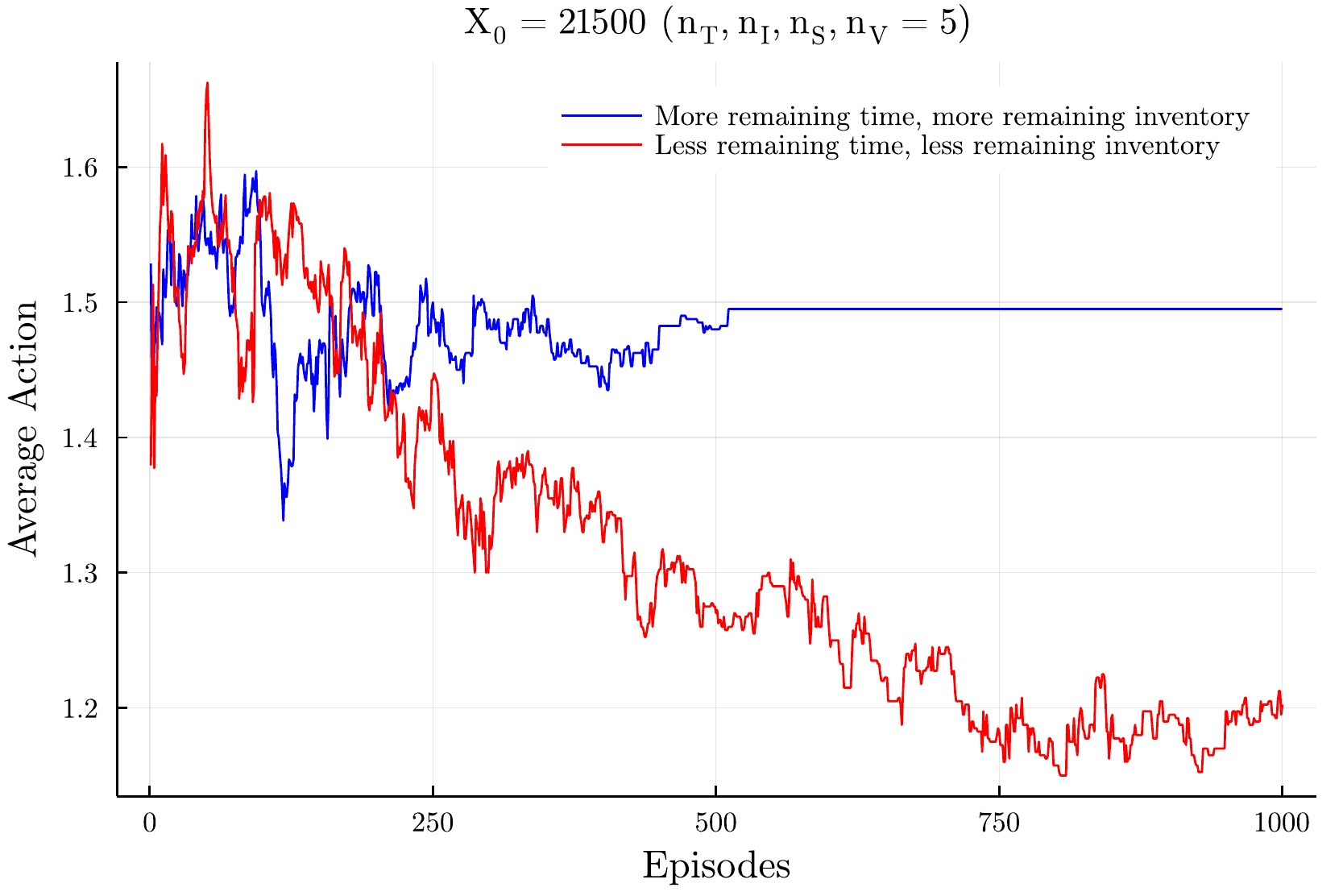}
            \caption[]{Small order, small state space}
            \label{fig:average-action-TI-V50-S5}
        \end{subfigure}
        \hfill
        \begin{subfigure}[b]{0.475\textwidth}   
            \centering 
            \includegraphics[width=\textwidth, height=6cm]{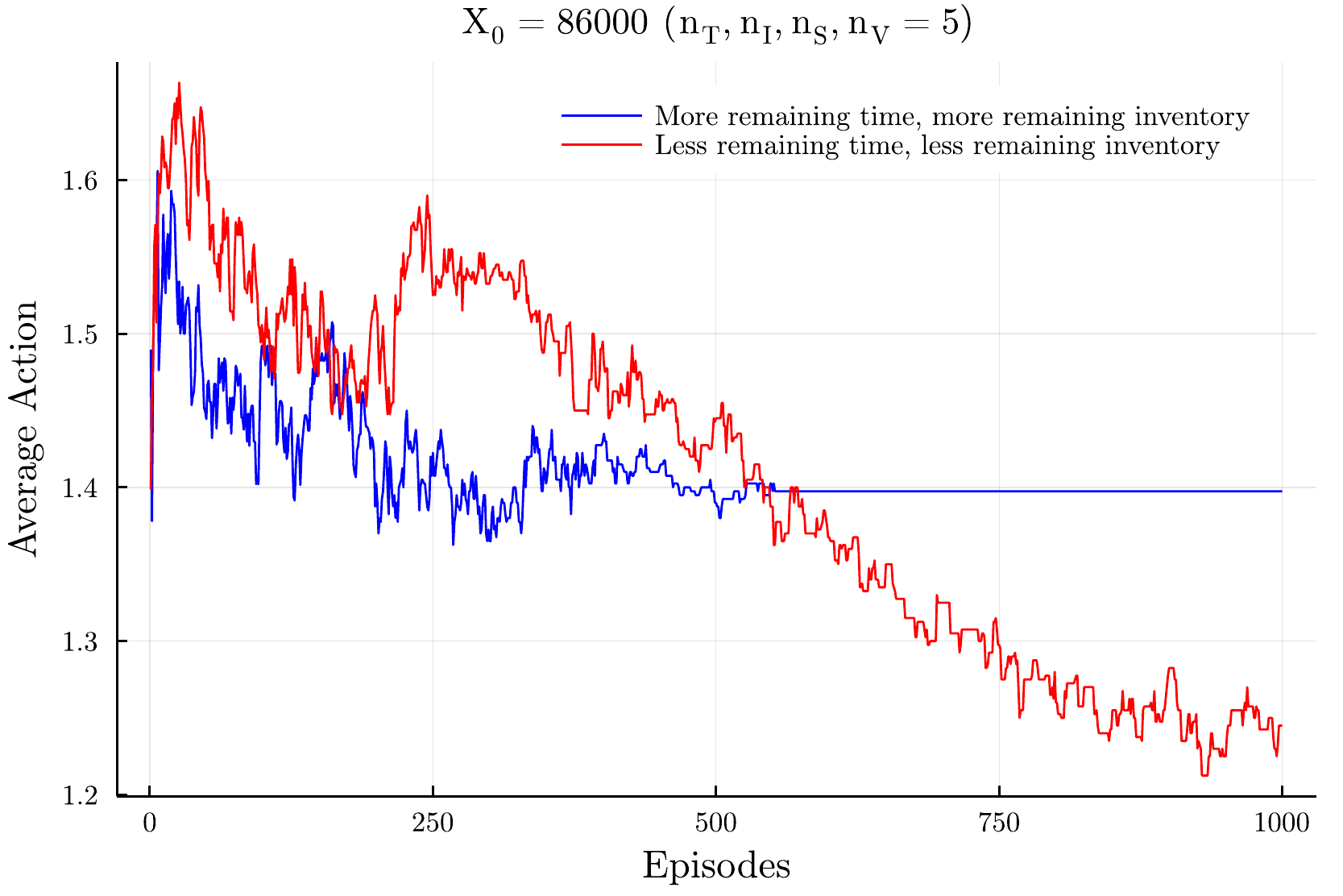}
            \caption[]{Large order, small state space}   
            \label{fig:average-action-TI-V200-S5}
        \end{subfigure}
        \vskip\baselineskip
        \begin{subfigure}[b]{0.475\textwidth}   
            \centering 
            \includegraphics[width=\textwidth, height=6cm]{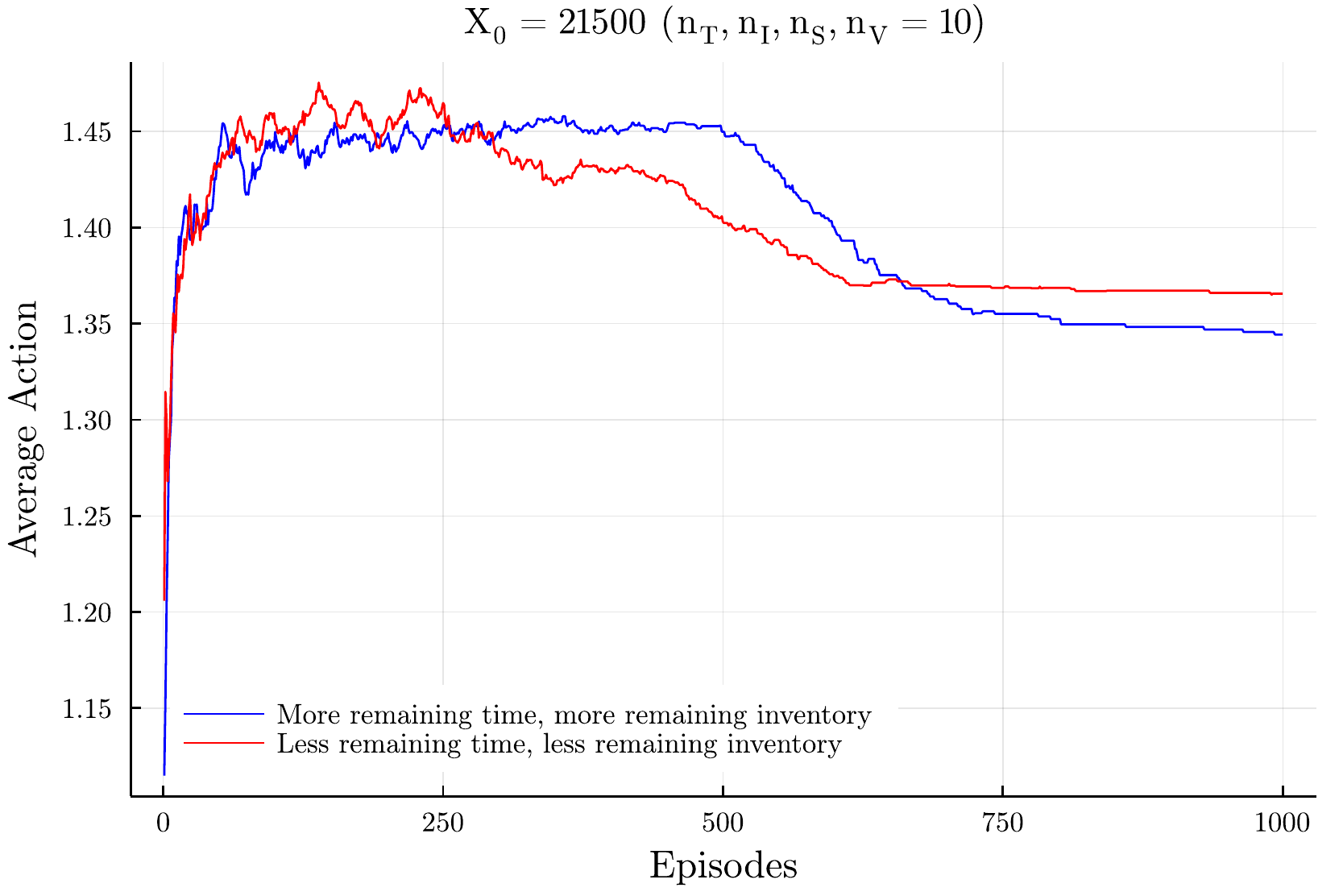}
            \caption[]{Small order, large state space}   
            \label{fig:average-action-TI-V50-S10}      
        \end{subfigure}
        \hfill
        \begin{subfigure}[b]{0.475\textwidth}   
            \centering 
            \includegraphics[width=8cm, height=6cm]{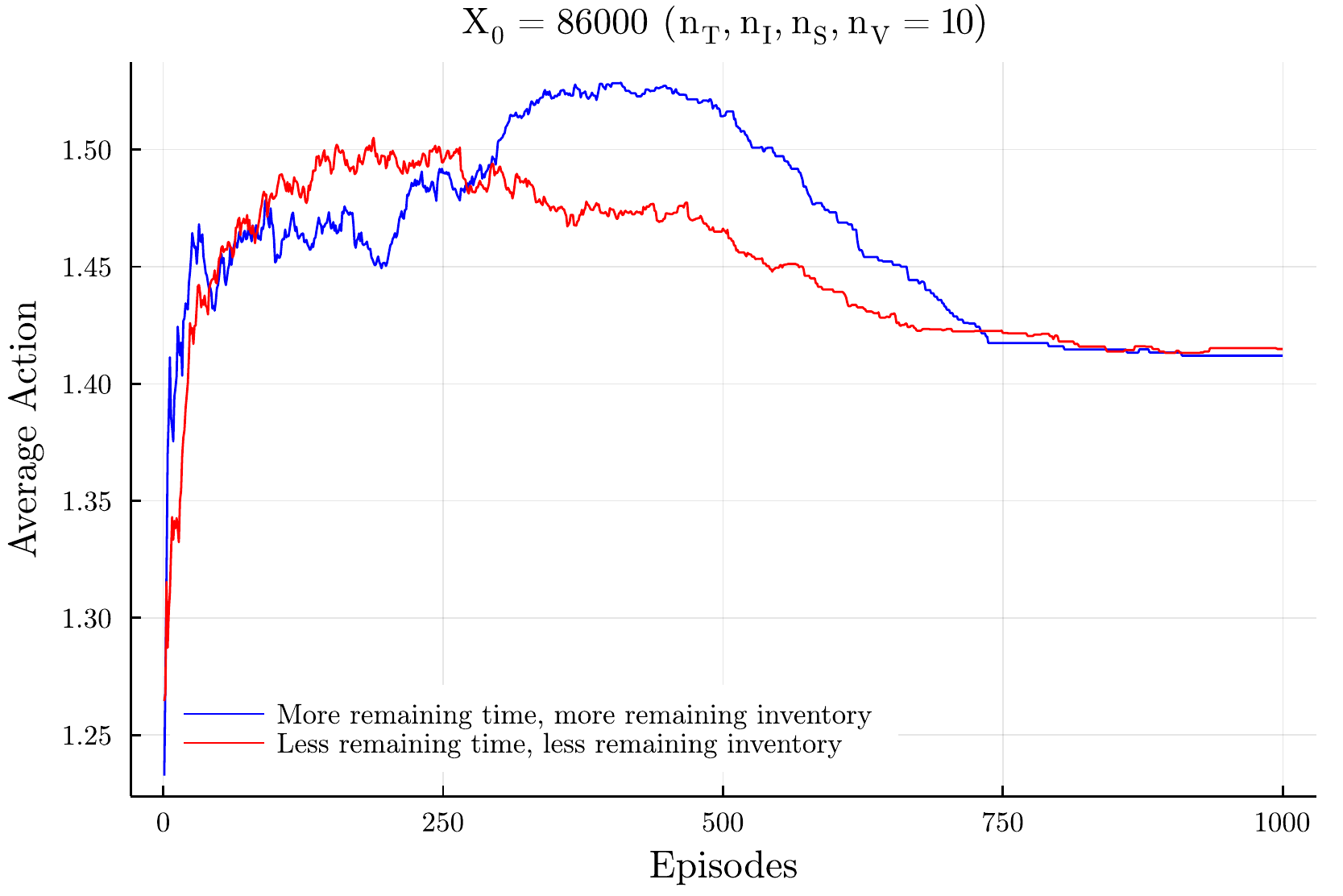}
              \caption[]{Large order, large state space}   
            \label{fig:average-action-TI-V200-S10}   
        \end{subfigure}
\caption{The average actions taken by the greedy policy when the agent has lots of inventory at the beginning of the simulation (Figure \ref{fig:average-action-TI-V50-S5}) and when the agent has less inventory at the end of the simulation (Figure \ref{fig:average-action-TI-V200-S5}) are plotted in the first row. The agent has learned to trade more when it is early in the simulation and the agent has a large amount of inventory. This effect is present when there is a low initial inventory  and when there is a high initial inventory. In the lower row are the plots, Figure \ref{fig:average-action-TI-V200-S10} and Figure \ref{fig:average-action-TI-V200-S10} the average action were taken by two of the larger state space agents for a different time and inventory state combinations. We find that all the agents have learned to trade, on average, more when there is less time remaining, and trade more when there is more inventory remaining. However, in the large state space, the effects cancel out whereas in the smaller state space the inventory effect dominates.}
        \label{rl-average-action-time-inventory-S10}
        \label{rl-average-action-time-inventory-S5}
    \end{figure}

To understand how the agents have learned to trade in different time and inventory states we chose to see how the agents' actions differ when they trade at the beginning, with a large amount of inventory, as compared to when they are trading at the end with less inventory. Figure \ref{rl-average-action-time-inventory-S5} plots the results for the two agents with the smaller state space. It is clear that the agent has learned to trade more inventory when it is early in the simulation and the agent has a large amount of inventory when compared with trading later when it has less inventory. This effect is present no matter how much initial inventory the agent was given. This finding corroborates the results seen in Figure \ref{rl-action-volume-V100-S5}. 

The time and inventory interaction results for the two agents with the larger state space are shown in Figure \ref{rl-average-action-time-inventory-S10}. We see that both agents have learned to decrease the amount of inventory traded as the agents are trained. However, the same effect, of trading more initially when the agent has a large inventory is not seen in the agents with the larger state space. The agents are trading roughly the same amounts in each state combination. All the agents have learned to trade, on average, more when there is less time remaining and trade more when there is less inventory remaining. However, in the smaller state space, we have that the time effect is much less than the inventory effect, which translates to the agents trading more when more time and inventory remain. However, for the agents with the larger state space, the time and inventory effects are roughly similar, therefore the effects are cancelled out in the interaction plot. 

\begin{figure}[h]
    \centering
    \begin{subfigure}[t]{0.49\textwidth}
        \centering
        \includegraphics[width=9cm, height=6cm]{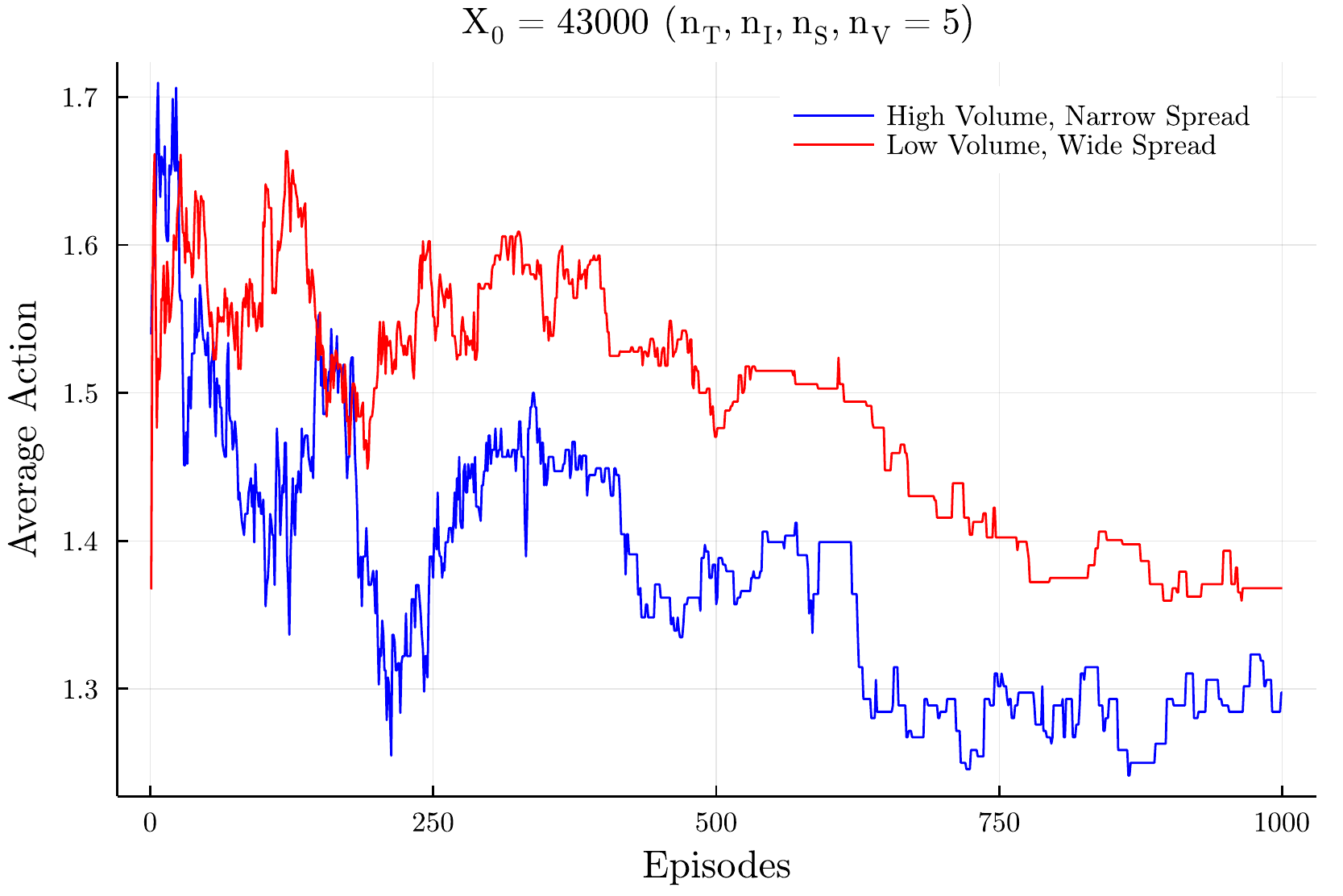}
        \caption{Plots the average action for the high (resp. low) volume and narrow (resp. wide) spread states over the training episodes. This figure plots the results for the agent with $X_{0}=43000$ and $n_{T},n_{I},n_{S},n_{V}=5$.}
        \label{app:rl-average-action-spread-volume-V100-S5}
    \end{subfigure}\hfill
     \begin{subfigure}[t]{0.49\textwidth}
        \centering
        \includegraphics[width=9cm, height=6cm]{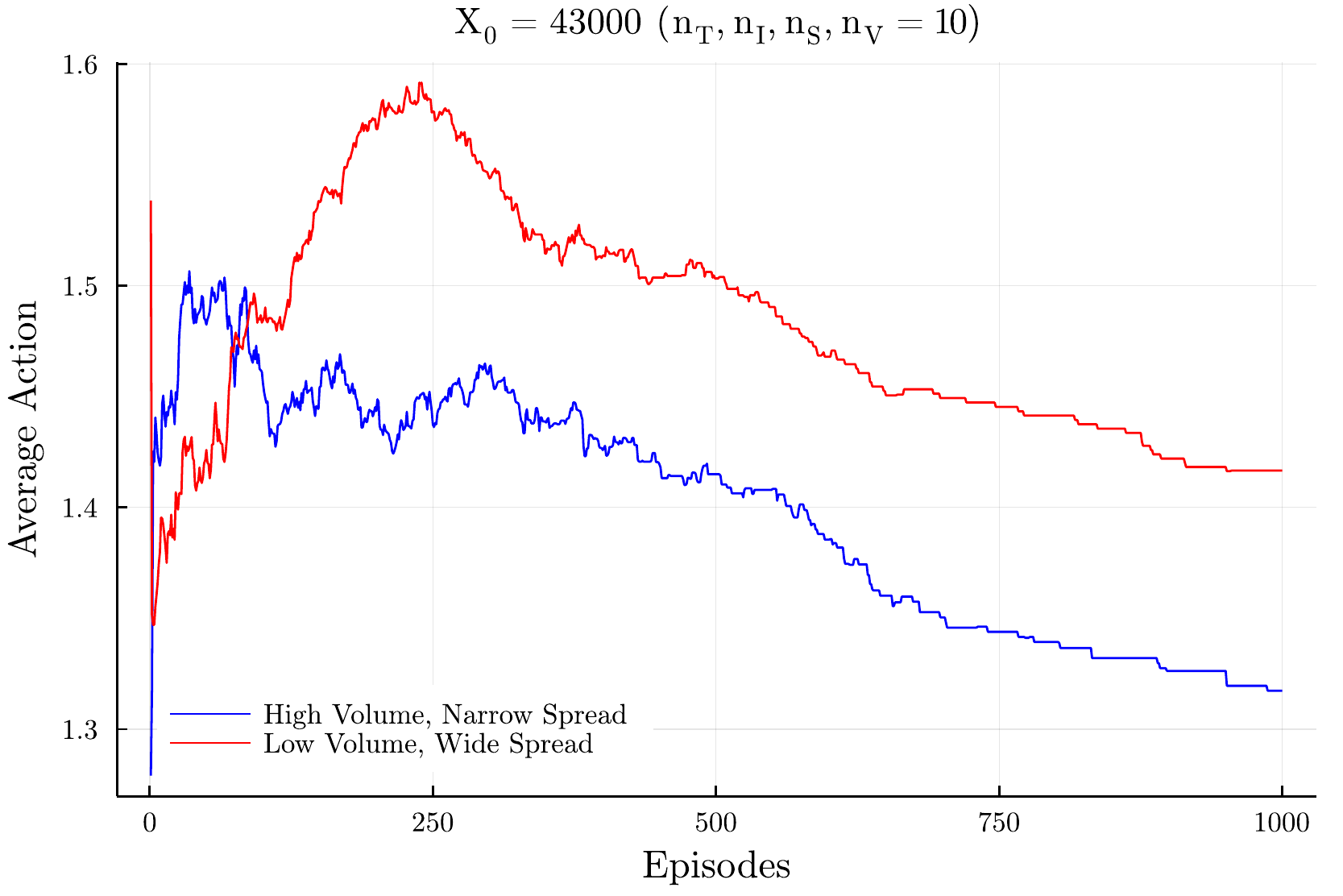}
        \caption{Plots the average action for the high (resp. low) volume and narrow (resp. wide) spread states over the training episodes. This figure plots the results for the agent with $X_{0}=43000$ and $n_{T},n_{I},n_{S},n_{V}=10$.}
        \label{app:rl-average-action-spread-volume-V100-S10}
    \end{subfigure}
    \caption{Plots the average action for the high (resp. low) volume and narrow (resp. wide) spread states over the training episodes. This figure plots the results for the agents with $X_{0}=43000$. The smaller state space agent is shown on the left (Figure \ref{app:rl-average-action-spread-volume-V100-S5}) and the larger state space agent is shown on the right (Figure \ref{app:rl-average-action-spread-volume-V100-S10})}
\end{figure}

\begin{figure}[h]
    \centering
    \begin{subfigure}[t]{0.49\textwidth}
        \centering
        \includegraphics[width=9cm, height=6cm]{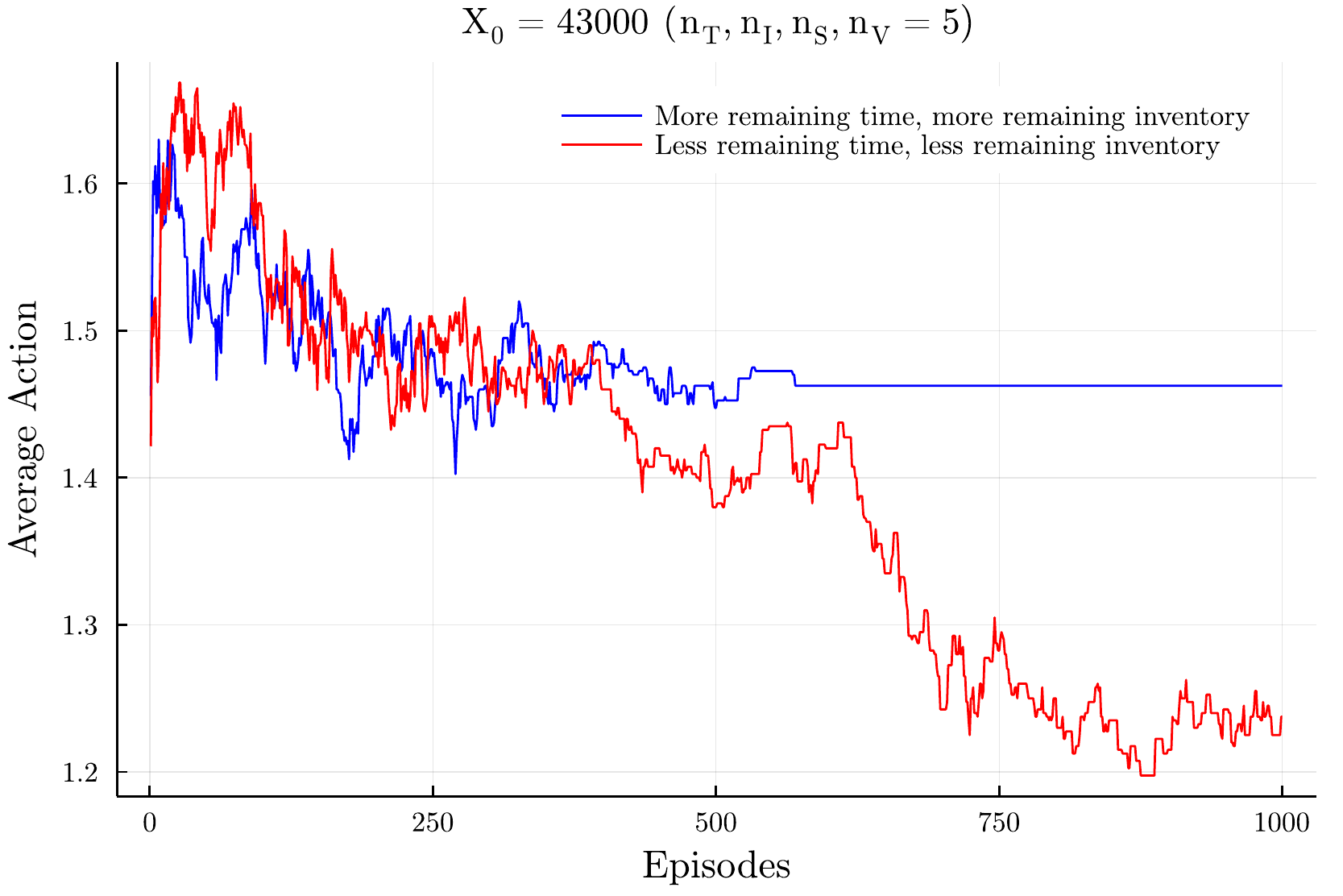}
        \caption{Plot the average action taken when the agent has lots of inventory at the beginning of the simulation and when the agent has less inventory at the end of the simulation. This figure plots the results for the agent with $X_{0}=43000$ and $n_{T},n_{I},n_{S},n_{V}=5$.}
        \label{app:rl-average-action-time-inventory-V100-S5}
    \end{subfigure}\hfill
     \begin{subfigure}[t]{0.49\textwidth}
        \centering
        \includegraphics[width=9cm, height=6cm]{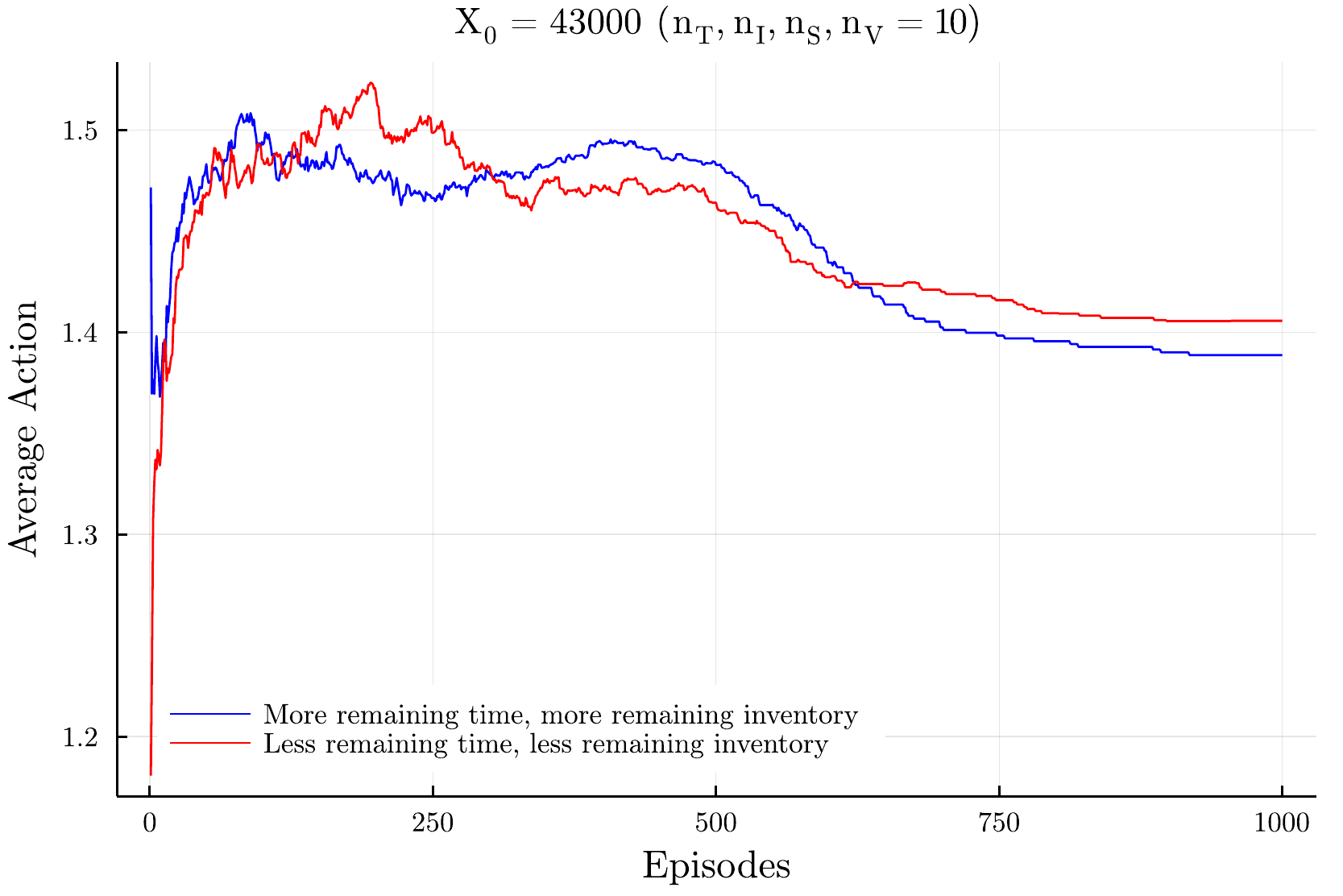}
        \caption{Plot the average action taken when the agent has lots of inventory at the beginning of the simulation and when the agent has less inventory at the end of the simulation. This figure plots the results for the agent with $X_{0}=43000$ and $n_{T},n_{I},n_{S},n_{V}=10$.}
        \label{app:rl-average-action-time-inventory-V100-S10}
    \end{subfigure}
    \caption{Plots the average action taken when the agent has lots of inventory at the beginning of the simulation and when the agent has less inventory at the end of the simulation. This figure plots the results for the agents with $X_{0}=43000$. The smaller state space agent is shown on the left (Figure \ref{app:rl-average-action-time-inventory-V100-S5}) and the larger state space agent is shown on the right (Figure \ref{app:rl-average-action-time-inventory-V100-S10})}
\end{figure}

\begin{figure}[h]
        \centering
        \begin{minipage}{.5\textwidth}
          \centering
              \includegraphics[width=9cm, height=6cm]{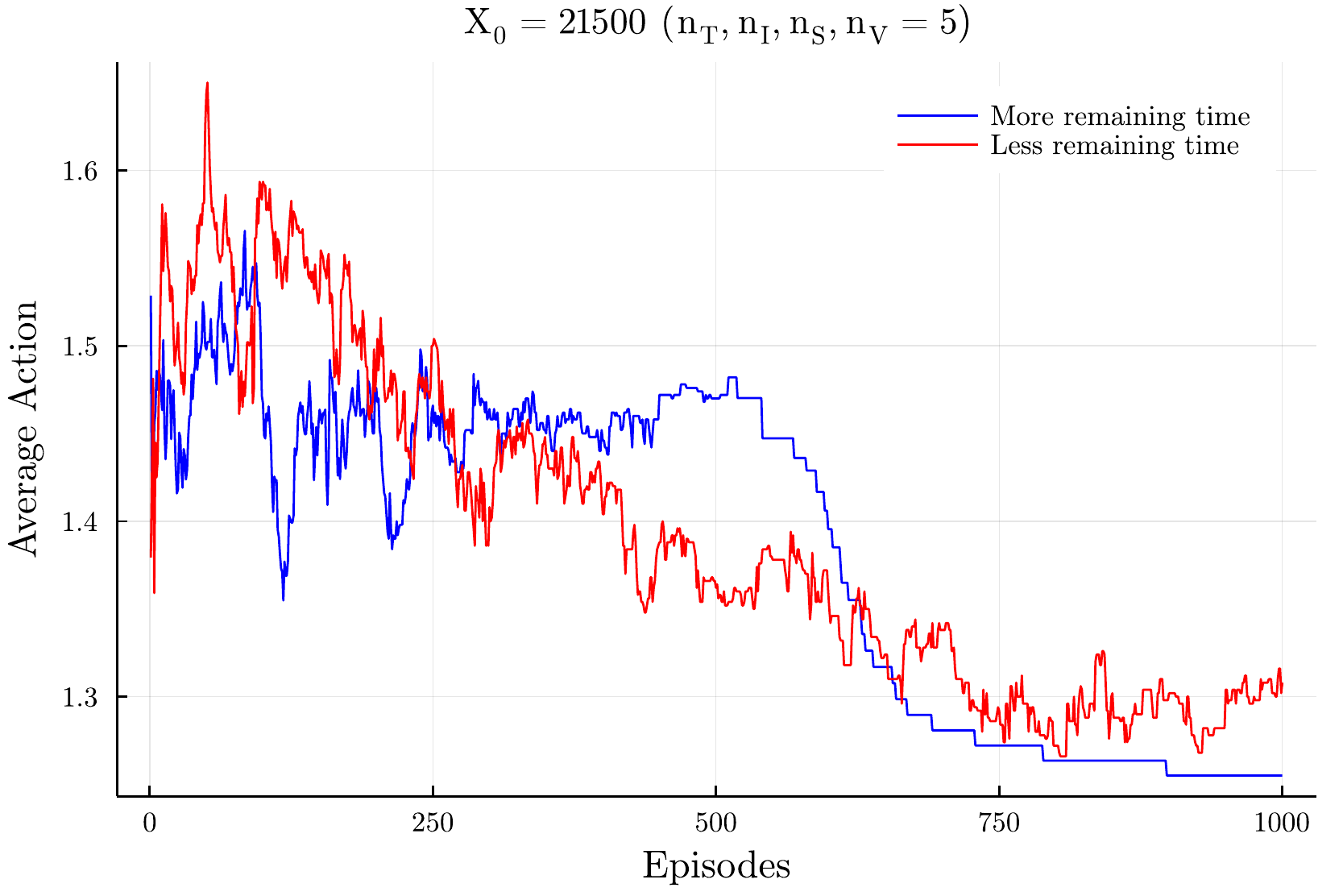}
        \end{minipage}%
        \begin{minipage}{.5\textwidth}
          \centering
              \includegraphics[width=9cm, height=6cm]{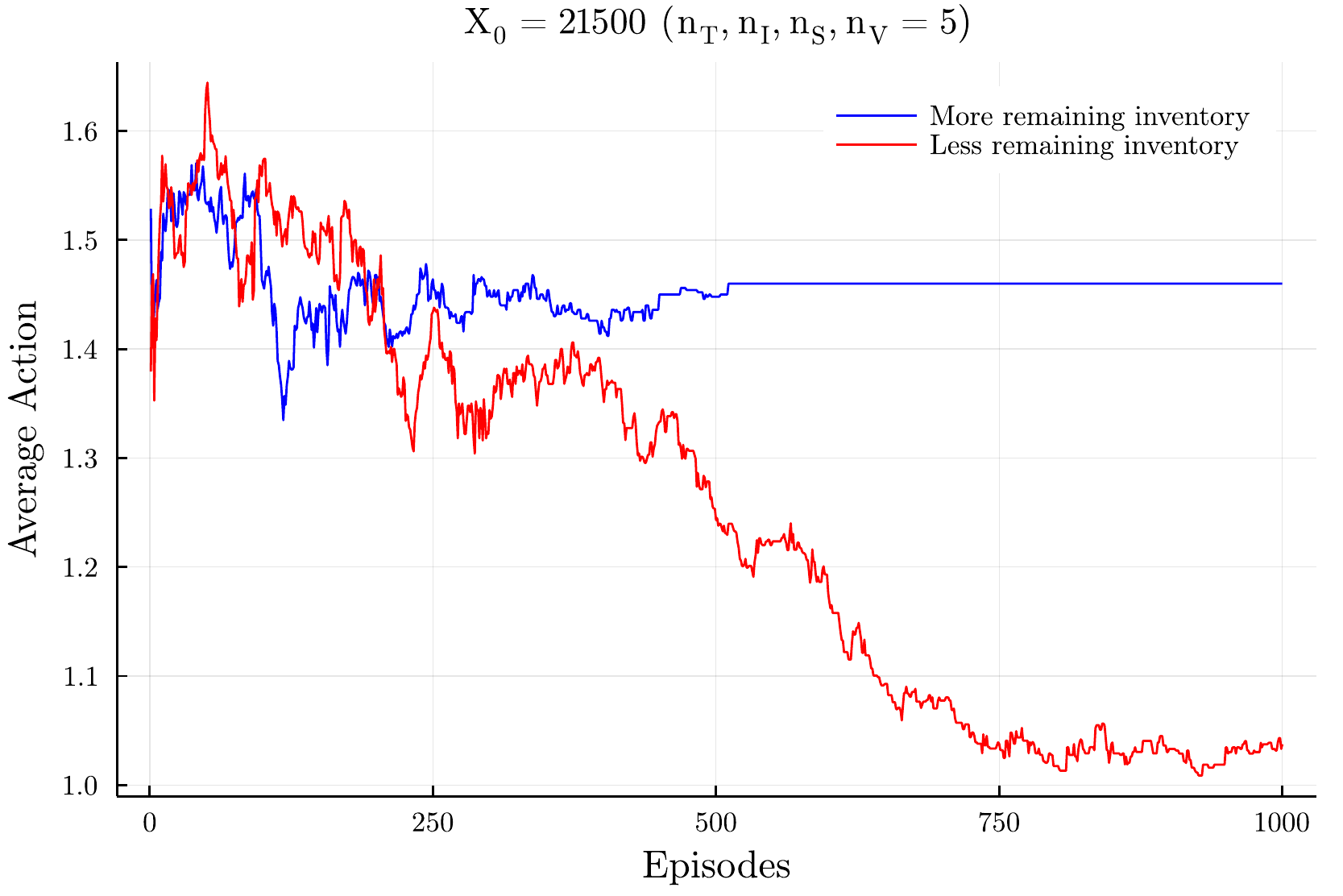}
        \end{minipage}
        \begin{minipage}{.5\textwidth}
          \centering
              \includegraphics[width=9cm, height=6cm]{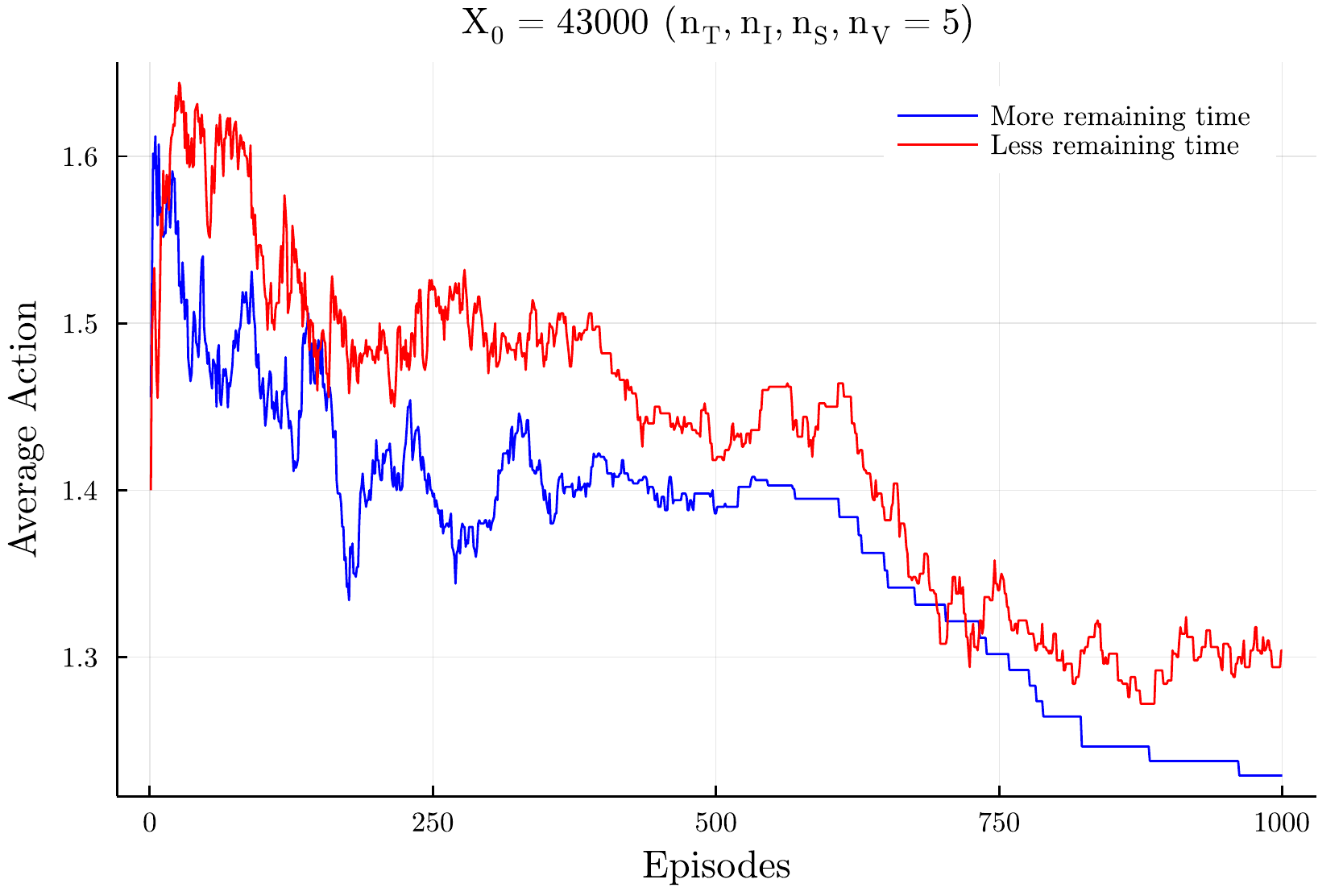}
        \end{minipage}%
        \begin{minipage}{.5\textwidth}
          \centering
              \includegraphics[width=9cm, height=6cm]{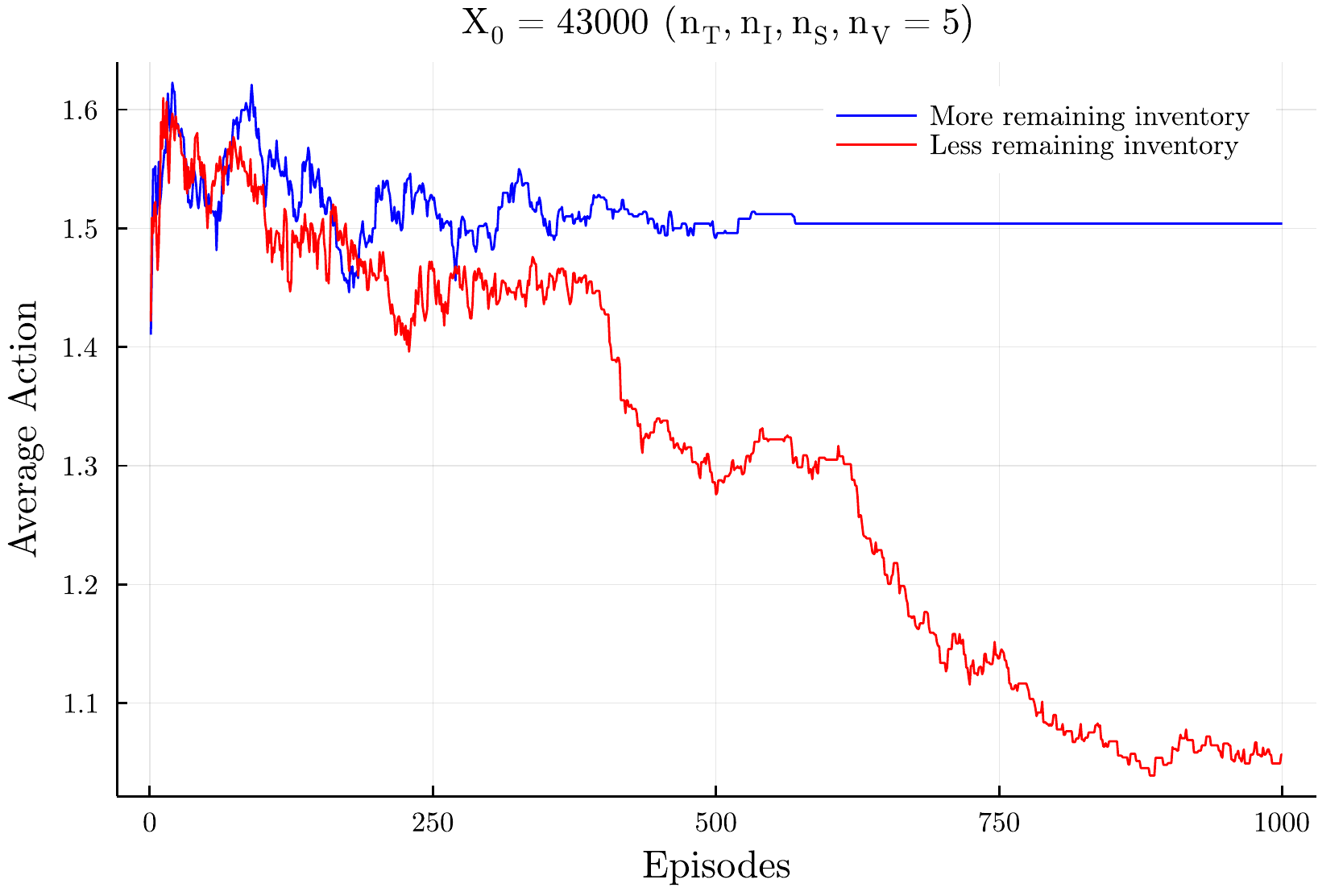}
        \end{minipage}
        \begin{minipage}{.5\textwidth}
          \centering
              \includegraphics[width=9cm, height=6cm]{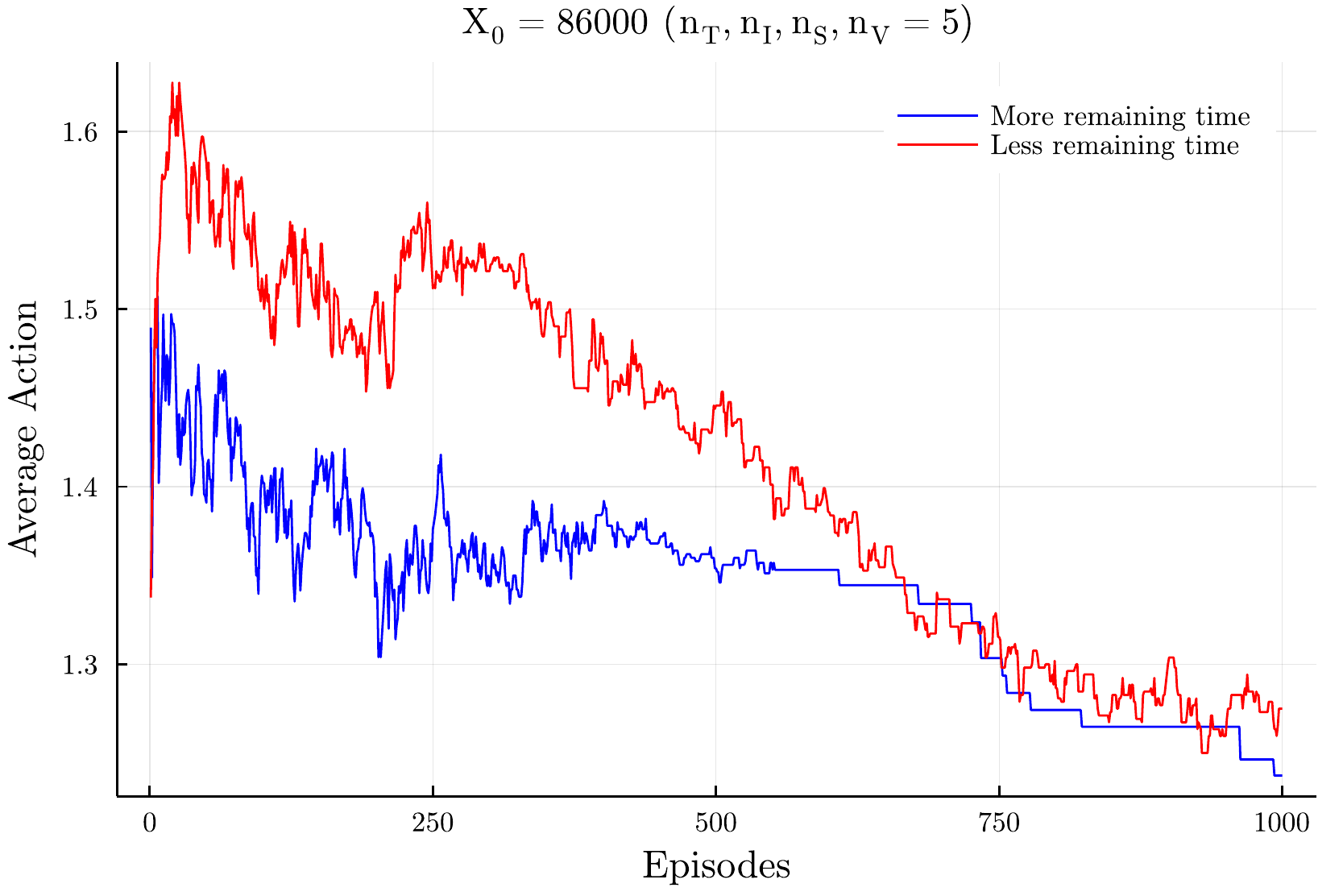}
        \end{minipage}%
        \begin{minipage}{.5\textwidth}
          \centering
              \includegraphics[width=9cm, height=6cm]{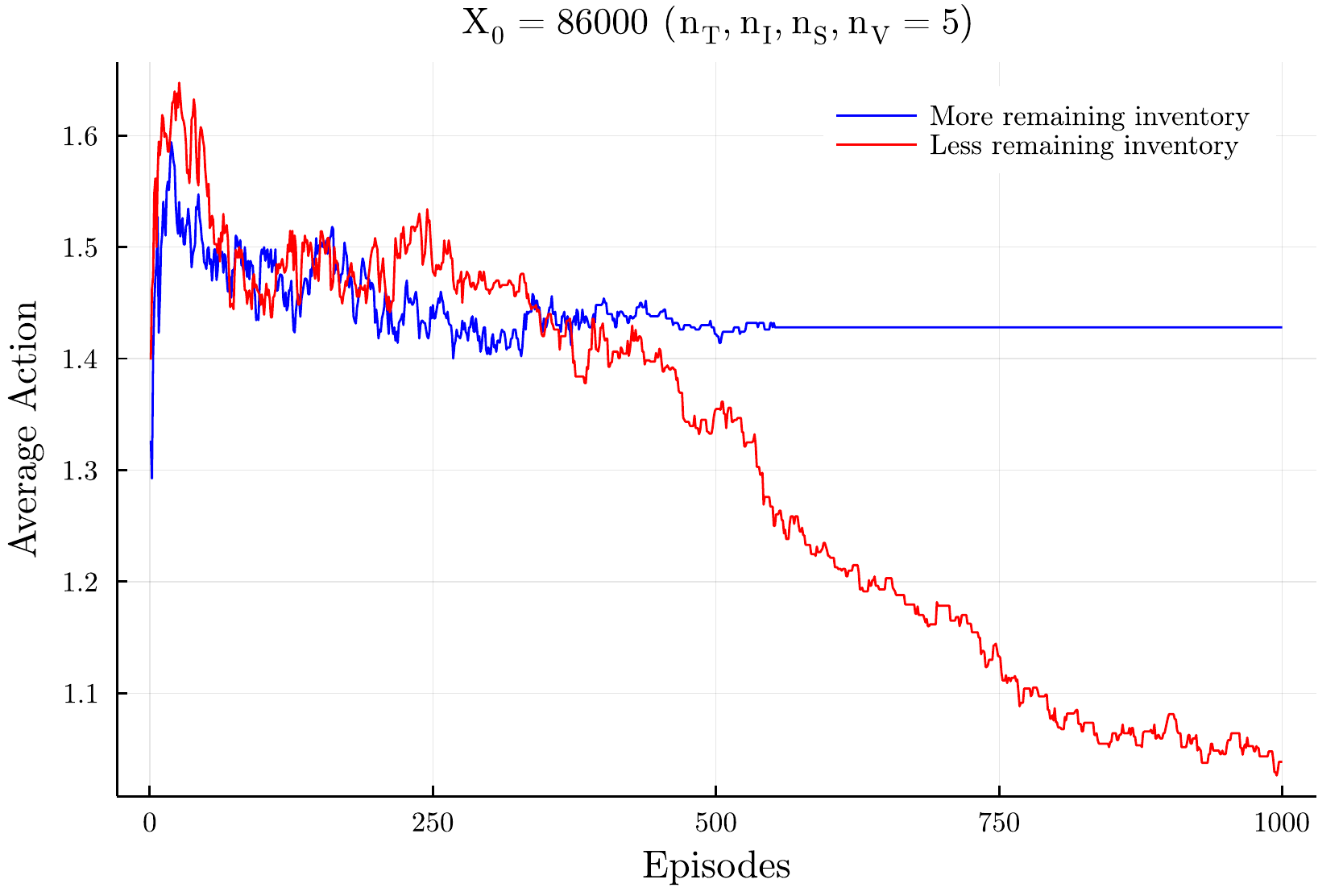}
        \end{minipage}
        \caption{Plots the average action taken by the RL agents with $n_{T},n_{I},n_{S},n_{V}=5$, when there is lots of time remaining and little time remaining as well as when the agent has a large amount of inventory remaining and a small amount of inventory remaining. Once again we see that the agents are learning to reduce the amount they trade as they get trained. Moreover, we see that all the agents have learned to trade, on average, more when there is less time remaining and trade more when there is less inventory remaining. It is clear that the amount of inventory remaining has a more significant effect when compared with the time remaining, as there is a larger difference between the average actions taken in the more and less, time and inventory remaining states.}
        \label{app:rl-average-action-marginals-time-inventory-s5}
\end{figure}

\begin{figure}[h]
        \centering
        \begin{minipage}{.5\textwidth}
          \centering
              \includegraphics[width=9cm, height=6cm]{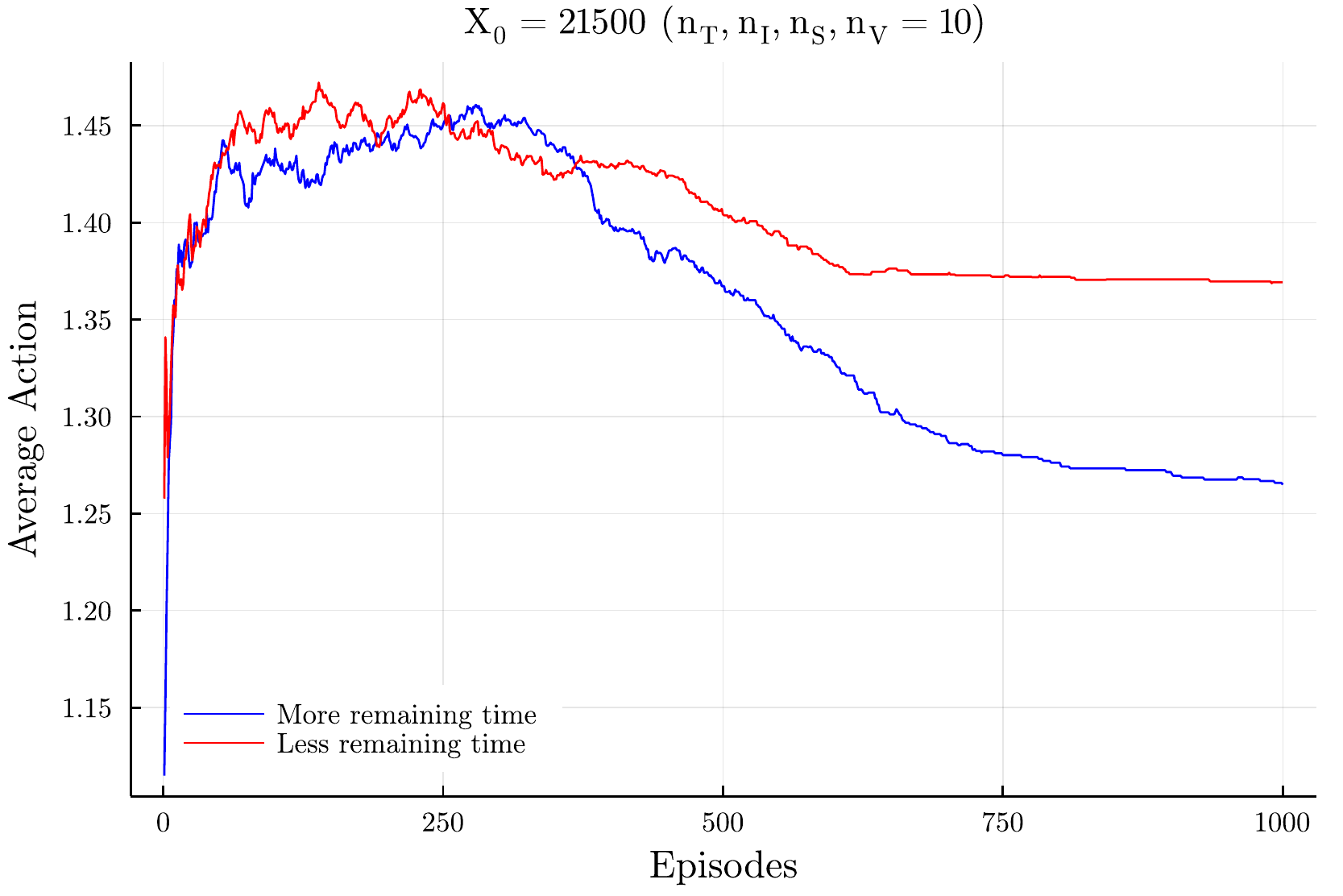}
        \end{minipage}%
        \begin{minipage}{.5\textwidth}
          \centering
              \includegraphics[width=9cm, height=6cm]{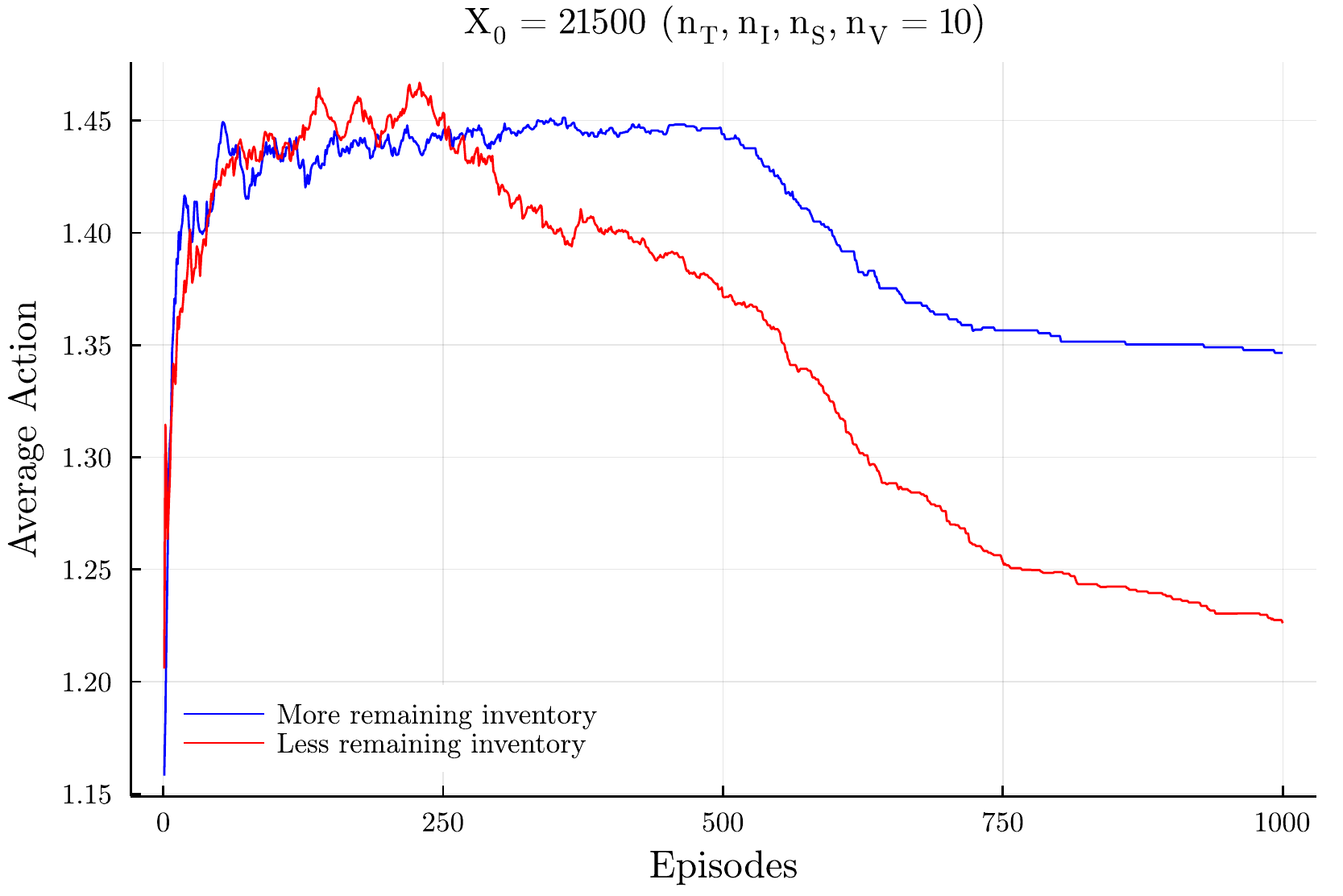}
        \end{minipage}
        \begin{minipage}{.5\textwidth}
          \centering
              \includegraphics[width=9cm, height=6cm]{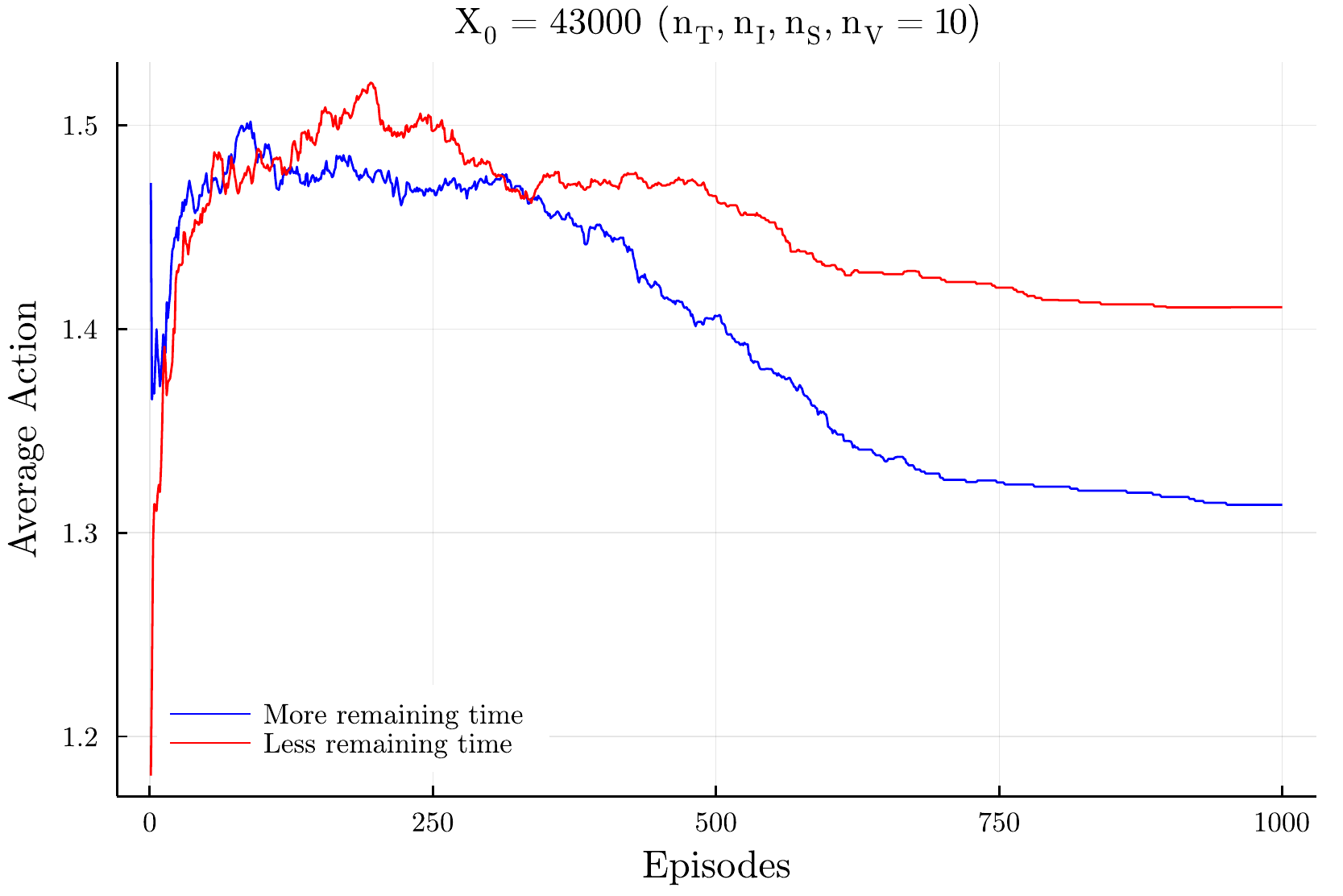}
        \end{minipage}%
        \begin{minipage}{.5\textwidth}
          \centering
              \includegraphics[width=9cm, height=6cm]{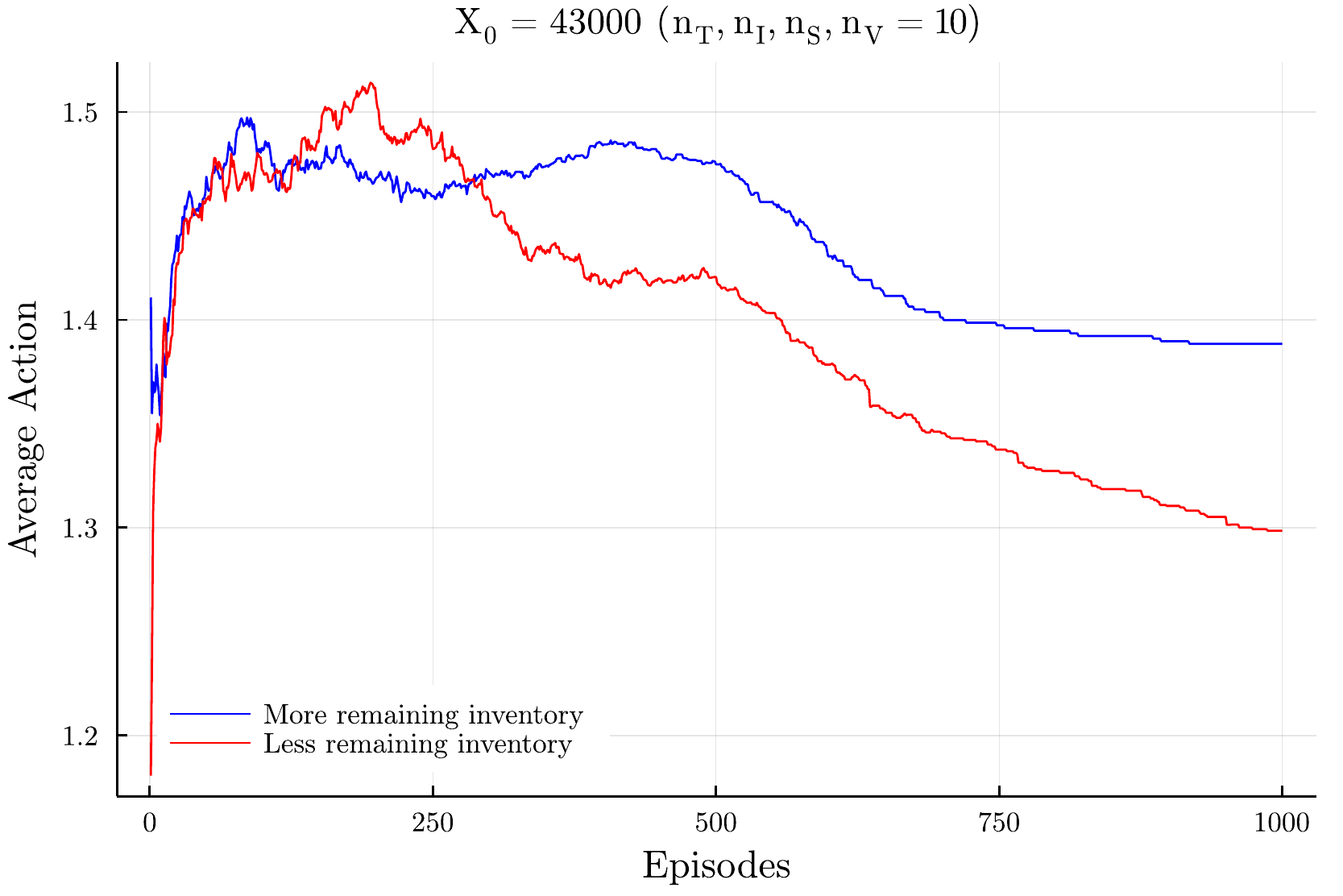}
        \end{minipage}
        \begin{minipage}{.5\textwidth}
          \centering
              \includegraphics[width=9cm, height=6cm]{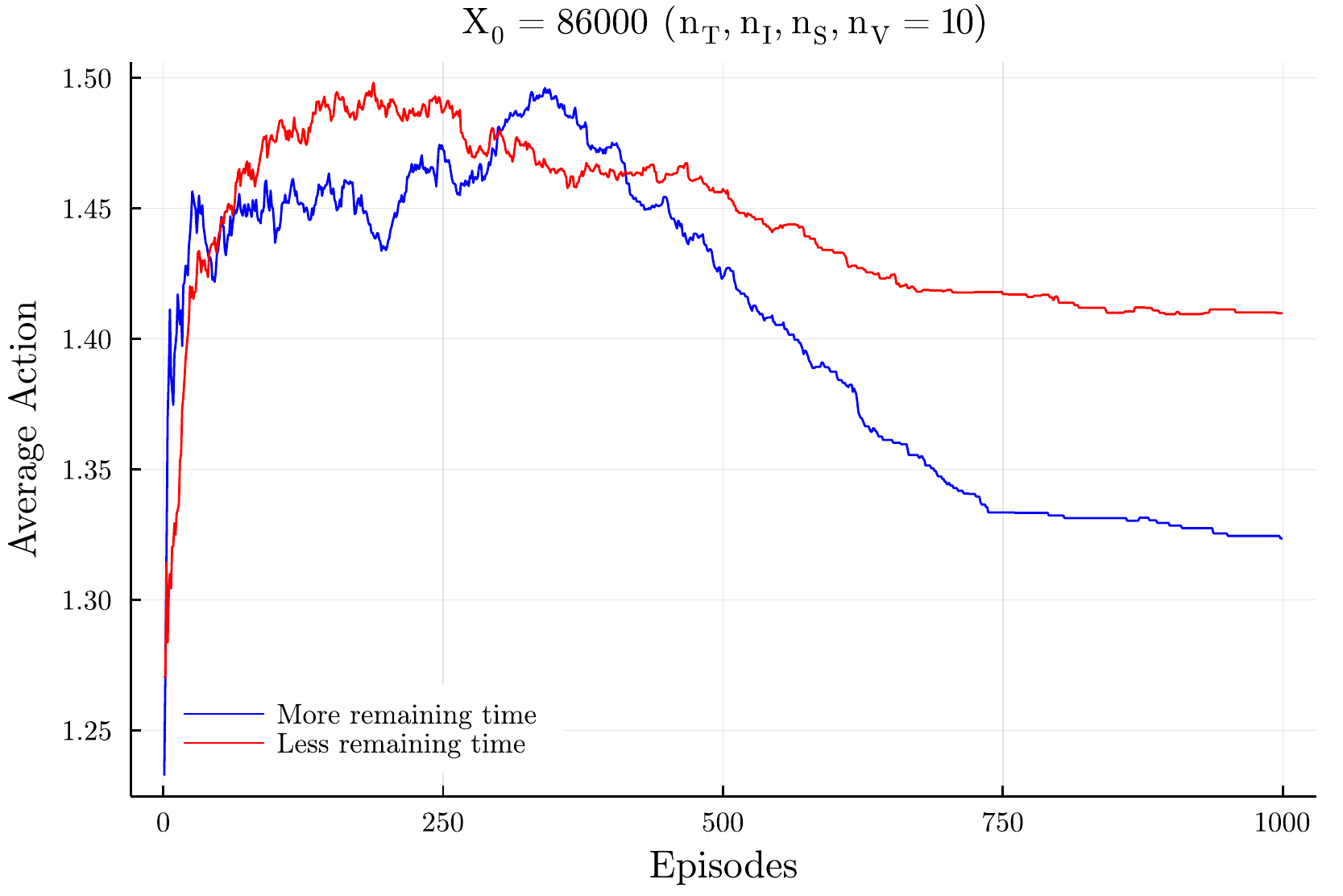}
        \end{minipage}%
        \begin{minipage}{.5\textwidth}
          \centering
              \includegraphics[width=9cm, height=6cm]{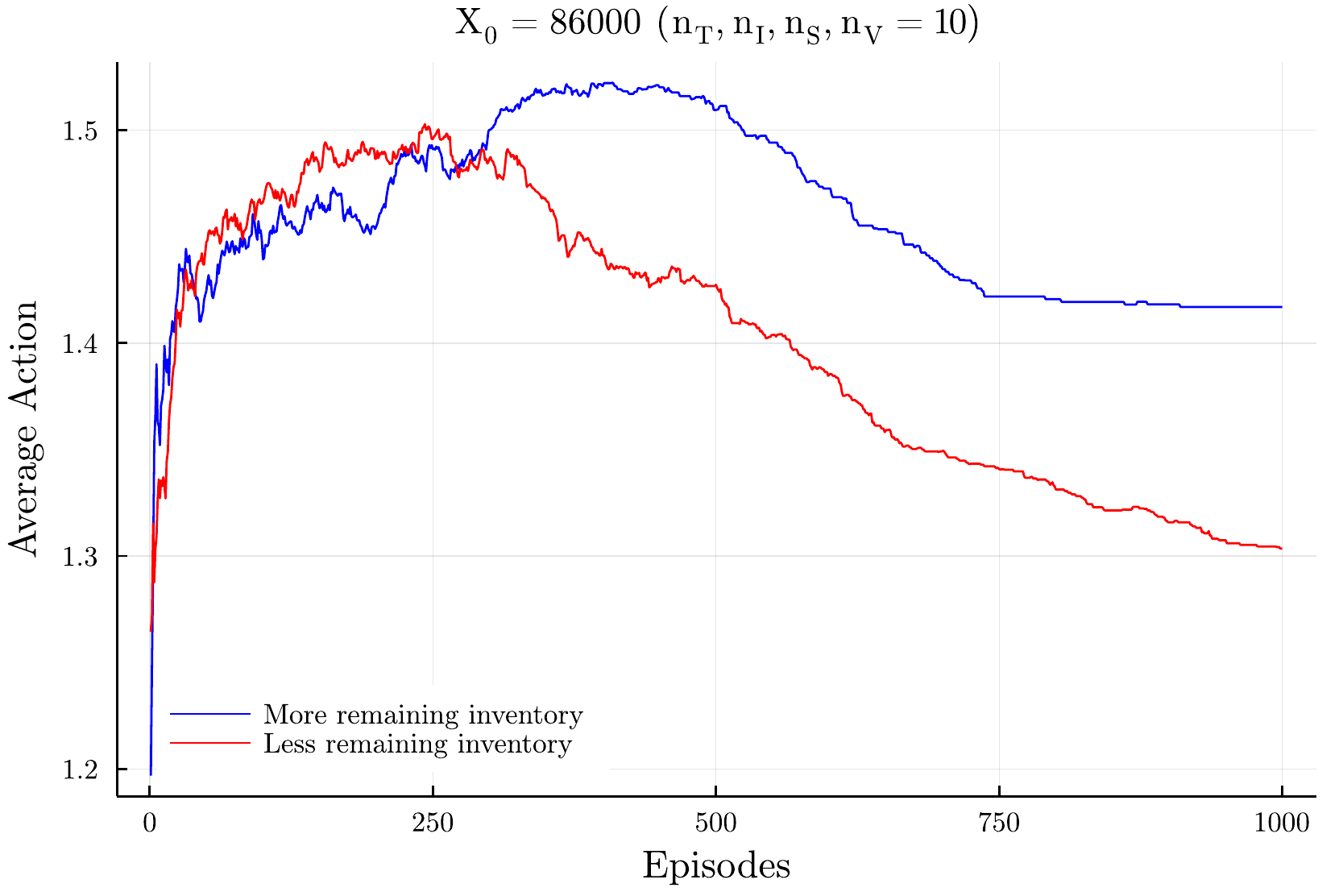}
        \end{minipage}
        \caption{Plots the average action taken by the RL agents with $n_{T},n_{I},n_{S},n_{V}=10$, when there is lots of time remaining and little time remaining as well as when the agent has a large amount of inventory remaining and a small amount of inventory remaining. Here, we see that the larger state space agents have also learned to trade less as the number of episodes increase. In contrast to the smaller states space agents (Figure \ref{app:rl-average-action-marginals-time-inventory-s5}), the time and inventory effects are roughly similar, therefore the effects are cancelled out in the interaction plot. }
        \label{app:rl-average-action-marginals-time-inventory-s10}
\end{figure}

\newpage
\subsection{Stylised fact dynamics} \label{app:RLAgentSimulationResults-StylisedFactDynamics}

Every 100 episodes, including the first and the last, all the raw messages of the simulation were recorded. The stylised fact moments were computed at each of these points so that we could see how adding in the RL agents affected the stylised facts of the ABM, and if these effects changed as the agents learned to trade. 

\begin{figure}[h!]
        \centering
        \begin{subfigure}[b]{0.475\textwidth}
            \centering
             \includegraphics[width=\textwidth]{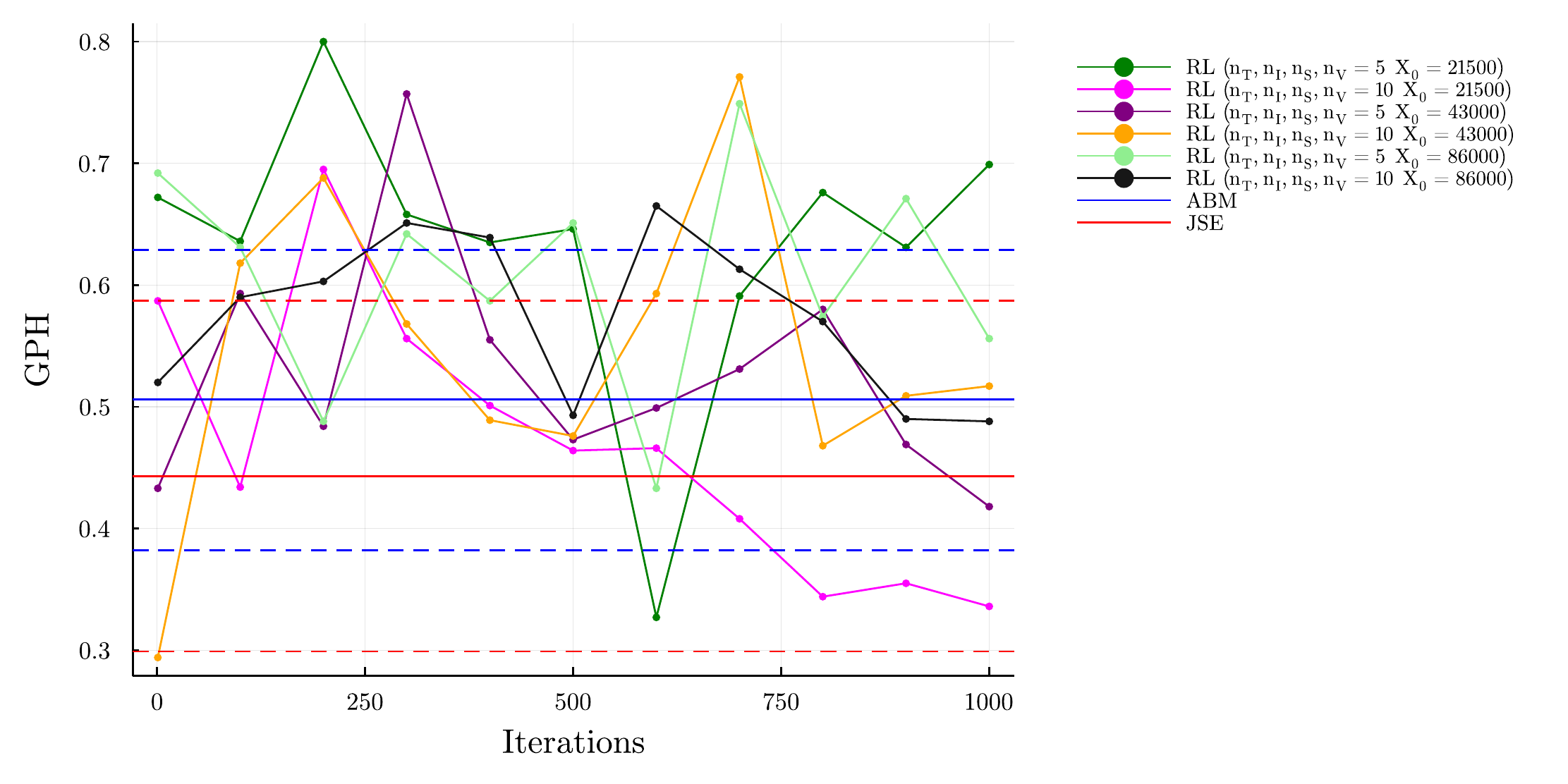}
            \caption{GPH}    
            \label{fig:GPH-modeldrift}
        \end{subfigure}
        \hfill
        \begin{subfigure}[b]{0.475\textwidth}  
            \centering 
            \includegraphics[width=\textwidth]{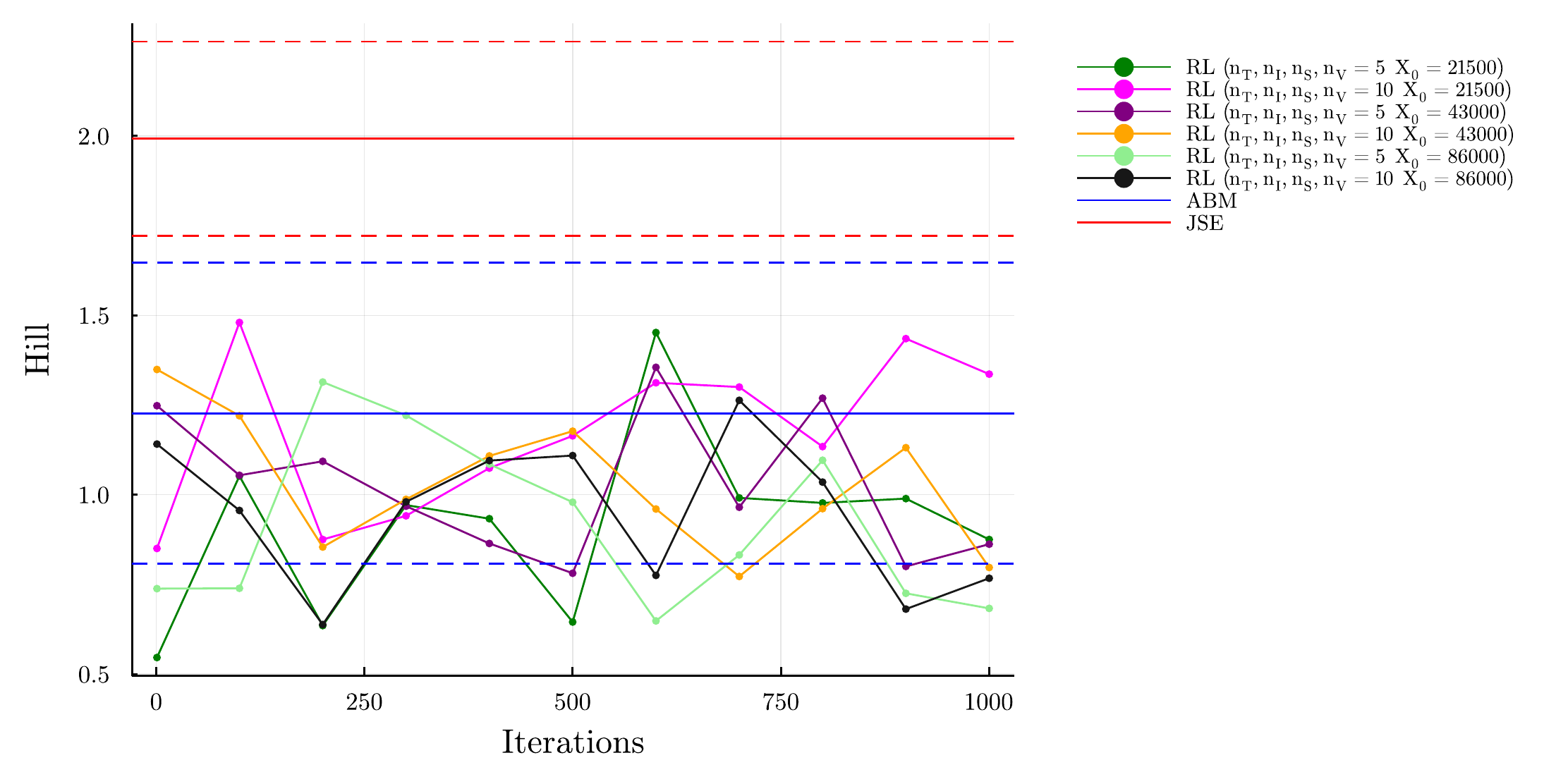}
            \caption{Hill}    
            \label{fig:Hill-modeldrift}
        \end{subfigure}
        \vskip\baselineskip
        \begin{subfigure}[b]{0.475\textwidth}   
            \centering 
          \includegraphics[width=\textwidth]{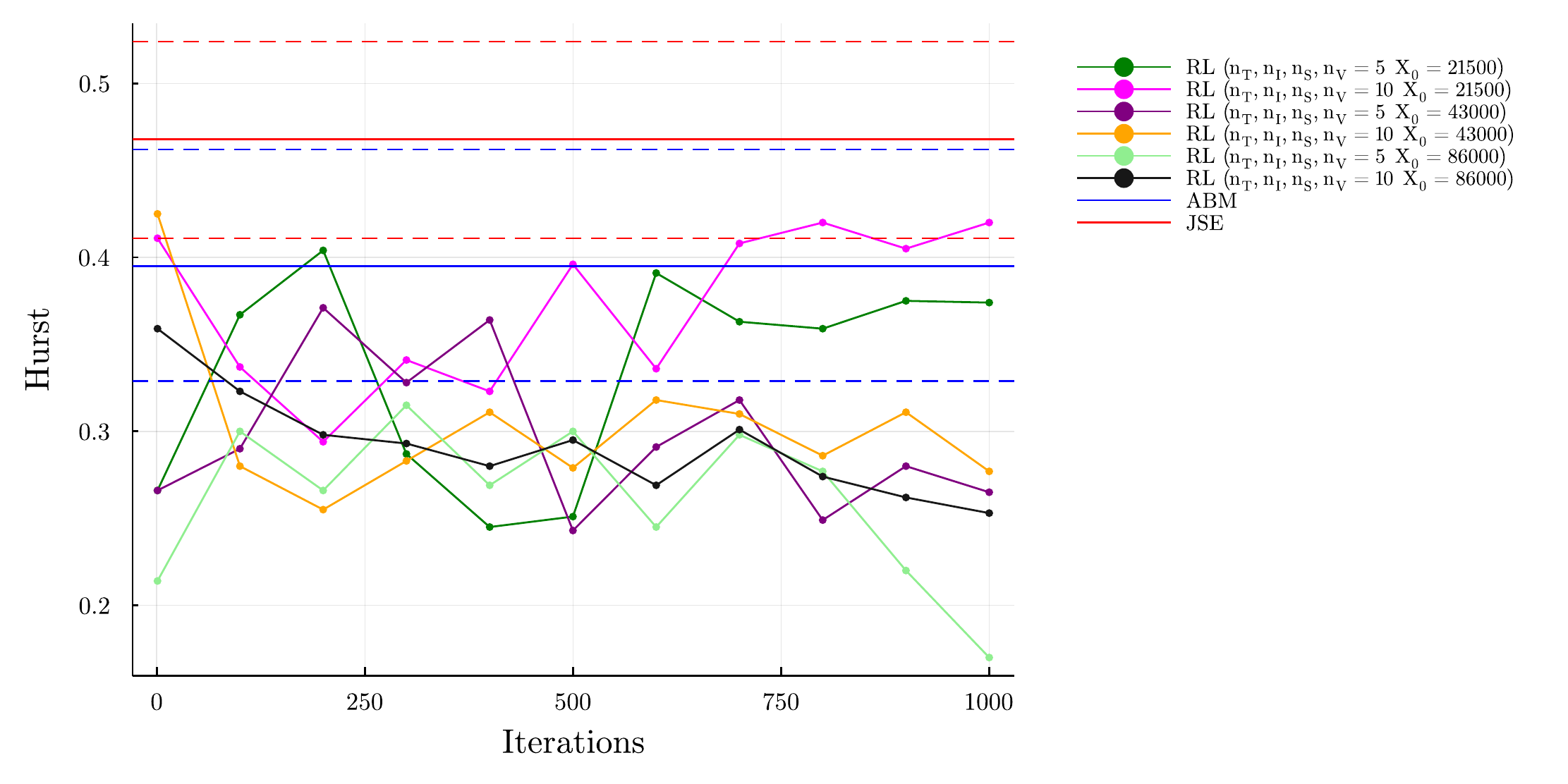}
            \caption{Hurst}    
            \label{fig:Hurst-modeldrift}
        \end{subfigure}
        \hfill
        \begin{subfigure}[b]{0.475\textwidth}   
            \centering 
            \includegraphics[width=\textwidth]{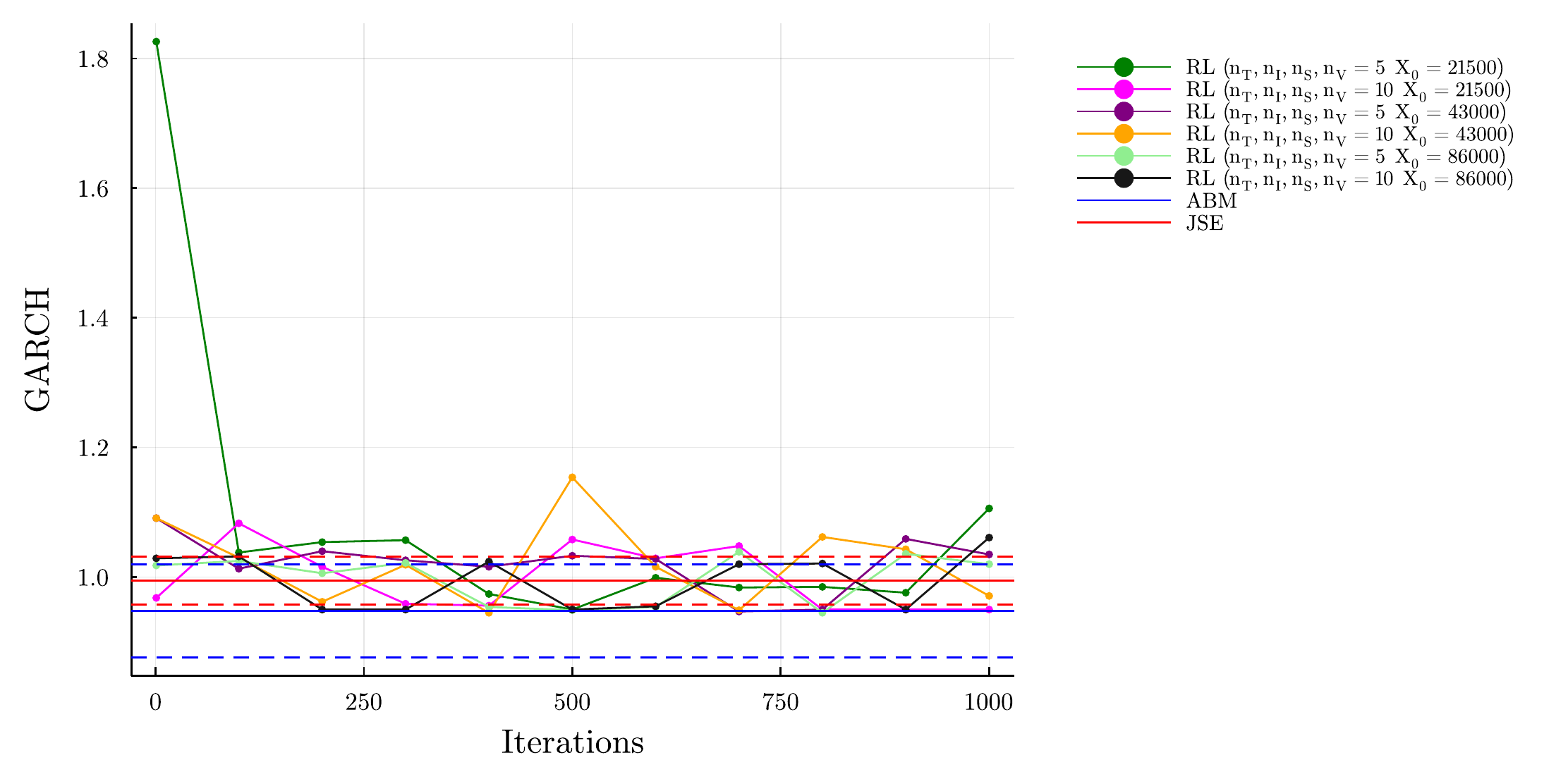}
            \caption{GARCH}    
            \label{fig:GARCH-modeldrift}
        \end{subfigure}
        \centering
        \begin{subfigure}{0.475\textwidth}
          \centering
              \includegraphics[width=\textwidth]{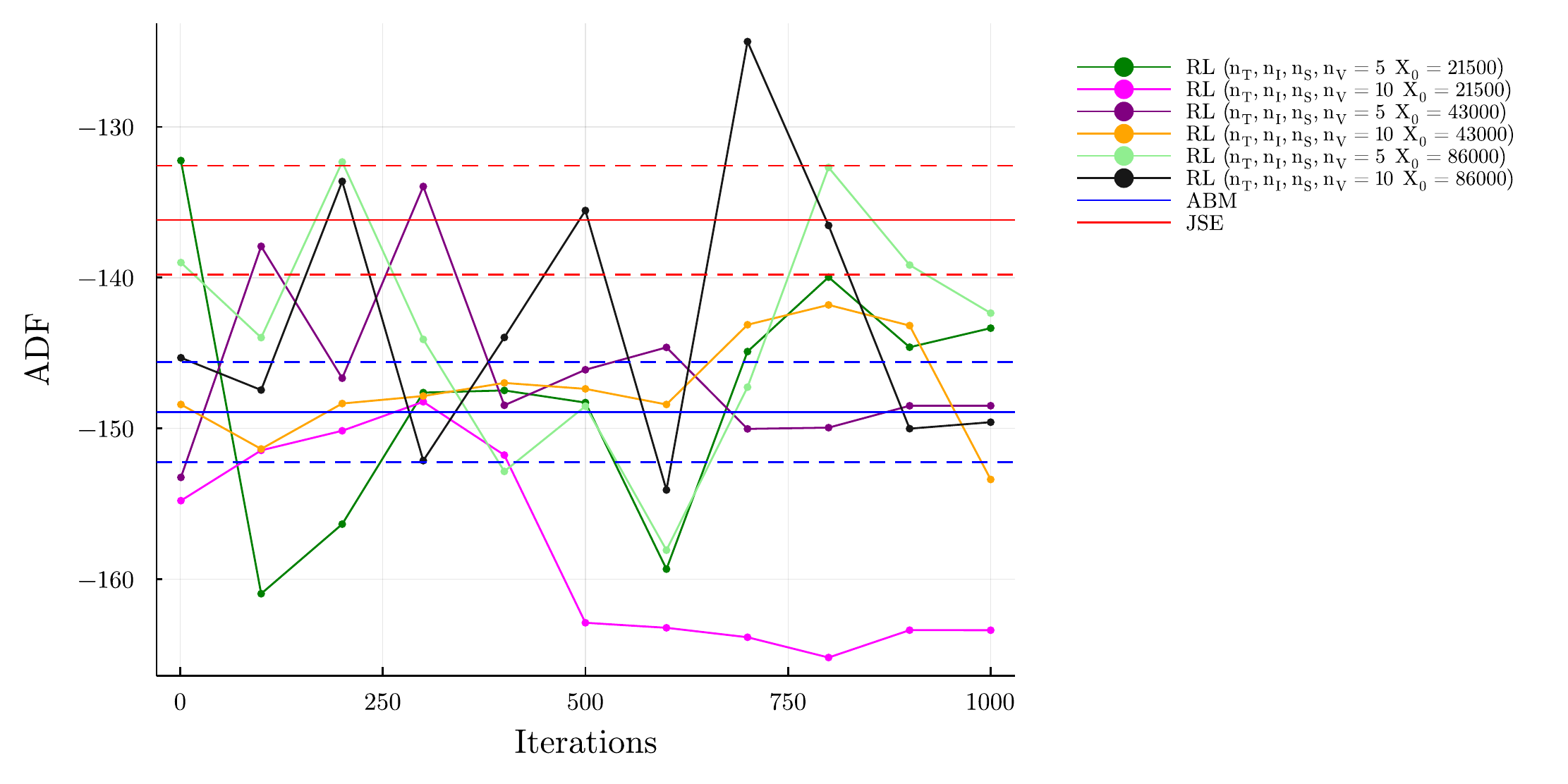}
              \caption{ADF}
        \end{subfigure}%
        \begin{subfigure}{0.475\textwidth}
          \centering
          \includegraphics[width=\textwidth]{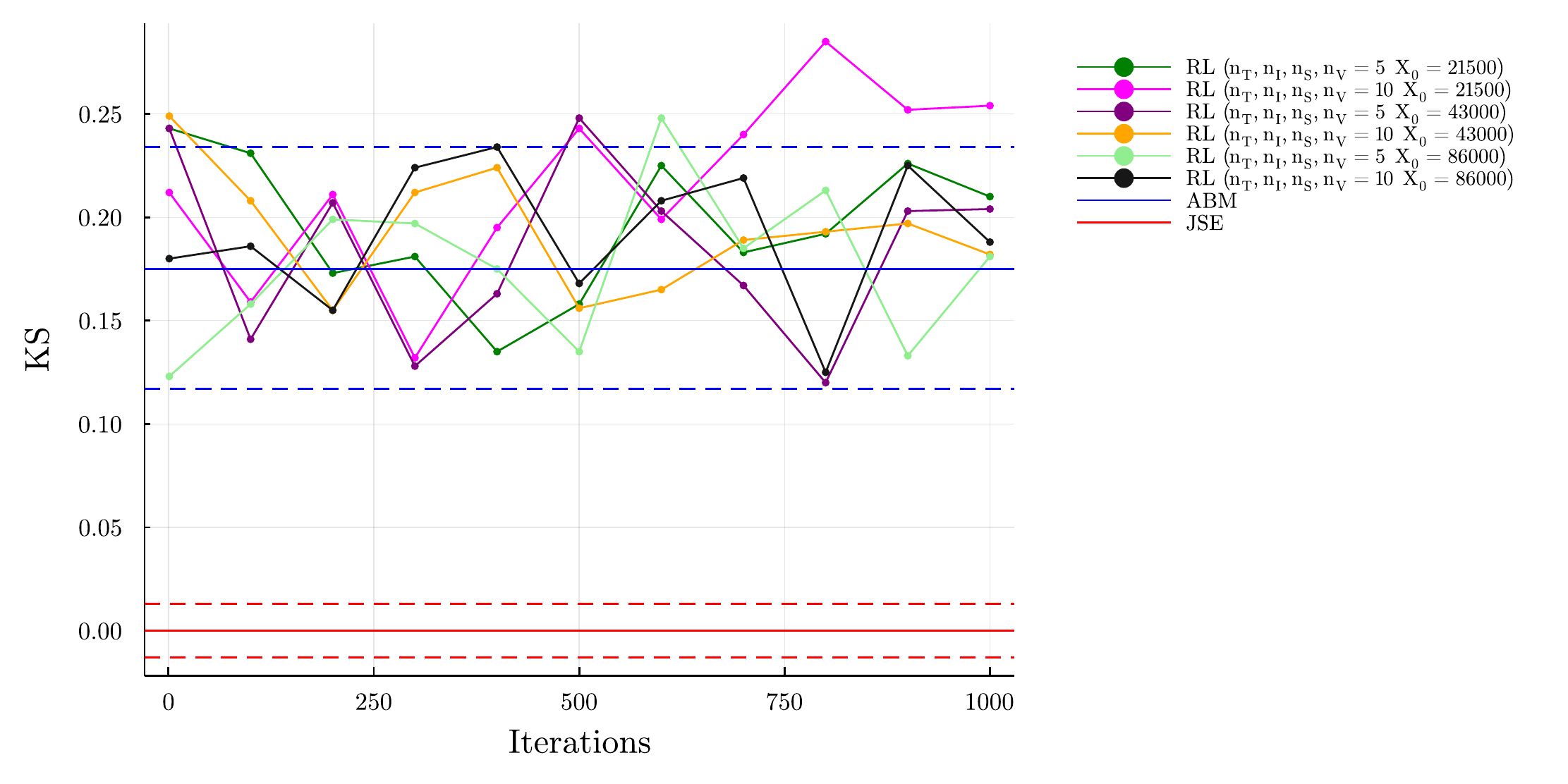}
              \caption{KS}
        \end{subfigure}
        \begin{subfigure}{0.475\textwidth}
          \centering
            \includegraphics[width=\textwidth]{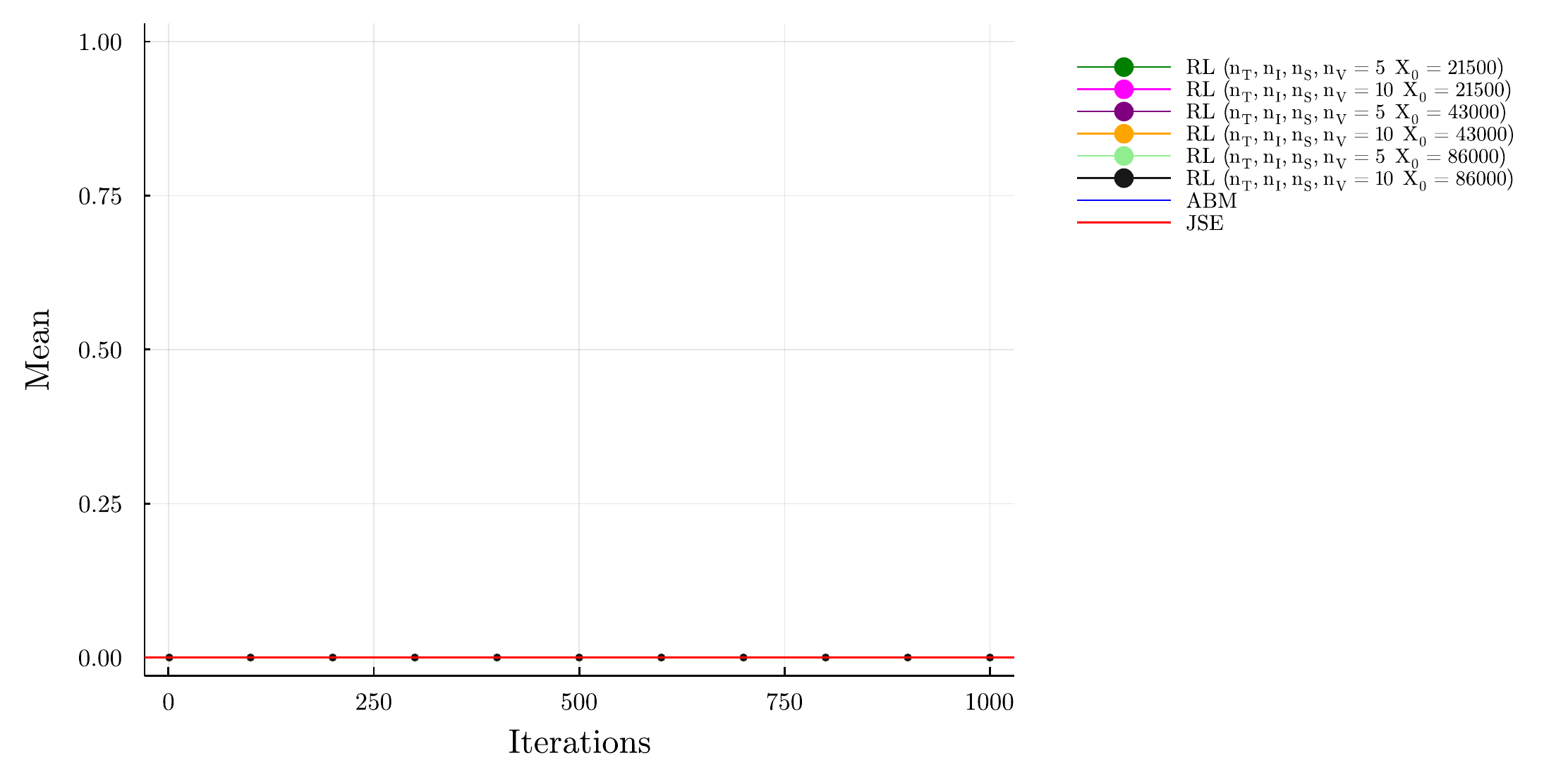}
              \caption{Mean}
        \end{subfigure}%
        \begin{subfigure}{0.475\textwidth}
          \centering
          \includegraphics[width=\textwidth]{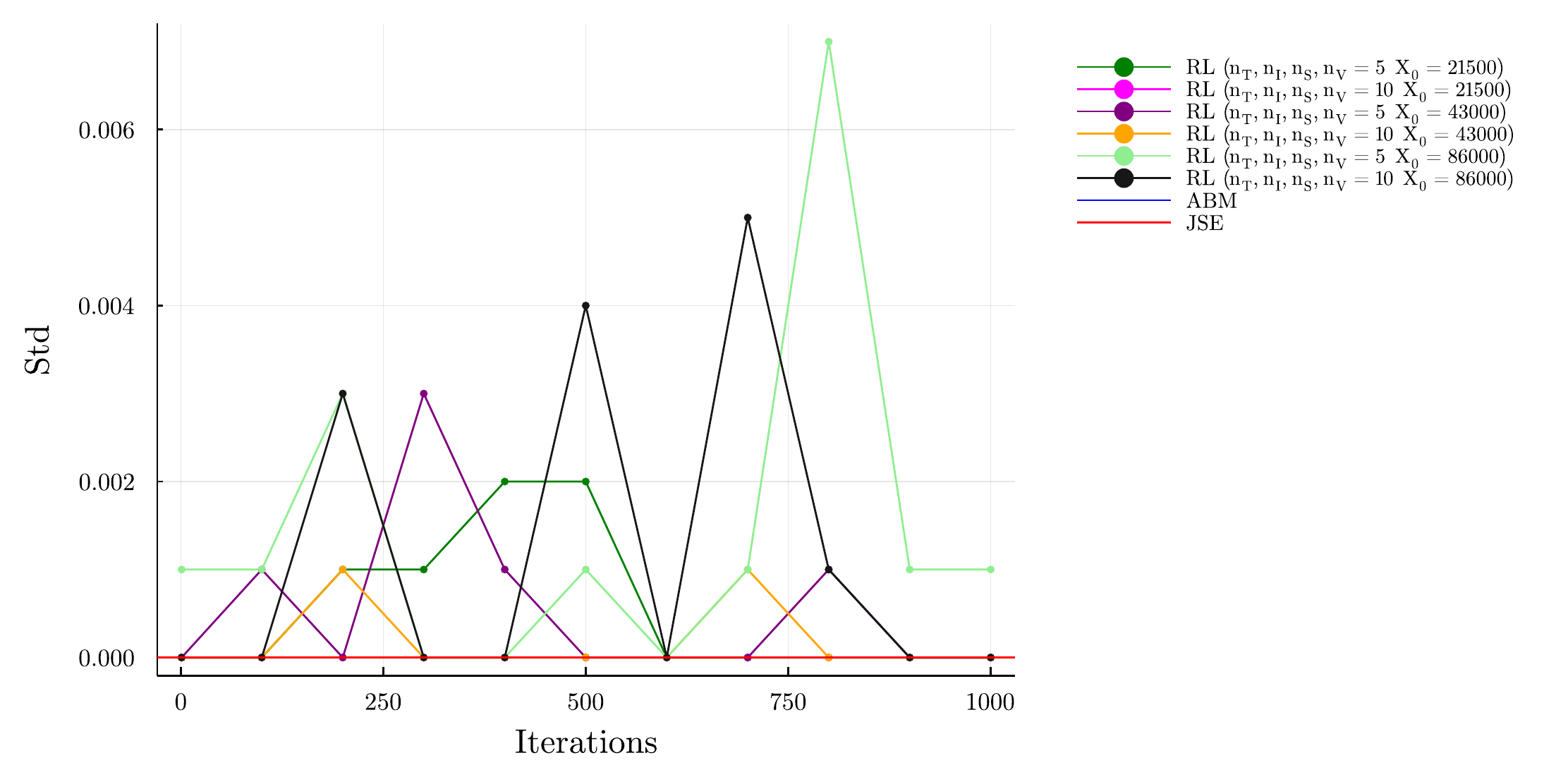}
              \caption{Std.}
        \end{subfigure}
     \caption{Here we compare the ABM and JSE moments, to those from the ABM+RL agent simulations.}
        \label{app:rl-stylised-fact-dynamics}  \label{rl-stylised-fact-dynamics}
\end{figure}

The moments were computed on both the JSE, and the ABM, and are plotted in Figure \ref{rl-stylised-fact-dynamics} along with their associated confidence intervals. This includes the moments for each of the RL agents over the episodes. The GPH statistic was roughly the same for the ABM and the JSE, and adding an RL agent to the simulation has not made a significant difference. This is also true for the GARCH moment, except for one erroneous point at the start of training for one of the agents. For the Hill estimator, the confidence intervals were close but did not overlap and the moments for the ABM's with the RL agents have Hill estimators that fall into the ABM confidence interval. 

The Hurst exponent moment was the only moment that showed slight differences between the ABM with and without the RL agent. For the RL simulations, the Hurst exponent is lower than that of the ABM and consistently falls below the lower confidence interval of the ABM. The only ones to fall into the ABM's confidence interval are the RL agents that had the smallest parent orders. Indicating that the larger the amount of inventory traded by the RL agent, the more this increases the amount of mean reversion in the simulation. As the episodes increase there is a slight downward trend in the Hurst exponent for the agents with larger parent orders. Figure \ref{rl-stylised-fact-dynamics} only shows four out of the eight moments.
The GPH statistic has remained similar for the ABM and the JSE, and adding an RL agent to the simulation has no significant impact. This is also true for the GARCH moment, except for an initialisation. For the Hill estimator, the confidence intervals were close but did not overlap and the moments for the ABMs with the RL agents have Hill estimators that fall into the ABM confidence interval. The Hurst exponent moment was the only moment that showed small differences between the ABM with and without the RL agent. For the RL simulations, the Hurst exponent is lower than that of the ABM and consistently falls below the lower confidence interval of the ABM. The only ones to fall into the original ABM's confidence interval are the RL agents that had the smallest parent order, indicating that the larger the amount of inventory traded by the RL agents the more this increases the amount of mean reversion in the simulation. This plot only shows four out of the eight moments.

\section{Pseudocode}\label{Pseudocode}

\begin{algorithm2e}
    \DontPrintSemicolon
    \SetKwInOut{Input}{Input}
    \SetKwInOut{Output}{Output}
    \SetKwFunction{Function}{HighFrequencyAgentAction}
    \SetKwProg{Fn}{Function}{:}{end}

    \Input{LOB state $\mathcal{L}_{k}$ and simulation parameters $\boldsymbol{\Theta}$}
    \Output{Limit order $o_{ik} = (p_{ik},\omega_{ik},D_{ik})$}
    \Fn{\Function{}}{
       \lIf{simulation is over}{
            \Return 
        }
        \lIf{simulation is not initialising}{
            cancel orders that have not been matched in $\phi$ time 
        }
        Sample $u \sim U(0,1)$\;
        \lIf{$u < (\rho_{t} + 1$)/2}{
            $D_{ik}=$ sell 
        }\lElse{
            $D_{ik}=$ buy
        }
        \uIf{$D_{ik}=$ sell}{
            $\alpha = 1 - \rho_{t}/\nu$\;
            Sample $\eta \sim \Gamma(s_{k}, e^{\rho_{t} / \kappa})$\;
            $p_{ik} = b_{k} + 1 + \lfloor\eta\rfloor$\;
            $\omega_{ik} \sim f(x;5,\alpha)$ Power law distribution with tail-index parameter $\alpha$ \cite{arxiv2021abm}
        }\Else{
            $\alpha = 1 + \rho_{t}/\nu$\;
            Sample $\eta \sim \Gamma(-s_{k}, e^{\rho_{t} / \kappa})$\;
            $p_{ik} = a_{k} - 1 - \lfloor\eta\rfloor$\;
            $\omega_{ik} \sim f(x;5,\alpha)$ Power law distribution with tail-index parameter $\alpha$ \cite{arxiv2021abm}
        }
        \uIf{initialising}{
            \lIf{$p_{k}$ is not inside initial spread}{Submit order  $o_{ik}$}
        }\Else{
            Submit order $o_{ik}$
        }
    
    }
    \caption{Liquidity Provider Action Algorithm}
    \label{Pseudocode-LP}
\end{algorithm2e}

\begin{algorithm2e}
    \DontPrintSemicolon
    \SetKwInOut{Input}{Input}
    \SetKwInOut{Output}{Output}
    \SetKwFunction{Function}{ChartistAgentAction}
    \SetKwProg{Fn}{Function}{:}{end}

    \Input{LOB state $\mathcal{L}_{k}$ and simulation parameters $\boldsymbol{\Theta}$}
    \Output{Market order $o_{ik} = (0,\omega_{ik},D_{ik})$}
    \Fn{\Function{}}{
        \lIf{simulation is over or initialising}{
            \Return 
        }
        $\Bar{m}_{i,k} = \Bar{m}_{i,k} + \lambda_{i} (\Bar{m}_{i,k} - m_{k})$\;
        \uIf{$\Bar{{m}}_{i,k} > m_{k} + 1/2 s_{k}$}{
            $D_{ik}$ = sell
        }\uElseIf{$\Bar{{m}}_{i,k} < m_{k} - 1/2 s_{k}$}{
            $D_{ik}$ = buy
        }\Else{
            \Return
        }
        $x_{m} = 20$\;
        \lIf{$|m_{k} - \Bar{m}_{i,k}| > \delta m_{k}$}{
            $x_{m} = 50$
        }
        \uIf{$D_{ik}=$ sell}{
            $\alpha = 1 - \rho_{t}/\nu$\;
            $\omega_{ik} \sim f(x;x_{m},\alpha)$ Power law distribution with tail-index parameter $\alpha$ \cite{arxiv2021abm}
        }\Else{
            $\alpha = 1 + \rho_{t}/\nu$\;
            $\omega_{ik} \sim f(x;x_{m},\alpha)$ Power law distribution with tail-index parameter $\alpha$ \cite{arxiv2021abm}
        }
        \uIf{there are LOs on the contra side and the order won't cause a volatility auction}{
            Submit order  $o_{ik}$
        }\Else{
            \Return
        }
    }
    \caption{Chartist Agent Action Algorithm}
    \label{Pseudocode-chartist}
\end{algorithm2e}

\begin{algorithm2e}
    \DontPrintSemicolon
    \SetKwInOut{Input}{Input}
    \SetKwInOut{Output}{Output}
    \SetKwFunction{Function}{FundamentalistAgentAction}
    \SetKwProg{Fn}{Function}{:}{end}

    \Input{LOB state $\mathcal{L}_{k}$ and simulation parameters $\boldsymbol{\Theta}$}
    \Output{Market order $o_{ik} = (0,\omega_{ik},D_{ik})$}
    \Fn{\Function{}}{
        \lIf{simulation is over or initialising}{
            \Return 
        }
        \uIf{$f_{i} < m_{k} - 1/2 s_{k}$}{
            $D_{ik}$ = sell
        }\uElseIf{$f_{i} > m_{k} + 1/2 s_{k}$}{
            $D_{ik}$ = buy
        }\Else{
            \Return
        }
        $x_{m} = 20$\;
        \lIf{$|m_{k} - \Bar{m}_{i,k}| > \delta m_{k}$}{
            $x_{m} = 50$
        }
        \uIf{$D_{ik}=$ sell}{
            $\alpha = 1 - \rho_{t}/\nu$\;
            $\omega_{ik} \sim f(x;x_{m},\alpha)$ Power law distribution with tail-index parameter $\alpha$ \cite{arxiv2021abm}
        }\Else{
            $\alpha = 1 + \rho_{t}/\nu$\;
            $\omega_{ik} \sim f(x;x_{m},\alpha)$ Power law distribution with tail-index parameter $\alpha$ \cite{arxiv2021abm}
        }
        \uIf{there are LOs on the contra side and the order won't cause a volatility auction}{
            Submit order  $o_{ik}$
        }\Else{
            \Return
        }
    }
    \caption{Fundamentalist Agent Action Algorithm}
    \label{Pseudocode-fundamentalist}
\end{algorithm2e}

\begin{algorithm2e}
    \DontPrintSemicolon
    \SetKwInOut{Input}{Input}
    \SetKwInOut{Output}{Output}
    \SetKwFunction{Function}{RLAgentAction}
    \SetKwProg{Fn}{Function}{:}{end}

    \Input{LOB state $\mathcal{L}_{k}$, simulation parameters $\boldsymbol{\Theta}$, RL agent parameters $\boldsymbol{\Theta}_{RL}$, and RL event messages from previous orders, previous state $S_{k-1}$, and previous action $A_{k-1}$ }
    \Output{Market order $o_{ik} = (0,\omega_{ik},D_{ik})$}
    \Fn{\Function{}}{
        \lIf{simulation is over or initialising}{
            \Return 
        }
        $v_{k}$, $R_{k}$ = ProcessMessages(messages) \tcp{previous trades volume traded and profit received}
        remainingInventory = remainingInventory - $v_{k}$\;
        remainingTime = $T_{0} - \mathrm{currentTime}$\;
        $s_{k}$, done = GetState($\mathcal{L}_{k}$, remainingTime, remainingInventory)\;
        \lIf{rl trading time is up and final trade submitted}{
            state = terminal state
        }
        \If{a previous trade was submitted}{
            $Q(S_{k-1},A_{k-1}) = Q(S_{k-1},A_{k-1}) + \alpha (R_{k} + \gamma \max\limits_{a} Q(S_{k},a) - Q(S_{k-1},A_{k-1}))$\;
        }
        $A_{k}$ = GetAction($Q,S_{k},\epsilon$) \tcp{Sample action from $\epsilon$-greedy policy}
        \If{rl trading time is up}{
            \uIf{still has inventory remaining}{
                $A_{k}$ = 2 \tcp{Submit largest trade possible}
            }\Else{
                $A_{k}$ = 0 \tcp{Ensure no trade}
            }   
        }
        $D_{ik}$ = sell\;
        $\omega_{ik}$ = min($A_{k} (X_{0}/N_{dp})$, remainingInventory)\;
        \uIf{there are LOs on the contra side and the order won't cause a volatility auction}{
            Submit order  $o_{ik}$
        }\Else{
            \Return
        }
        $S_{k-1}$ = $S_{k}$\;
        $A_{k-1}$ = $A_{k}$
    }
    \caption{Learning Agent Action Algorithm}
    \label{Pseudocode-rlagent}
\end{algorithm2e}

\begin{algorithm2e}
    \DontPrintSemicolon
    \SetKwInOut{Input}{Input}
    \SetKwInOut{Output}{Output}
    \SetKwFunction{Function}{Simulate}
    \SetKwProg{Fn}{Function}{:}{end}
    \SetKwFor{Loop}{Loop}{}{EndLoop}

    \Input{Simulation parameters $\boldsymbol{\Theta}$, RL agent parameters $\boldsymbol{\Theta}_{RL}$, and CoinTossX trading gateway}
    \Fn{\Function{}}{
        Start the CoinTossX trading session\;
        Initialise the LOB $\mathcal{L}_{0}$ with the initial spread $s_{0}$, best bid and offer $b_{0}$ and $a_{0}$, as well as the mid-price $m_{0}$\;
        Connect to the UDP socket with the IP and port of the CoinTossX market data feed\;
        Start the asynchronous task that will listen for and store CoinTossX messages\;
        Initialise the LPs, LTs and the RL agents (if it is an RL simulation)\;
        Initialise the LOB using the initial LOB parameters and the LPs\;
        \Loop(\tcp*[f]{Event loop}){}{
            \uIf{there are messages from CoinTossX}{
                \For{message in messages}{
                    UpdateLOBState($\mathcal{L}_{k}$,message)\;
                    \lIf{there are RL traders}{store the RL trade messages}
                }
                Pass updated LOB to all the agents for them to trade off\;
            }\Else{
                break
            }
        }
        Close asynchronous task resources, the UDP socket and close the {\tt CoinTossX} trading session\;
        Write required data to files\;
    }
    \caption{Simulation Loop Algorithm}
    \label{Pseudocode-simulation}
\end{algorithm2e}

\begin{algorithm2e}
    \DontPrintSemicolon
    \SetKwInOut{Input}{Input}
    \SetKwInOut{Output}{Output}
    \SetKwFunction{Function}{TrainRLAgent}
    \SetKwProg{Fn}{Function}{:}{end}
    \SetKwFor{Loop}{loop}{}{end}

    \Input{Calibrated simulation parameters $\boldsymbol{\Theta}$ and RL agent parameters $\boldsymbol{\Theta}_{rl}$}
    \Output{Training results for an RL agent}
    \Fn{\Function{}}{
        Initialise $Q_{0}$\;
        Login to CoinTossX\;
        \For{i in $\{1,\ldots,\mathrm{number\;\;of\;\;episodes}\}$}{
            Simulate the e-ABM with the RL agent\;
            Update $\epsilon$ using the $\epsilon$ decay function in Figure \ref{rl-epsilon-decay}\;
            $Q_{i-1} = Q_{i}$ \tcp{Store previous Q-matrix for next iteration}
            Manual garbage collection\;
        }
        Logout of CoinTossX\;
        Close CoinTossX\;
        Write the RL results to files\;
    }
    \caption{Learning Agent Training Algorithm}
    \label{Pseudocode-trainrl}
\end{algorithm2e}
\newpage
\bibliographystyle{elsarticle-harv}
\bibliography{MDAPTG.bib}